\documentclass[11pt]{article}
\usepackage{graphicx}
\usepackage[margin=1.25in]{geometry}
\usepackage[usenames,dvipsnames]{color}
\usepackage{url}
\usepackage[colorlinks = true,
            linkcolor = blue,
            urlcolor  = blue,
            citecolor = blue,
            anchorcolor = blue]{hyperref}
\usepackage[square,comma,sort&compress,numbers]{natbib}
\usepackage{caption}
\usepackage{subcaption}
\usepackage{booktabs}
\usepackage{placeins}

\newenvironment{Abstract}{\begin{quotation} \begin{center}
                       ABSTRACT
     \end{center}\bigskip  }{\end{quotation}}

\def\beq{\begin{equation}}
\def\eeq#1{\label{#1}\end{equation}}
\def\eeqn{\end{equation}}

\newenvironment{Eqnarray}%
   {\arraycolsep 0.14em\begin{eqnarray}}{\end{eqnarray}}
\def\beqa{\begin{Eqnarray}}
\def\eeqa#1{\label{#1}\end{Eqnarray}}
\def\eeqan{\end{Eqnarray}}

\let\bar=\overbar

\def\lsim{\mathrel{\raise.3ex\hbox{$<$\kern-.75em\lower1ex\hbox{$\sim$}}}}
\def\gsim{\mathrel{\raise.3ex\hbox{$>$\kern-.75em\lower1ex\hbox{$\sim$}}}}

\def\del{\partial}
\def\Dslash{\not{\hbox{\kern-4pt $D$}}}
\def\dslash{\not{\hbox{\kern-2pt $\del$}}}
\def\pslash{\not{\hbox{\kern-2pt $p$}}}
\def\ETmiss{\not{\hbox{\kern-4pt $E$}}_T}

\def\Dlr{\mathrel{\raise1.5ex\hbox{$\leftrightarrow$\kern-1em\lower1.5ex\hbox{$D$}}}}

\def\MSB{{\bar{M \kern -2pt S}}}
\def\msb{{\bar{\scriptsize M \kern -1pt S}}}

\def\drb{{\bar{\scriptsize D \kern -1pt R}}}

\newcommand\snowmass{\begin{center}\rule[-0.2in]{\hsize}{0.01in}\\\rule{\hsize}{0.01in}\\
\vskip 0.1in Submitted to the  Proceedings of the US Community Study\\ 
on the Future of Particle Physics (Snowmass 2021)\\ 
\rule{\hsize}{0.01in}\\\rule[+0.2in]{\hsize}{0.01in} \end{center}}

\usepackage{authblk}

\title{Summarizing experimental sensitivities of collider experiments to dark matter models and comparison to other experiments}

\author[1]{Antonio~Boveia}
\affil[1]{The Ohio State University}
\author[2,3]{Caterina~Doglioni}
\affil[2]{University of Manchester and Lund University}
\affil[3]{Lund University}
\author[4]{Boyu~Gao}
\affil[4]{Duke University}
\author[3]{Josh~Greaves}
\author[5]{Philip~Harris}
\affil[5]{MIT}
\author[6]{Katherine~Pachal}
\affil[6]{TRIUMF}

\author[7]{Etienne~Dreyer}
\affil[7]{Weizmann Institute of Science}
\author[8]{Giuliano~Gustavino}
\affil[8]{CERN}
\author[9]{Robert~Harris}
\affil[9]{Fermi National Accelerator Laboratory}
\author[10]{Daniel~Hayden}
\affil[10]{Michigan State University}
\author[11]{Tetiana~Hrynova}
\affil[11]{LAPP Annecy}
\author[4]{Ashutosh~Kotwal}
\author[10]{Jared~Little}
\affil[12]{University of Wisconsin-Madison}
\author[12]{Kevin~Black}
\author[12]{Tulika~Bose}
\author[12]{Yuze~Chen}
\author[12]{Sridhara~Dasu}
\author[12]{Haoyi~Jia}
\author[12]{Deborah~Pinna}
\author[12]{Varun~Sharma}
\author[12]{Nikhilesh~Venkatasubramanian}
\author[12]{Carl~Vuosalo}

\begin{document}




\maketitle




\medskip

 \begin{Abstract}
   \noindent Comparisons of the coverage of current and proposed dark matter searches can help us to understand the context in which a discovery of particle dark matter would be made. In some scenarios, a discovery could be reinforced by information from multiple, complementary types of experiments; in others, only one experiment would see a signal, giving only a partial, more ambiguous picture; in still others, no experiment would be sensitive and new approaches would be needed. In this whitepaper, we present an update to a similar study performed for the European Strategy Briefing Book performed within the dark matter at the Energy Frontier (EF10) Snowmass Topical Group We take as a starting point a set of projections for future collider facilities and a method of graphical comparisons routinely performed for LHC DM searches using simplified models recommended by the LHC Dark Matter Working Group and  also used for the BSM and dark matter chapters of the European Strategy Briefing Book. These comparisons can also serve as launching point for cross-frontier discussions about dark matter complementarity.
\end{Abstract}

\snowmass

\def\thefootnote{\fnsymbol{footnote}}
\setcounter{footnote}{0}

\tableofcontents

\section{Introduction}

Collider and accelerator experiments searching for dark matter can constrain a wide variety of models of dark matter production in collisions of known matter. Underground and above-ground direct-detection experiments can constrain the interaction rate of dark matter in our solar neighborhood with nuclei and electrons, and indirect-detection instruments can constrain self-annihilation of dark matter beyond our local neighborhood. Nevertheless, putting together the information learned from these different types of particle physics experiments is challenging, since it requires a model, and we have few clues as to what the correct model of dark matter might be.

One model framework long used to connect dark matter searches is supersymmetry. Nevertheless, supersymmetry is not the result of a top-down attempt to provide a theory of dark matter---indeed, it need not provide a dark matter candidate at all. Supersymmetry also has many unknown parameters, many of which have no observable effect on the phenomenology of DM at the experiments searching for it. Recently, collider experiments, as well as some direct-detection (e.g. Refs~\cite{PICO:2017tgi}) and indirect-detection experiments (e.g. Ref.~\cite{Balazs:2017hxh}) have adopted simplified models~\cite{Alwall:2008ag} for this purpose. Fewer parameters, allow focus on low-energy phenomenology at the experiments, at the expense of theoretical completeness and unambiguous connection to the observed galactic abundance of DM. With fewer parameters, one can construct slices of parameter space to make rough comparisons of the coverage from the different types of experiments.

In this whitepaper, we use the simplified models from Ref. ~\cite{Abercrombie:2015wmb} as benchmarks to \textbf{visualize projected sensitivities of future hadron collider experiments to DM production} (Sections \ref{sec:colliderinputs} and \ref{sec:futurecollider} ). 
This whitepaper also contains a section dedicated to a  \textbf{dark matter search at muon colliders} (Section \ref{subsec:muonCollider}), presented in terms of a dark matter candidate produced directly through electroweak interactions \cite{Han:2020uak}.

We also use the simplified models framework of Ref.~\cite{Abercrombie:2015wmb} to \textbf{display scenarios of projected sensitivities from proposed collider, direct-detection and indirect-detection searches} (Sections \ref{sec:complementarity_DD} and \ref{sec:complementarity_ID}). These projections have been prepared within the EF10 topical group starting from a selection of projections provided by studies of future collider facilities. The simplified models used for these scenarios consist of a subset of those recommended by the LHC Dark Matter Working Group and currently used by LHC searches, as well as in the BSM and dark matter chapters of the recent European Strategy Briefing Book~\cite{EuropeanStrategyBook}. It is important to note from the start that due to the model dependence and different technology readiness of different projects, these plots are not meant to be used to compare sensitivities of different experiments in a realistic way\footnote{For this reason, they will be discussed with the relevant Cosmic Frontier TG conveners how they can be turned into sketches to appear in a subsequent complementarity whitepaper.}.  

In a parallel development to LHC simplified models, a set of simplified models has also emerged from CERN's Physics Beyond Colliders initiative to target light dark sector particles~\cite{Beacham:2019nyx,Lanfranchi:2020crw}. Owing to the flexibility of the LHC dark sector searches, it is possible to map the LHC simplified models onto these light dark sector models. In this white paper, we also discuss the connection between LHC light dark matter models and present an approach to \textbf{map our existing dark matter bounds and searches to that of the light dark sector models} (Section \ref{sec:lightDarkSectorPortals}). 

\section{Benchmarks and inputs from future collider facilities}
\label{sec:colliderinputs}

We use the DM simplified models in~\cite{Abercrombie:2015wmb} since they are simple descriptions of collider phenomenology that capture common features across many full models, while ignoring the differences among these models at energies higher than collider scales~\cite{doi:10.1146/annurev-nucl-101917-021008}. 
In the simplified models, the interaction between Standard Model (SM) particles and DM is mediated by a new particle called a mediator, where the mediator – quark coupling controls the interaction strength between mediator and SM quarks. These models can also be used as a stepping stone to connect collider searches to accelerator searches since the searches can be reinterpreted using different portal models~\cite{Beacham:2019nyx}, as discussed in Section \ref{sec:lightDarkSectorPortals}. 

In this whitepaper, we initially focus on models of a new $s-$channel mediator between Standard Model (SM) and DM particles with a mass $M_{med}$, with independent (axial-)vector couplings to SM particles (e.g. couplings to quarks $g_{q}$ and to leptons $g_{DM}$).
The mediator also couples to the DM particle of mass $m_{DM}$ with a coupling $g_{\chi}$. 
Due to the presence of both SM and DM couplings, the main search avenues for this kind of models at colliders is by looking for an excess of missing energy from the mediator decaying to DM particles, or by looking for localized (resonant) excesses signaling the presence of a mediator decaying into pairs of DM particles. 
We consider mainly vector and axial-vector models for the summary plots in this whitepaper, but we also touch upon models where the mediator is a new scalar or pseudoscalar particle. 

The inputs received from the collider community in the context of Snowmass 2021 and used to prepare the plots in this whitepaper are: 
\begin{itemize}
    \item MET+jet search (also called monojet search) using the ATLAS detector at the upgraded LHC (HL-LHC) ~\cite{hllhc-monojet};
    \item MET+jet and MET+hadronically decaying vector boson at the Future Hadron Collider in the Future Circular Collider (FCC) complex~\cite{Harris:2015kda};
    \item Dijet resonance search at the HL-LHC and at the Future Hadron Collider~\cite{Harris:2022kls};
    \item Dilepton resonance search at the HL-LHC using the ATLAS detector~\cite{ATL-PHYS-PUB-2018-044}. The CMS version of this search is also available and will be included in version 2 of this paper if the theory cross-sections are made available~\cite{CMS-PAS-FTR-21-005}.
\end{itemize}

The inputs above only covers hadron colliders, and as a consequence the focus of a number of the plots in this whitepaper is on the coupling between the mediator particle and quarks. Lepton colliders such as ILC would also have a significant impact in constraining parameter space of dark matter models where the mediator couples to leptons \cite{Kalinowski:2021tyr}, and we  aim to include those constrains in v2.  
Inputs from a monophoton analysis at a future muon collider are also included in a dedicated section (Sec. \ref{subsec:muonCollider}), where they are described in more detail.



\section{Future collider reach to DM mediator vector models}
\label{sec:futurecollider}







To compare how existing collider searches may discover or constrain a given model the LHC Dark Matter Working Group~\cite{Abercrombie:2015wmb,BOVEIA2020100365,ALBERT2019100377} has prompted the LHC collaborations to make summary plots in terms of the  parameters of the Lagrangian of simplified dark matter models mentioned in the previous section~\cite{CMSSummary,ATL-PHYS-DMSUM-JHEP-2019}. In this whitepaper, we start from the same recommendations to understand how future collider experiments would extend current LHC results. 

When comparing visible and invisible searches at colliders, we focus on the vector model from~\cite{Abercrombie:2015wmb}. 
Projections of the performance of these searches can be displayed as exclusion comparisons in the DM mass--mediator mass plane, as well as on the mediator--quark coupling and mediator mass plane. 

In order to give a representative idea of collider constraints for this model under different choices of parameters (quark and lepton couplings, DM mass and mediator mass), we use methods from work done in synergy between the LHC DM WG and the Snowmass community and resulted in a Snowmass whitepaper~\cite{Albert:2022xla}. 
This work allows us to easily derive collider limits with lower mediator - quark coupling values, rather than only for the fixed coupling values proposed by the LHC DM WG ~\cite{BOVEIA2020100365,ALBERT2019100377} and used in the European Strategy Briefing Book~\cite{EuropeanStrategyBook}. 

\subsection{Comparing visible and invisible searches at hadron colliders}
\label{sec:visibleInvisible}

\subsubsection{DM mass vs mediator mass plane}

Projections of expected exclusion limits for HL-LHC and FCC-hh searches are interpreted in terms of vector model parameters (couplings and mass of mediator, as well as DM mass), and then scaled to a representative set of coupling values using the methods in~\cite{Albert:2022xla}. 

For HL-LHC, all three of monojet, dijet, and dilepton limits were available. 
For FCC-hh, only monojet and dijet limits were available. 
Therefore, the set of couplings explored in the HL-LHC scenario involved varying quark couplings ($g_q$) and lepton couplings ($g_l$), while for FCC-hh only scans of $g_q$ are given here since the impact of varying $g_l$ is not significant for the monojet and dijet final states. 
The DM couplings $g_{\chi}=1.0$ are also varied when combining all analyses at the end of this section. 

\begin{figure}[htb!]
     \centering
     \begin{subfigure}[b]{0.49\textwidth}
         \centering
         \includegraphics[width=\textwidth]{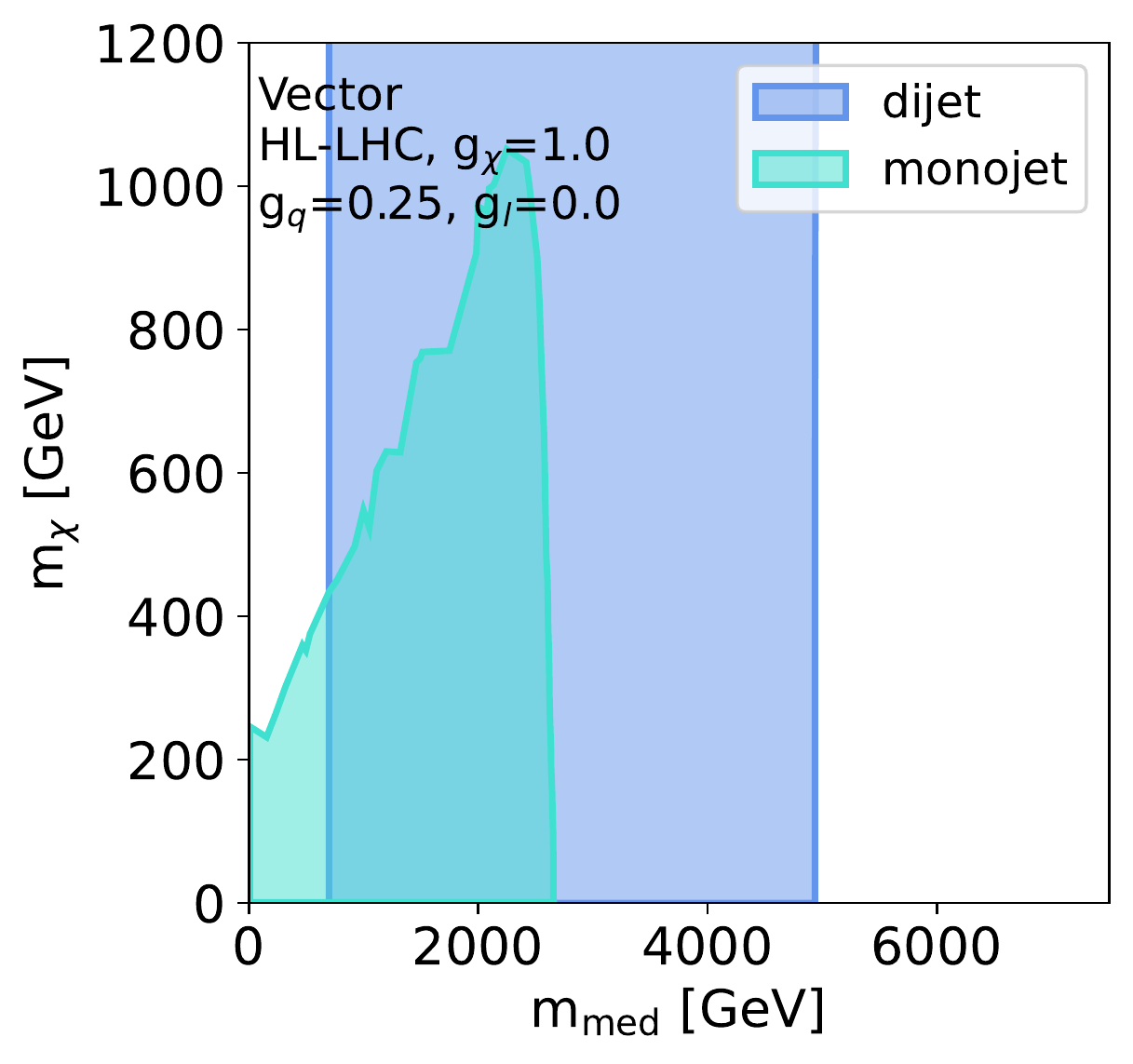}
         \caption{``V1'': $g_q=0.25$, $g_{\chi}=1.0$, $g_l=0.0$}
         \label{subfig:vector-hl-lhc-v1}
     \end{subfigure}
     \hfill
     \begin{subfigure}[b]{0.49\textwidth}
         \centering
         \includegraphics[width=\textwidth]{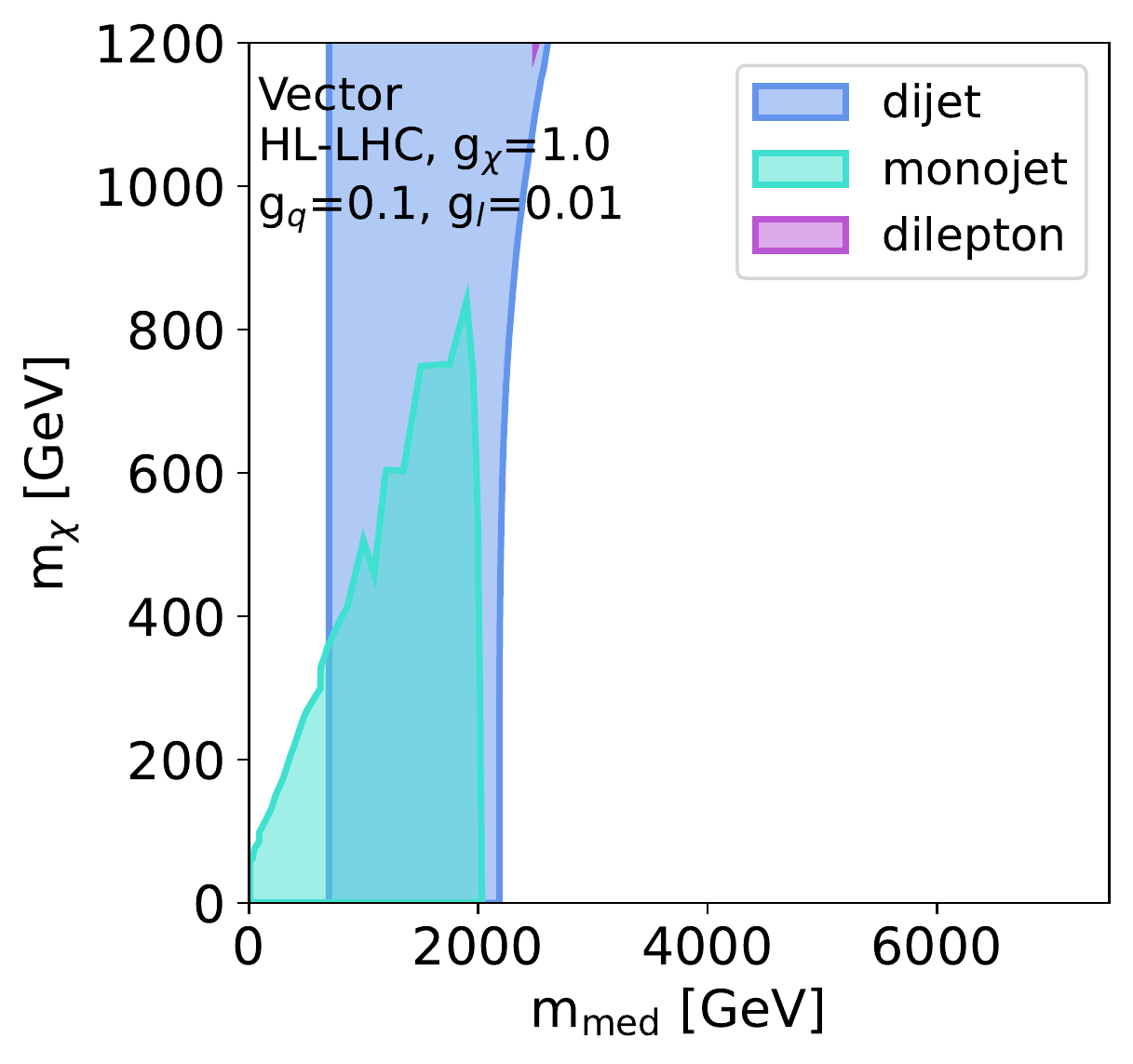}
         \caption{``V2'': $g_q=0.1$, $g_{\chi}=1.0$, $g_l=0.01$}
         \label{subfig:vector-hl-lhc-v2}
     \end{subfigure}

     \begin{subfigure}[b]{0.49\textwidth}
         \centering
         \includegraphics[width=\textwidth]{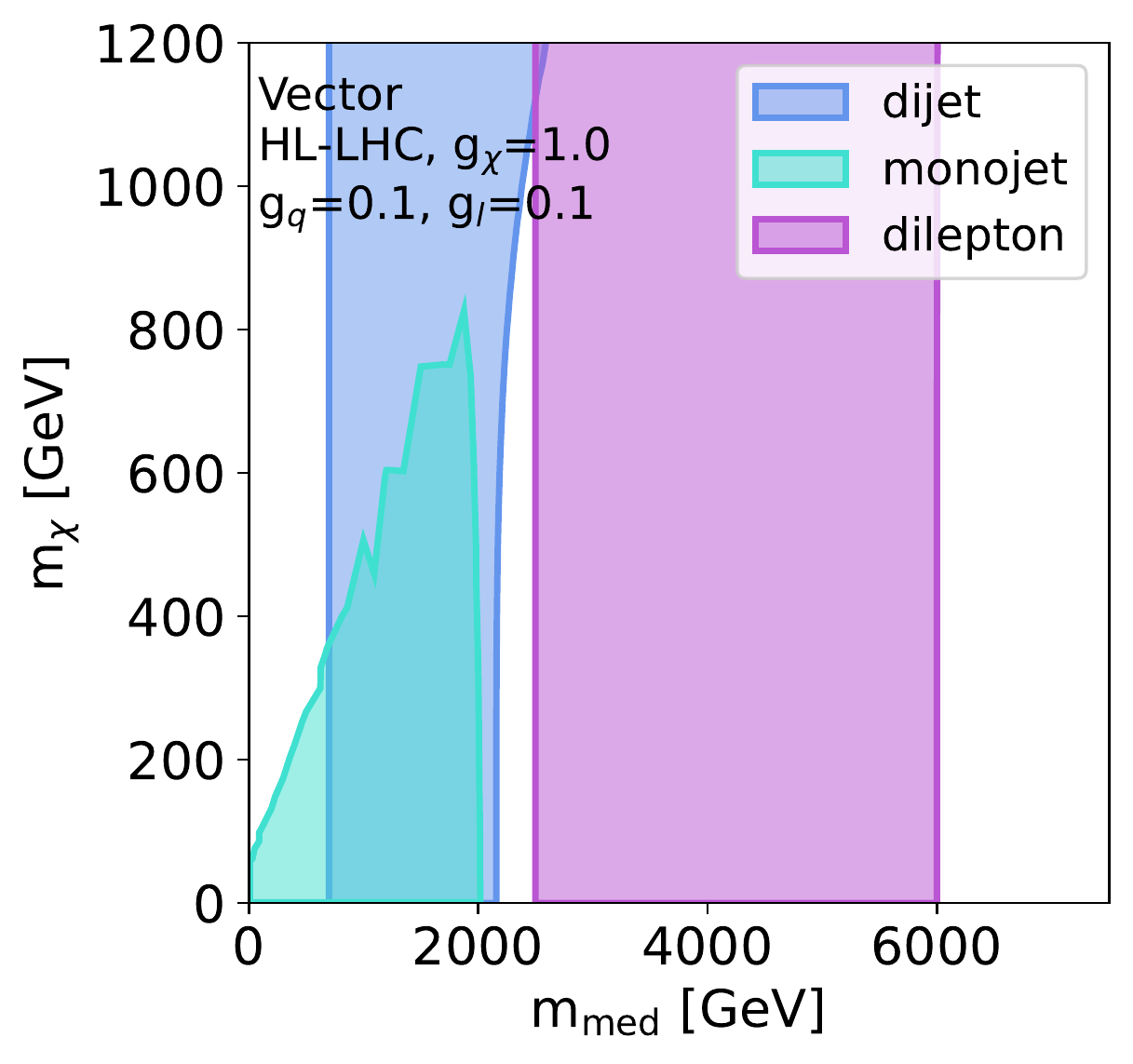}
         \caption{New scenario with varied SM couplings: $g_q=0.1$, $g_{\chi}=1.0$, $g_l=0.1$}
         \label{subfig:vector-hl-lhc-biggerlepton}
     \end{subfigure}
     \hfill
     \begin{subfigure}[b]{0.49\textwidth}
         \centering
         \includegraphics[width=\textwidth]{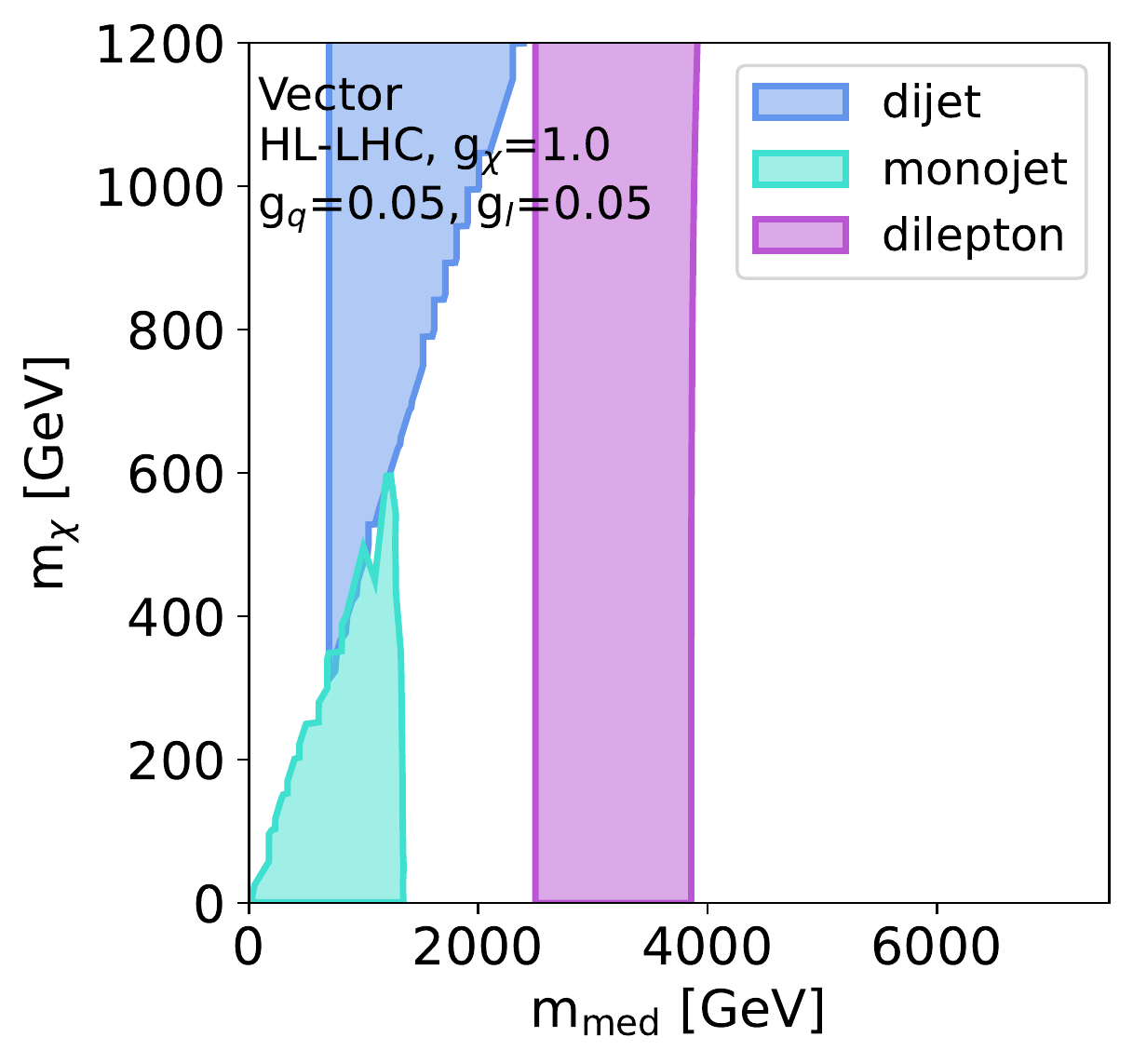}
         \caption{New scenario with varied SM couplings: $g_q=0.05$, $g_{\chi}=1.0$, $g_l=0.05$}
         \label{subfig:vector-hl-lhc-allsmall}       
     \end{subfigure}
        \caption{HL-LHC projected exclusions for individual analyses in the vector model and with a range of couplings.}
        \label{fig:hl-lhc-massmass-separate}
\end{figure}

Figure~\ref{fig:hl-lhc-massmass-separate} shows the three projected analysis limits for HL-LHC in the vector model for a representative selection of coupling points, including the DMWG benchmark points ``V1'' ($g_q=0.25$, $g_{\chi}=1.0$, $g_l=0.0$) and ``V2'' ($g_q=0.1$, $g_{\chi}=1.0$, $g_l=0.01$), alongside alternatives where the coupling to quarks is reduced to $g_q=0.1$ and where the coupling to leptons is increased to $g_q=0.1$ with respect to the previous scenarios, with dark matter coupling fixed to unity. Equivalent plots for the axial-vector model are given in Appendix~\ref{apppendix}. 

These plots depict the complementarity between visible and invisible searches. The monojet search is most sensitive when the mediator can decay to DM directly. 
The dijet and dilepton searches can also probe this kind of dark interaction when the dark matter is too heavy to be produced, as the mediator can still decays back into quarks or into leptons. However, an excess in the monojet search compatible with an excess in the dijet and dilepton searches would be needed to be able to associate visible and invisible decays to the same model interpretation. 

Figure~\ref{fig:hl-lhc-massmass-separate} shows that the effect of reducing the quark couplings affects all searches (since it controls the mediator production), but it is more prominent for dijet searches since  $g_q$ enters both in the mediator production and decay. The exclusion area covered by dijet searches reduces exclusively to the off-shell region when $g_q=0.05$. 
Variations in $g_l$ within the studied range have little impact on the monojet and dijet analyses but a strong impact on the dilepton analysis.

The lower mass edge of the dijet and dilepton contours in Figure~\ref{fig:hl-lhc-massmass-separate} correspond to the mediator masses at which the projected exclusions are provided by analyses that focus on high mass processes. 
The lower mass edge is also responsible for the lack of dilepton contour appearing in Figure~\ref{subfig:vector-hl-lhc-v2}: the ATLAS projected dilepton contour begins only at $m_\mathrm{med} = 2.5$~TeV, above which point sensitivity is limited for the dark matter masses under consideration. 
These bounds however do not represent a hard limit on $M_{\mathrm{med}}$ for dijet and dilepton searches, especially if non-standard analysis workflows are employed (see e.g. \cite{ATLAS:2016xiv, Khachatryan:2016ecr}). 

\begin{figure}[htb!]
     \centering
     \begin{subfigure}[b]{0.49\textwidth}
         \centering
         \includegraphics[width=\textwidth]{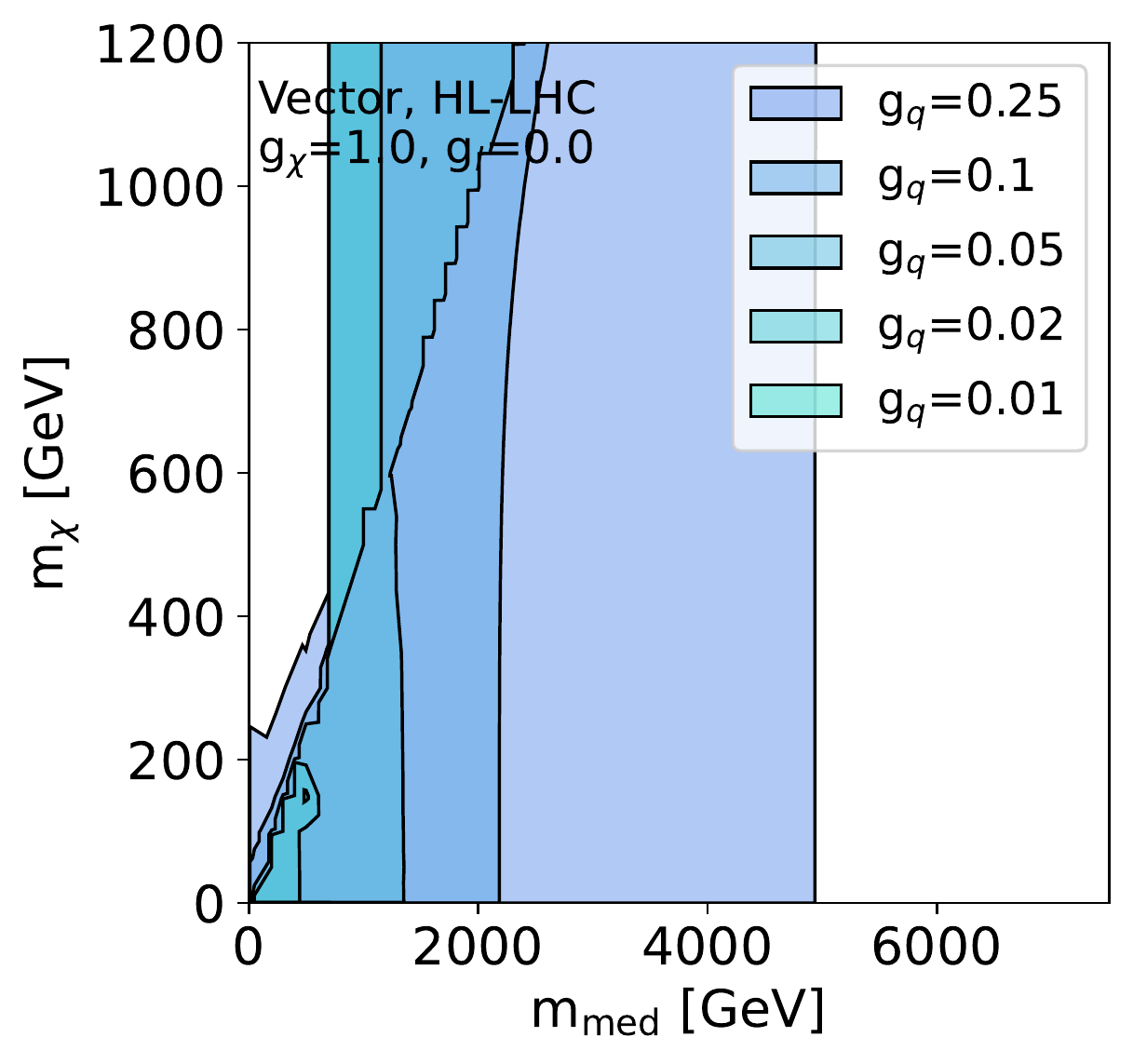}
         \caption{$g_\chi$ fixed to 1.0}
         \label{subfig:vector-hl-lhc-gqvariations1}
     \end{subfigure}
     \hfill
     \begin{subfigure}[b]{0.49\textwidth}
         \centering
         \includegraphics[width=\textwidth]{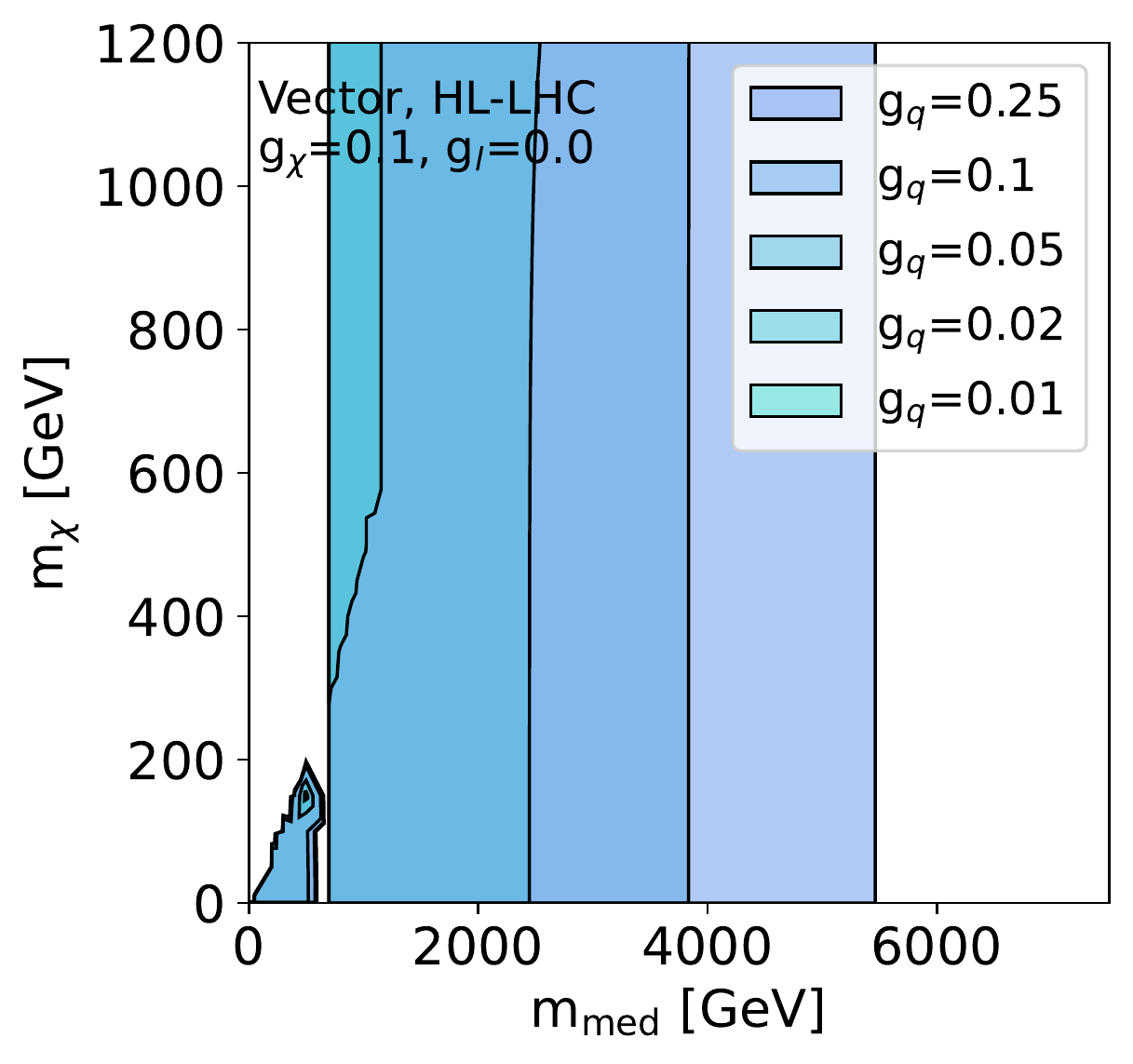}
         \caption{$g_l$ fixed to 0.1}
         \label{subfig:vector-hl-lhc-gqvariations2}
     \end{subfigure}
        \caption{Effects of varying $g_q$ coupling on HL-LHC projected exclusions, combined across analyses. No coupling to leptons is allowed; $g_\chi$ is set to 1.0 in (\subref{subfig:vector-hl-lhc-gqvariations1}) and 0.1 in (\subref{subfig:vector-hl-lhc-gqvariations2}).}
        \label{fig:hl-lhc-massmass-combined-gqvary}
\end{figure}

\begin{figure}[htb!]
     \begin{subfigure}[b]{0.49\textwidth}
         \centering
         \includegraphics[width=\textwidth]{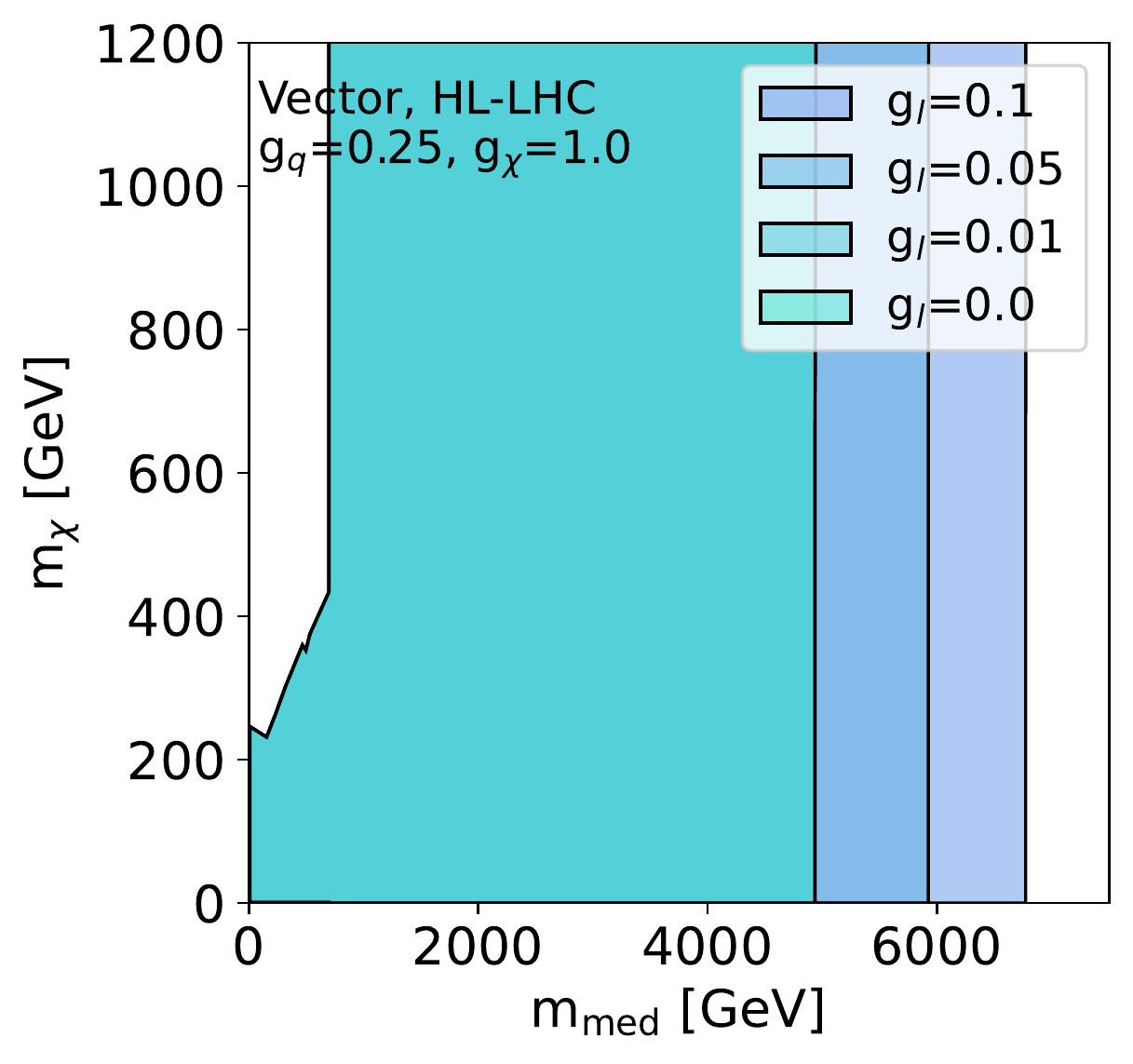}
         \caption{$g_q$ fixed to 0.25}
         \label{subfig:vector-hl-lhc-glvariations1}
     \end{subfigure}
     \hfill
     \begin{subfigure}[b]{0.49\textwidth}
         \centering
         \includegraphics[width=\textwidth]{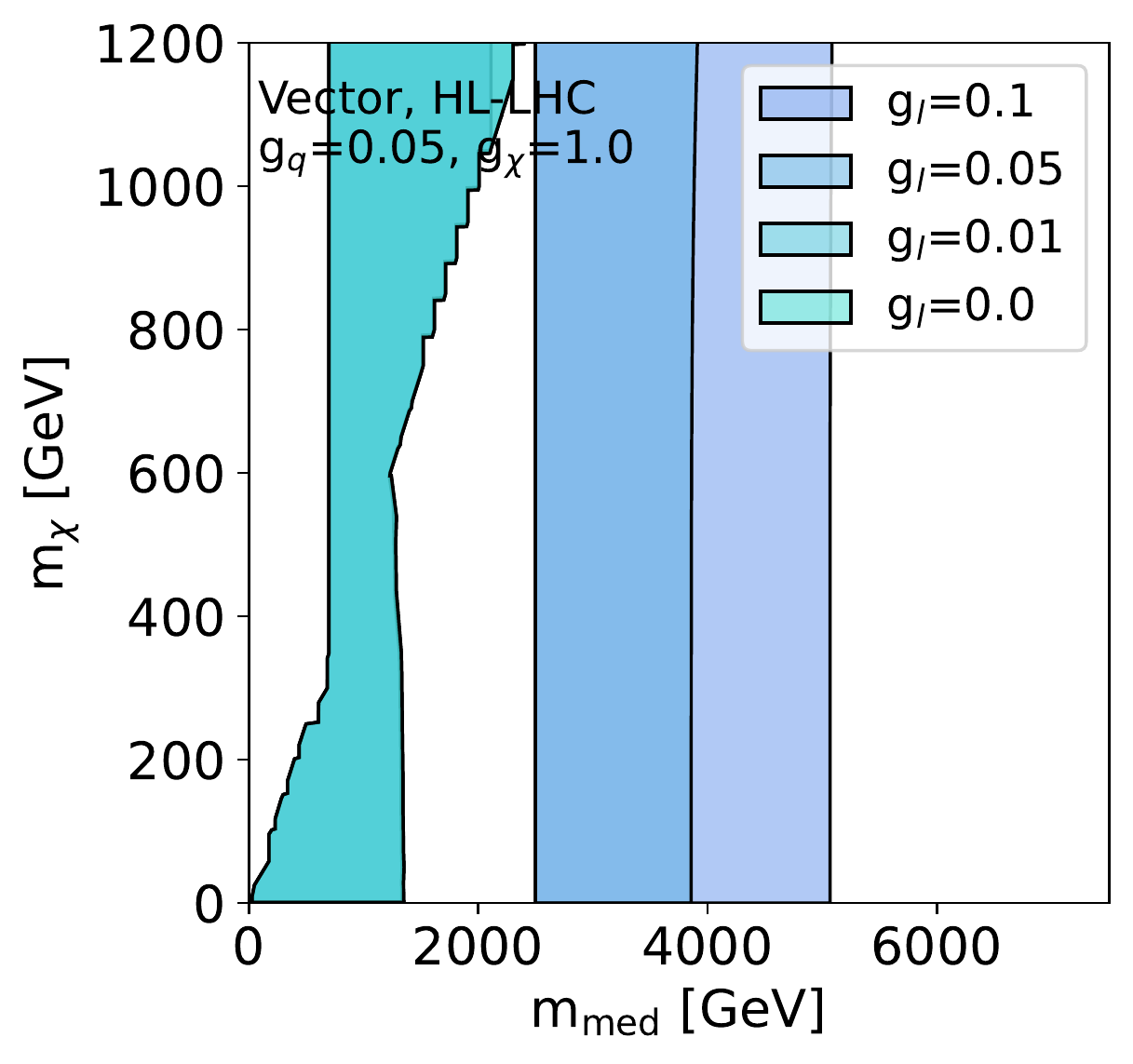}
         \caption{$g_q$ fixed to 0.05}
         \label{subfig:vector-hl-lhc-glvariations2}       
     \end{subfigure}
        \caption{Effects of varying $g_l$ coupling on HL-LHC projected exclusions, combined across analyses. The coupling to DM is fixed to 1.0; $g_q$ is set to 0.25 in (\subref{subfig:vector-hl-lhc-glvariations1}) and 0.05 in (\subref{subfig:vector-hl-lhc-glvariations2}).}
        \label{fig:hl-lhc-massmass-combined-glvary}
\end{figure}

\begin{figure}[htb!]
     \begin{subfigure}[b]{0.49\textwidth}
         \centering
         \includegraphics[width=\textwidth]{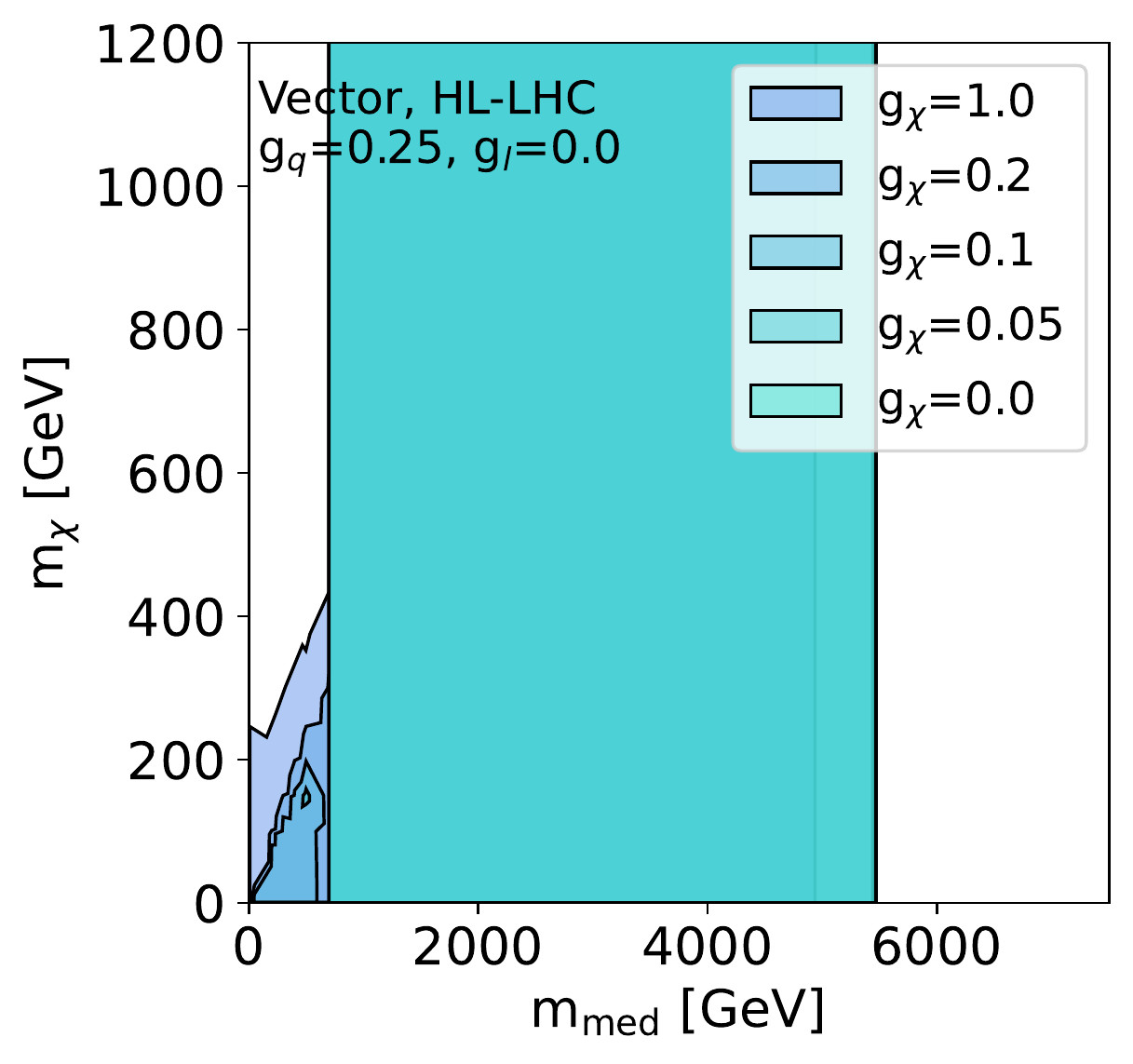}
         \caption{$g_q$ fixed to 0.25, $g_l$ fixed to 0.0}
         \label{subfig:vector-hl-lhc-gdmvariations1}
     \end{subfigure}
     \hfill
     \begin{subfigure}[b]{0.49\textwidth}
         \centering
         \includegraphics[width=\textwidth]{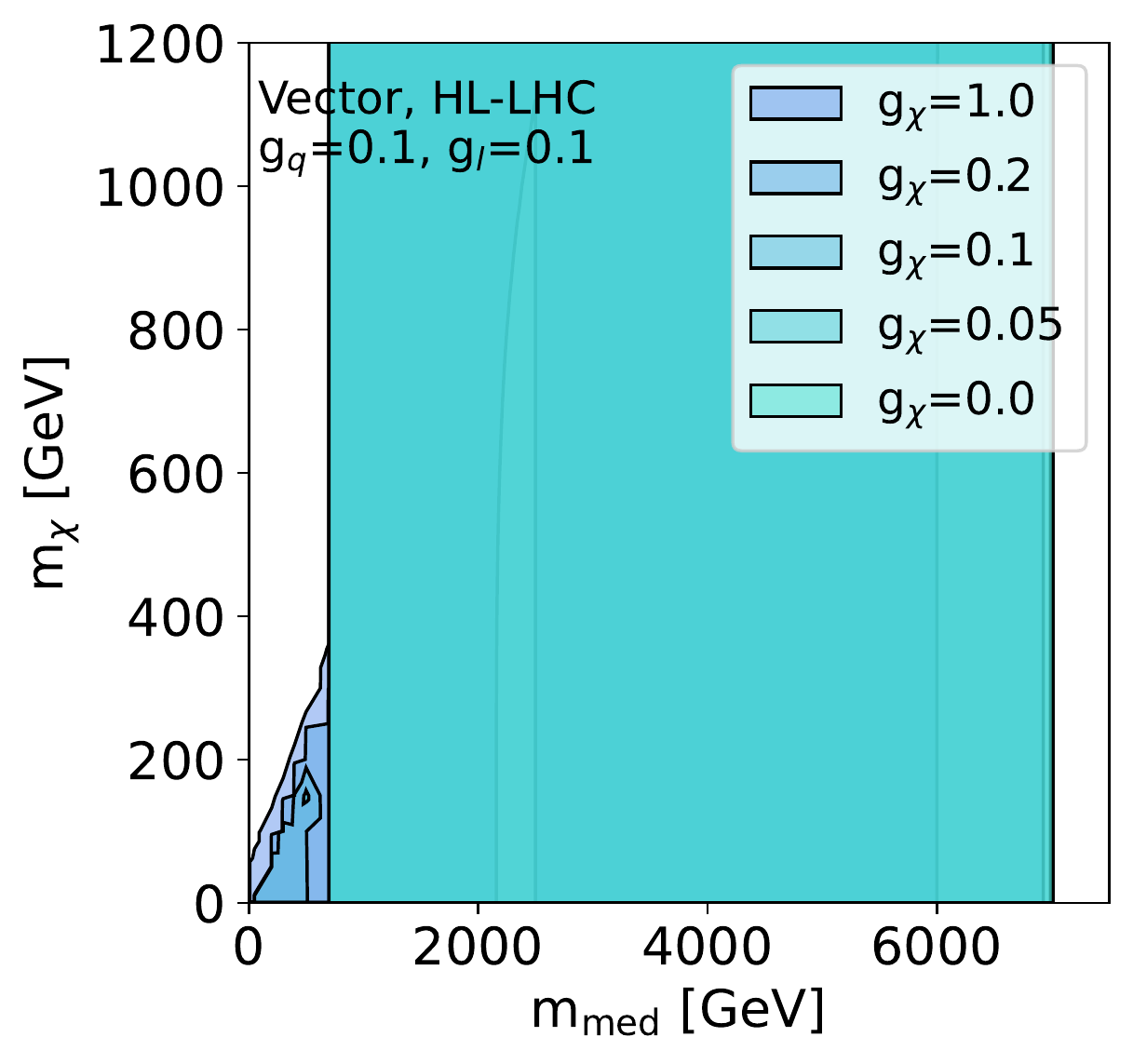}
         \caption{$g_q$ fixed to 0.1, $g_l$ fixed to 0.1}
         \label{subfig:vector-hl-lhc-gdmvariations2}       
     \end{subfigure}
        \caption{Effects of varying $g_\chi$ coupling on HL-LHC projected exclusions, combined across analyses. Other couplings are fixed. }
        \label{fig:hl-lhc-massmass-combined-gdmvary}
\end{figure}

In Figures~\ref{fig:hl-lhc-massmass-combined-gqvary} through~\ref{fig:hl-lhc-massmass-combined-gdmvary}, curves of different shades illustrate the effect on the overall excluded region of varying the couplings, including also variations of $g_\chi$.
A single contour is shown for each set of couplings, corresponding to the total excluded region from all three input HL-LHC analyses. 
With the other two couplings held constant, the remaining coupling is varied and the effects on the exclusion region shown. 
At low DM and mediator mass, some artifacts due to the limited signal grid available are visible. 

Figure \ref{fig:hl-lhc-massmass-combined-gdmvary} shows the effects of only varying $g_\chi$ coupling on HL-LHC projected exclusions, combined across analyses. As $g_\chi$ decreases, the monojet limits weaken but the visible final state limits improve because their relative branching ratios increase.
In these figures, the relative impact of monojet and dijet analyses changes with the value of $g_\chi$, but the sensitivity of HL-LHC for this model generally vanishes when $g_q$ is below 0.02.

\FloatBarrier

\begin{figure}[htb!]
     \centering
     \begin{subfigure}[b]{0.49\textwidth}
         \centering
         \includegraphics[width=\textwidth]{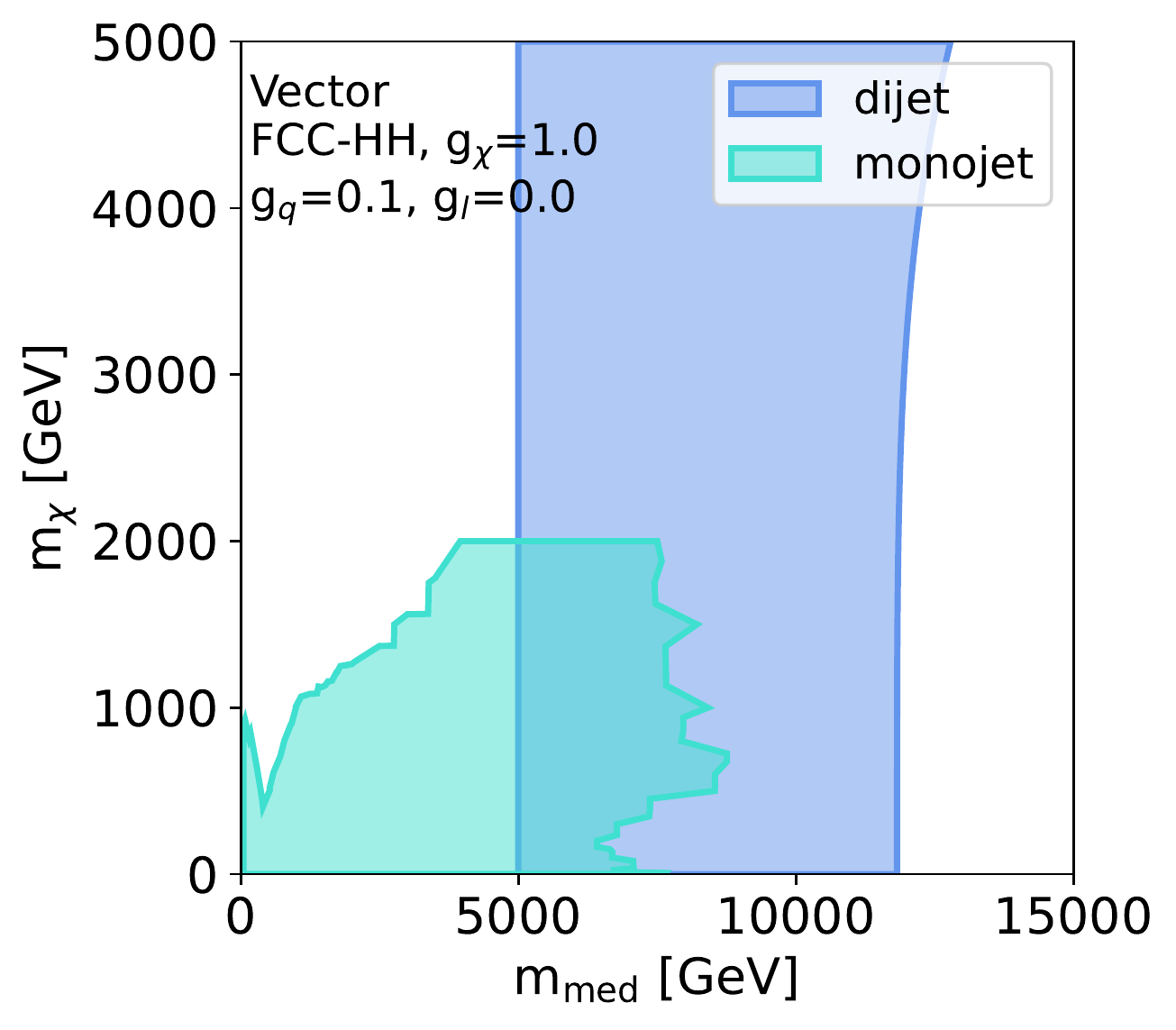}
         \caption{$g_q=0.1$, $g_{\chi}=1.0$, $g_l=0.0$}
         \label{subfig:vector-fcc-v2}
     \end{subfigure}
     \hfill
     \begin{subfigure}[b]{0.49\textwidth}
         \centering
         \includegraphics[width=\textwidth]{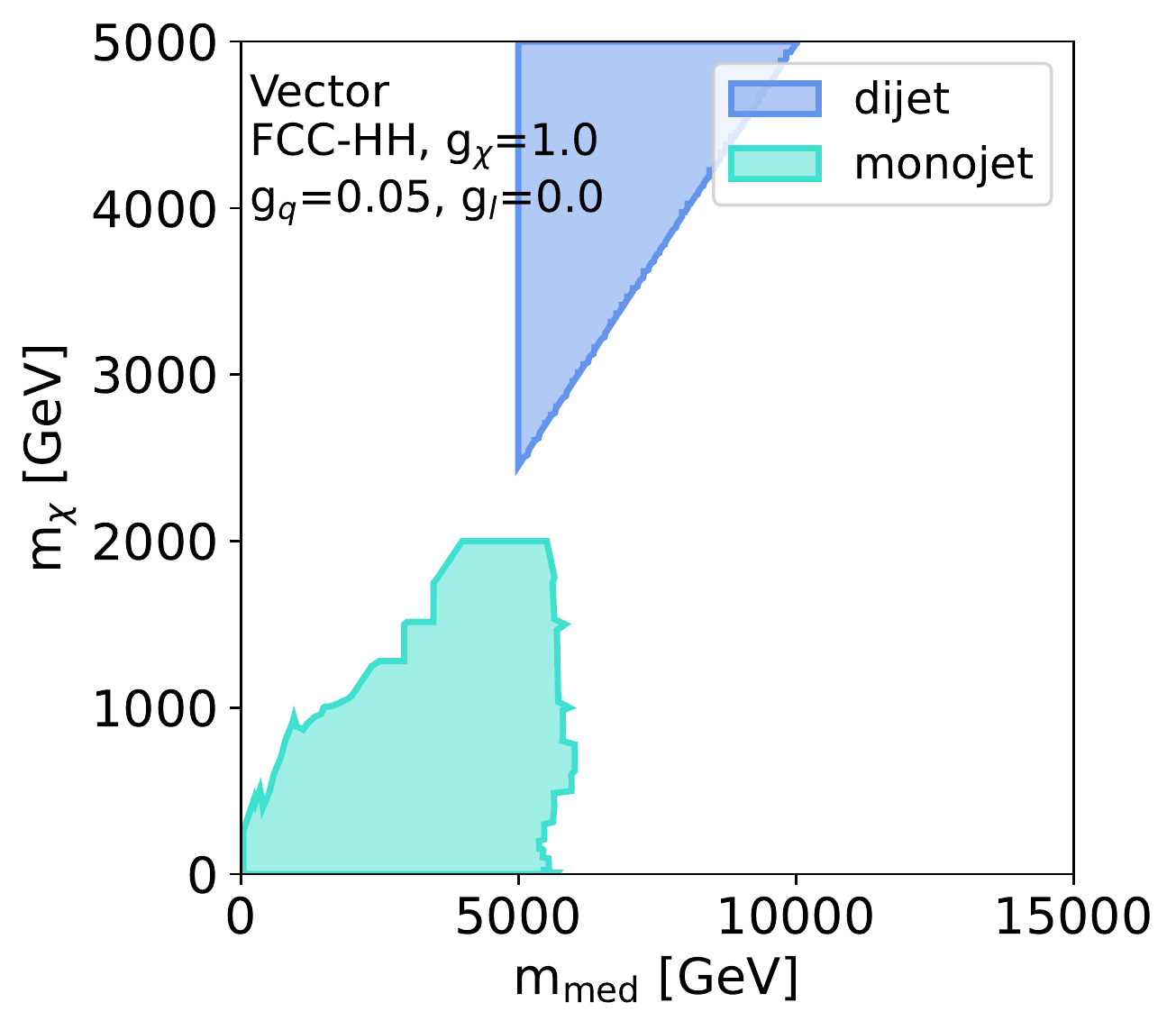}
         \caption{$g_q=0.05$, $g_{\chi}=1.0$, $g_l=0.0$}
         \label{subfig:vector-fcc-v3}
     \end{subfigure}

     \begin{subfigure}[b]{0.49\textwidth}
         \centering
         \includegraphics[width=\textwidth]{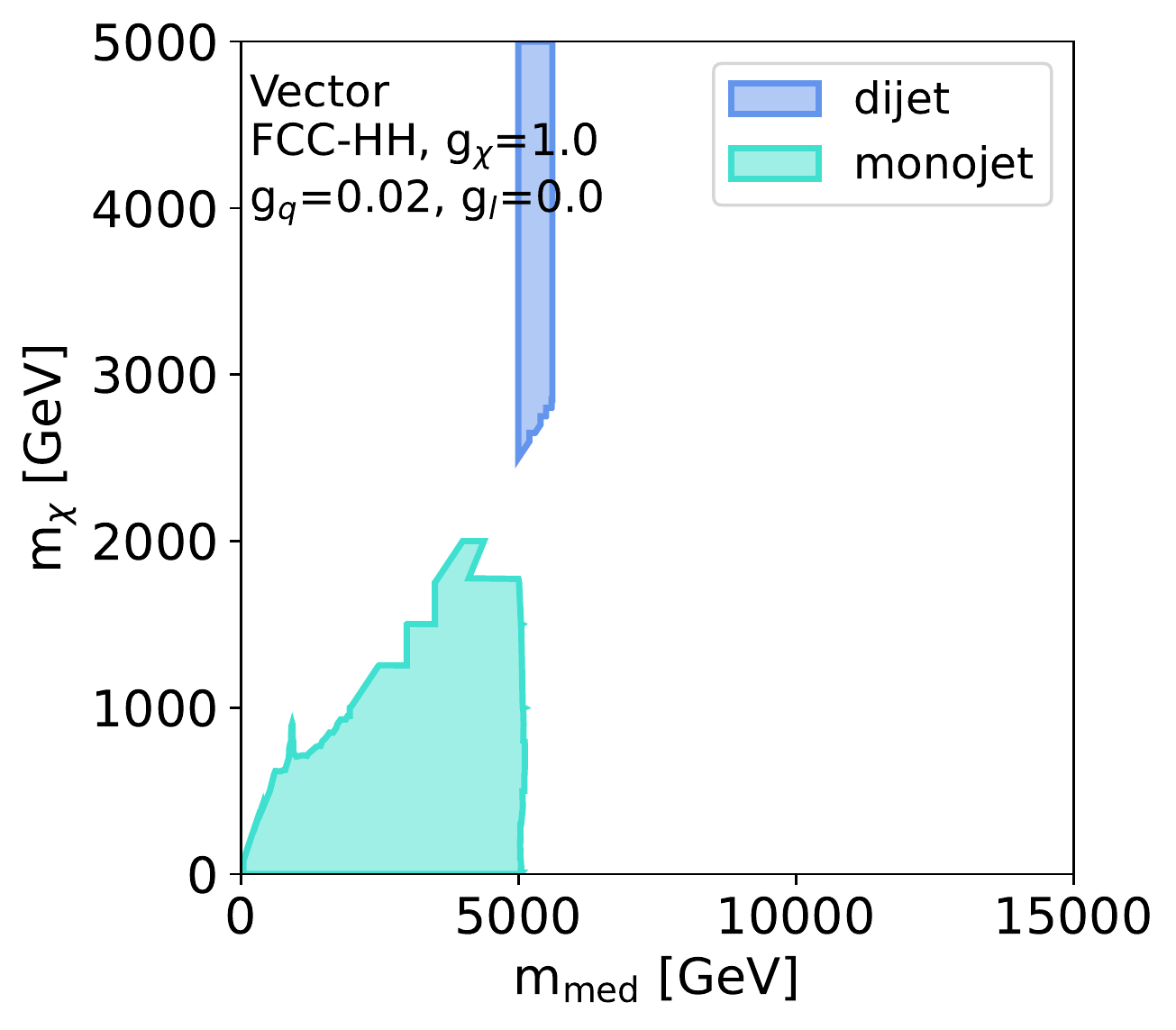}
         \caption{$g_q=0.25$, $g_{\chi}=1.0$, $g_l=0.0$}
         \label{subfig:vector-fcc-v1}
     \end{subfigure}
     \hfill
     \begin{subfigure}[b]{0.49\textwidth}
         \centering
         \includegraphics[width=\textwidth]{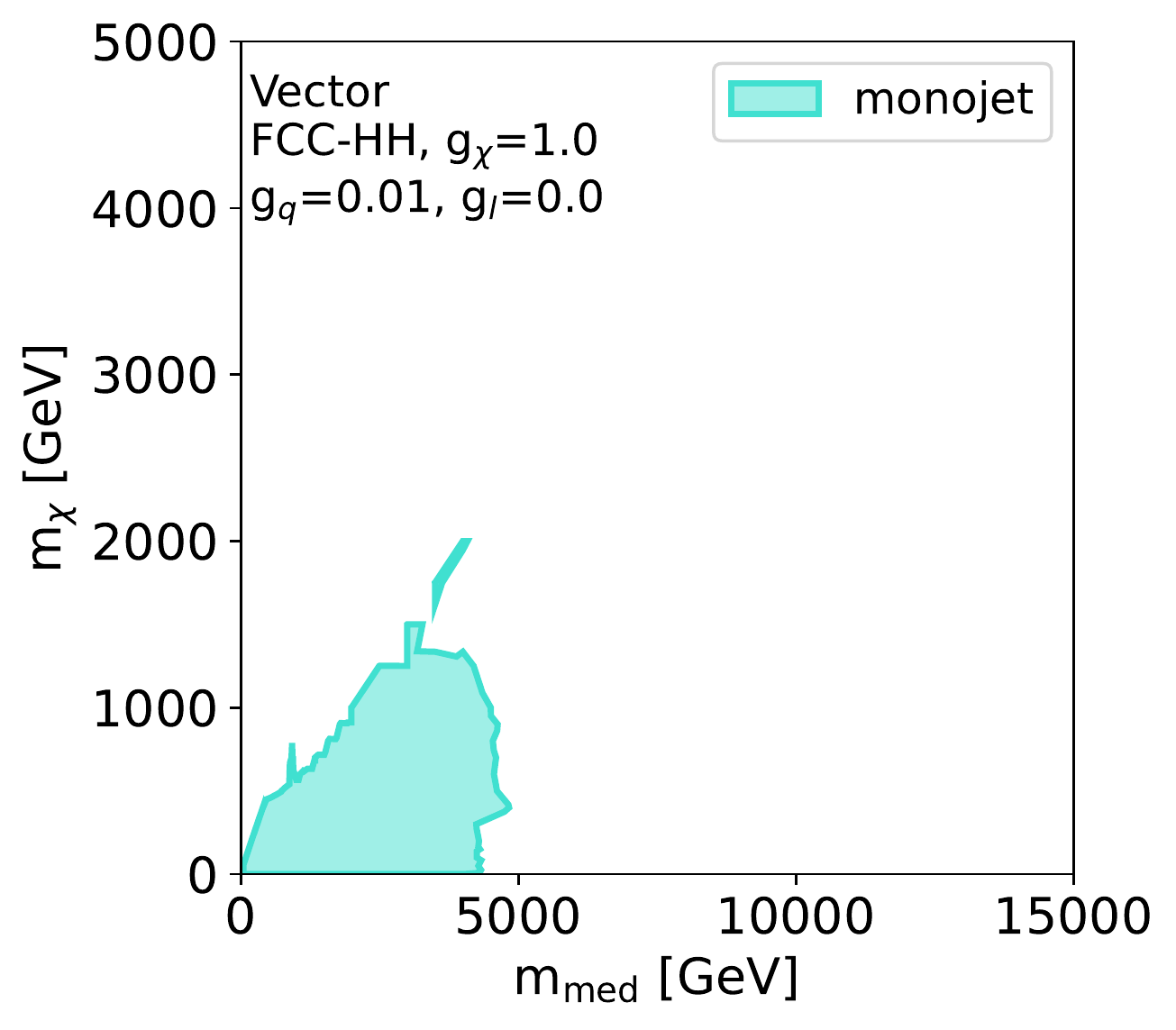}
         \caption{$g_q=0.01$, $g_{\chi}=1.0$, $g_l=0.0$}
         \label{subfig:vector-fcc-v4}       
     \end{subfigure}
        \caption{FCC-hh projected exclusions for individual analyses in the vector model and with a range of couplings. The truncation at the top of the monojet exclusion in subfigures~\subref{subfig:vector-fcc-v2} and~\subref{subfig:vector-fcc-v3} is due to a limitation in the input exclusion grid rather than a physical effect of the analysis at FCC-hh.}
        \label{fig:fcc-hh-massmass-separate}
\end{figure}

For the future hadron collider (FCC-hh), only dijet and monojet projections are available as of this writing. Figure~\ref{fig:fcc-hh-massmass-separate} shows these two analysis limits for 100 TeV center of mass energy at 30 $ab^{-1}$ in the vector model for a selection of interesting coupling points. 
In the absence of a dilepton prediction, the limits show little change with $g_{l}$ so the selected points differ in $g_q$ only. 
Note that the highlighted $g_q$ values are significantly smaller than for the HL-LHC projections: this reflects the significantly greater exclusion power of FCC-hh, such that the scenarios with $g_q = 0.25$, developed as benchmarks for LHC Run 2, are too strongly excluded to remain of much interest.

\begin{figure}[htb!]
     \centering
     \begin{subfigure}[b]{0.49\textwidth}
         \centering
         \includegraphics[width=\textwidth]{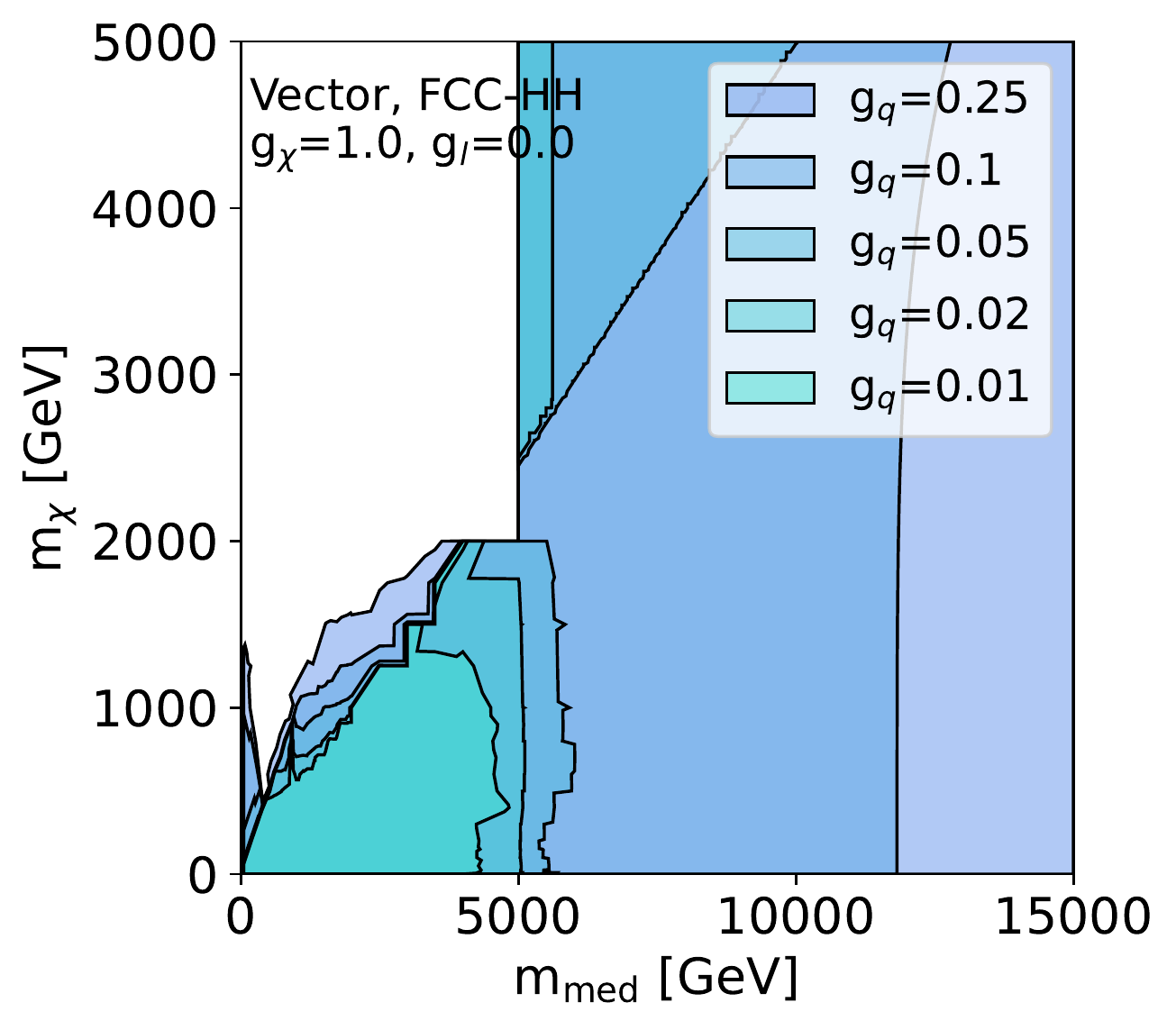}
         \caption{Effects on a vector model of varying $g_q$ with fixed $g_l=0.0$ and $g_\chi=1.0$}
         \label{subfig:vector-fcc-gqvariations1}
     \end{subfigure}
     \hfill
    \begin{subfigure}[b]{0.49\textwidth}
         \centering
         \includegraphics[width=\textwidth]{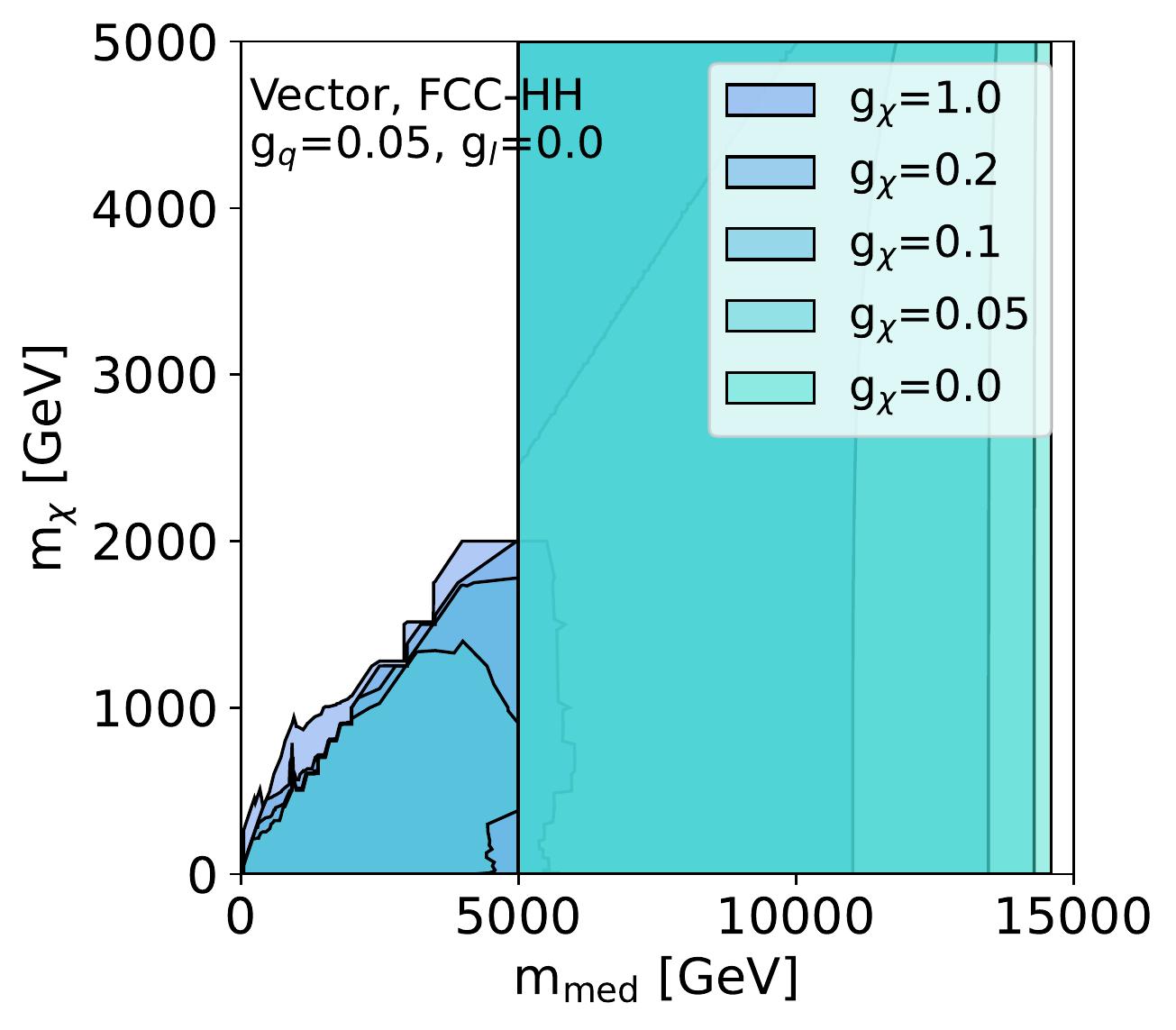}
         \caption{Effects on a vector model of varying $g_\chi$ with fixed $g_q=0.05$ and $g_l=0.0$}
         \label{subfig:vector-fcc-gqvariations3}
     \end{subfigure}
    \caption{FCC-hh projected exclusions, combined across analyses, for varying coupling values.}
    \label{fig:fcc-hh-massmass-combined}
\end{figure}

In Figure~\ref{fig:fcc-hh-massmass-combined}, combined exclusion contours are drawn for the FCC-hh monojet and dijet limits. In the absence of an effect from varying $g_l$, only variations in $g_q$ and $g_\chi$ are interesting. 

\FloatBarrier

\subsubsection{Coupling vs mediator mass plane}

Given the previous section, in particular the apparent insensitivity of collider constraints to changes in coupling values at low masses, it is useful to extend the projections to coupling values lower than 0.05 to explore where the collider searches would lose sensitivity to these mdoels.  
This can be done using plots that display search results in the coupling-mediator mass plane.
In general, searches can easily constrain larger couplings that explicitly feature in the mediator production and decay processes, and the same constraints eventually vanish for small coupling values. 
For couplings which do not appear in the production and decay processes that are the target of the analysis (e.g. $g_\chi$ in the dijet search), constraints can still be set depending on the relative strengths of all three couplings.

\begin{figure}[htb!]
    \centering
     \begin{subfigure}[b]{0.49\textwidth}
         \includegraphics[width=\textwidth]{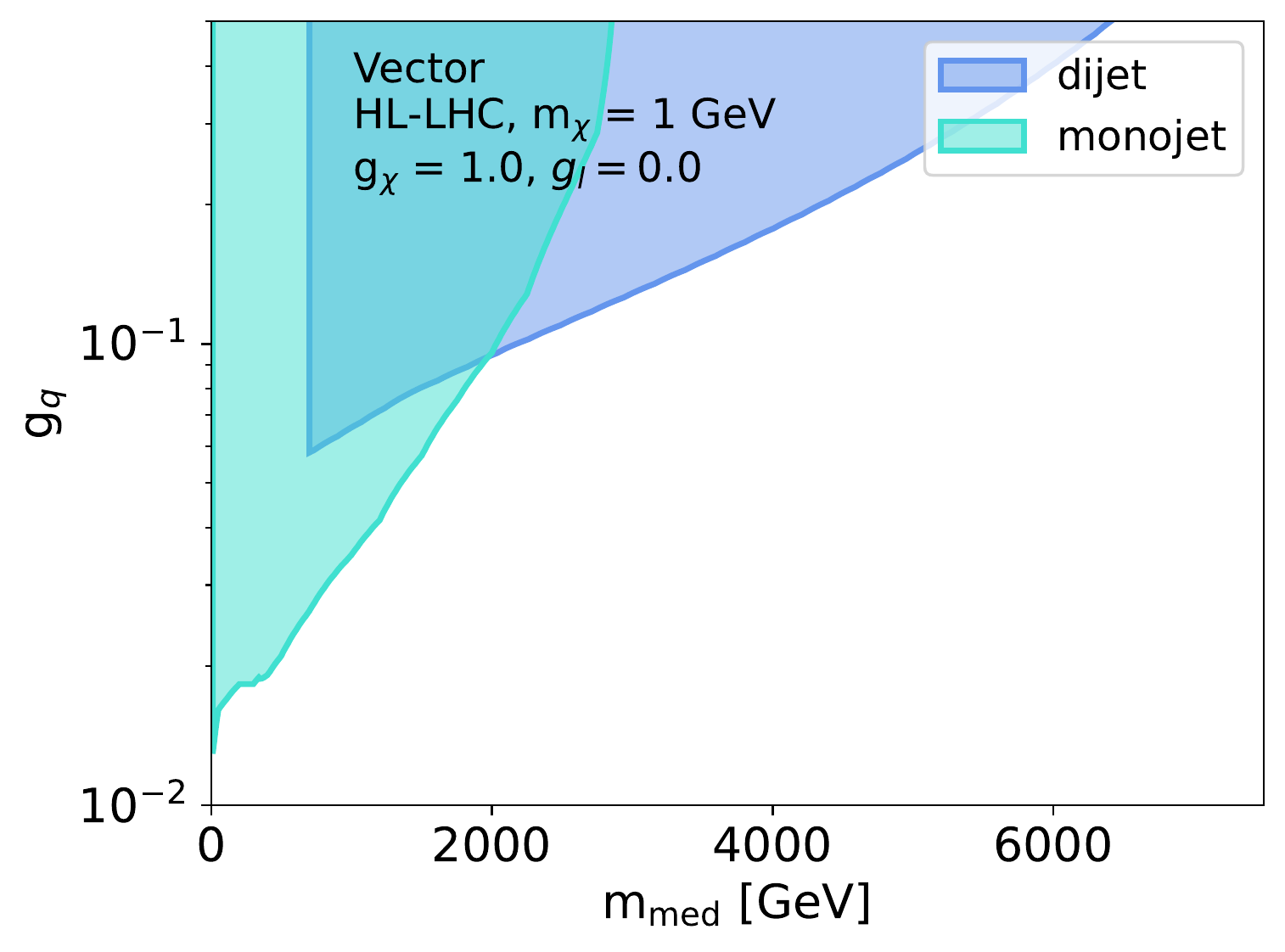}
         \caption{}
         \label{subfig:gqscan-dmLight}
     \end{subfigure}
     \begin{subfigure}[b]{0.49\textwidth}
         \includegraphics[width=\textwidth]{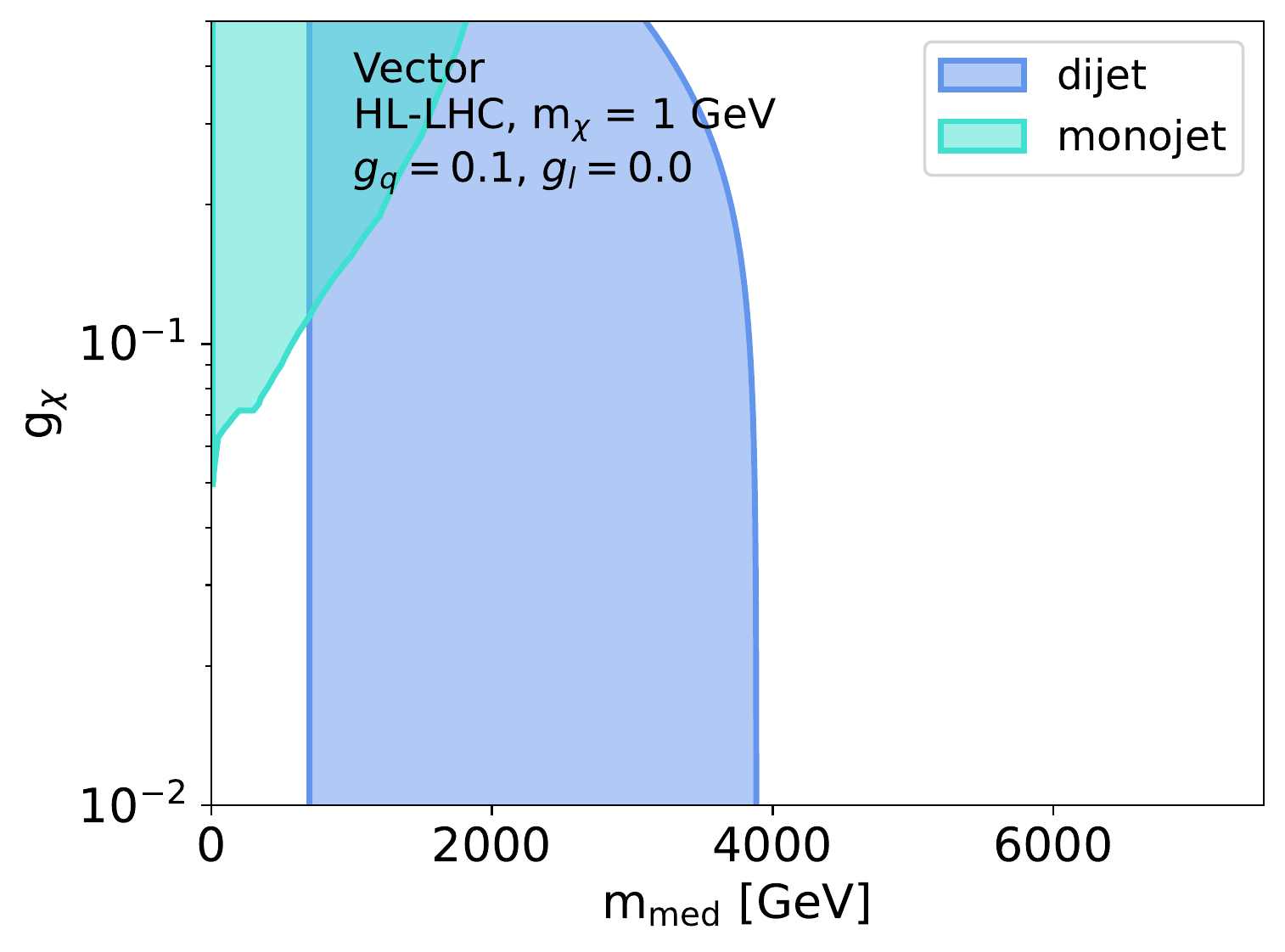}
         \caption{}
         \label{subfig:gchiscan-dmLight}
     \end{subfigure}     
     
    \begin{subfigure}[b]{0.49\textwidth}
         \includegraphics[width=\textwidth]{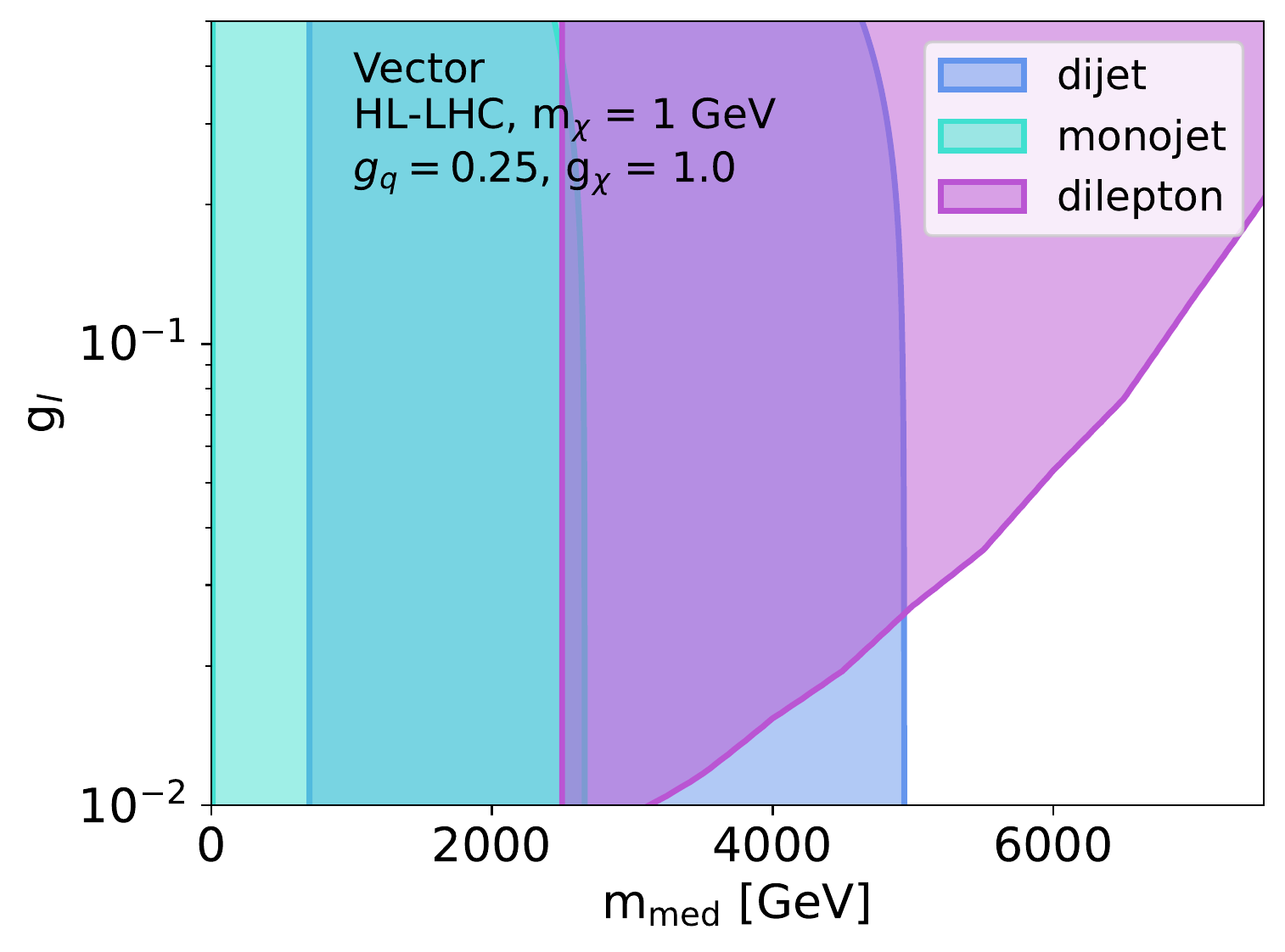}
         \caption{}
         \label{subfig:glscan-dmLight}
     \end{subfigure}    
         \caption{Projected exclusion limits on the couplings $g_q$ (\subref{subfig:gqscan-dmLight}), $g_\chi$ (\subref{subfig:gchiscan-dmLight}), and $g_l$ (\subref{subfig:glscan-dmLight}) for a vector mediator at the HL-LHC. The result is shown as a function of the mediator mass $m_{med}$; the mass of the DM candidate is fixed to 1 GeV in all cases. The coupling on the $y$ axis is varied while the other two couplings are fixed: in~(\subref{subfig:gqscan-dmLight}), $g_\chi$=1.0 and $g_l$=0.0; in~(\subref{subfig:gchiscan-dmLight}), $g_q$=0.1 and $g_l$=0.0; and in~\subref{subfig:glscan-dmLight}, $g_q$=0.25 and $g_\chi$=1.0.}
         \label{fig:couplinglimits-hl-lhc-allanalyses}
\end{figure}

Figure~\ref{fig:couplinglimits-hl-lhc-allanalyses} shows the interplay between analyses at the HL-LHC by illustrating the excluded couplings as a function of mediator mass for all three couplings, with representative fixed values for the couplings not being tested. 

\begin{figure}[htb!]
    \centering
     \begin{subfigure}[b]{0.49\textwidth}
         \includegraphics[width=\textwidth]{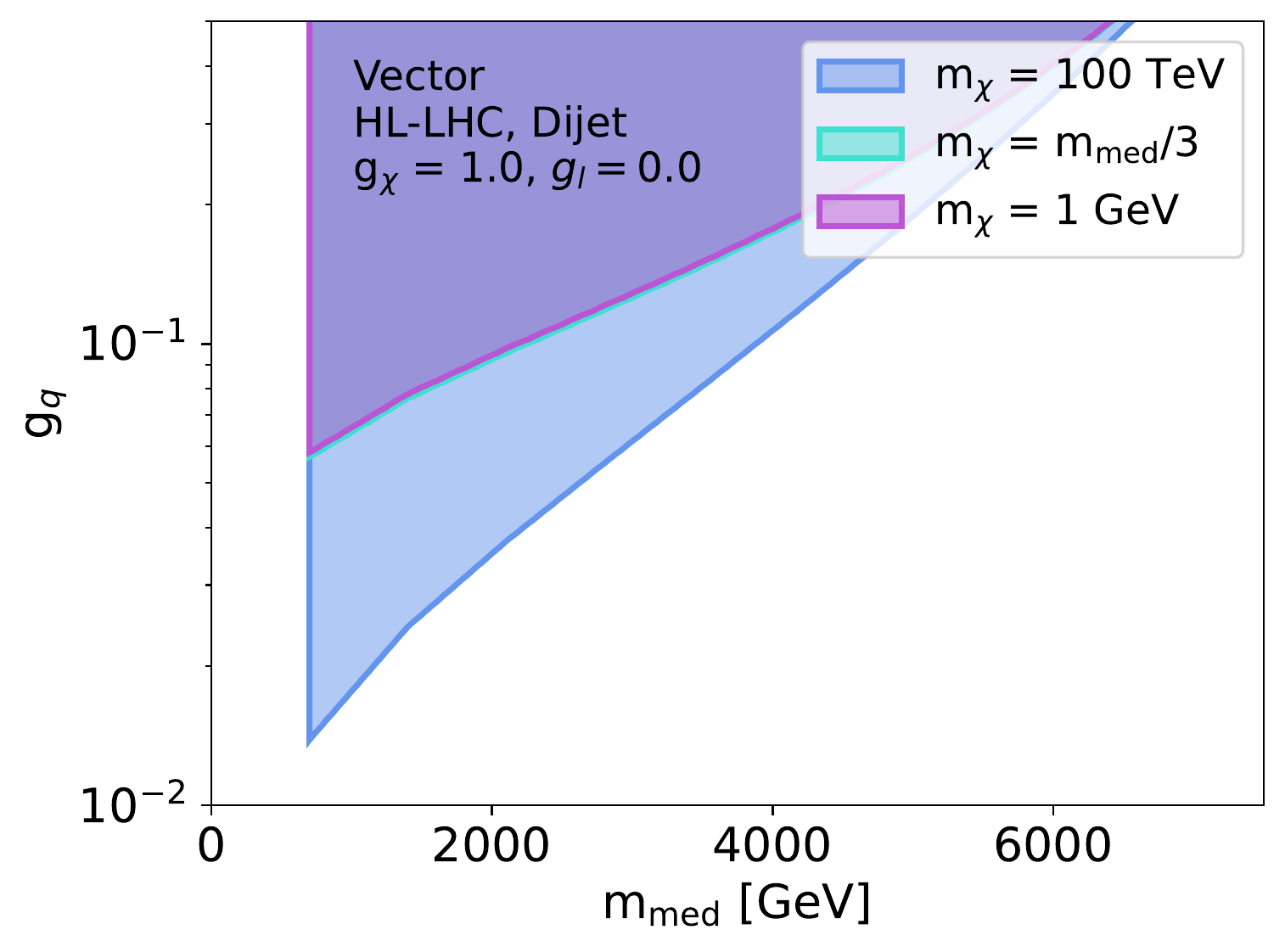}
         \caption{}
         \label{subfig:gqscan-comparedm}
     \end{subfigure}
     \begin{subfigure}[b]{0.49\textwidth}
         \includegraphics[width=\textwidth]{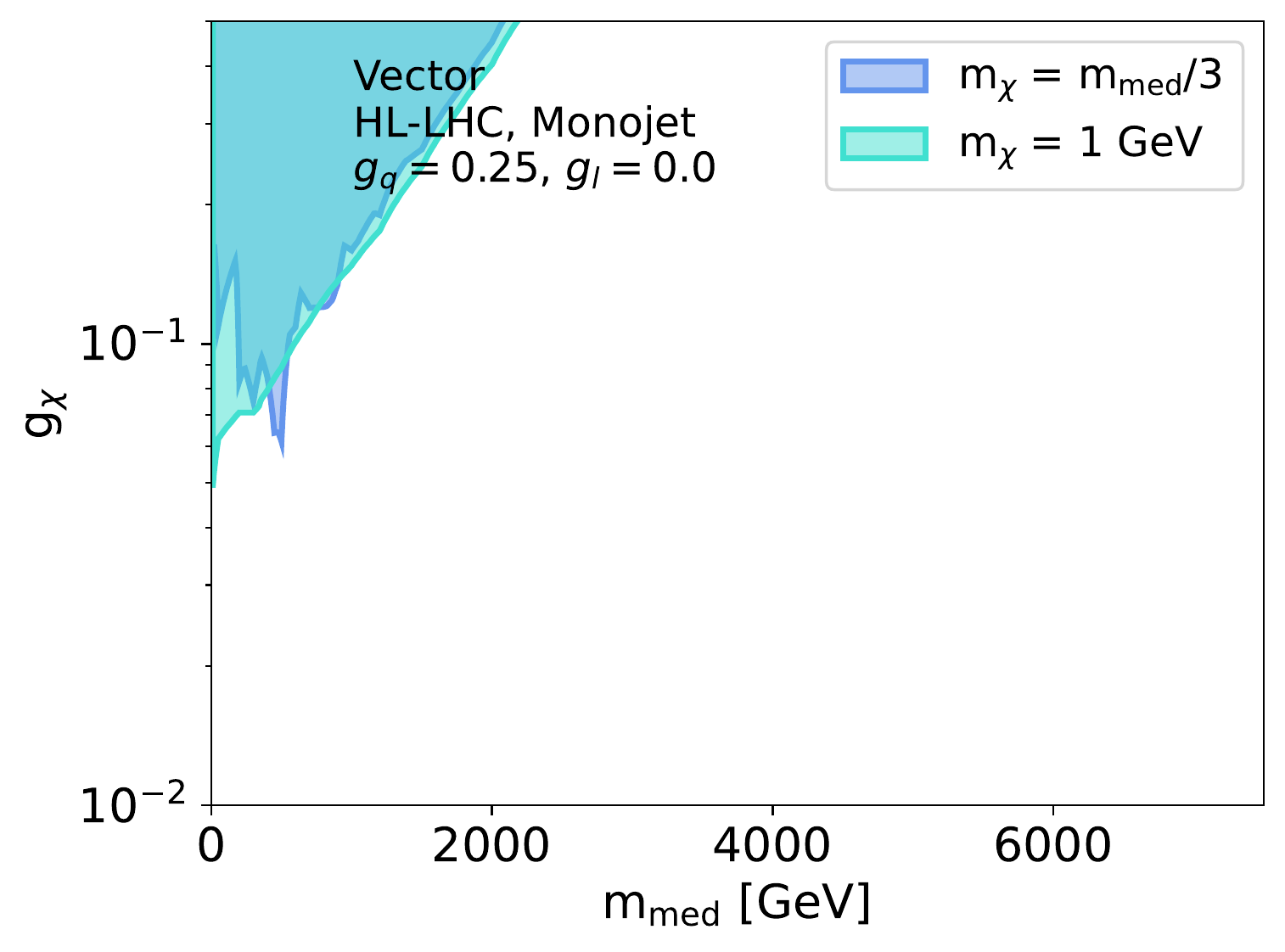}
         \caption{}
         \label{subfig:gchiscan-comparedm}
     \end{subfigure}     
         \caption{An illustration of the impact of the choice of DM particle mass on the coupling exclusion limits. Limits set by the dijet analysis on $g_q$ with $g_\chi=1.0$, $g_l=0.0$ are shown in (\subref{subfig:gqscan-comparedm}). Limits set by the monojet analysis on $g_\chi$ with $g_q=0.25$, $g_l=0.0$ are shown in (\subref{subfig:gchiscan-comparedm}. All limits displayed refer to vector mediators and HL-LHC projected exclusions. }
         \label{fig:couplinglimits-hl-lhc-dmmass}
\end{figure}

While Figure~\ref{fig:couplinglimits-hl-lhc-allanalyses} uses a light dark matter particle with $m_\chi=1$ GeV, alternate choices of dark matter mass can also affect these limits to varying degrees. Figure~\ref{fig:couplinglimits-hl-lhc-dmmass} illustrates this impact on the monojet and dijet analysis coupling limits by setting the mass of dark matter to $m_\chi=1$ GeV, $m_\chi=m_\mathrm{med}/3$, or 100 TeV (effectively decoupling it).
The difference between constraints is small when comparing a dark matter mass of 1~GeV and a DM mass corresponding to $1/3$ of the mediator mass. 
Decoupling dark matter entirely has a larger effect on the strength of the dijet limit, while eliminating any constraints from monojet by construction.

\FloatBarrier

\begin{figure}[htb!]
    \centering
     \begin{subfigure}[b]{0.49\textwidth}
         \includegraphics[width=\textwidth]{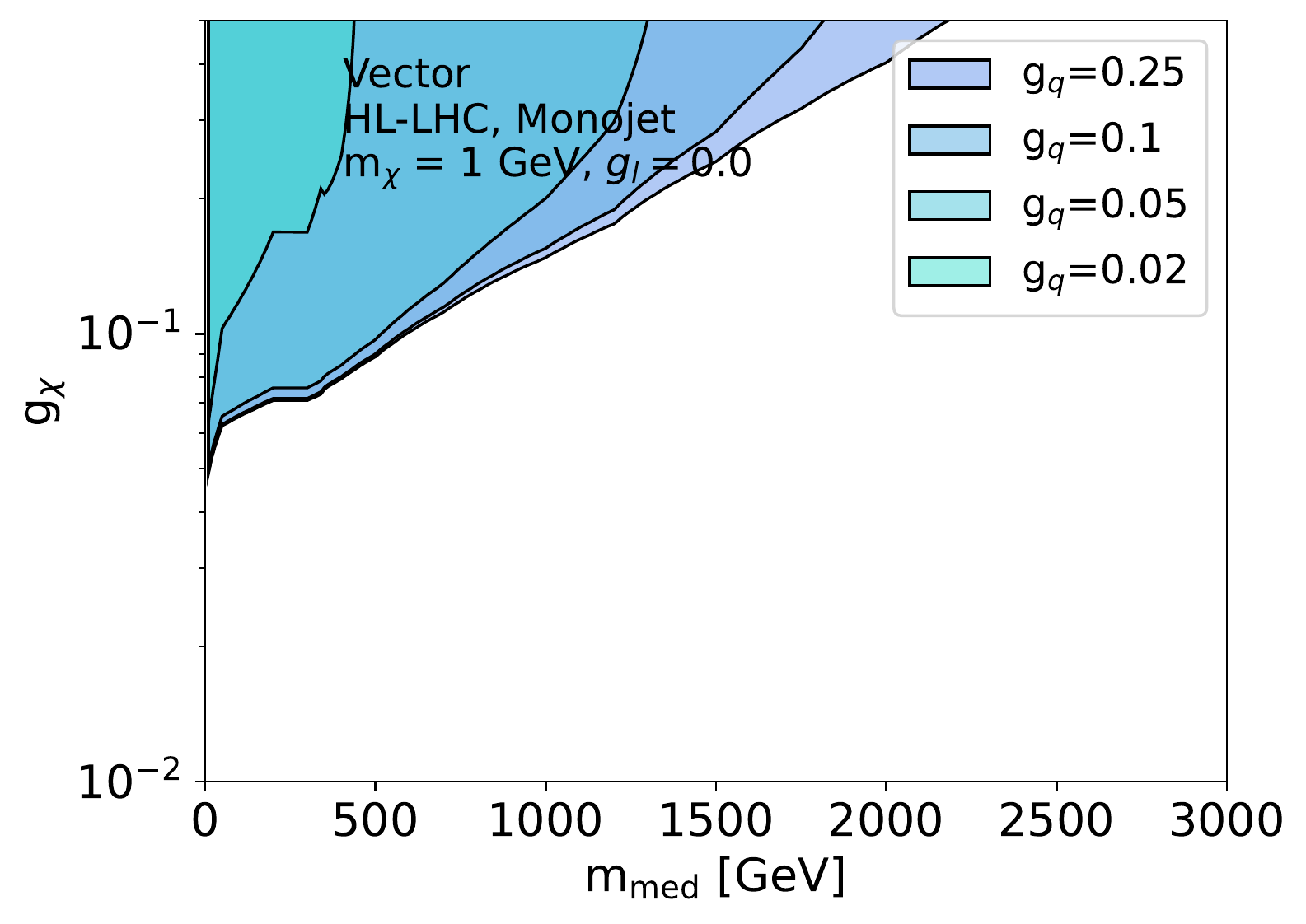}
         \caption{Monojet limits on $g_\chi$}
         \label{subfig:gchiscan-fixedmdm-monojet}
     \end{subfigure}
     \begin{subfigure}[b]{0.49\textwidth}
         \includegraphics[width=\textwidth]{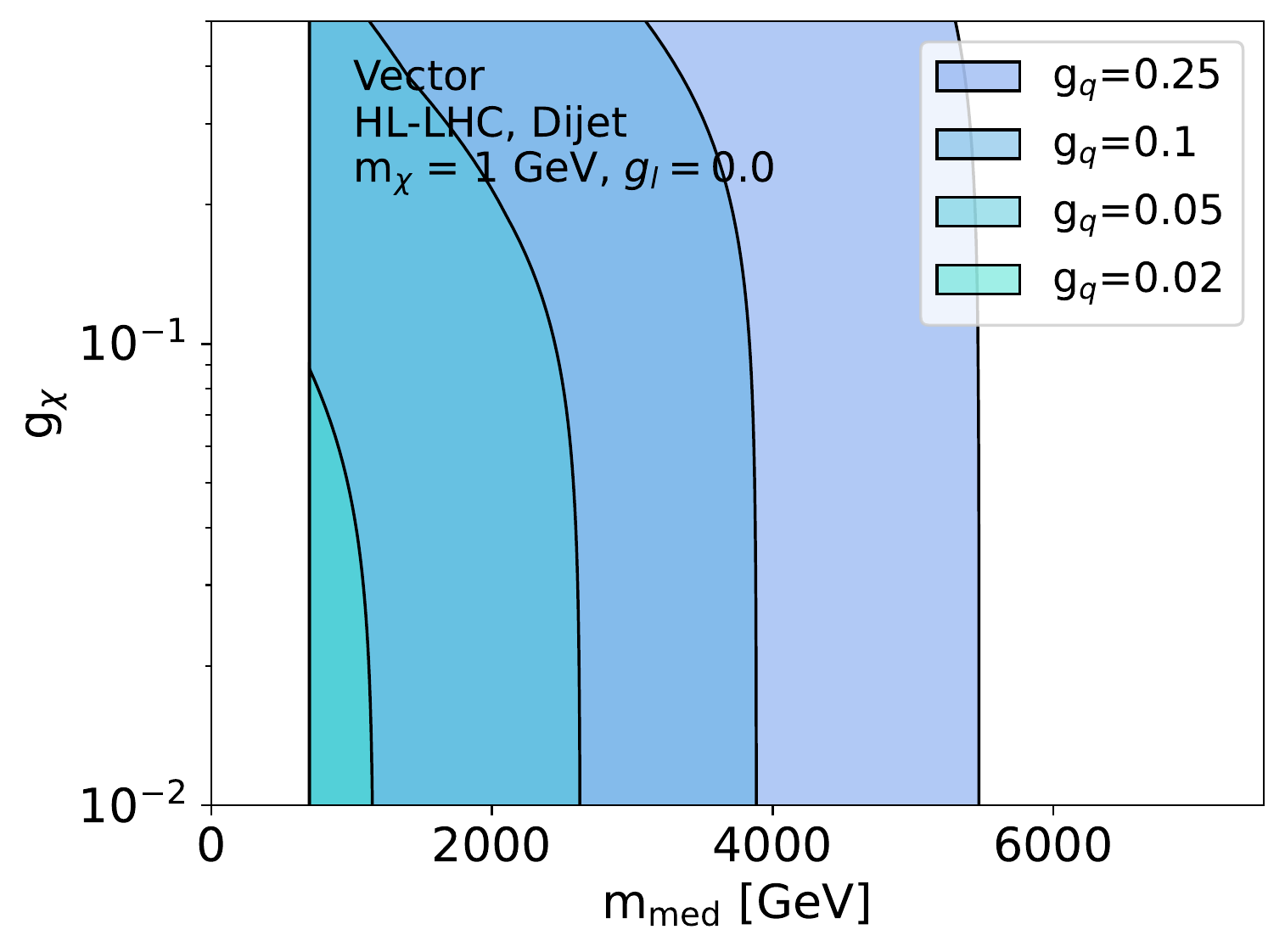}
         \caption{Dijet limits on $g_\chi$}
         \label{subfig:gchiscan-fixedmdm-dijet}
     \end{subfigure}     
     
     \begin{subfigure}[b]{0.49\textwidth}
         \includegraphics[width=\textwidth]{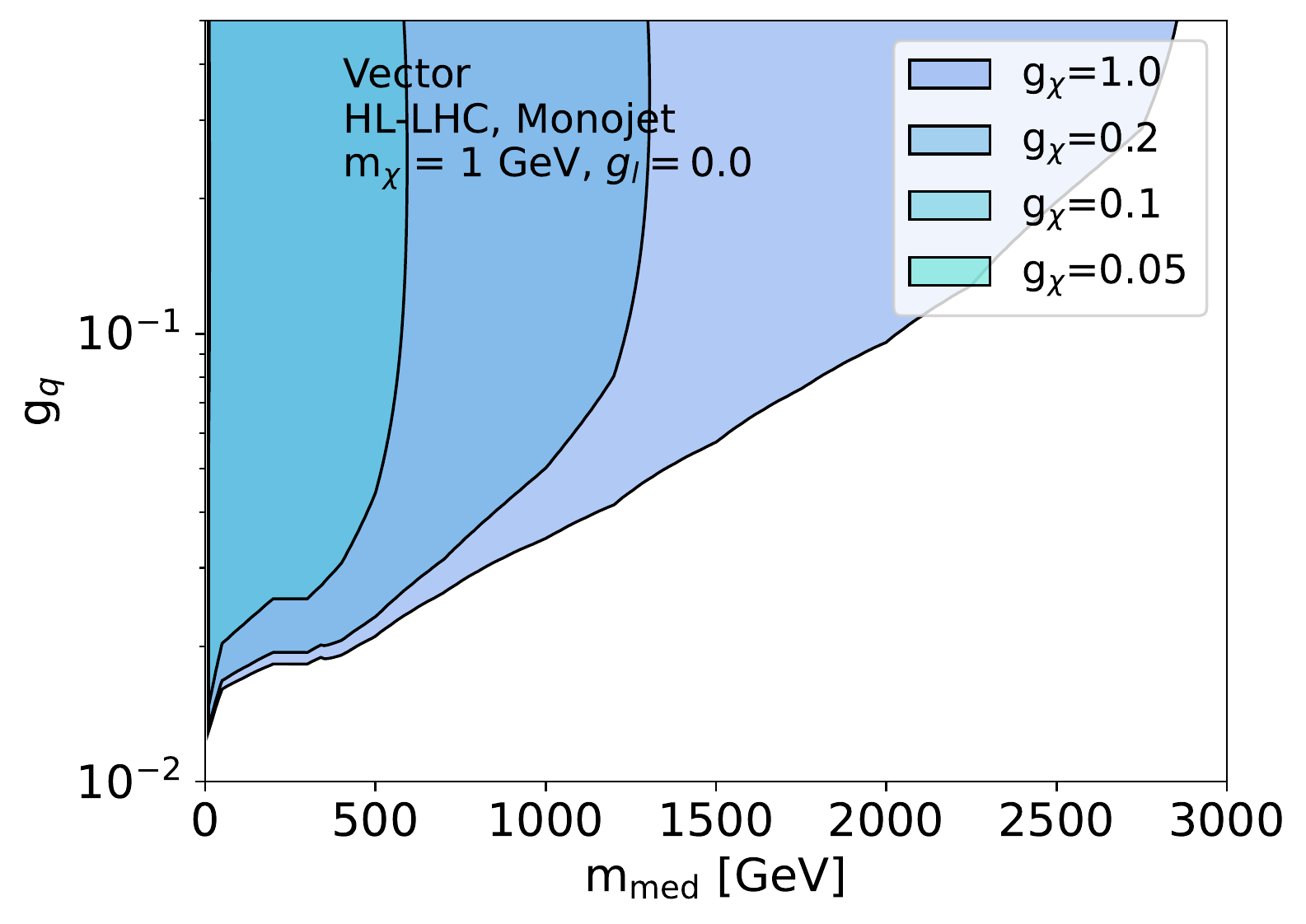}
         \caption{Monojet limits on $g_q$}
         \label{subfig:gqscan-fixedmdm-monojet}
     \end{subfigure}
     \begin{subfigure}[b]{0.49\textwidth}
         \includegraphics[width=\textwidth]{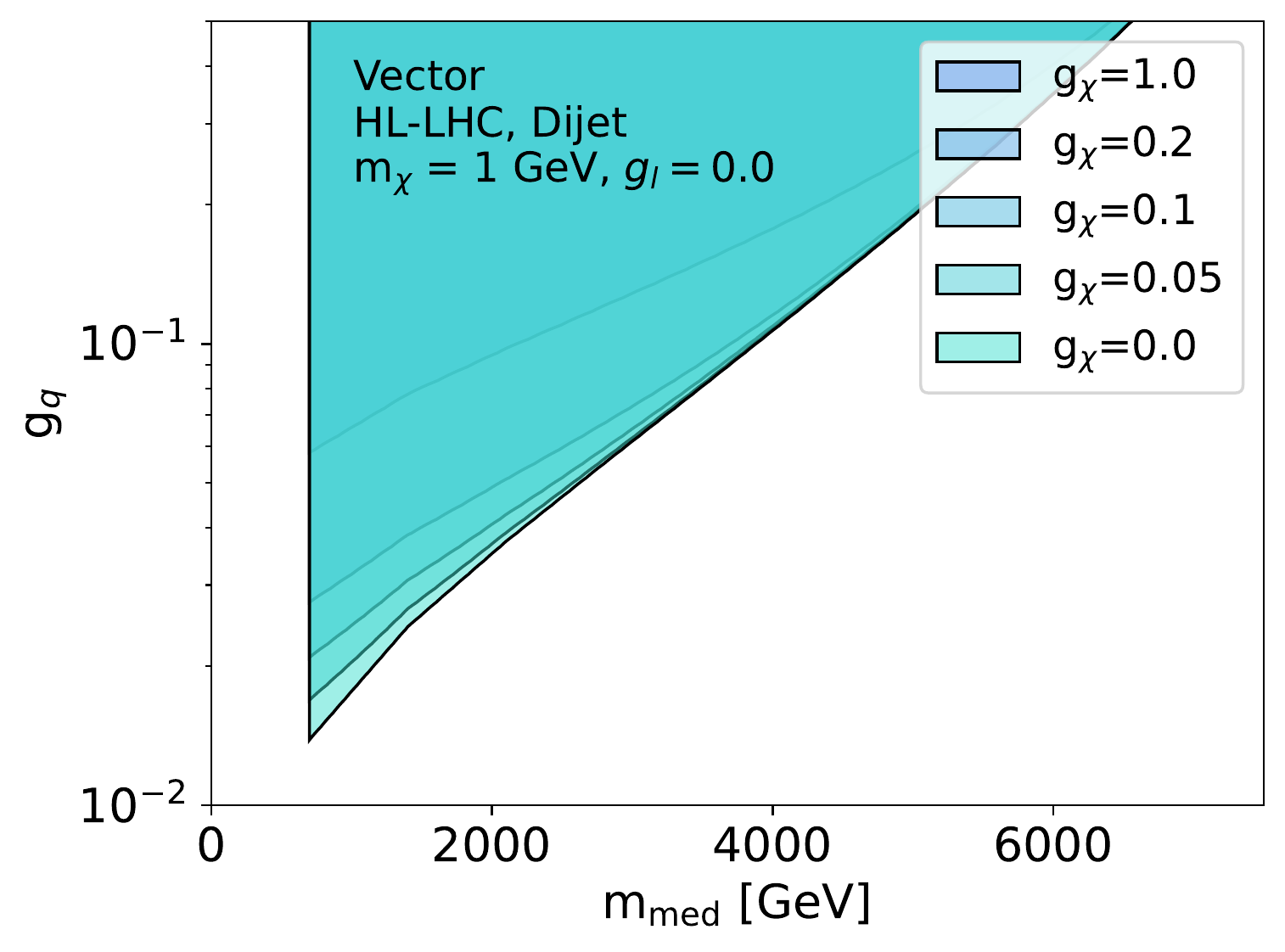}
         \caption{Dijet limits on $g_q$}
         \label{subfig:gqscan-fixedmdm-dijet}
     \end{subfigure}     
         \caption{Limits on $g_\chi$ (\subref{subfig:gchiscan-fixedmdm-monojet},~\subref{subfig:gchiscan-fixedmdm-dijet}) and $g_q$ (\subref{subfig:gqscan-fixedmdm-monojet},~\subref{subfig:gqscan-fixedmdm-dijet}) set by the HL-LHC monojet and dijet analyses as a function of vector mediator mass. The mediator has no coupling to leptons. The mass of the dark matter particle is fixed to $m_\chi=1$~GeV.}
         \label{fig:couplinglimits-hl-lhc-fixedmdm-couplingscan}
\end{figure}

Like the limits at fixed coupling values from the previous section, these mass-coupling limits can also be explored through comparisons across scanned couplings. Figure~\ref{fig:couplinglimits-hl-lhc-fixedmdm-couplingscan} illustrates how changing either $g_q$ or $g_\chi$ affects the limit on the remaining coupling. 
The dark matter mass is held fixed to 1~GeV and there is no coupling to leptons. 
The monojet analysis constrains $g_\chi$ most strongly at higher values (where the invisible signature is most easily produced), while the dijet analysis constrains $g_\chi$ most strongly from smaller values where the relative branching ratio to dijets is higher. For limits on $g_q$, in contrast, both analyses rely on the coupling to quarks for production and thus both constrain it at large coupling values. 
Also in this case, the power of the monojet signature decreases with decreasing $g_\chi$ while the power of the dijet signature increases.

\FloatBarrier

\begin{figure}[htb!]
    \centering
     \begin{subfigure}[b]{0.49\textwidth}
         \includegraphics[width=\textwidth]{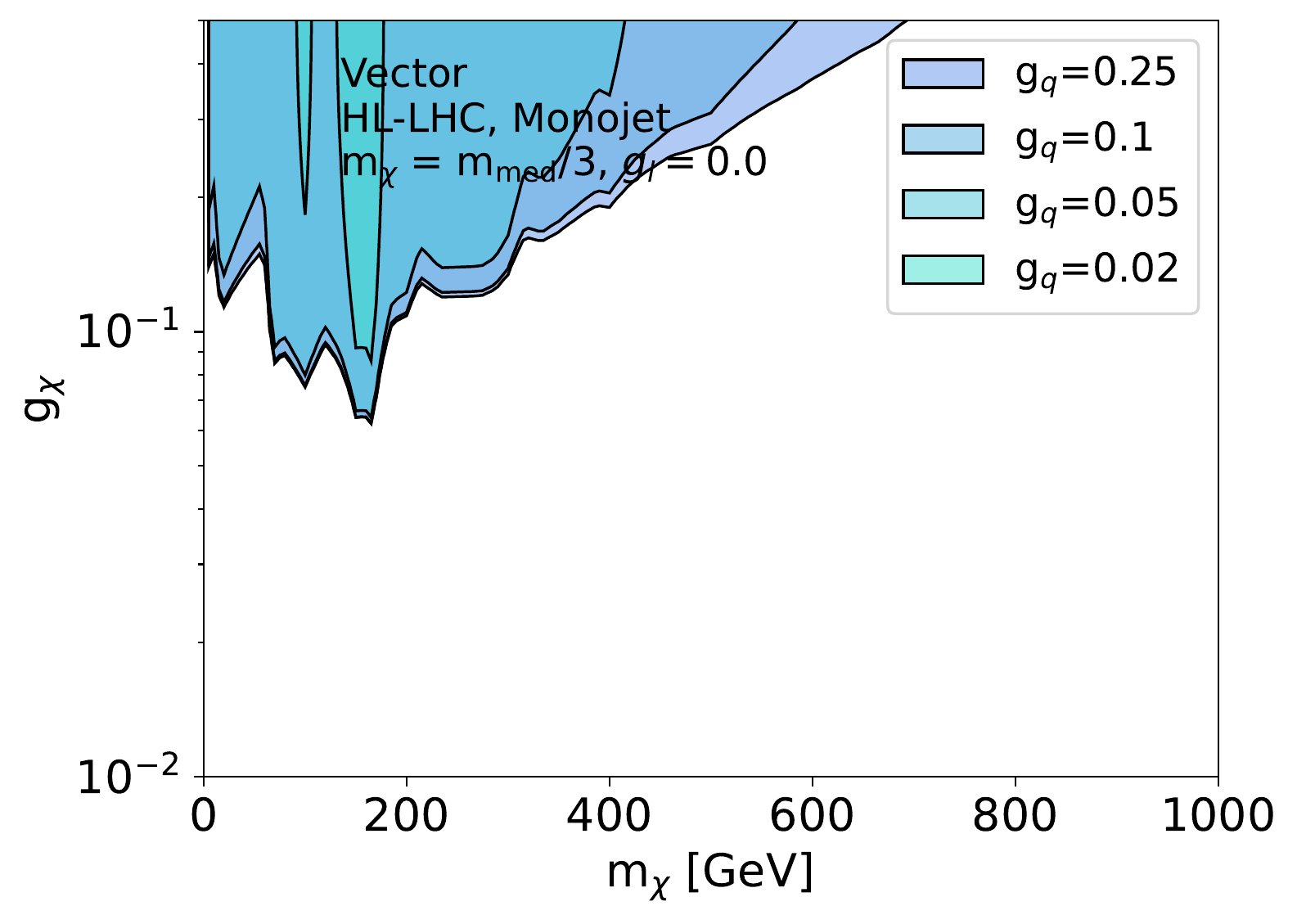}
         \caption{}
         \label{subfig:gchiscan-fixedmmed-monojet}
     \end{subfigure}
     \begin{subfigure}[b]{0.49\textwidth}
         \includegraphics[width=\textwidth]{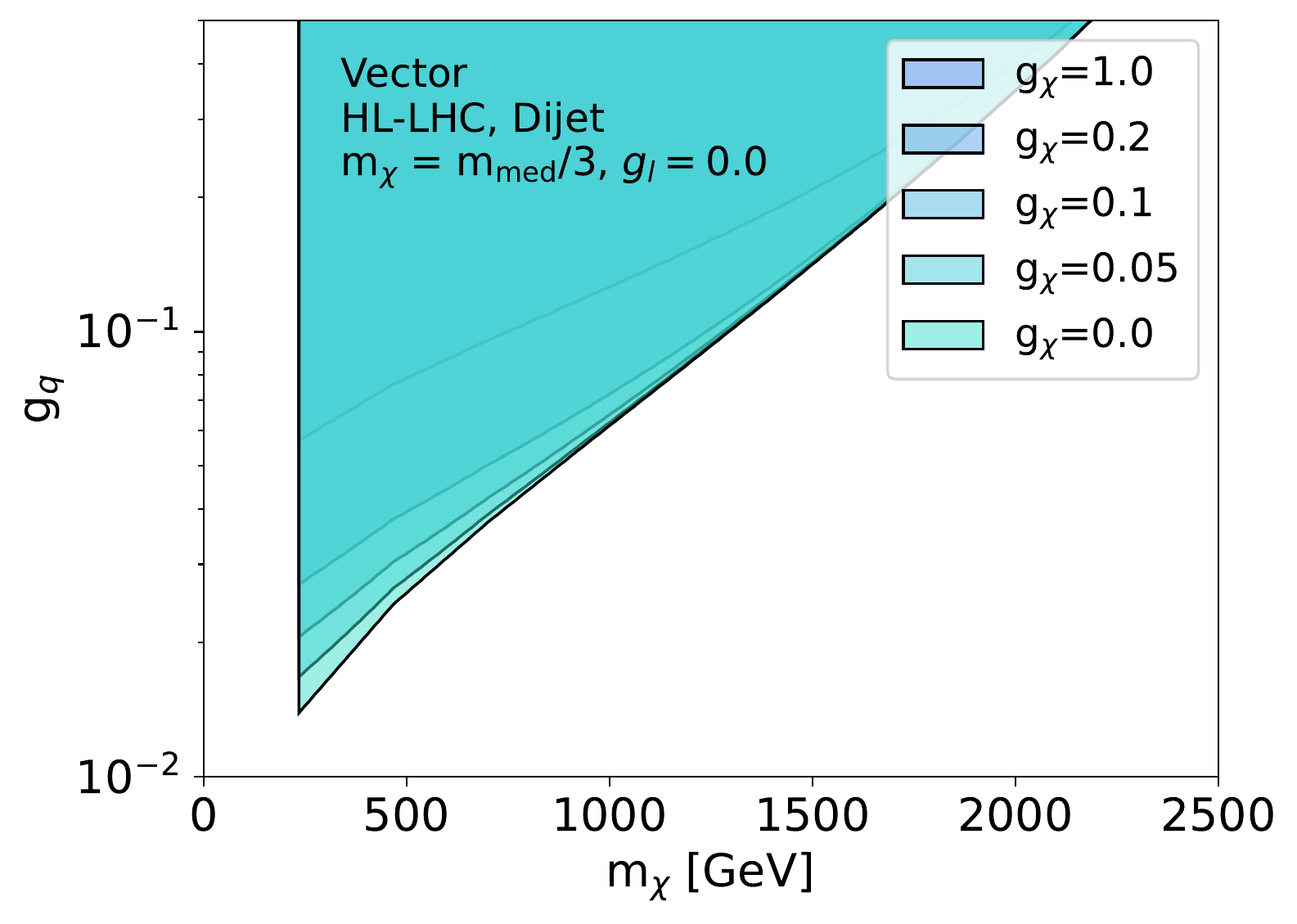}
         \caption{}
         \label{subfig:gqscan-fixedmmed-dijet}
     \end{subfigure}     
         \caption{Coupling limits as a function of $m_\chi$ with mediator mass fixed to $m_\mathrm{med} = 3 m_\chi$. Monojet constraints on $g_\chi$ with varying $g_q$ values are shown in (\subref{subfig:gchiscan-fixedmmed-monojet}) while dijet constraints on $g_q$ with varying $g_\chi$ values are shown in (\subref{subfig:gqscan-fixedmmed-dijet}).}
         \label{fig:couplinglimits-hl-lhc-fixedmmed}
\end{figure}

Fixing the mediator mass and scanning the DM mass provides an alternative version of these coupling limits, which can be shown instead as a function of dark matter mass. Figure~\ref{fig:couplinglimits-hl-lhc-fixedmmed} shows the limits on $g_\chi$ and $g_q$ obtained from monojet and dijet HL-LHC analyses respectively with $m_\mathrm{med}$ fixed to three times the dark matter mass. These limits correspond directly to the equivalent limits as a function of $m_\mathrm{med}$ with the same mass hypothesis, but will be particularly useful for complementarity statements with other experiments where limits are often stated as a function of dark matter mass.

\begin{figure}[htb!]
    \centering
     \begin{subfigure}[b]{0.49\textwidth}
         \includegraphics[width=\textwidth]{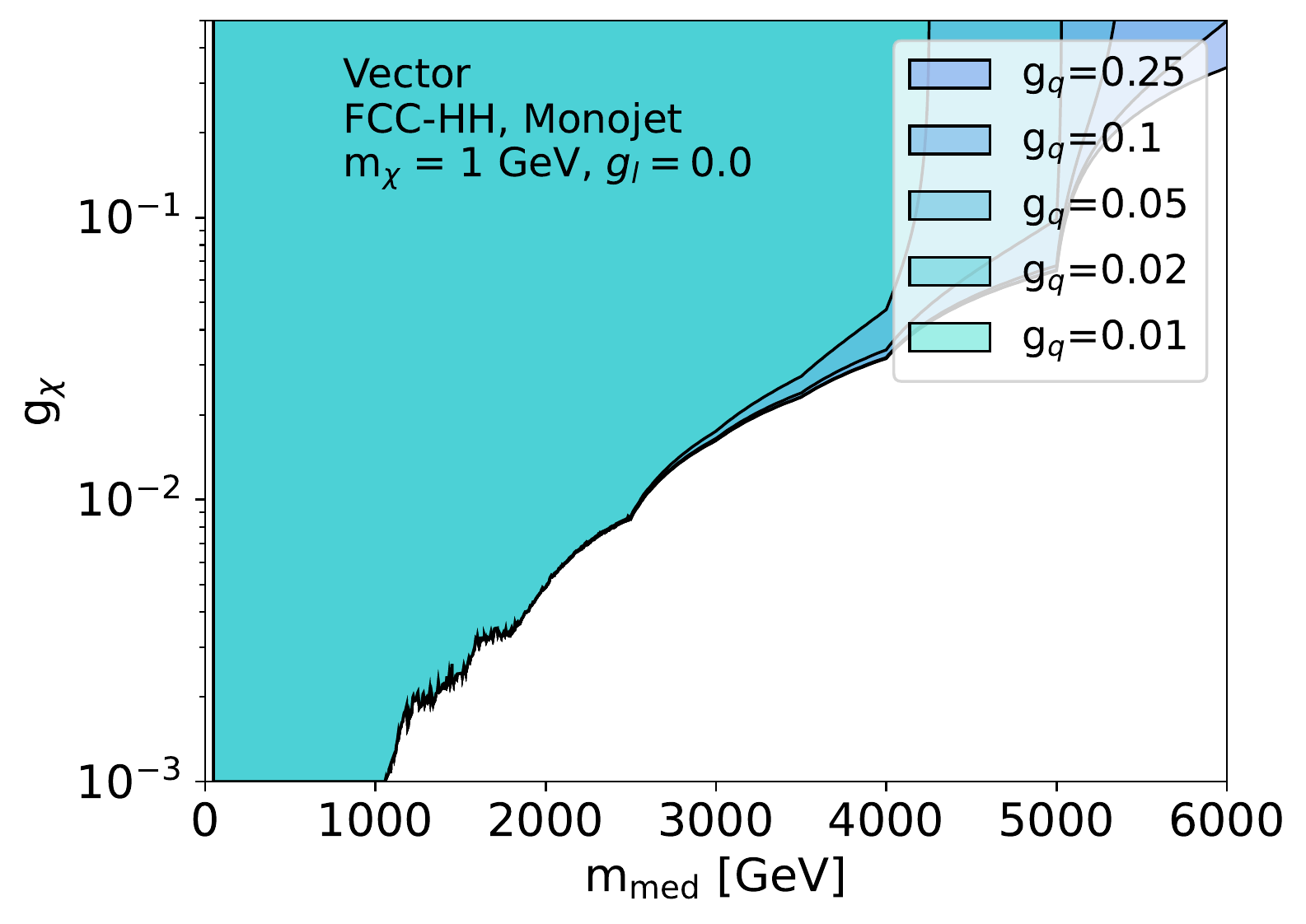}
         \caption{Monojet limits on $g_\chi$}
         \label{subfig:gchiscan-fixedmdm-monojet-fcc}
     \end{subfigure}
     \begin{subfigure}[b]{0.49\textwidth}
         \includegraphics[width=\textwidth]{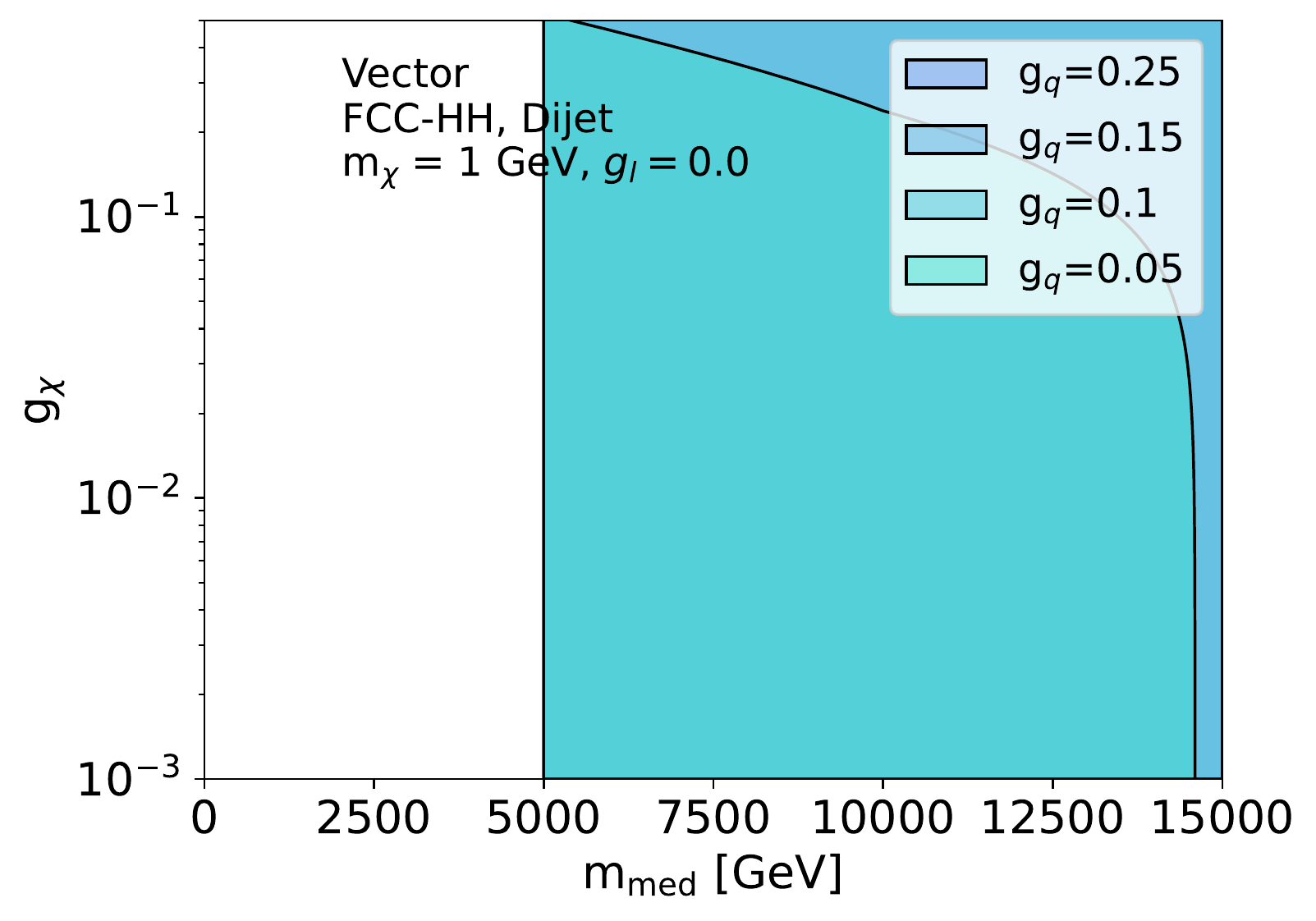}
         \caption{Dijet limits on $g_\chi$}
         \label{subfig:gchiscan-fixedmdm-dijet-fcc}
     \end{subfigure}     
     
     \begin{subfigure}[b]{0.49\textwidth}
         \includegraphics[width=\textwidth]{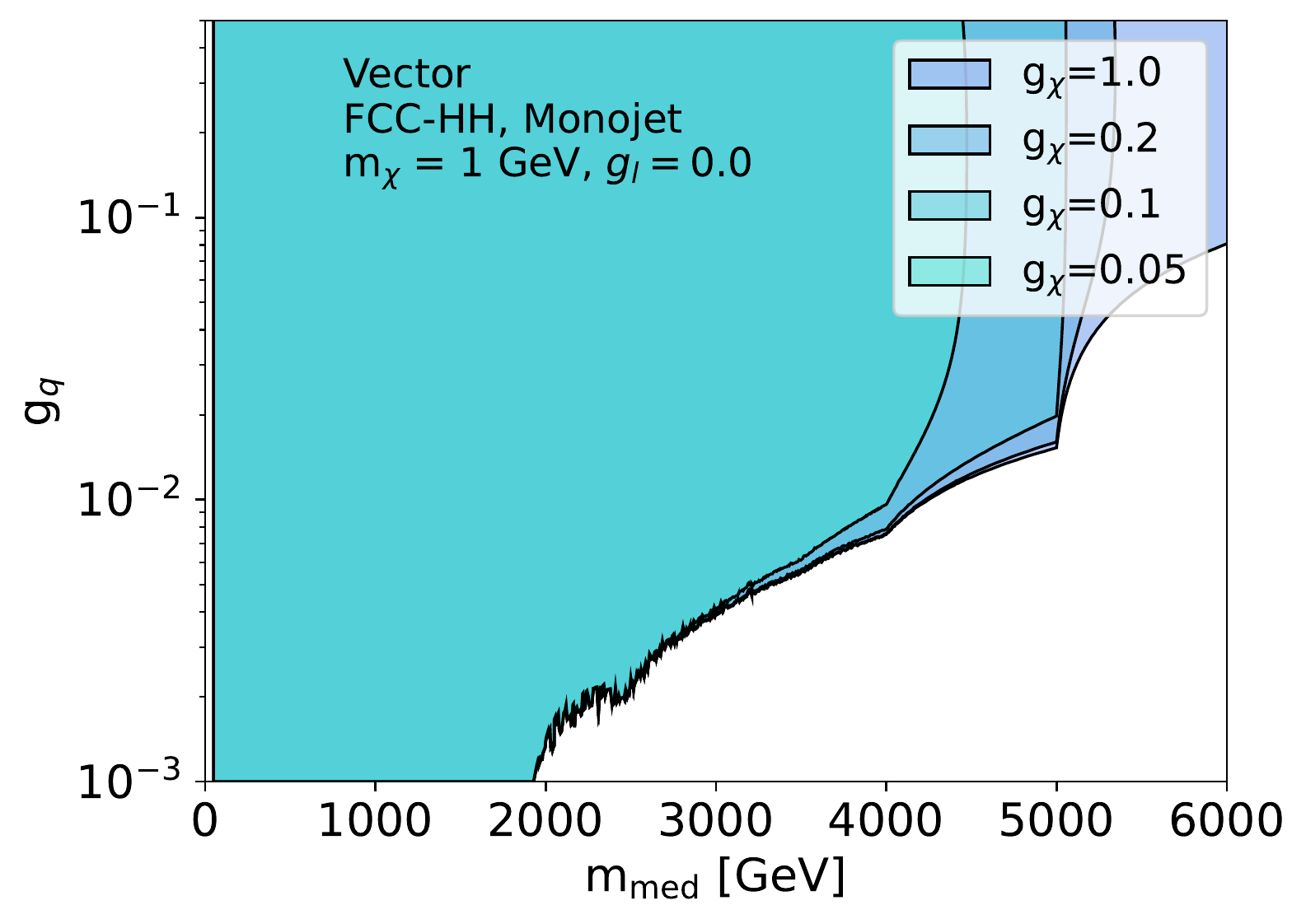}
         \caption{Monojet limits on $g_q$}
         \label{subfig:gqscan-fixedmdm-monojet-fcc}
     \end{subfigure}
     \begin{subfigure}[b]{0.49\textwidth}
         \includegraphics[width=\textwidth]{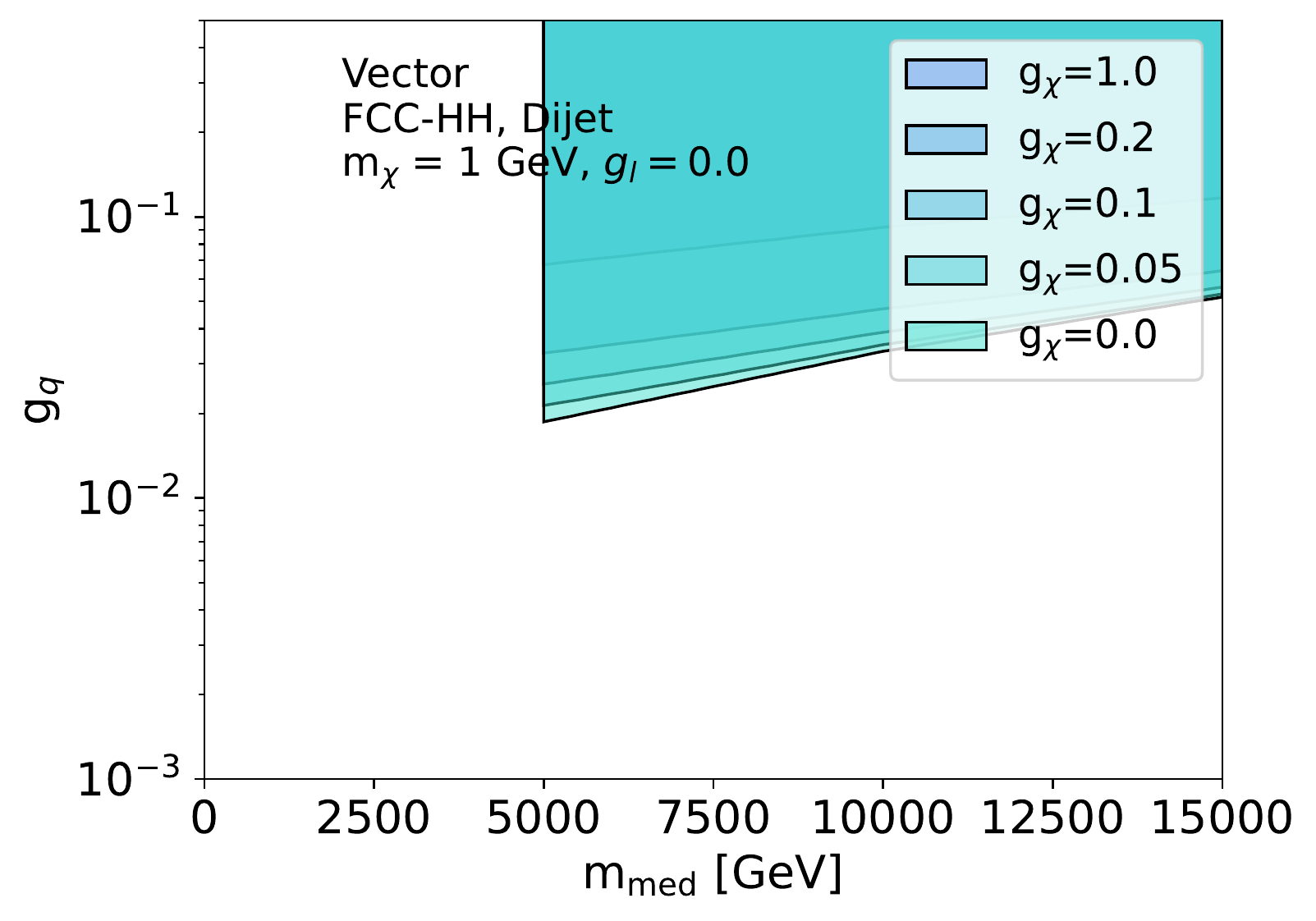}
         \caption{Dijet limits on $g_q$}
         \label{subfig:gqscan-fixedmdm-dijet-fcc}
     \end{subfigure}       
         \caption{Coupling limits as a function of mediator mass at FCC-hh with $m_\chi = 1$ GeV.}
         \label{fig:couplinglimits-fcc-hh-various}
\end{figure}

Equivalent plots can be produced for FCC-hh and show similar trends, but reach smaller coupling values. Example plots are shown as a function of mediator mass for light dark matter in Figure~\ref{fig:couplinglimits-fcc-hh-various}. The results are similar for $m_\chi = m_\mathrm{med}/3$; these can be found in the Appendix.

\clearpage

\subsection{Dark matter projections for a muon collider}
\label{subsec:muonCollider}

One of the most studied DM candidates is a Weakly Interacting Massive Particle (WIMP), which is predicted to have weak scale interactions with SM particles, allowing for various experiments to investigate its signatures. Among the WIMP candidates, one interesting scenario assumes the DM particle to be the lightest member of an electroweak (EW) multiplet and it can be either neutral or a charged~\cite{Han:2020uak}. If the sole responsible for the measured thermal relic abundance, the DM mass scale is set in the range 1-23 TeV. The model independent mono-X signal searches at the LHC are expected to saturate at few hundred GeV, even after the high-luminosity phase of the LHC program completes~\cite{Low:2014,Han:2018,Vidal:2019}.
Great discovery potential for this type of new physics is provided by muon colliders. In fact, these operate at much higher center of mass energies with respect to hadron colliders, and they can probe new physics up to the highest possible energy thresholds.

In this study we report on prospects for WIMP discovery at muon colliders with center of mass energies between 3 and 30 TeV, with integrated luminosities ranging from 1 to 10 ab$^{-1}$. The following studies are based on ~\cite{Han:2020uak}, with the extension of considering the full detector simulation for both signal and background processes to include acceptance effects and resolutions, as well as an optimized variable selection to further increase the signal sensitivity of the search. In particular, we explore the possibility of discoveries in the mono-photon final state, considering only the most significant SM backgrounds to these processes. Although there are several multiplets of the SM group, with varying cross sections for production that could be considered, we limit ourselves to color-singlet-electroweak-doublet case in this initial study. 

Signal events are generated using the FeynRules based model for Madgraph5~\cite{Han:2020uak}. The Madgraph5, Pythia8 and Delphes programs are setup for various muon collider configurations and used to produce both signals for various dark matter masses and centers of mass energies of the collider. 

\begin{figure}[htb!]
    \centering
    \begin{subfigure}{0.5\textwidth}
        \centering
        \includegraphics[width=3in]{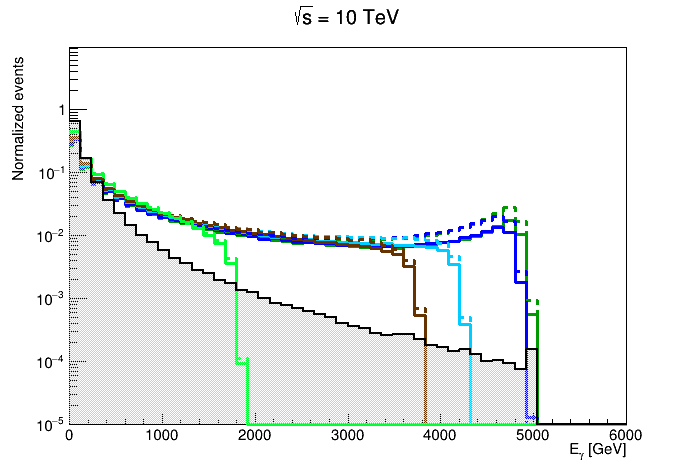}
        \caption{}
        \label{fig:monop_COM10_egamma}
    \end{subfigure}\hfill
    \begin{subfigure}{0.5\textwidth}
        \centering
        \includegraphics[width=3in]{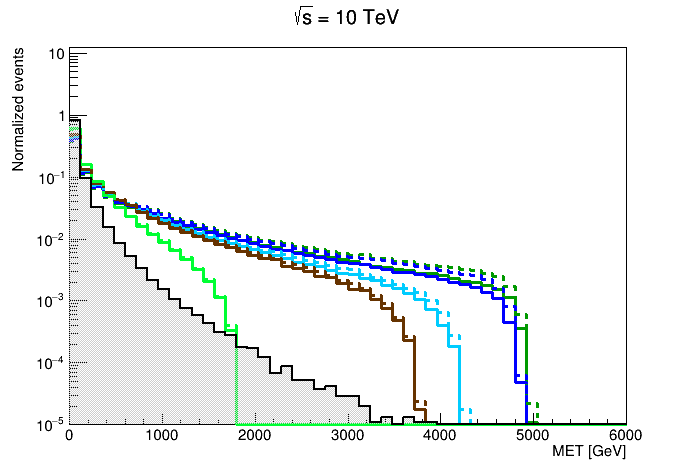}
        \caption{}
        \label{fig:monop_COM10_MET}
    \end{subfigure}\hfill
    \begin{subfigure}{0.5\textwidth}
        \centering
        \includegraphics[width=3in]{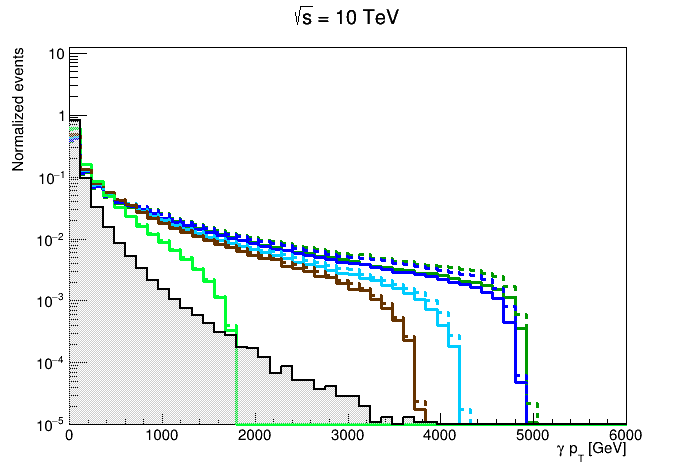}
        \caption{}
        \label{fig:monop_COM10_gammapT}
    \end{subfigure}\hfill
   \begin{subfigure}{0.5\textwidth}
        \centering
        \includegraphics[width=3in]{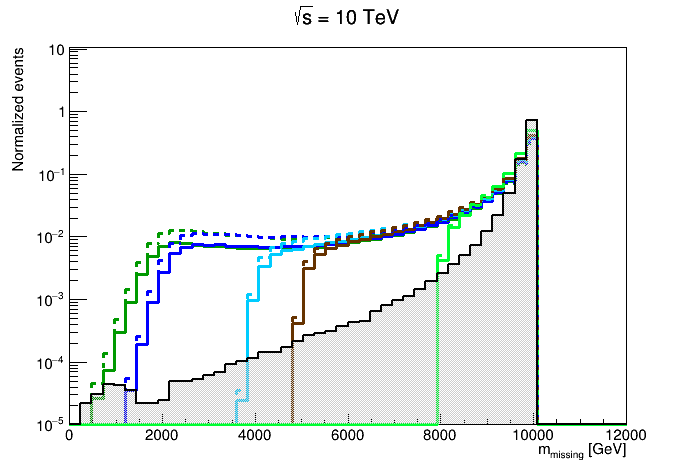}
        \caption{}
        \label{fig:monop_COM10_mm}
    \end{subfigure}\hfill
    \begin{subfigure}{0.5\textwidth}
        \centering
        \includegraphics[width=3in]{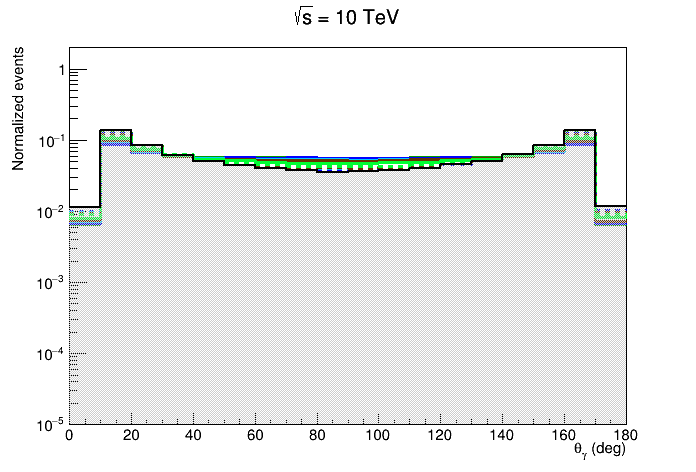}
        \caption{}
        \label{fig:monop_COM10_theta}
    \end{subfigure}\hfill
     \begin{subfigure}{0.5\textwidth}
        \centering
        \includegraphics[width=3in]{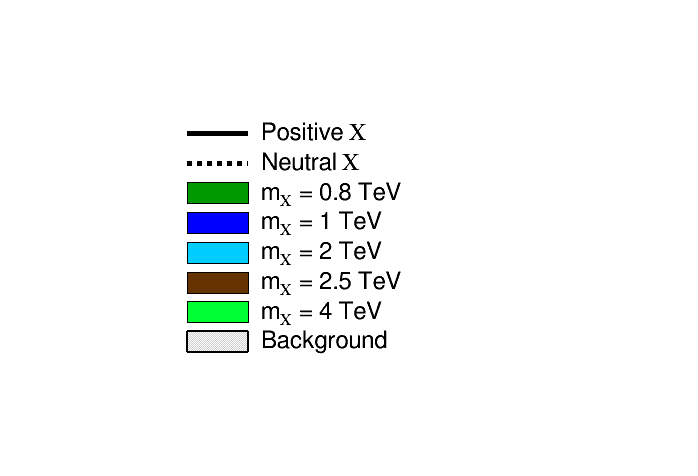}
        \label{fig:monop_COM10_legend}
    \end{subfigure}
    \caption{Normalized distributions for the photon energy E$_\gamma$ (a), MET (b), photon transverse momentum $\gamma \mathrm{p}_T$ (c), missing mass m$_{missing}$ (d), theta of the photon $\theta_{\gamma}$ (e) for different dark matter masses with both charged and neutral DM particles for a center of mass energy of 10~TeV after the requirement that at least one photon is present in the final state.
    }
    \label{fig:monop-10}
\end{figure}

Since the dark matter particle exit the detector without interacting, the signal events are expected to have large missing transverse momentum (MET), 

\[
\mathrm{MET} = \sqrt{\Big[\sum_i{{p_x^i}\Big]^2 + \Big[\sum_i{p_y^i}\Big]^2}},
\]

and missing mass,

\[
m_{missing}^2 = {\Big(p_{\mu^+} + p_{\mu^-} - \sum_i{p_i}\Big)}^2,
\]

\noindent where $p_i$ are the four-vectors of the observed particles in the final states and $p_{\mu^+}$ and $p_{\mu^-}$ are the momenta of the colliding muons. 
Because any high-energy particles that exit the detector without interaction cause large MET and $m_{missing}$, we look for signal in that 2-dimensional plane. Other discriminating variables are investigated for the mono-photon scenario, in addition to  MET and (m$_{missing}$), in order to identify a selection to discriminate between signal and SM background events. Some of these distributions are presented for signal and background, for a 10 TeV center of mass energy of a muon collider and various dark matter particle masses in Figure~\ref{fig:monop-10} for the mono-photon analysis. 

We report on the significance of measurements of mono-photon event signatures of DM production over the SM backgrounds in terms of a figure of merit (FOM) defined as $FOM = \frac{\mathrm{s}}{\sqrt{\mathrm{b}}}$, where s is the number of expected signal events and b is the number of expected SM background events after the identified selection is applied. The FOM for different centers of mass energies of the collider and dark matter masses are listed in Table~\ref{tab:mono-photon_FOM}. 
\begin{table}[htb!]
    \centering
    \resizebox{\columnwidth}{!}{
    \begin{tabular}{|l|c|c|c|c|}
        \hline
        &&&&\\
        $M_\chi$ / $\sqrt{s}~(\int{d{\cal L}})$ & 3 TeV (1 ab$^{-1}$) & 6 TeV (4 ab$^{-1}$) & 10 TeV (10 ab$^{-1}$) & 30 TeV (10 ab$^{-1}$) \\ 
        &&&&\\\hline
        &&&&\\
        0.8 TeV & 1.02 & 1.01 & 0.81 & 0.23\\
        &&&&\\
        1.0 TeV  & 0.72 & 0.91 & 0.75 & 0.22\\
        &&&&\\
        2.0 TeV & n/a & 0.47 & 0.55 & 0.18\\
        &&&&\\
        2.5 TeV & n/a & 0.22 & 0.47 & 0.17\\
        &&&&\\
        4.0 TeV & n/a & n/a & 0.21 & 0.15\\
        &&&&\\
        10.0 TeV & n/a & n/a & n/a & 0.07\\
        &&&&\\\hline
    \end{tabular}
    }
    \caption{Significance figure of merit for observing mono-photon signals due to dark matter WIMP production above anticipated Standard Model backgrounds, as a function of dark matter mass ($M_\chi$) considering  both neutral and charged DM particles, for muon colliders operating at various centers of mass energies and integrated luminosity after selections based on discriminating variables.}
    \label{tab:mono-photon_FOM}
\end{table}

While this is not a full study leading to a projected constraint, we still find that the mono-photon channel would be sensitive to color-singlet-electroweak-doublet DM candidates. Therefore, we conclude that a mono-photon analysis at a high energy muon collider could have a significant impact in the search for the thermal dark matter. 
In order to target DM masses up to ~2 TeV, muon colliders operating at low center of mass energies will be the preferred option, while to discover or constrain DM candidates above 2 TeV, higher energies and higher luminosities will be needed. 

More details on the presented study can be found in Ref.~\cite{Black:2022qlg} where prospects for WIMP discovery in the mono-Z (leptonic decays only) channel are also reported. The mono-Z channel, with the current presented analysis, would need a 10~TeV and 10-ab$^{-1}$ muon collider machine to go significantly beyond the HL-LHC performance. Despite being currently less sensitive with respect to the mono-photon signature, it could provide important complementary in the search for dark matter an muon colliders. In addition, possible future improvements in the analyses strategy as well as including additional final states could boost the sensitivity reach for both channels.


\section{Comparisons of collider projections to other experiments}
\label{sec:complementarity}

To understand how current and future collider searches within the Energy Frontier may complement Cosmic Frontier and Intensity Frontier searches, we update comparisons done for the recent European Strategy Briefing Book~\cite{EuropeanStrategyBook} which extrapolate the projected constraints from collider searches on simplified models to the direct-detection (scattering) and indirect-detection (annihilation) cross section plots typically provided by the non-collider searches.

In Sections \ref{sec:complementarity_DD} and \ref{sec:complementarity_ID}, we plot collider projections   together with current results from direct-detection (DD) and indirect-detection (ID) experiments, using variables commonly employed to display indirect- and direct-detection results. The translation of collider limits is done using the procedures within Refs.~\cite{BOVEIA2020100365,ALBERT2019100377}. 
The goal for the EF10 Topical Group whitepaper is to update the plots in the European Strategy Briefing Book~\cite{EuropeanStrategyBook} with varying assumptions on the mediator coupling strengths to quarks and to DM. 

In Section \ref{sec:lightDarkSectorPortals}, we reframe LHC and HL-LHC monojet limits in terms of Intensity Frontier dark photon benchmarks. 

\subsection{Collider and direct-detection results}
\label{sec:complementarity_DD}

\subsubsection{Vector mediator}
\label{sub:complementarity_DD_vector}

\paragraph{Collider results shown in the direct-detection plane}

Assumptions on theoretical models that are made when deriving one set of constraints from collider data, such as the presence and coupling strength of a mediator particle, do not feature in astrophysical searches. 
Visually displaying and highlighting the consequences of varying these assumptions can help paint a more complete picture of the complementarity of collider DM searches with direct and indirect detection. 
This is why in this section we display a sample of collider constraints, overlaid to direct-detection constraints, using different mediator-quark coupling values. 



\begin{figure}[!htp]
     \centering
     \begin{subfigure}[b]{0.45\textwidth}
         \centering
         \includegraphics[width=\textwidth]{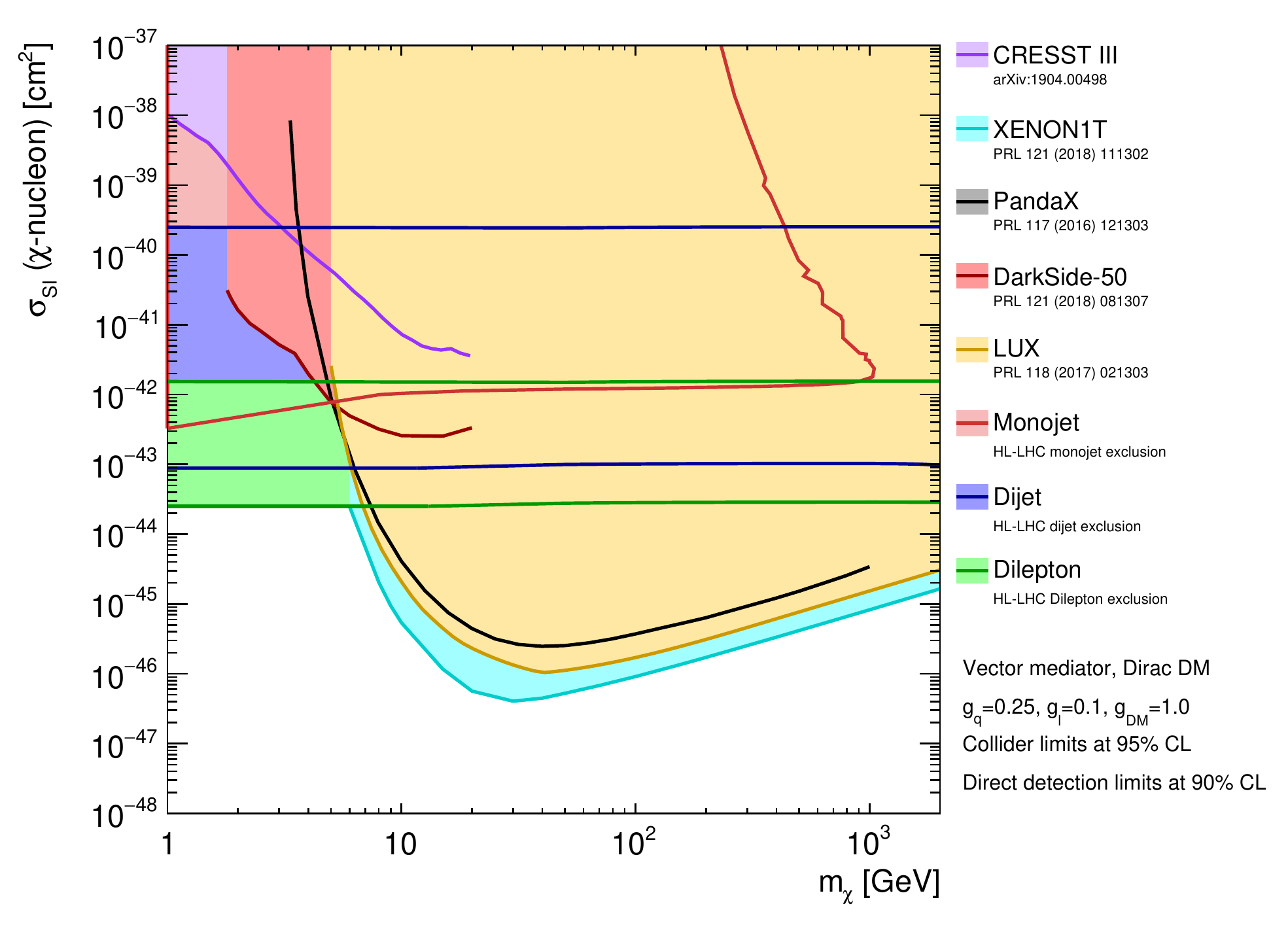}
         \caption{$g_q=0.25$, $g_{\chi}=1.0$, $g_l=0.1$}
         \label{subfig:dd-hl-lhc-vec-1}
     \end{subfigure}
     \begin{subfigure}[b]{0.45\textwidth}
         \centering
         \includegraphics[width=\textwidth]{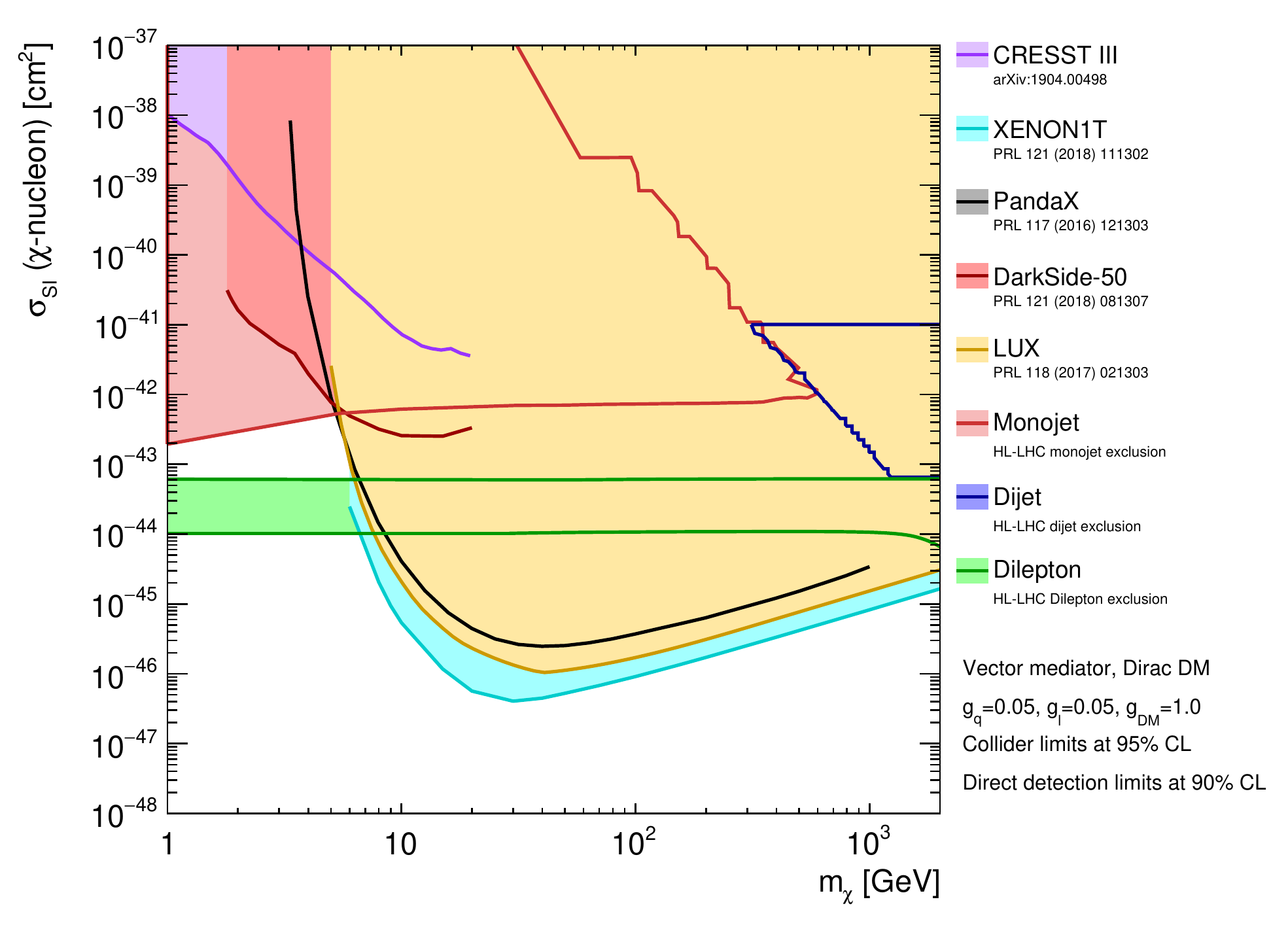}
         \caption{$g_q=0.05$, $g_{\chi}=1.0$, $g_l=0.05$}
         \label{subfig:dd-hl-lhc-vec-2}       
     \end{subfigure}
        \caption{Comparison of projected exclusions from HL-LHC with constraints from current direct-detection experiments on the spin-independent DM–nucleon scattering cross section in the context of the vector simplified model.}
        \label{fig:hl-lhc-dd-separate-si}
\end{figure}

\begin{figure}[!htp]
     \centering
     \begin{subfigure}[b]{0.45\textwidth}
         \centering
         \includegraphics[width=\textwidth]{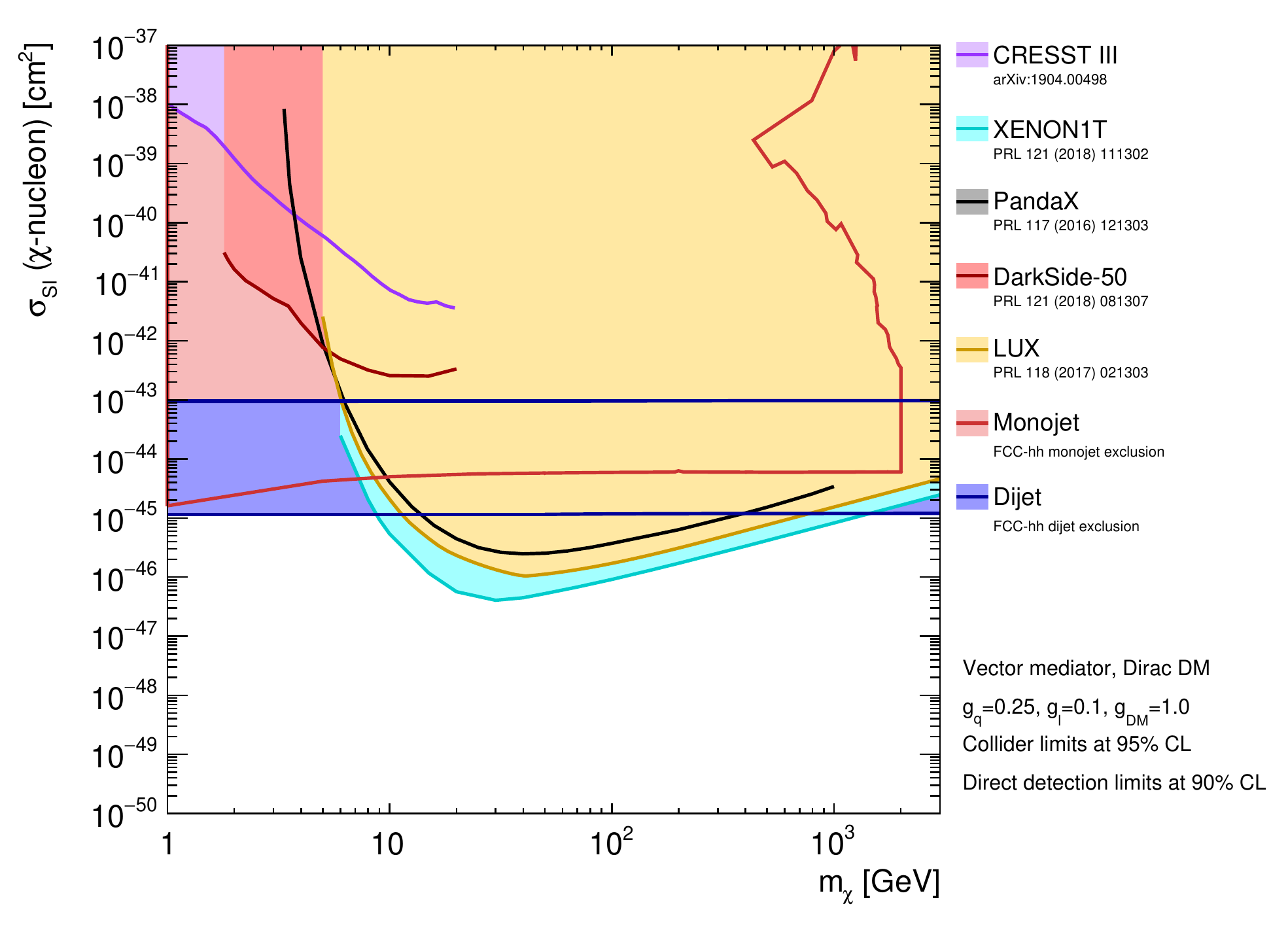}
         \caption{$g_q=0.25$, $g_{\chi}=1.0$, $g_l=0.1$}
         \label{subfig:dd-fcc-hh-vec-1}
     \end{subfigure}
     \begin{subfigure}[b]{0.45\textwidth}
         \centering
         \includegraphics[width=\textwidth]{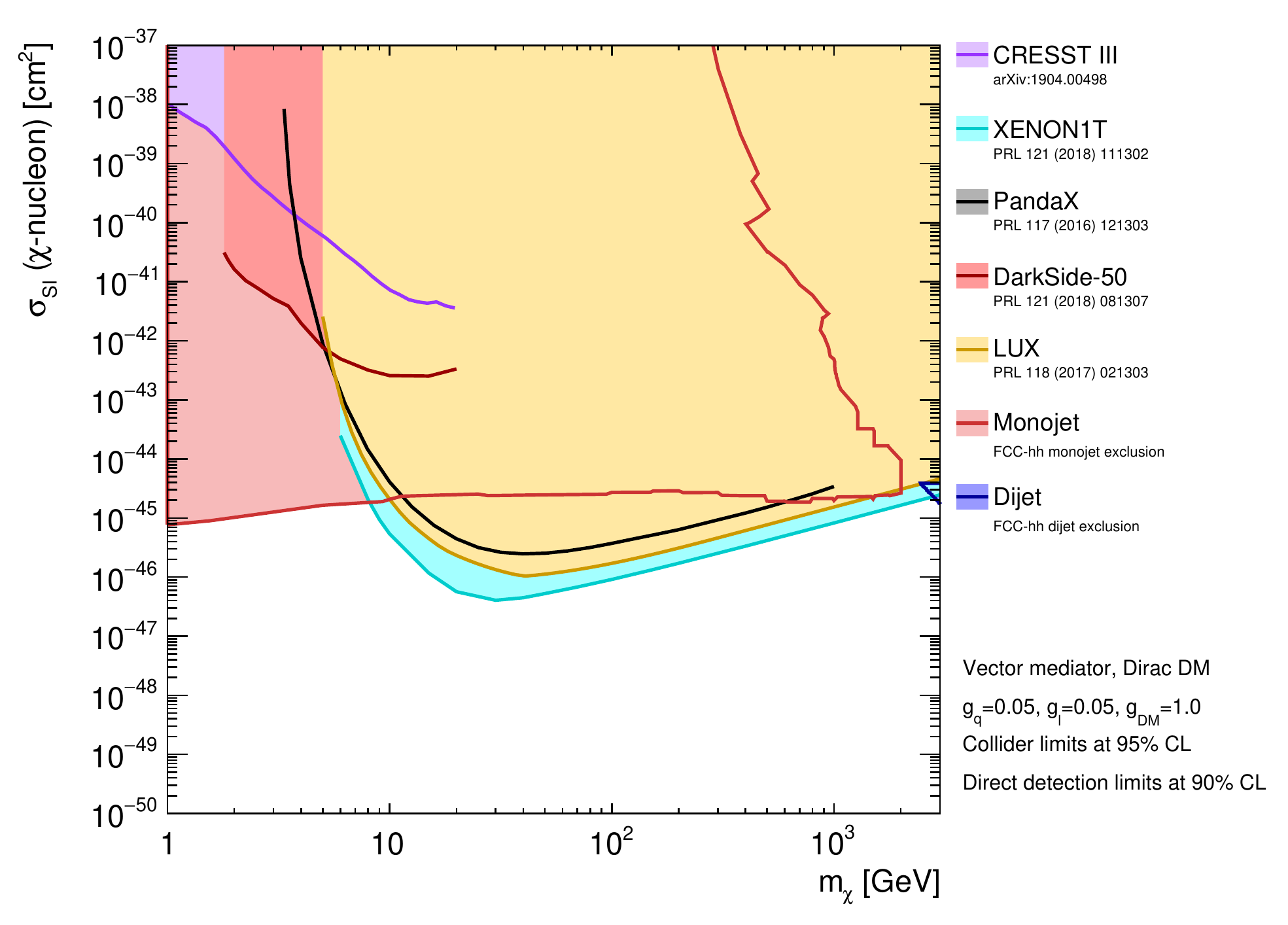}
         \caption{$g_q=0.05$, $g_{\chi}=1.0$, $g_l=0.05$}
         \label{subfig:dd-fcc-hh-vec-2}
     \end{subfigure}
        \caption{Comparison of projected exclusions from FCC-hh with constraints from current DD experiments on the spin-independent DM–nucleon scattering cross section in the context of the vector simplified model.}
        \label{fig:fcc-hh-dd-separate-si}
\end{figure}

The exclusion limits for HL-LHC and FCC-hh in the vector model shown in the previous section can be converted into limits on the DM-nucleon interaction cross section~\cite{BOVEIA2020100365}. 
The limits from HL-LHC are displayed for representative couplings in comparison to existing direct-detection experimental limits for the vector model in Figure~\ref{fig:hl-lhc-dd-separate-si}. 
Figure~\ref{fig:fcc-hh-dd-separate-si} displays the exclusion limits corresponding to the FCC-hh projections for the same sets of couplings. 

While in the present version of this whitepaper, future collider constraints are compared to existing direct-detection limits, we anticipate replacing thse curves with projected exclusions from upcoming direct-detection experiments, in collaboration with the Cosmic Frontier, in a future update of this document.
The direct-detection results currently shown for comparison in the vector mediator case correspond to XENON1T~\cite{XENON:2018voc}, DarkSide-50~\cite{DarkSide:2018bpj}, and a dedicated low-mass search at XENON1T using the Migdal effect (labelled XENON1T MIGD in plots)~\cite{XENON:2019zpr}. 

As these figures show, the increasing limit strength from HL-LHC to FCC-hh corresponds to lower minimum values of $\sigma_{\mathrm{SI}}$ and $\sigma_{\mathrm{SD}}$ which can be probed at these accelerators. 

Spin-dependent limits for axial-vector mediators can be found in Appendix~\ref{apppendix}, and overlaid to results from LUX~\cite{LUX:2017ree} and XENON1T~\cite{PhysRevLett.122.141301} for DM-neutron interactions and PICO-60~\cite{PhysRevD.100.022001} for DM-proton interactions.
     
\FloatBarrier 

\paragraph{Further considerations on comparing monojet and dijet collider constraints to direct-detection constraints: varying couplings}

An interesting study of $\sigma_{\mathrm{SI}}$ and $\sigma_{\mathrm{SD}}$ limits in the context of the DM WG vector model can be made by comparing the reinterpreted exclusions for the same analysis results as couplings of the model are changed, in parallel to what has been shown in Section \ref{sec:visibleInvisible} for the mass-mass plane. 

In the mass-mass plane, decreasing the coupling to quarks $g_q$ for a fixed analysis result causes the upper limit in $M_{\mathrm{med}}$ to worsen, i.e. to move to smaller mediator masses. 
On the other hand, constraints on $\sigma_{\mathrm{SI}}$ or $\sigma_{\mathrm{SD}}$ corresponding to a fixed $M_{\mathrm{med}}, M_{\mathrm{DM}}$ get stronger as $g_q$ decreases. 
Intuitively, this happens because collider sensitivity to a smaller coupling corresponds to a stronger limit in the direct-detection plane.  However, there is interplay between constraints on different parameters entering the translation between these two planes, and the overall results depend on the strength of exclusion at each point in the mass-mass plane and it is not easy to predict. 

\begin{figure}[htp]
     \centering
     \begin{subfigure}[b]{0.49\textwidth}
         \centering
         \includegraphics[width=\textwidth]{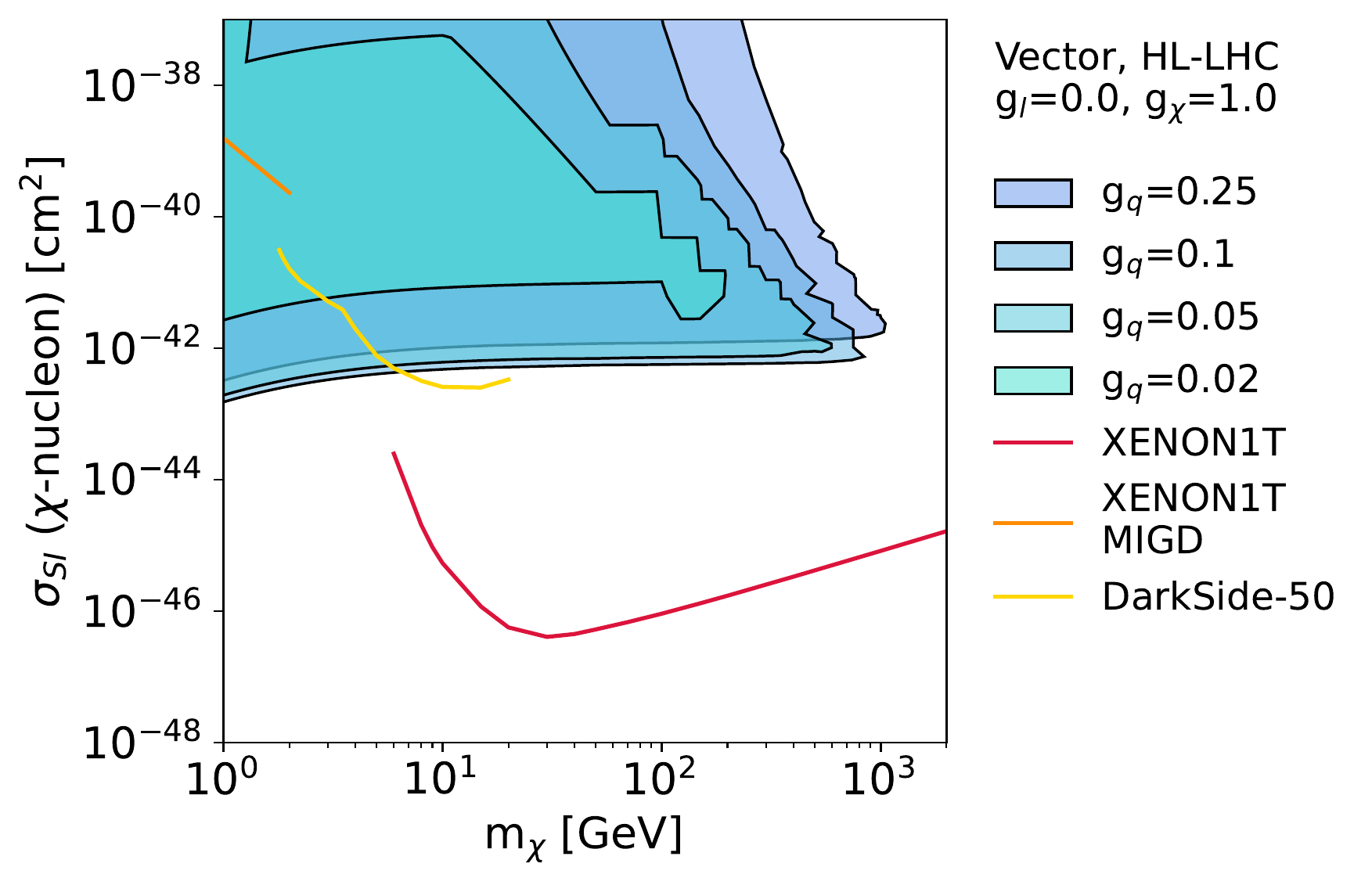}
         \caption{Monojet analysis}
         \label{subfig:couplingscan-dd-monojet}
     \end{subfigure}
     \hfill
     \begin{subfigure}[b]{0.49\textwidth}
         \centering
         \includegraphics[width=\textwidth]{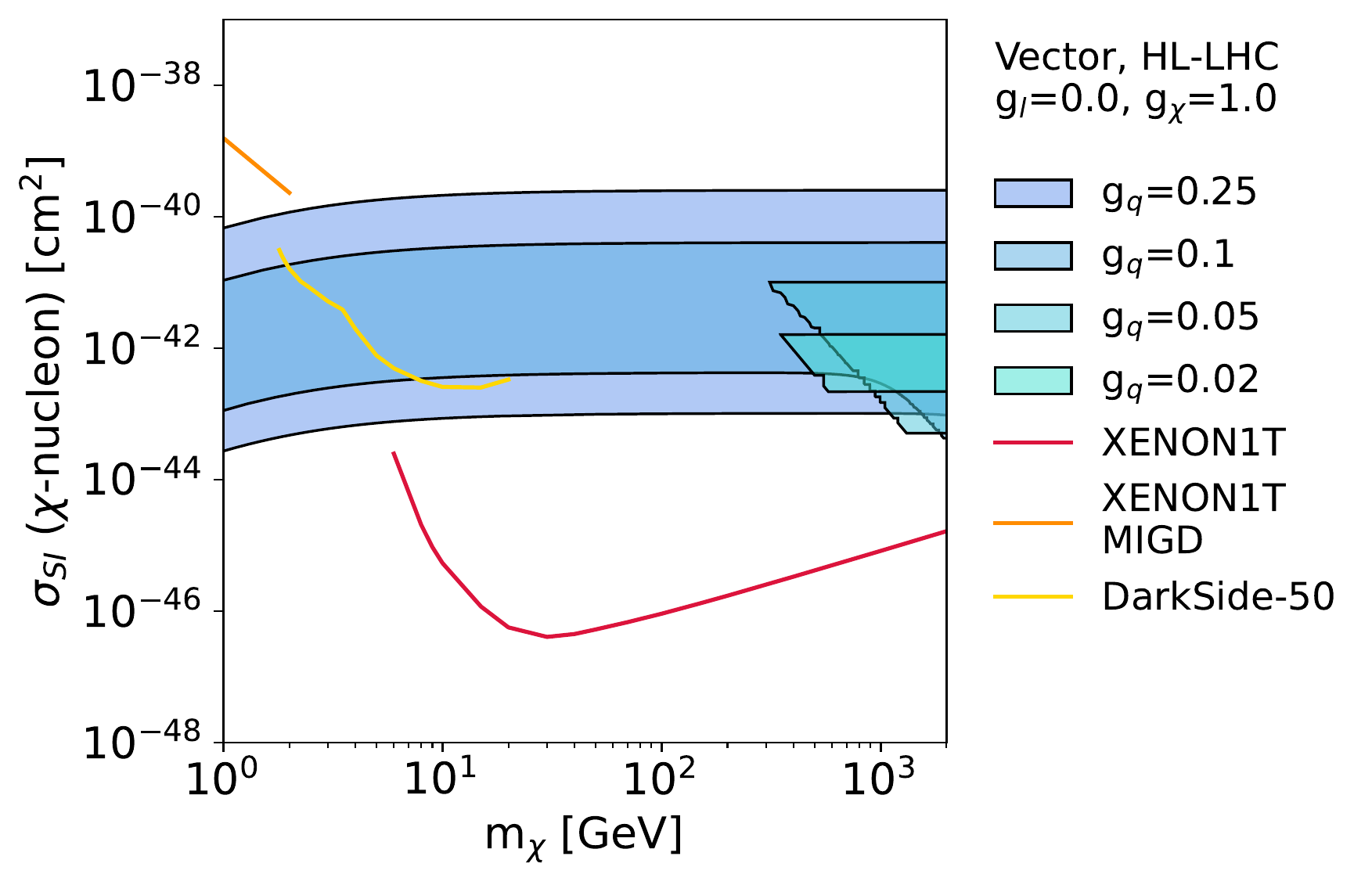}
         \caption{Dijet analysis}
         \label{subfig:couplingscan-dd-dijet}
     \end{subfigure}
    \caption{Effects on the HL-LHC exclusion limits in $\sigma_{\mathrm{SI}}$ for the monojet (\subref{subfig:couplingscan-dd-monojet}) and dijet (\subref{subfig:couplingscan-dd-dijet}) signatures  when varying the $g_q$ coupling. The dark matter coupling is held fixed to $g_{\mathrm{DM}}=1$; there is no coupling to leptons. Limits from existing direct-detection experiments are shown for context.}
    \label{fig:couplingscan-dd}
\end{figure}

In practice, the limit set by dijet and dilepton analyses is found to weaken noticeably as $g_q$ decreases, but while the high-$\sigma_{\mathrm{SI/SD}}$ monojet exclusion limit shrinks visibly, the lowest $\sigma_{\mathrm{SI/SD}}$ values reached by monojet are fairly stable as $g_q$ changes. 
In Figure~\ref{fig:couplingscan-dd}, this approximate independence of the monojet lower limit from $g_q$, and the greater coupling dependence of the dijet limit, are illustrated for HL-LHC projections and a vector mediator. 

\FloatBarrier

\begin{figure}[htp]
     \centering
     \begin{subfigure}[b]{0.49\textwidth}
         \centering
         \includegraphics[width=\textwidth]{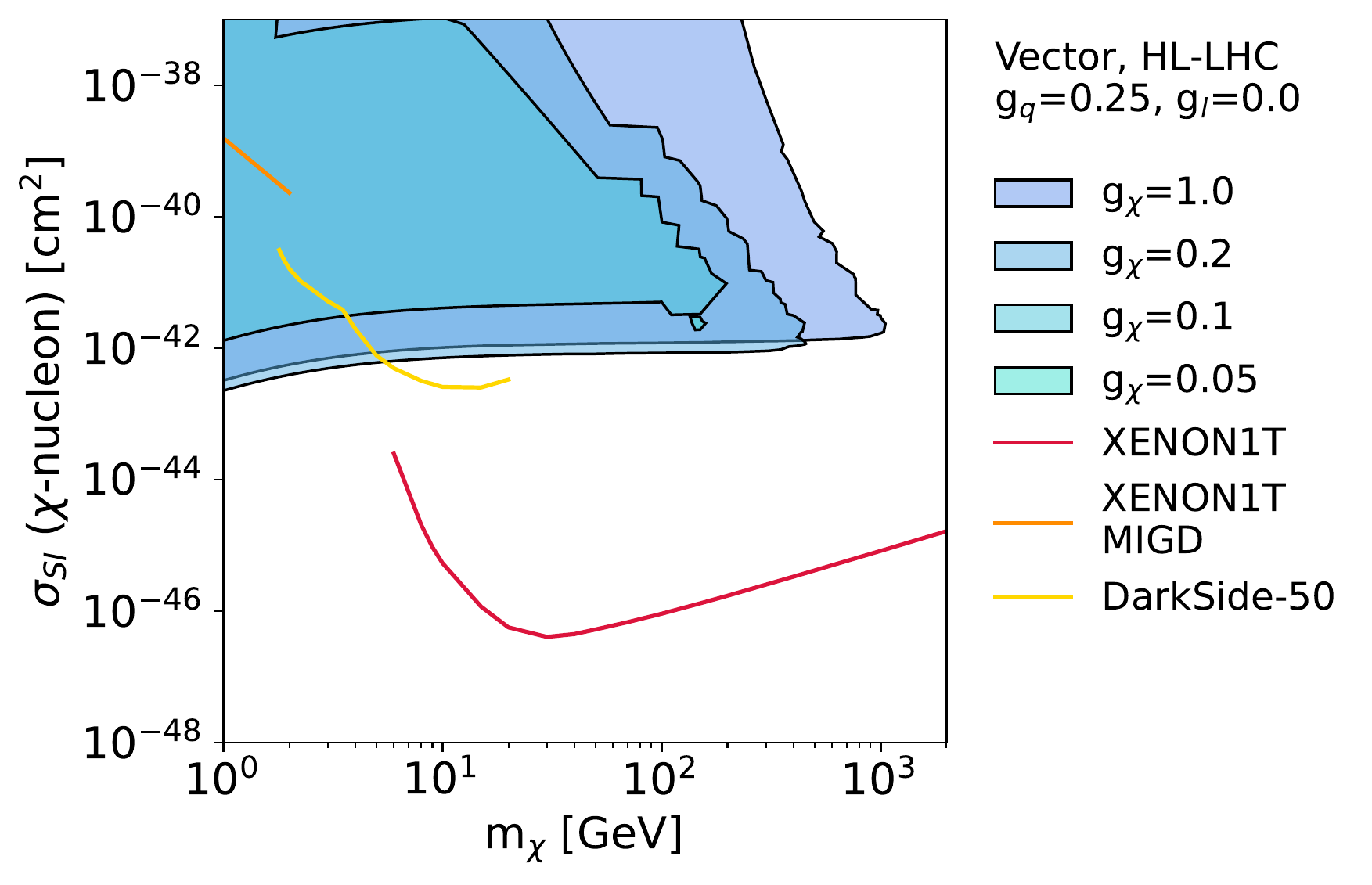}
         \caption{Monojet analysis}
         \label{subfig:couplingscan-dd-gdm-monojet}
     \end{subfigure}
     \hfill
     \begin{subfigure}[b]{0.49\textwidth}
         \centering
         \includegraphics[width=\textwidth]{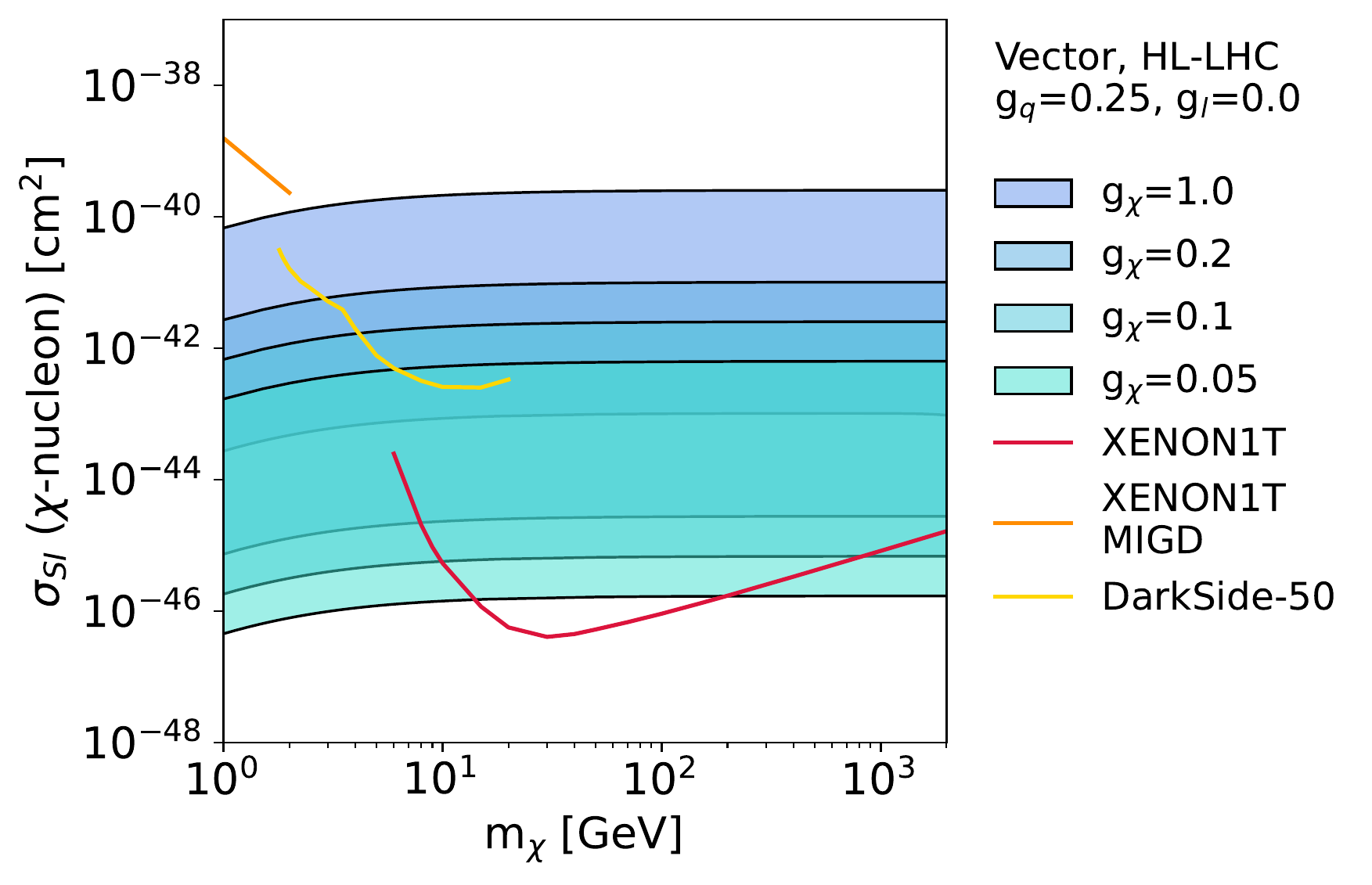}
         \caption{Dijet analysis}
         \label{subfig:couplingscan-dd-gdm-dijet}
     \end{subfigure}
    \caption{Effects on the HL-LHC exclusion limits in $\sigma_{\mathrm{SI}}$ for the monojet (\subref{subfig:couplingscan-dd-gdm-monojet}) and dijet (\subref{subfig:couplingscan-dd-gdm-dijet}) signatures  when varying the $g_\chi$ coupling. The coupling to quarks is held fixed to $g_q=0.25$; there is no coupling to leptons. Limits from existing direct-detection experiments are shown for context.}
    \label{fig:couplingscan-dd-gdm}
\end{figure}

Varying the dark matter-mediator coupling $g_\chi$ while holding $g_q$ fixed also has interesting consequences. 
As one would expect from the mass-mass contours, decreasing $g_\chi$ while $g_q$ and $g_l$ are held constant increases the power of the exclusions for visible final states and moves them towards smaller values of $\sigma_\mathrm{SI/SD}$. 
For the monojet contours, the effect of decreasing $g_\chi$ is essentially the same as the effect of decreasing $g_q$, since each vertex enters the production process once and each coupling has an equivalent role in the conversion from mass-mass exclusion to DM-nucleon interaction cross section. 
Figure~\ref{fig:couplingscan-dd-gdm} illustrates these effects. Varying $g_l$ has no visible effect on the monojet exclusion, as would be naively expected, and is not shown. 

Combining this result with the result in Figure~\ref{fig:couplingscan-dd}, it can be seen that the lower edge of the monojet contour is fairly robust to changes in any coupling in this simplified model. 
This consideration is valid when the couplings are varied and the DM and mediator masses are left floating with respect to each other -- we will see in the  following that this is not the case when the ratio between these two masses is fixed. 

Once again, equivalent plots corresponding to axial-vector results can be found in Appendix~\ref{apppendix}.

\FloatBarrier

\paragraph{Further considerations on comparing monojet and dijet collider constraints to direct-detection constraints: fixing the DM/mediator mass ratio}

The space mapped out by the five free parameters of the vector mediator simplified model can be sliced in numerous ways, each illustrating different aspects of experimental sensitivity, as shown by the juxtaposition of the exclusion limits in the $m_\mathrm{med}$-$m_\chi$ plane and the $m_\mathrm{med}$-coupling plane. 

A similar complementary set of exclusion limits in the $\sigma_{\mathrm{ID/DD}}$ plane can be created by keeping the ratio between mediator mass $m_\mathrm{med}$ and DM mass $m_\mathrm{chi}$ fixed, and allowing the variation in a coupling to map out the variation in $\sigma_{\mathrm{SI/SD}}$.

\begin{figure}[htp]
     \centering
     \begin{subfigure}[b]{0.49\textwidth}
         \centering
         \includegraphics[width=\textwidth]{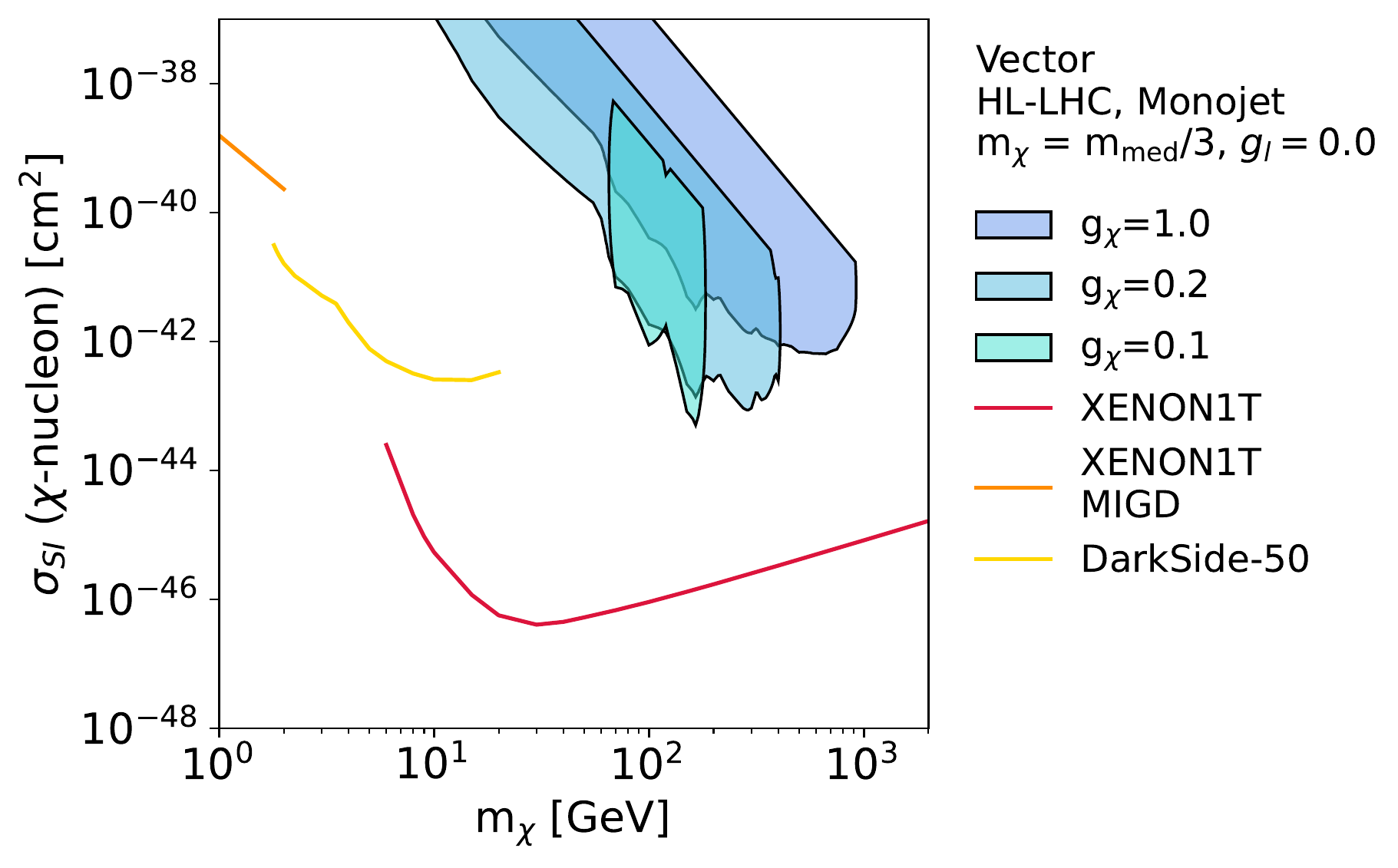}
         \caption{Monojet analysis}
         \label{subfig:couplingscan-dd-gdm-monojet-fixedmmed}
     \end{subfigure}
     \hfill
     \begin{subfigure}[b]{0.49\textwidth}
         \centering
         \includegraphics[width=\textwidth]{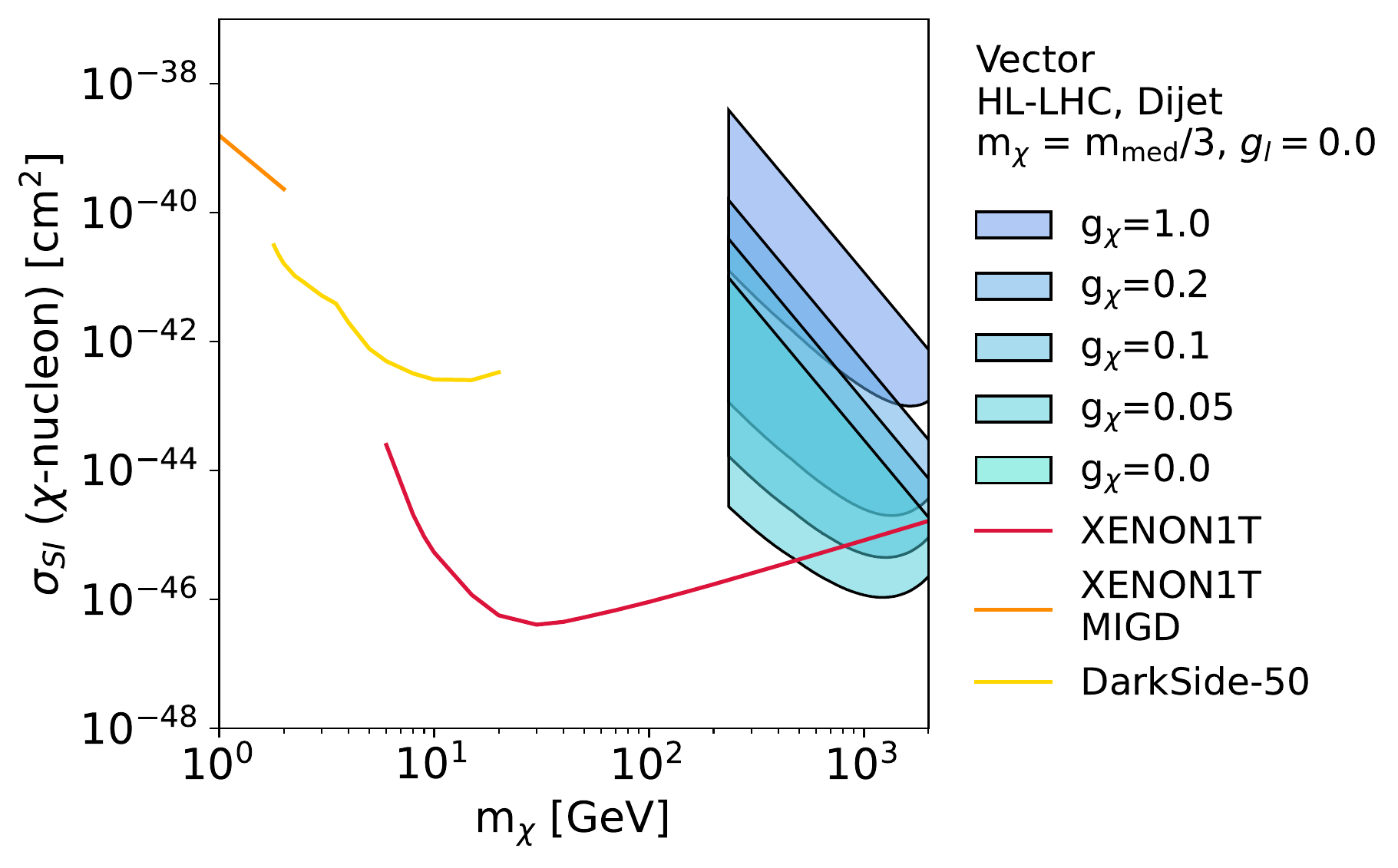}
         \caption{Dijet analysis}
         \label{subfig:couplingscan-dd-gdm-dijet-fixedmmed}
     \end{subfigure}
    \caption{Effects on the HL-LHC exclusion limits in $\sigma_{\mathrm{SI}}$ for the monojet (\subref{subfig:couplingscan-dd-gdm-monojet-fixedmmed}) and dijet (\subref{subfig:couplingscan-dd-gdm-dijet-fixedmmed}) signatures when varying the $g_\chi$ coupling. The mass of the mediator is fixed to $m_\mathrm{med} = 3 m_\chi$; there is no coupling to leptons. Limits from existing direct-detection experiments are shown for context.}
    \label{fig:couplingscan-dd-gdm-fixedmmed}
\end{figure}

Exclusion limits for a vector mediator in the $m_\chi$-$\sigma_{\mathrm{SI}}$ plane are shown with fixed $m_\mathrm{med} = 3 m_\chi$ in Figure~\ref{fig:couplingscan-dd-gdm-fixedmmed}.

In this plane, the constraints from colliders are significantly less powerful. 
This arises from the conversion between couplings, masses, and $\sigma_{\mathrm{SI/SD}}$, wherein $\sigma_{\mathrm{SI/SD}}$ is proportional to $g_q^2$, $g_\chi^2$, and $m_\mathrm{med}^{-4}$. 
The consequence of this dependence is that with fixed couplings, a collider search can improve the limit on $\sigma_{\mathrm{SI/SD}}$ by a factor of 16 by doubling the upper limit on the mediator mass, whereas with mediator mass fixed, a similar factor 2 improvement in the limit on $g_q$ or $g_\chi$ only improves the upper limit on mediator mass by a factor of 4.
This also means that colliders only need to access mediator masses a factor 10 higher at fixed couplings to have the same impact as accessing couplings a factor 100 smaller at fixed $m_\mathrm{med}$, and the former is much easier. 

All presented visualisations of the limits in this simplified model are equally valid, and they present a strong message on the complementarity between direct-detection and collider searches as they highlight regions of the model space where each method can provide unique power.

\clearpage

\subsubsection{Scalar mediator}
\label{sub:complementarity_DD_scalar}

\begin{figure}[htp]
     \centering
     \begin{subfigure}[b]{0.49\textwidth}
         \centering
         \includegraphics[width=\textwidth]{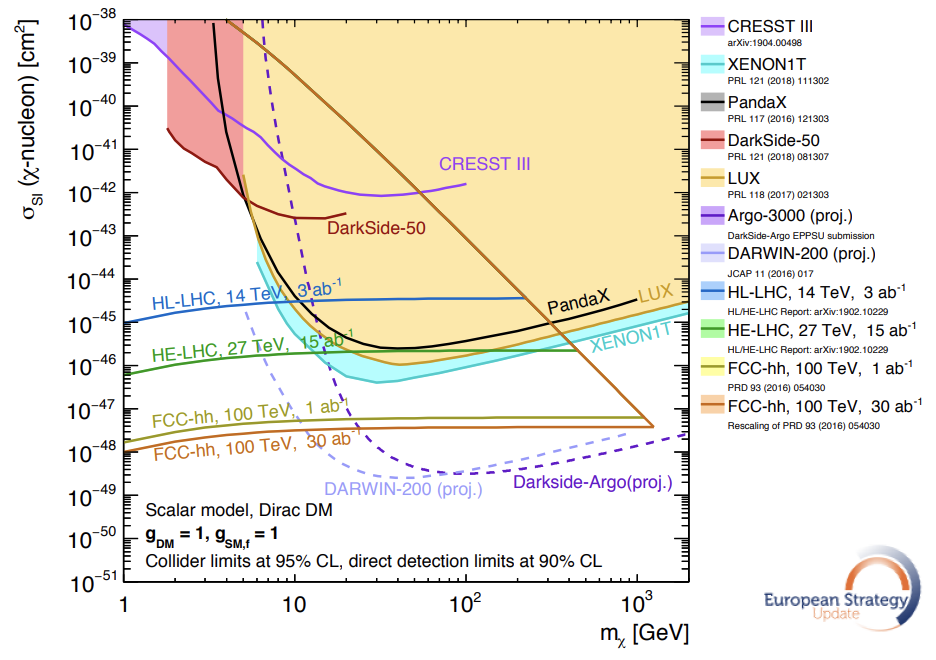}
         \caption{Scalar mediator model}
         \label{subfig:ScalarMediator}
     \end{subfigure}
     \hfill
     \begin{subfigure}[b]{0.49\textwidth}
         \centering
         \includegraphics[width=\textwidth]{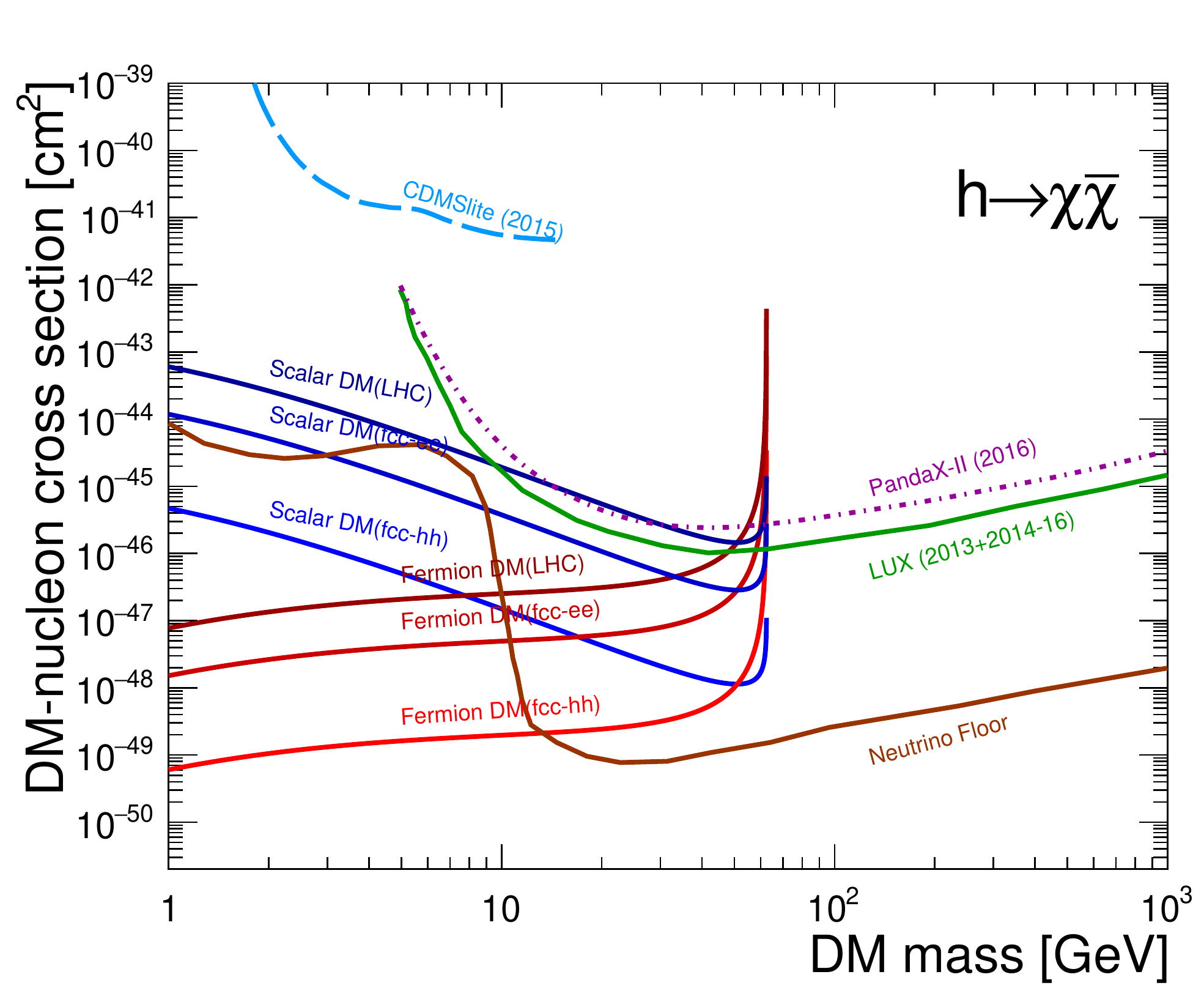}
         \caption{Higgs portal model}
         \label{subfig:HiggsPortal}
     \end{subfigure}
    \caption{(Left) Comparison of Direct-Detection bounds for projections of the scalar simplified model, from \cite{Ellis:2691414}. (Right) direct-detection bounds for the projections of sensitivity for Higgs to Invisible decays from \cite{L.Borgonovi:2642471}.}
    \label{fig:DDColliderSpin0}
\end{figure}

Spin-0 simplified models used at the LHC include a Yukawa coupling between the spin-0 mediator and the standard model fermions \cite{Abercrombie:2015wmb}. 
Unlike the case with the spin-1 mediators, for a scalar mediator that does not mix with the Higgs boson, the visible final states are typically found to be less sensitive than the invisible final states, due to the small cross section, and due to the high backgrounds in the hadronic final states. 
This Yukawa coupling also leads to a suppression in the direct-detection cross sections, since the couplings to light particles are suppressed, and the coupling to gluons is only present at loop level. 
As a result the sensitivity of the LHC with respect to direct detection tends to be stronger than when comparing spin-1 mediator models. 
This is shown in Fig.\ref{subfig:ScalarMediator}. 
For v2 of this paper, this plot may be updated including constraints from HL-LHC monojet searches (the current constraints are from heavy flavor + MET searches). We do not yet have a procedure to obtain plots that display the sensitivity of collider results to models with scalar mediator couplings much smaller than unity. 

Other spin-0 models that are frequently considered are the Higgs portal model (where the mediator is the Higgs boson itself) and the Dark Higgs boson model also discussed later in~\ref{sec:darkhiggs} (see e.g. Refs \cite{Patt:2006fw, Djouadi:2011aa}).
The Dark Higgs boson model includes a dark Higgs boson which mixes with the SM Higgs and allows for decays to dark matter. If the dark Higgs is very massive (decoupled), this model reduces to a simple Higgs portal. 
Both models can be constrained using Higgs to invisible searches and Higgs coupling measurements. 

Figure~\ref{subfig:HiggsPortal} shows the comparison between direct-detection results and collider results from the Higgs portal model. 
For v2 of this paper, this plot will be updated with the most recent constraints. 
For both generic scalar portal and Higgs portal, collider have a similar order-of-magnitude sensitivity with respect to direct-detection experiments.
In case of discovery this is a desirable feature, as both kind of experiments are complementary. 
Collider constraints on these models can also probe cross-sections below the expected rate of solar neutrino neutral current interactions. 


\FloatBarrier

\subsection{Collider and indirect-detection results}
\label{sec:complementarity_ID}

To highlight the complementarity between collider  and indirect-detection searches for DM, we use the simplified model with a pseudoscalar mediator and Dirac DM within the ATLAS/CMS Dark Matter Forum whitepaper (see ~\cite{Harris:2014hga,Buckley:2014fba,Haisch:2015ioa,Abercrombie:2015wmb} and references therein).

We do not yet have a procedure to obtain plots that display the sensitivity of collider results to models with pseudoscalar mediator couplings much smaller than unity. However, we update the European Strategy plots with the discussion of a feature in the translation of collider plots in the ID plane (cross-section times average velocity, $\langle \sigma v_{rel} \rangle$) that leads to a reduction in the exclusion area in the case of collider results. 

The reason for this reduction is an ambiguity between how single points in the $\langle \sigma v_{rel} \rangle$ plane map to multiple points in the collider DM mass - quark coupling plane, according to the translation formulas in~\cite{BOVEIA2020100365}. If any of the mass-coupling plane points is not within reach of collider searches, we conservatively suggest to use a different display convention for this point, leading to an area that is only partially excluded by collider searches, especially in the case of projections of exclusion limits.

\begin{figure}[!htb]
         \centering
         \includegraphics[width=\textwidth]{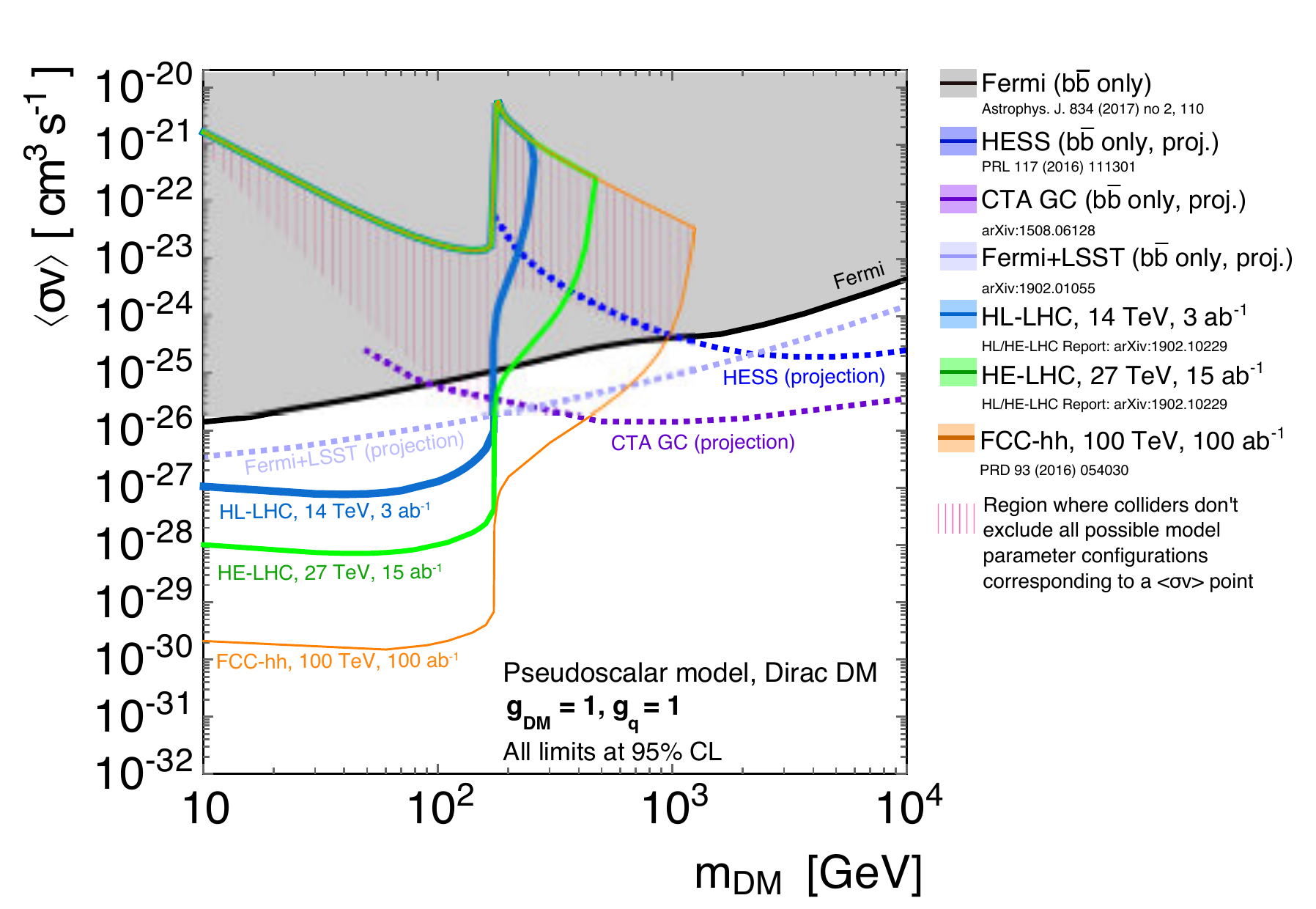}
         \caption{Comparison of a selection of projected exclusions from future colliders with constraints from current and future indirect-detection experiments in the context of a simplified model where a pseudoscalar particle with unit couplings mediates the interaction between SM fermions and Dirac fermionic DM. The shaded area represents the region where colliders don't exclude all possible points mapping to the value in the plane. All limits are shown at 95\% CL.
         Projections, curves and caption from \cite{EuropeanStrategyBook} (2020)}
         \label{fig:IDCollider}
\end{figure}

A plot displaying both collider and indirect-detection results and showing the areas in the $\langle \sigma v_{rel} \rangle$ plane that are not completely excluded by collider searches is in Fig. \ref{fig:IDCollider}.

\subsection{Reframing collider results in the context of light dark sector portals}
\label{sec:lightDarkSectorPortals}

Recently, a new set of simplified models have emerged as search targets of light dark matter-focused experiments \cite{Beacham:2019nyx}. 
These simplified models have a number of characteristics in common with the simplified models used by the LHC Dark Matter working group, and presented in LHC dark matter searches \cite{Abercrombie:2015wmb}. 
In this section, we will compare some selected benchmarks from the two sets of models, show how they can be mapped onto each other, and highlight the differences coming mainly from including or neglecting interference between the mediator and the Z or Higgs boson. 

\subsubsection{Vector portal (dark photon)}

One of the most common benchmarks for experiments at accelerators is the \textit{vector portal} model, where the vector mediator is also called \textit{dark photon} \cite{Holdom:1985ag} or \textit{dark boson} ($Z'$).  
This model has much in common with the vector dark matter mediator model (called \texttt{DMSimp} or \texttt{LHC DM} in the following), but unlike the model in \cite{Abercrombie:2015wmb}, its couplings to SM particles are fixed, owing to the mediator's kinetic mixing with the SM photon.

The dark photon simplified model considered here includes a mixing term between the dark mediator $Z'$ and the Z boson \cite{Curtin_2014,Curtin_2015}. 
The term added to the Lagrangian contains a mixing angle $\theta_a$ between the two:

\begin{equation}
\mathcal{L} \subseteq g_{DM} \cos{\theta_a}Z'_{\bar{\chi}\chi} + g_{DM} \sin{\theta_a}Z_{\bar{\chi}\chi}
\end{equation}

The implementation of this dark photon model comes from Refs.\cite{Curtin_2014,Curtin_2015}, where the model is called \textit{HAHM}, or Hidden Abelian Higgs Model. We turn off the hidden Higgs boson coupling to isolate the effect of the Z mixing and to effectively reduce it to the "minimal" dark photon model in \cite{Beacham:2019nyx}. 

We start from the results of the CMS search for new phenomena in events with energetic jets and large missing transverse momentum \cite{CMS_dmsimp} using LHC Run-2 data at 13 TeV center-of-mass energy. 
In the CMS publication these are interpreted in terms of the \texttt{DMSimp} vector mediator model parameters; in this whitepaper we reinterpret this search in terms of the dark photon mixing parameter. 

This reinterpretation is done using both the \texttt{DMSimp} LHC DM and the HAHM dark photon model implementations, to highlight the differences due to the Z mixing that is only present in the latter. 

The CMS constraints only start at 100 GeV in mediator mass, so we extend these down to 10 GeV. 
We generate events using Madgraph+PYTHIA \cite{Alwall:2011uj,Sjostrand:2014zea}, then employ Delphes \cite{deFavereau:2013fsa} for fast detector simulation, and finally pass the events through the selection and statistical analysis implemented in MadAnalysis5 \cite{MA5_2013}, to compare with CMS data and obtain a 95\% CL limit.  

In order to display the dark photon parameter constraints, we follow the choice in \cite{Beacham:2019nyx}. 
We plot the constraints on the plane $y$ - dark photon mass, with 
$y = \epsilon^2 \alpha_D (m_{DM}/m_{med})^4$ where $\epsilon$ is the dark photon - SM photon mixing parameter, $\alpha_D$ is the coupling constant between the dark photon and DM set as  $\alpha=\frac{g_{DM}^2}{4\pi}$ with $g_{DM}=1.0$, and $m_{DM}$ and $m_{med}$ are the DM mass and the dark photon mass respectively. 
The ratio between mediator and DM mass is fixed to 1/3, allowing the mediator to decay to DM on-shell.  

In the case of the \texttt{DMSimp} constraints, obtained in terms of the mediator-quark coupling constant, they are mapped to constraints on the variable $y$ by using the following approximate relationship between the coupling $g_q$ and the dark photon mixing parameter $\epsilon$: 

\begin{equation}\epsilon=g_q\frac{2(\Delta_z-1)}{e \cos{\theta_w}} \quad\textrm{from}\quad g_q=\frac{e\sin{\theta_a}}{2\tan{\theta_w}} \quad\textrm{where} \quad\sin{\theta_a}\approx\epsilon\frac{\sin{\theta_w}}{\Delta_z-1} \label{eq:eps_gq}
\end{equation}

Here, $\theta_a$ is the mixing between the dark photon and the SM Z boson, $\theta_w$ is the weak mixing angle and $\Delta_z=\frac{M_{Z'}}{M_Z}$ is the ratio of the dark photon and of the Z mass. For this scaling, we neglect the interference term. 

The HAHM dark photon model is calculated at leading order (LO), while DMSimp results are calculated at LO and then scaled by approximately $30\%$ to better match the next to leading order (NLO) constraints in the CMS publication. 


We also conservatively project the recasted LHC constraints to the HL-LHC by scaling the cross section limit by the square root of the integrated luminosities ratio (3000/139), while we neglect the change in the center-of-mass energy from 13 to 14 TeV.

\begin{figure}[!hptb]
     \centering
     \includegraphics[width=0.9\textwidth]{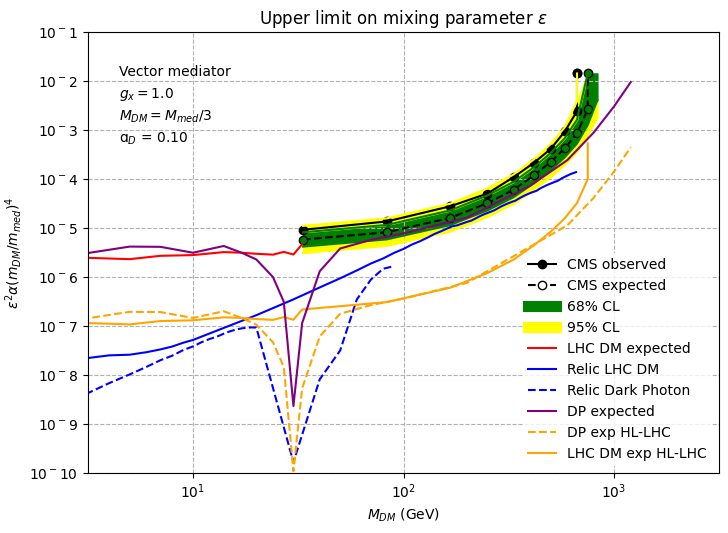}
     \caption{Comparison of two vector-mediated models, \textit{DMsimp} and \textit{HAHM}, corresponding to a simplified vector-mediated DM model at the LHC, and a dark photon-mediated model including Z interference, respectively. The mass ratio between mediator and DM mass is fixed to $\frac{1}{3}$, allowing the mediator to decay to DM. 
     Observed and expected limits at $95\%$ CL from the CMS analysis [CMS\_EXO\_20\_004] \cite{CMS_dmsimp} for the monojet final state, at $13 \textrm{ TeV}$ using $137\textrm{ fb}^{-1}$ of data, are plotted as a yellow and green band. The red and purple lines represent the constraints on the dark photon parameters from the \texttt{DMSimp} LHC DM and HAHM dark photon model implementation, respectively.
     The blue lines represent the minimum parameter combinations which reproduce the observed thermal relic density for each model, with the expected deviation for the dark photon model around the Z resonance. The orange lines forecast the increased sensitivity of this search for these two models at the HL-LHC, estimated by the effect on the cross section of scaling up the luminosity. \cite{josh_thesis}}
     \label{fig:dark_photon}
\end{figure}

The results of this reinterpretation are shown in Fig.\ref{fig:dark_photon}, where it is clear that the largest difference between the two models is the presence of the interference with the Z boson. 

The parameter combination corresponding to the relic density for both models are also shown. 
However, in both cases of underproduction or overproduction of DM, either model could still provide a valid DM candidate as part of a richer dark sector with other processes at play to satisfy the DM abundance seen today. 

It is also foreseen (from \cite{Curtin_2015}) that electron-positron colliders will also be able to place strong constraints on this model for dark photons around the Z boson mass using fits to electroweak precision observables. 

\subsubsection{Dark Higgs portal}
\label{sec:darkhiggs}

Dark Higgs portal models include an additional scalar dark matter mediator that has the same properties as the Higgs boson except for the mass. 
This scalar mediator is also often referred as the "Dark Higgs". 

The Lagrangian for the Dark Higgs model introduces a Dark Higgs($S$) to Higgs($H$) coupling that induces a mixing between the Dark Higgs and the Higgs, the additional terms in the Lagrangian are given by 
\begin{equation}
\mathcal{L} \subseteq -y_{DM} S\bar{\chi}\chi + (\mu S + \lambda S^2)H^{\dagger}H
\end{equation}

The mixing originates from the fact the $\mu$ interaction term yielding the observed $S$ and $H$ to be mixed states. The Higgs boson can therefore decay to dark sector particles, and the Dark Higgs boson can decay to standard model particles. 

The mixing of the Dark Higgs with the Higgs boson is typically characterized by an angle yielding the terms that we can write in terms of the eigenstates $h_{1},h_{2}$ of the mixed model between $S$ and $H$:

\begin{equation}
\mathcal{L} \subseteq y_{DM} \left(  h_{1} \sin \theta +  h_{2} \cos \theta \right)\bar{\chi}\chi +\\
\left(\cos \theta h_{1} - \sin \theta h_{2} \right) \left(2\frac{M_{W}^2}{v} W^{+}_{\mu}W^{-\mu} + \frac{M_{Z}^{2}}{v}Z^{+}_{\mu}Z^{-\mu} - \sum_{f} \frac{m_{f}}{v} \bar{f}f\right)
\end{equation}

where $h_{1}$ is our nominal Higgs and $h_{2}$ is aligned with the Dark Higgs. 

The mixing implies that the standard model couplings of the Higgs boson will deviate from unity by $\cos \theta$. Furthermore, it also introduces the possibility of Higgs decays to invisible particles, with a contribution to the Higgs boson given by
\begin{equation}
    \Gamma(h_{1}\rightarrow\chi\bar{\chi}) = \frac{y^{2}_{DM}\sin^{2}\theta m_{h_{1}}}{8\pi} \left(1-\frac{4m^{2}_{\chi}}{m^2_{h_{1}}}\right)^{3/2}
\end{equation}

Additionally, there is the possibility to produce the Dark Higgs $h_{2}$ through the same standard model production modes as the Higgs boson. 
However, its production is suppressed by a factor of $\sin^{2} \theta$. Furthermore, provided we assume a large dark matter coupling, the dark Higgs will decay immediately into invisible particles, leading to an additional increase in the overall Higgs to invisible cross section. 

Constraints on the dark Higgs model primarily come from constraints on the Higgs couplings, and the Higgs to invisible bounds. Higgs couplings are modified by the mixing angle $\cos \theta$. The Higgs to invisible bound is driven by the Higgs to invisible branching ratio $\Gamma(h_{1})$, which is proportional to $\sin^{2}\theta$

\begin{figure}[htp]
     \centering
         \centering
         \includegraphics[width=0.45\textwidth]{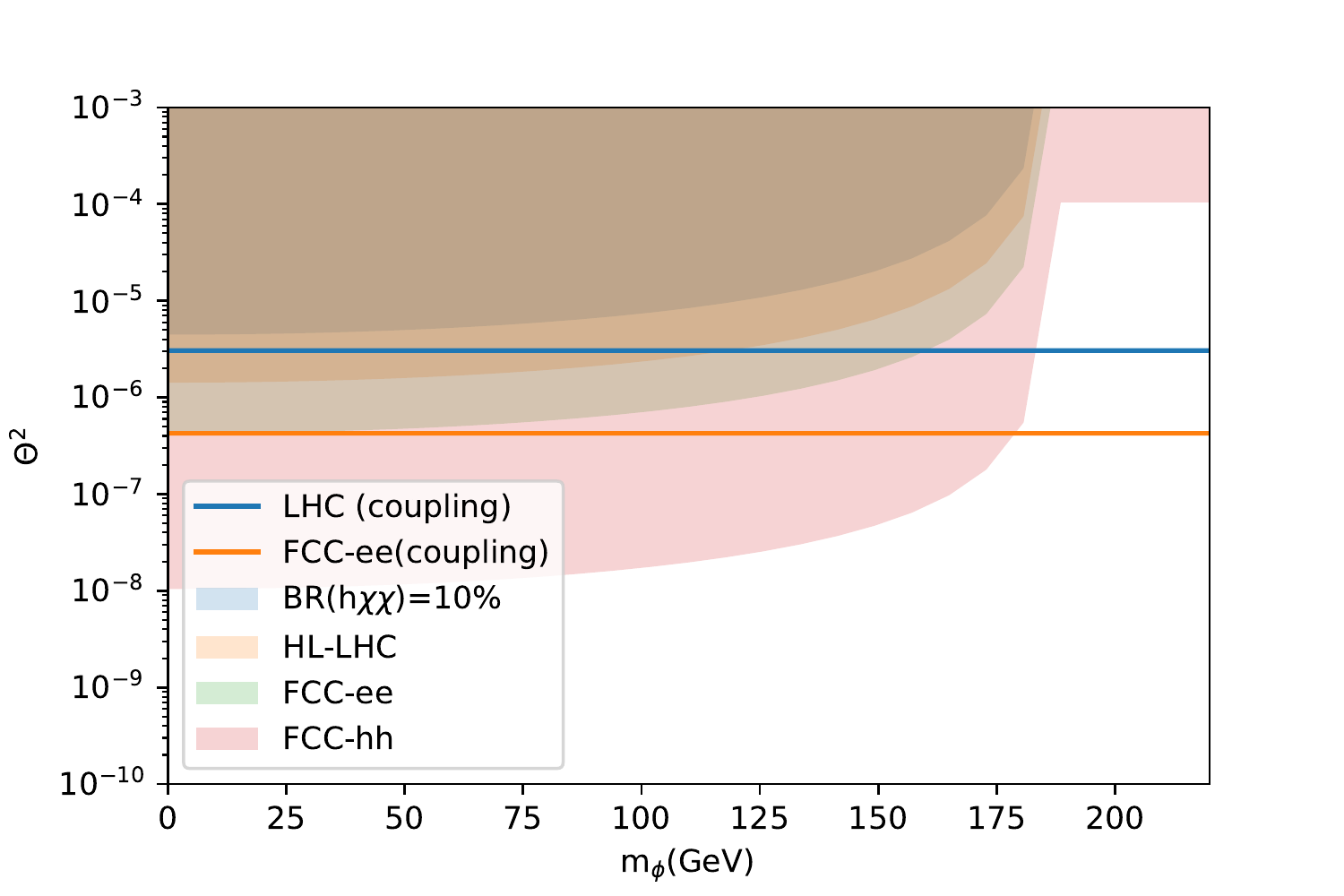}
         \includegraphics[width=0.45\textwidth]{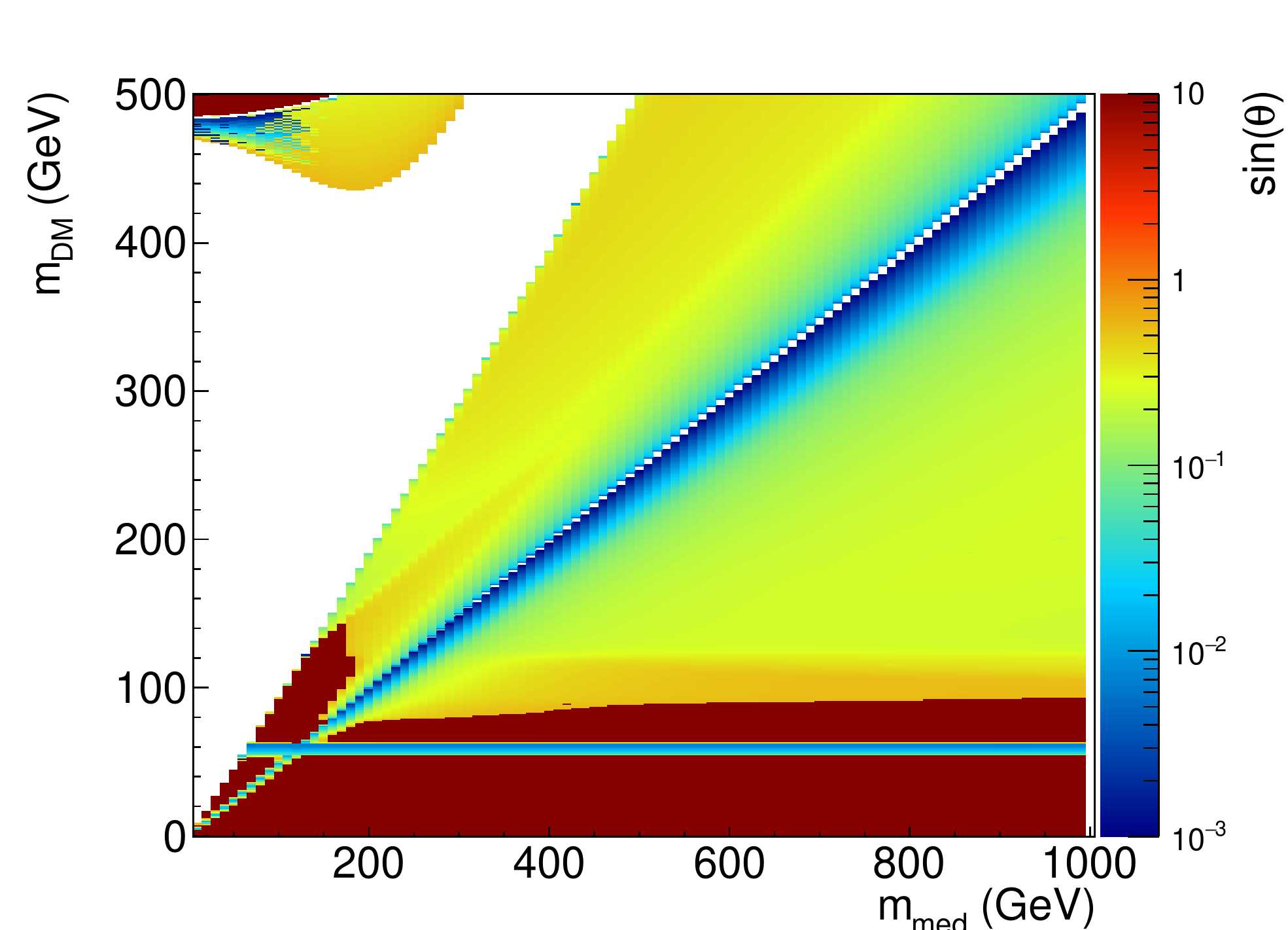}
         \caption{(Left) minimium mixing angle for the Higgs to invisible search when directly applying this search to the singlet mixing model. The solid lines indicate the constraints coming from indirecrt bounds on the Higgs couplings.  (Right) minimum allowed mixing angle for a model containing a Dark Higgs that mixes with the standard model Higgs boson.  }
         \label{fig:darkhiggs}
\end{figure}

Figure~\ref{fig:darkhiggs} presents the constraints on the the mixing angle $\theta^{2}$ from the Higgs to invisible decays as a function of the scalar mass, assuming a dark matter mass to mediator ratio of $m_{\chi}/m_{\phi}=\frac{1}{3}$.

Additionally, in figure~\ref{fig:darkhiggs}, we present the value of $sin(\theta)$ that would lead to the minimum allowed relic density for this model. For couplings below this value, a modified model is needed to get the right relic density.  
From the figure, we observe that small dark Higgs mass, and small dark matter mass there is no viable solutions to get the right relic density. However, for large dark matter mass, and singlet mass, there are possible values of $\sin \theta$ that would lead to a consistent relic density. Nevertheless, these solutions lead to large mixing angles, that will likely be excluded with HL-LHC running.

Lastly, the Dark Higgs portal can be compared to direct-detection results by noting that the Dark Higgs portal is a complete model of the standard Higgs to invisible portal presented in figure~\ref{fig:DDColliderSpin0}. As a consequence, we observe that collider bounds largely complement direct-detection dark matter bounds and lead to stronger results for light dark matter mass.

\subsubsection{Axion Portal}

The Axion portal very closely resembles the Pseudoscalar portal that is a standard interpretation for LHC dark matter experiments \cite{Abercrombie:2015wmb}.

This portal assumes coupling of the mediator to the Standard Model of either: 
\begin{equation}
\mathcal{L} \subseteq \frac{c_{\gamma}}{4\Lambda} a F_{\mu\nu}\tilde{F}^{\mu\mu} \rm{~or}
\mathcal{L} \subseteq \frac{4\pi\alpha_{s}c_{g}}{\Lambda} a G_{i,\mu\nu}\tilde{G}_{i}^{\mu\mu},
\end{equation}
for photon and gluon interactions respectively.

LHC Dark matter results are reported in terms of a single pseudoscalar mediator,$\phi$, that assumes Yukawa couplings to fermions. 
The Lagrangian is given by 
\begin{equation}
\mathcal{L} \subseteq -ig_{q}\frac{\phi}{\sqrt{2}}\sum_{\rm q\in quarks} y_{q} \bar{q}\gamma^{5}q
\end{equation}
As a consequence, the translation of the axion, $a$ interacton with the pseudoscalar coupling $g_{q}$ can be written as
\begin{equation}
\frac{c_{g}}{\Lambda} = g_{q}/v
\end{equation}

\begin{figure}[htp]
     \centering
         \centering
         \includegraphics[width=0.6\textwidth]{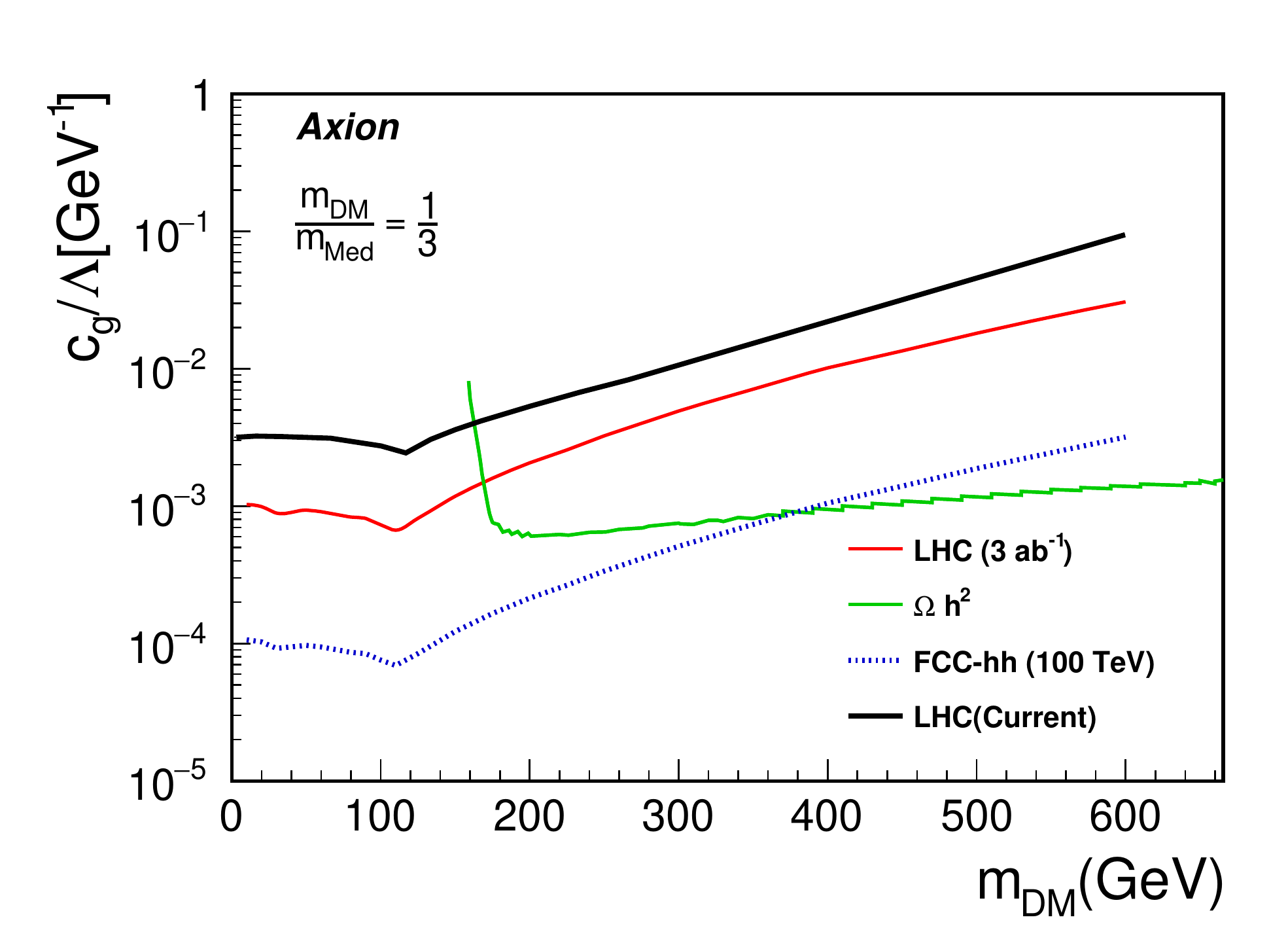}
          \caption{ Recast of Pseudoscalar simplified model bounds to the axion portal using the gluon effective coupling \cite{Curtin:2018mvb}.}
         \label{fig:axion}
\end{figure}

Figure~\ref{fig:axion} shows the LHC and FCC-hh bounds for the axion portal from a reinterpretation of the pseudoscalar mediator results. 
Typical bounds from the LHC missing energy searches are at a values of roughly $10^{-3}$. Furthermore, bounds from future collider searches push this further down by an order of magnitude\cite{Curtin:2018mvb}.  

\section{Conclusions}




Plots summarizing the constraints from different searches for dark matter can map the complementary and contrasting coverage of these searches within a given theoretical model, as well as help to illustrate the qualitative relationships between these probes.

We have presented an update to a similar, and less extensive array of plots prepared for the European Strategy Briefing Book~\cite{EuropeanStrategyBook}, now incorporating projections for future facilities that have been prepared for the Snowmass process, including the key mono-jet, dijet, and dilepton searches; illustrating the coupling dependence of the coverage of key searches; and discussing the expected performance of some general-purpose searches at a proposed muon collider. We also compare projected sensitives from the proposed collider experiments to direction and indirection detection experiments, and compare existing dark matter searches to light dark sector models in the context of CERN's Physics Beyond Colliders initiative.

The main conclusions from these studies are as follows \textbf{This list will be updated in v2 following the update of the summary plots.}.

\begin{itemize}
    \item the complementarity between visible and invisible searches continues at future hadron colliders, reaching higher mediator and dark matter masses at higher-energy colliders such as the FCC-hh. The exact reach is dependent primarily on the interaction strength between the mediator and the SM.   
    \item Searches that probe a specific coupling (e.g. dilepton resonances for lepton-mediator couplings) dominate the sensitivity with respect to other searches for models with large couplings. Models with smaller couplings are also constrained indirectly by other searches. 
    \item A muon collider operating at lower center-of-mass energy (up to 6 TeV) For minimal WIMP models, targeting new particles below 2 TeV, muon colliders operating at low center of mass energies are the preferred option, while probing particles above 2 TeV require higher energies and higher luminosities. 
    \item Collider searches for invisible particles (such as the mono-jet search) the ability to constrain smaller mediator-SM couplings leads to stronger limits in the DM-nucleon plane used to display direct detection searches. 
    When the coupling sensitivity limit approaches, the constraints gradually disappear. 
    The sensitivity of collider searches involving visible particles (such as the dijet search) decreases faster with decreasing coupling, since the same coupling is involved in both production and decay. 
    \item The sensitivity of collider searches depends strongly on the ratio of the DM to mediator mass. Searches at high energy colliders  are most sensitive in the region where the mediator can be produced directly from SM particle collision, and when the mediator is much heavier than the DM particle. Since in these benchmark models different mass hierarchy hypotheses are equally plausible, more mass ratios (as well as couplings) should be tested and displayed. 
    \item Light Dark Sector portal models are closely related to the simplified dark matter models utilized by LHC analysis. The main difference between the models is that the light dark sector models include interference effects with either the Higgs, for scalar mediators, or the Z boson for vector mediators. For invisible searches, a recast of the existing invisible searches can be performed, the Higgs invisible for a scalar mediator, and the monojet for vector mediator. The recast can be done utilizing existing analyses results. Once recasted, with the dark sector portal models, it is further possible to close a large region of the relic density benchmarks with future LHC running. 

\end{itemize}

\section{Acknowledgements} 

This work has received funding from the European Union's Horizon 2020 research and innovation programme under grant agreement No 824064 (ESCAPE, the European Science Cluster of Astronomy \& Particle Physics ESFRI Research Infrastructures) as well as under grant agreement No 101017536 (EOSC Future). 
Research by C. Doglioni is part of projects that have received funding from the European Research Council under the European Union’s Horizon 2020 research and innovation program (grant agreement 679305 and 101002463) and from the Swedish Research Council.
K.~Pachal's research is supported by TRIUMF, which receives federal funding via a contribution agreement with the National Research Council (NRC) of Canada.
Research for P.~Harris is supported by Department of Energy Early Career Award grant de-sc0021943.
Research for R. Harris is supported by the Fermi National Accelerator Laboratory, managed and operated by Fermi Research Alliance, LLC under Contract No. DE-AC02-07CH11359 with the U.S. Department of Energy.
Research by A.~Boveia is supported by the US Department of Energy (Grant DE-SC0011726).
Research by authors from the University of Wisconsin-Madison is supported by the research grants from the US Department of Energy (Award number DE-SC0017647) and the University of Wisconsin.
\clearpage







\bibliography{Refs}
\bibliographystyle{JHEP}

\appendix

\section{Appendix}
\label{apppendix}

The following figures contain the equivalent plots to all vector mediator plots shown in Sections~\ref{sec:futurecollider} and \ref{sec:complementarity} for axial-vector mediators instead. Results are comparable in most cases. The biggest differences occur for direct-detection comparison plots, where vector versus axial-vector mediators correspond to different interactions with nucleons.

\begin{figure}[!hp]
     \centering
     \begin{subfigure}[b]{0.49\textwidth}
         \centering
         \includegraphics[width=\textwidth]{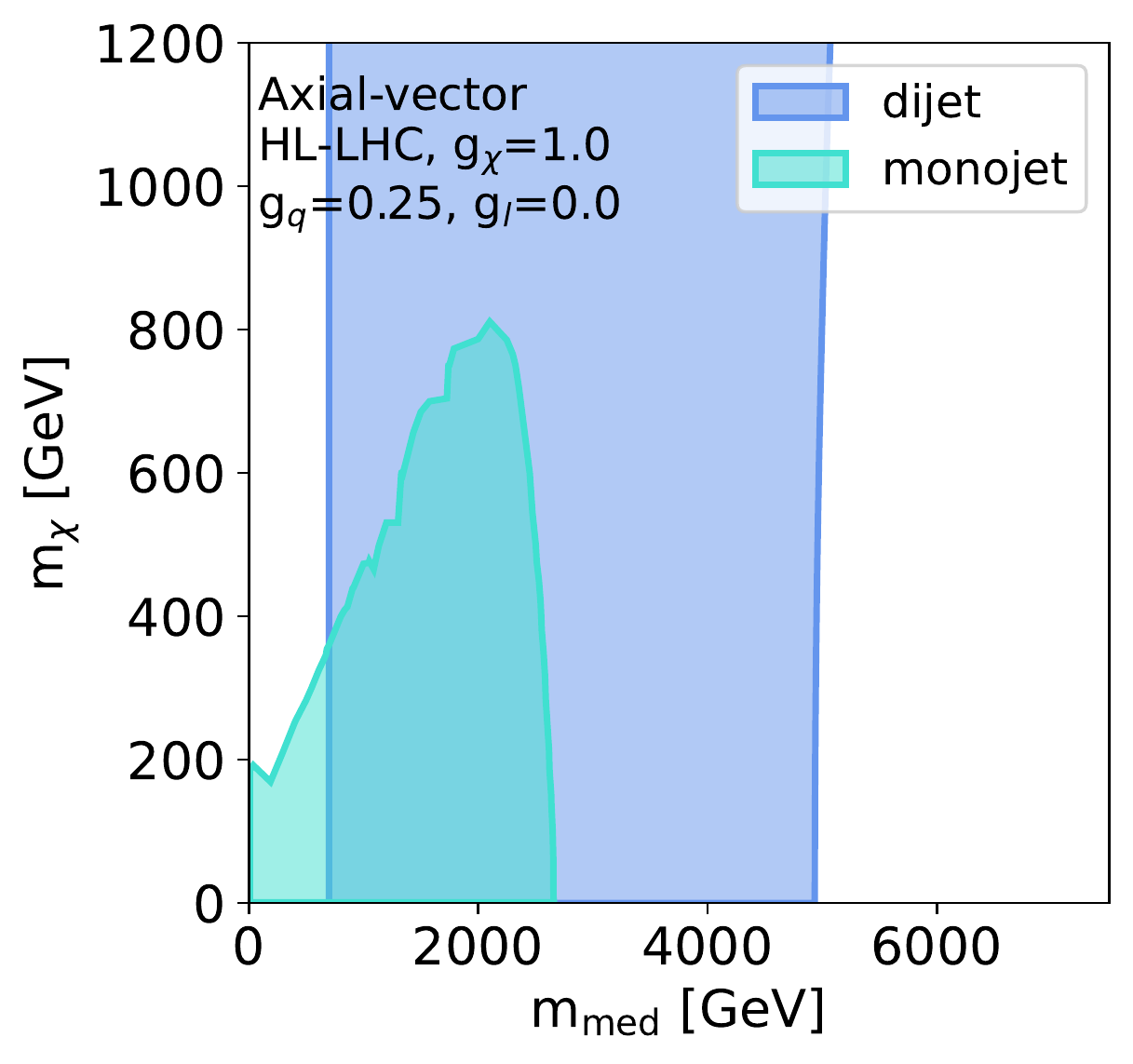}
         \caption{``A1'' scenario: couplings are $g_q=0.25$, $g_{\chi}=1.0$, $g_l=0.0$}
         \label{subfig:axial-hl-lhc-v1}
     \end{subfigure}
     \hfill
     \begin{subfigure}[b]{0.49\textwidth}
         \centering
         \includegraphics[width=\textwidth]{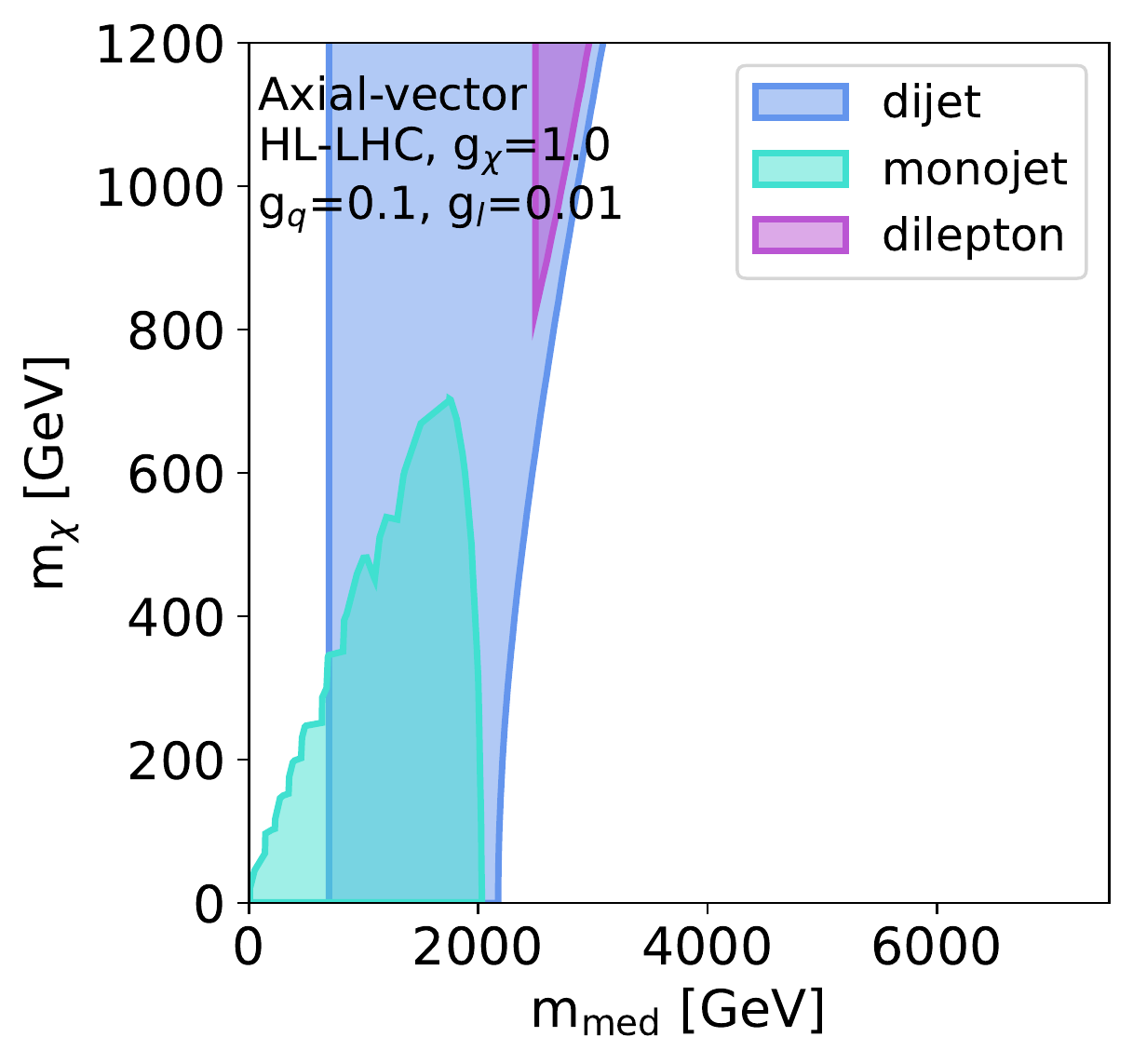}
         \caption{New scenario with varied SM couplings: $g_q=0.1$, $g_{\chi}=1.0$, $g_l=0.01$}
         \label{subfig:axial-lhc-v2}
     \end{subfigure}

     \begin{subfigure}[b]{0.49\textwidth}
         \centering
         \includegraphics[width=\textwidth]{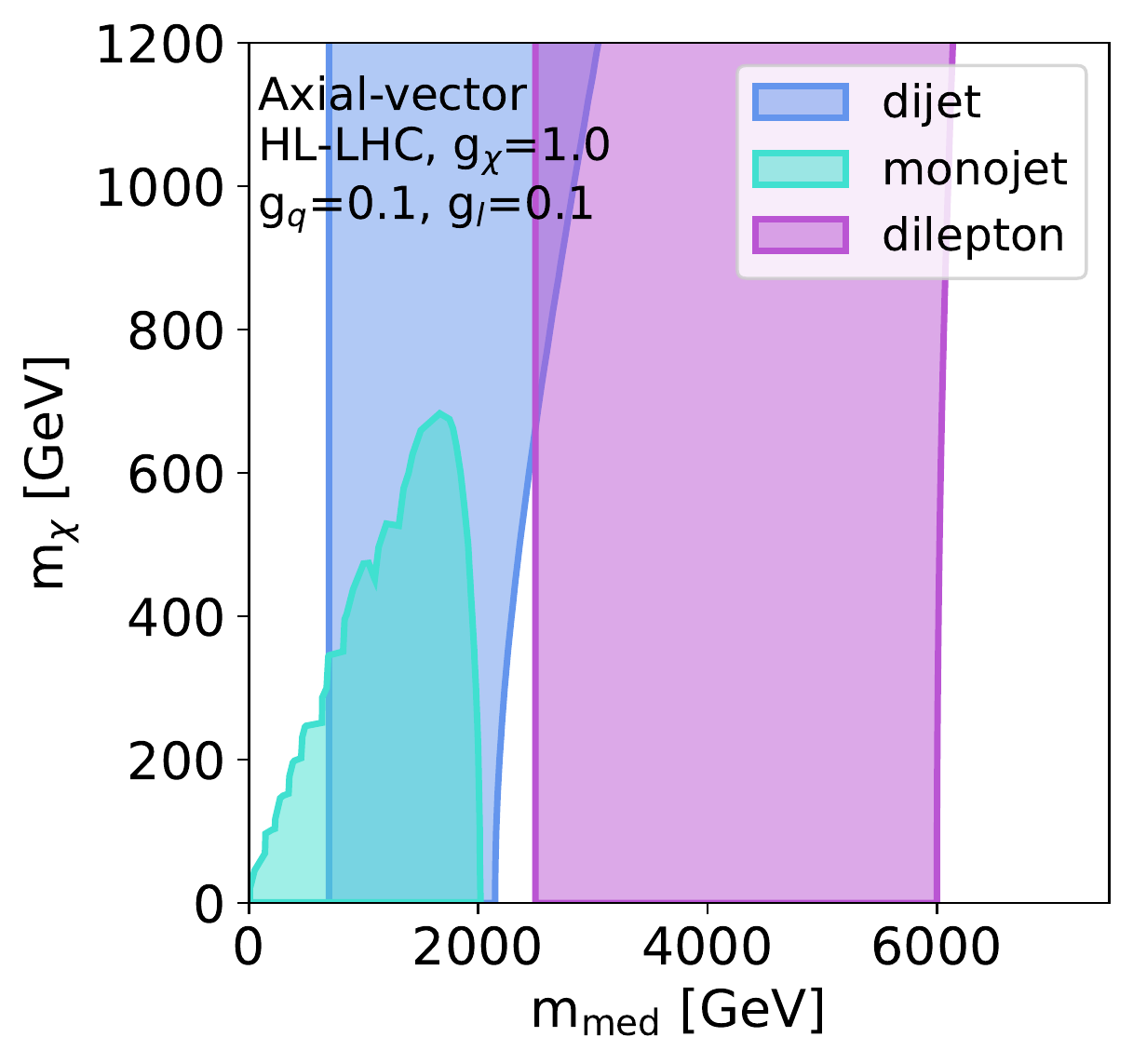}
         \caption{``A2'' scenario: couplings are $g_q=0.1$, $g_{\chi}=1.0$, $g_l=0.1$}
         \label{subfig:axial-hl-lhc-v3}
     \end{subfigure}
     \hfill
     \begin{subfigure}[b]{0.49\textwidth}
         \centering
         \includegraphics[width=\textwidth]{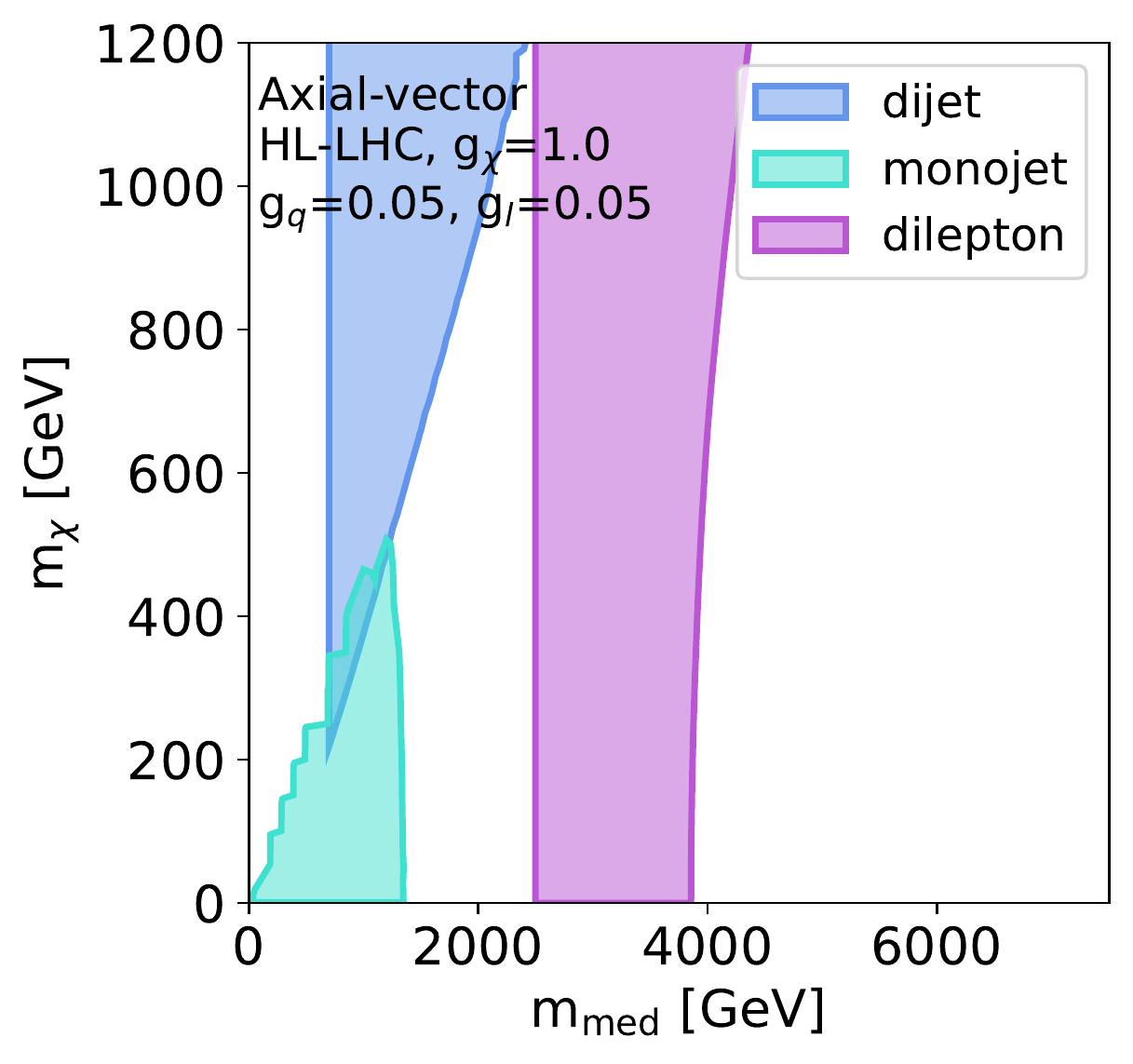}
         \caption{New scenario with varied SM couplings: $g_q=0.05$, $g_{\chi}=1.0$, $g_l=0.05$}
         \label{subfig:axial-hl-lhc-v4}       
     \end{subfigure}
        \caption{HL-LHC projected exclusions for individual analyses in the axial-vector model and with a range of couplings.}
        \label{fig:hl-lhc-massmass-separate-axial}
\end{figure}

\begin{figure}[!hp]
     \centering
     \begin{subfigure}[b]{0.49\textwidth}
         \centering
         \includegraphics[width=\textwidth]{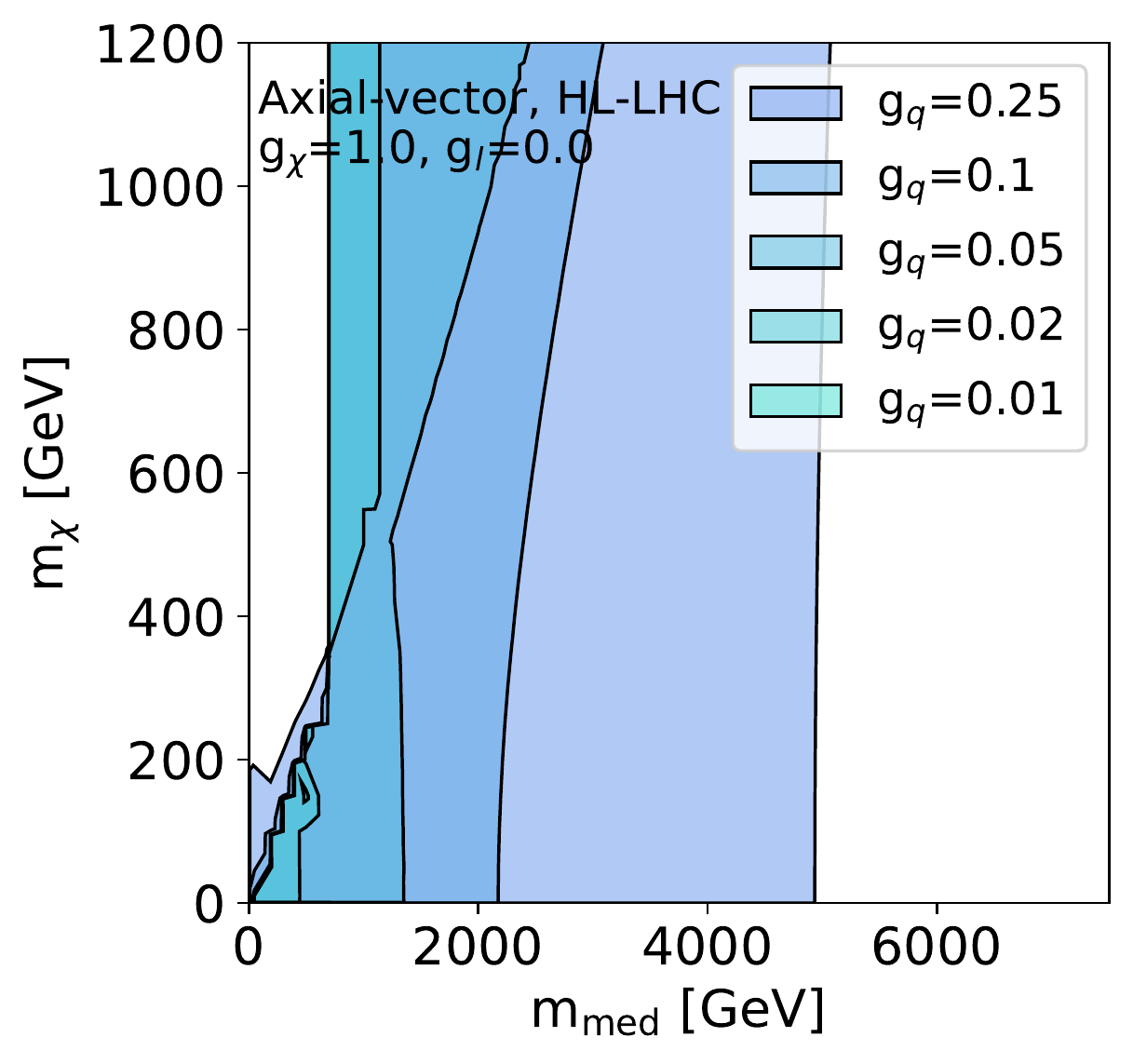}
         \caption{No coupling to leptons}
         \label{subfig:axial-hl-lhc-gqvariations1}
     \end{subfigure}
     \hfill
     \begin{subfigure}[b]{0.49\textwidth}
         \centering
         \includegraphics[width=\textwidth]{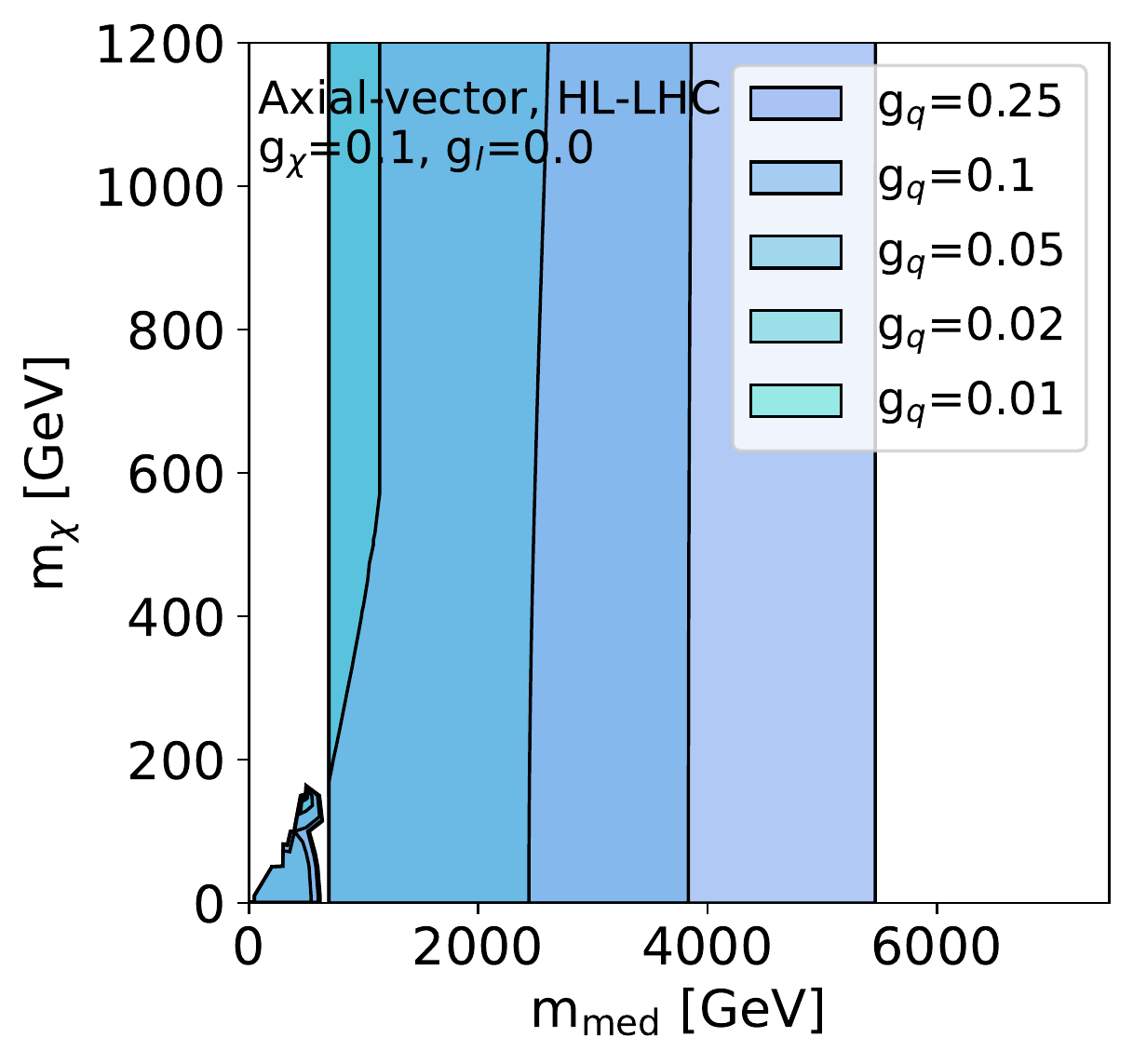}
         \caption{$g_l$ fixed to 0.05}
         \label{subfig:axial-hl-lhc-gqvariations2}
     \end{subfigure}
        \caption{Effects of varying $g_q$ coupling on HL-LHC projected exclusions in the axial-vector model, combined across analyses. No coupling to leptons is allowed; $g_\chi$ is set to 1.0 in (\subref{subfig:axial-hl-lhc-gqvariations1}) and 0.1 in (\subref{subfig:axial-hl-lhc-gqvariations2}). Decreasing $g_q$ decreases the strength of all limits since $g_q$ a coupling to protons is required to produce the mediator in all cases, but the effects are strongest on the dijet analysis where $g_q$ enters the cross section twice.}
        \label{fig:hl-lhc-massmass-combined-gqvary-ax}
\end{figure}

\begin{figure}[htp]
     \begin{subfigure}[b]{0.49\textwidth}
         \centering
         \includegraphics[width=\textwidth]{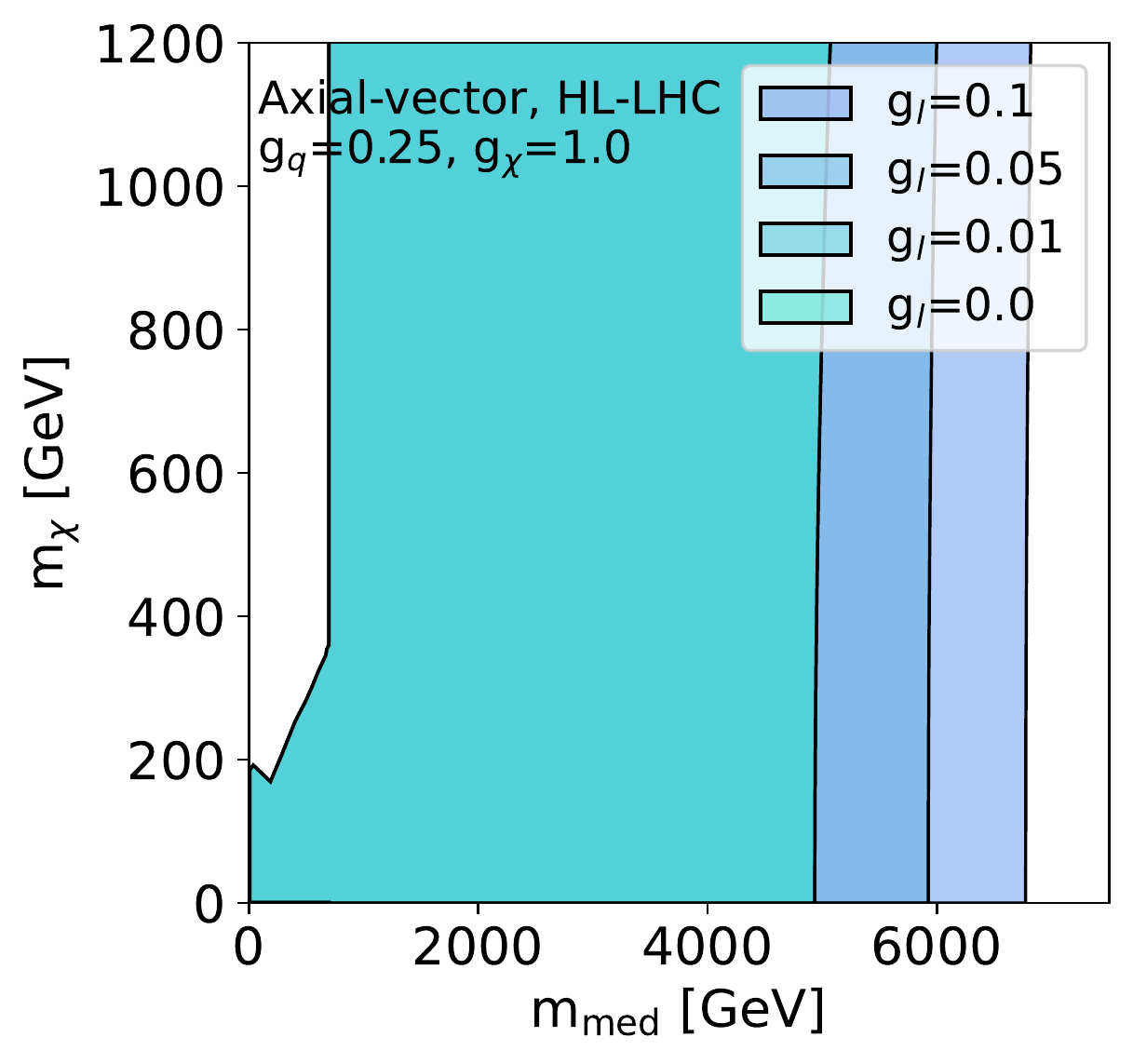}
         \caption{$g_q$ fixed to 0.15}
         \label{subfig:axial-hl-lhc-glvariations1}
     \end{subfigure}
     \hfill
     \begin{subfigure}[b]{0.49\textwidth}
         \centering
         \includegraphics[width=\textwidth]{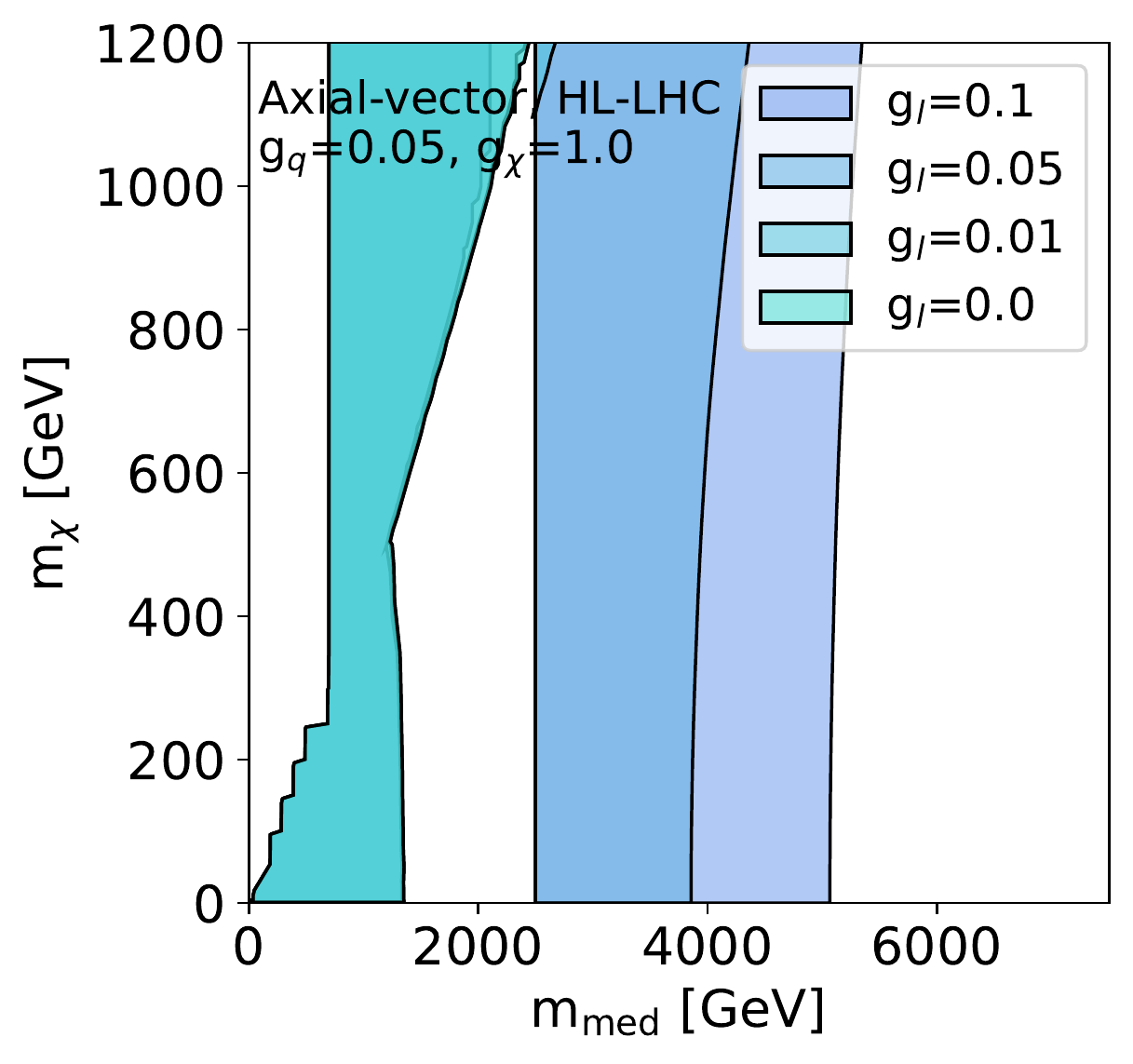}
         \caption{$g_q$ fixed to 0.05}
         \label{subfig:axial-hl-lhc-glvariations2}       
     \end{subfigure}
        \caption{Effects of varying $g_l$ coupling on HL-LHC projected exclusions in the axial-vector model, combined across analyses. Other couplings are fixed. Variations in $g_l$ within the studied range have little impact on the monojet and dijet analyses but a strong impact on the dilepton analysis.}
        \label{fig:hl-lhc-massmass-combined-glvary-ax}
\end{figure}

\begin{figure}[htp]
     \begin{subfigure}[b]{0.49\textwidth}
         \centering
         \includegraphics[width=\textwidth]{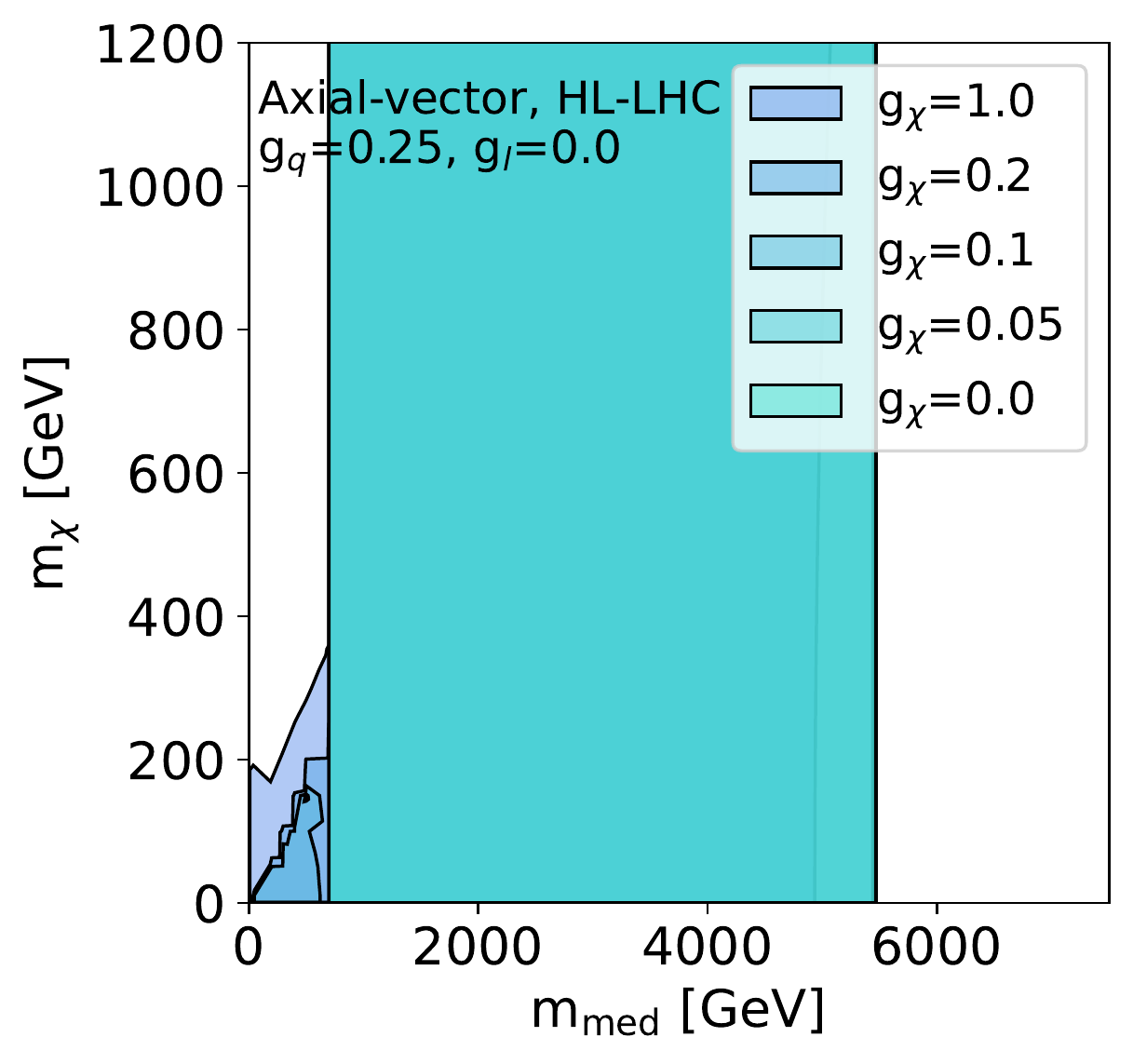}
         \caption{$g_q$ fixed to 0.25, $g_l$ fixed to 0.0}
         \label{subfig:axial-hl-lhc-gdmvariations1}
     \end{subfigure}
     \hfill
     \begin{subfigure}[b]{0.49\textwidth}
         \centering
         \includegraphics[width=\textwidth]{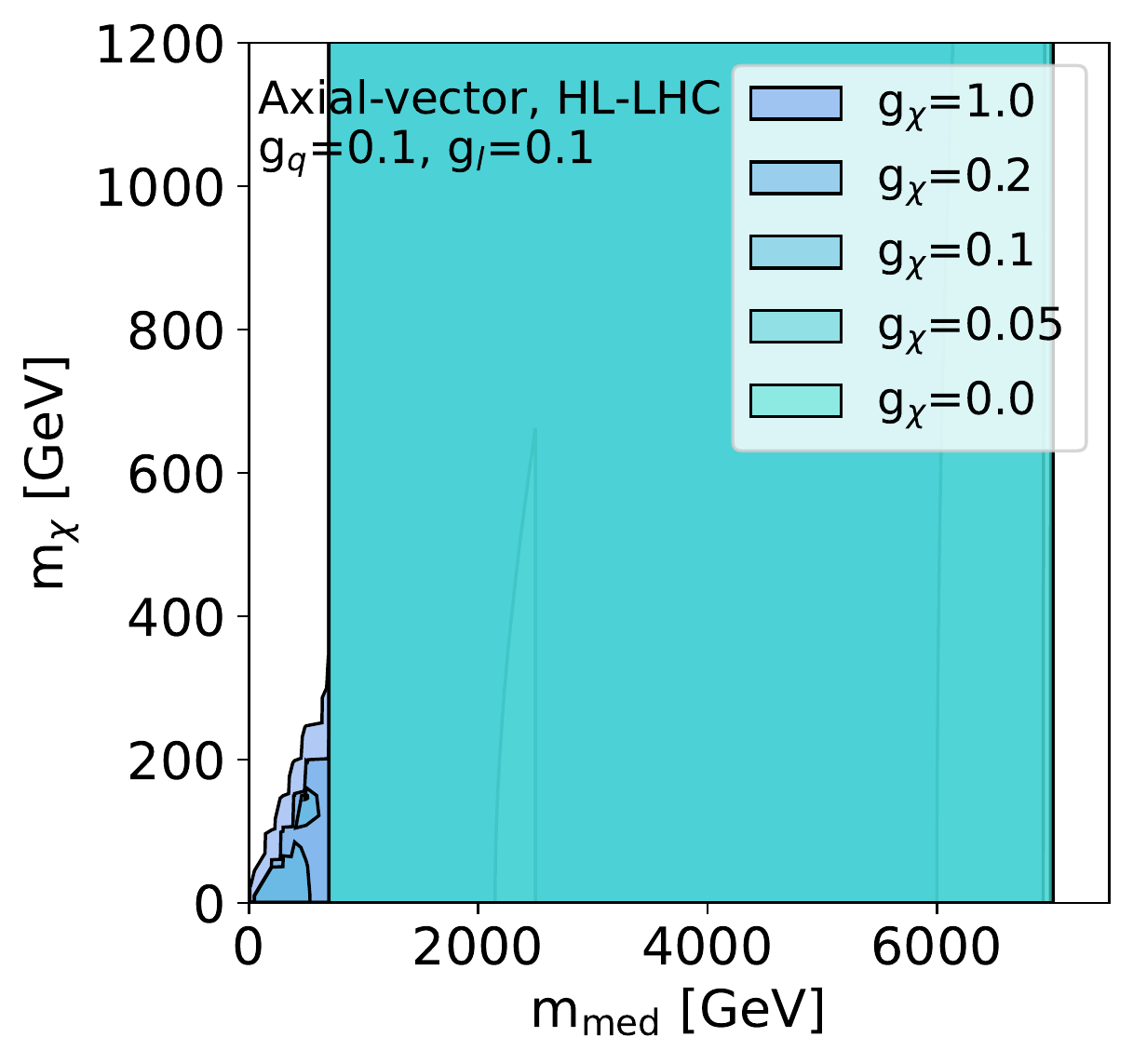}
         \caption{$g_q$ fixed to 0.1, $g_l$ fixed to 0.1}
         \label{subfig:axial-hl-lhc-gdmvariations2}       
     \end{subfigure}
        \caption{Effects of varying $g_\chi$ coupling on HL-LHC projected exclusions in the axial-vector model, combined across analyses. Other couplings are fixed. As $g_\chi$ decreases, the monojet limits weaken but visible final state limits improve because their relative branching ratios increase.}
        \label{fig:hl-lhc-massmass-combined-gdmvary-ax}
\end{figure}

\begin{figure}[htp]
     \centering
     \begin{subfigure}[b]{0.49\textwidth}
         \centering
         \includegraphics[width=\textwidth]{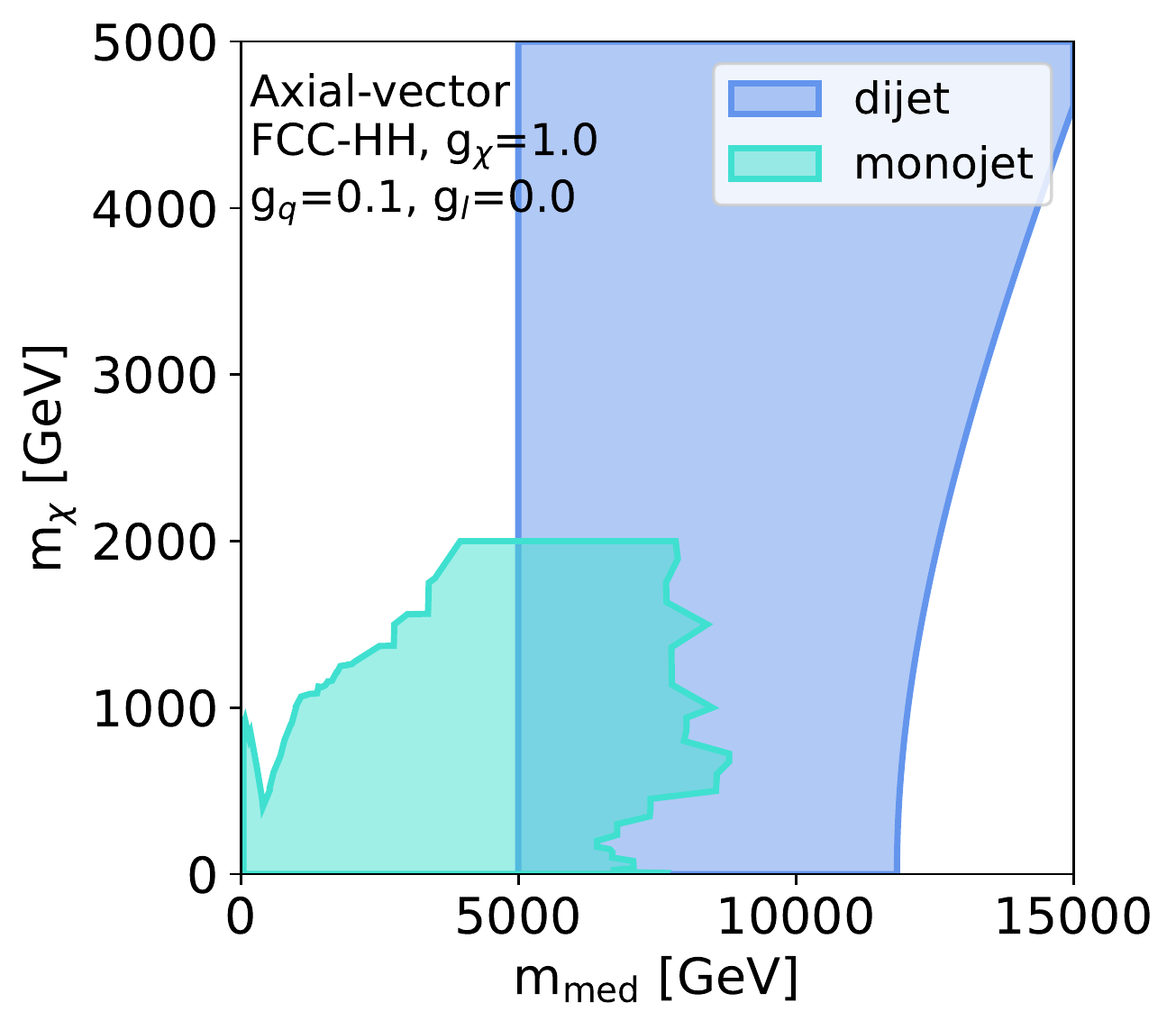}
         \caption{$g_q=0.1$, $g_{\chi}=1.0$, $g_l=0.0$}
         \label{subfig:axial-fcc-v1}
     \end{subfigure}
     \hfill
     \begin{subfigure}[b]{0.49\textwidth}
         \centering
         \includegraphics[width=\textwidth]{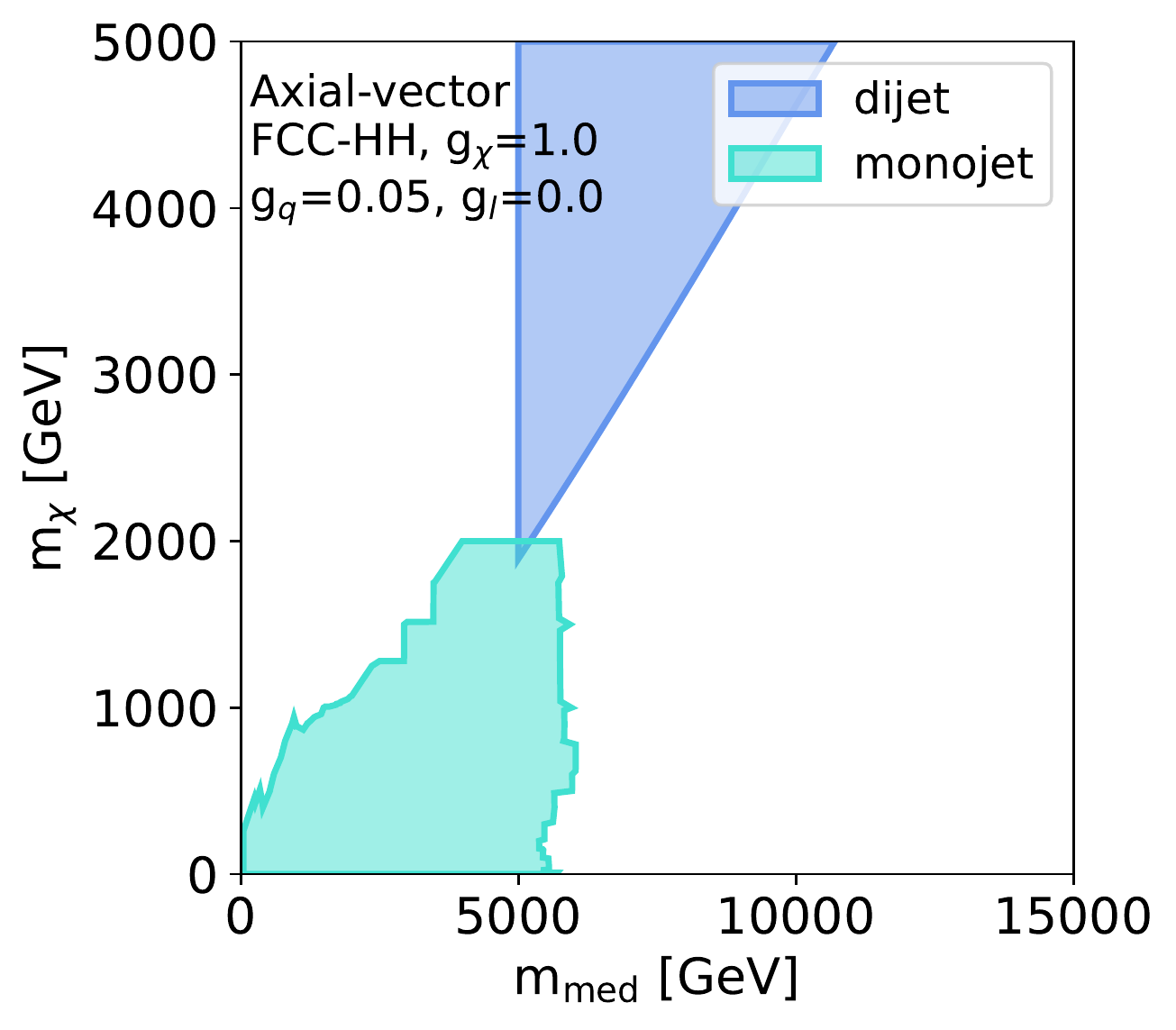}
         \caption{$g_q=0.05$, $g_{\chi}=1.0$, $g_l=0.0$}
         \label{subfig:axial-fcc-v2}
     \end{subfigure}

     \begin{subfigure}[b]{0.49\textwidth}
         \centering
         \includegraphics[width=\textwidth]{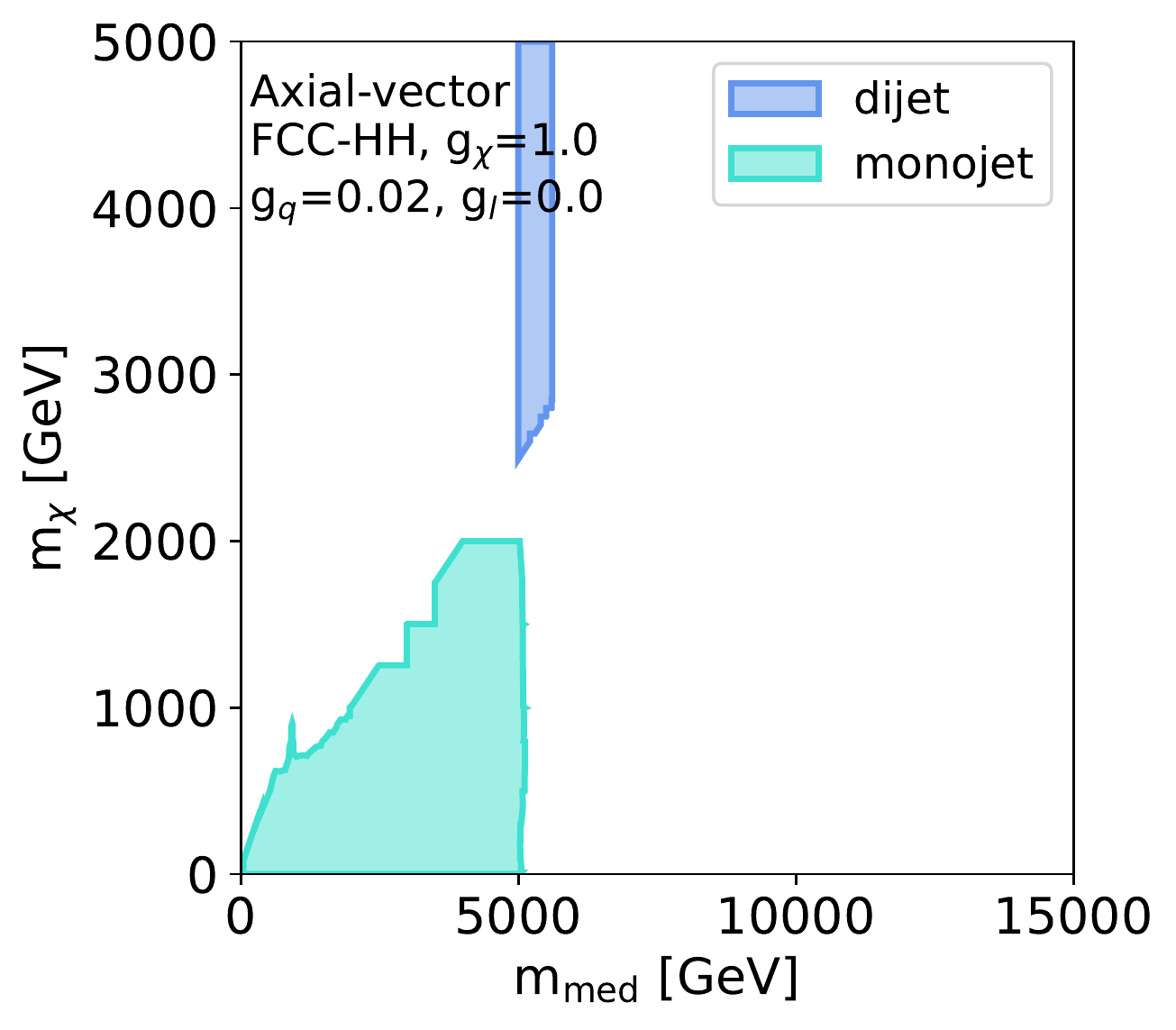}
         \caption{$g_q=0.02$, $g_{\chi}=1.0$, $g_l=0.0$}
         \label{subfig:axial-fcc-v3}
     \end{subfigure}
     \hfill
     \begin{subfigure}[b]{0.49\textwidth}
         \centering
         \includegraphics[width=\textwidth]{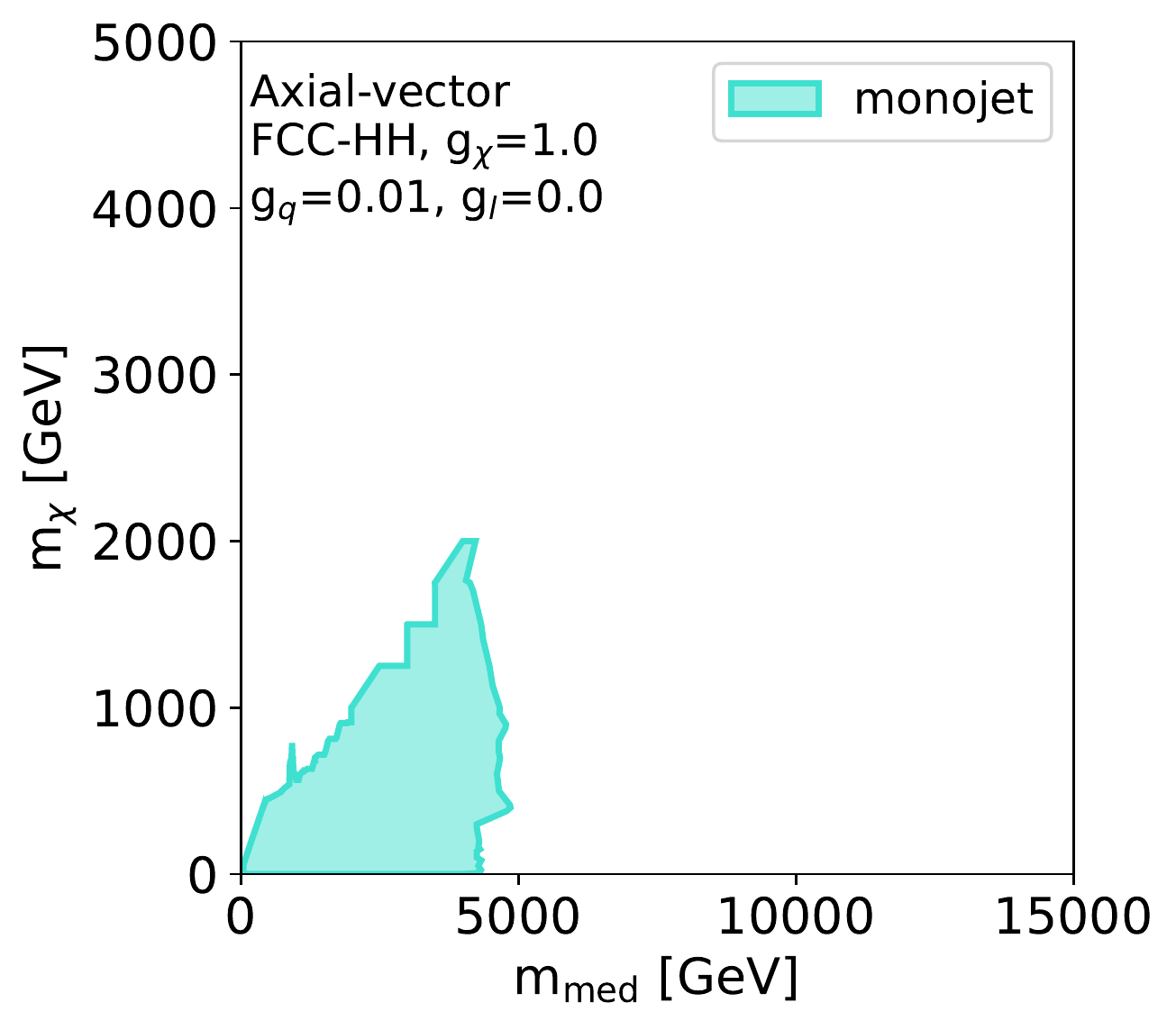}
         \caption{$g_q=0.01$, $g_{\chi}=1.0$, $g_l=0.0$}
         \label{subfig:axial-fcc-v4}       
     \end{subfigure}
        \caption{FCC-hh projected exclusions for individual analyses in the axial-vector model and with a range of couplings. The truncation at the top of the monojet exclusion in subfigures~\subref{subfig:axial-fcc-v1} and~\subref{subfig:axial-fcc-v2} is due to a limitation in the input exclusion grid rather than a physical effect of the analysis at FCC-hh.}
        \label{fig:fcc-hh-massmass-separate-ax}
\end{figure}

\begin{figure}[htb!]
     \centering
     \begin{subfigure}[b]{0.49\textwidth}
         \centering
         \includegraphics[width=\textwidth]{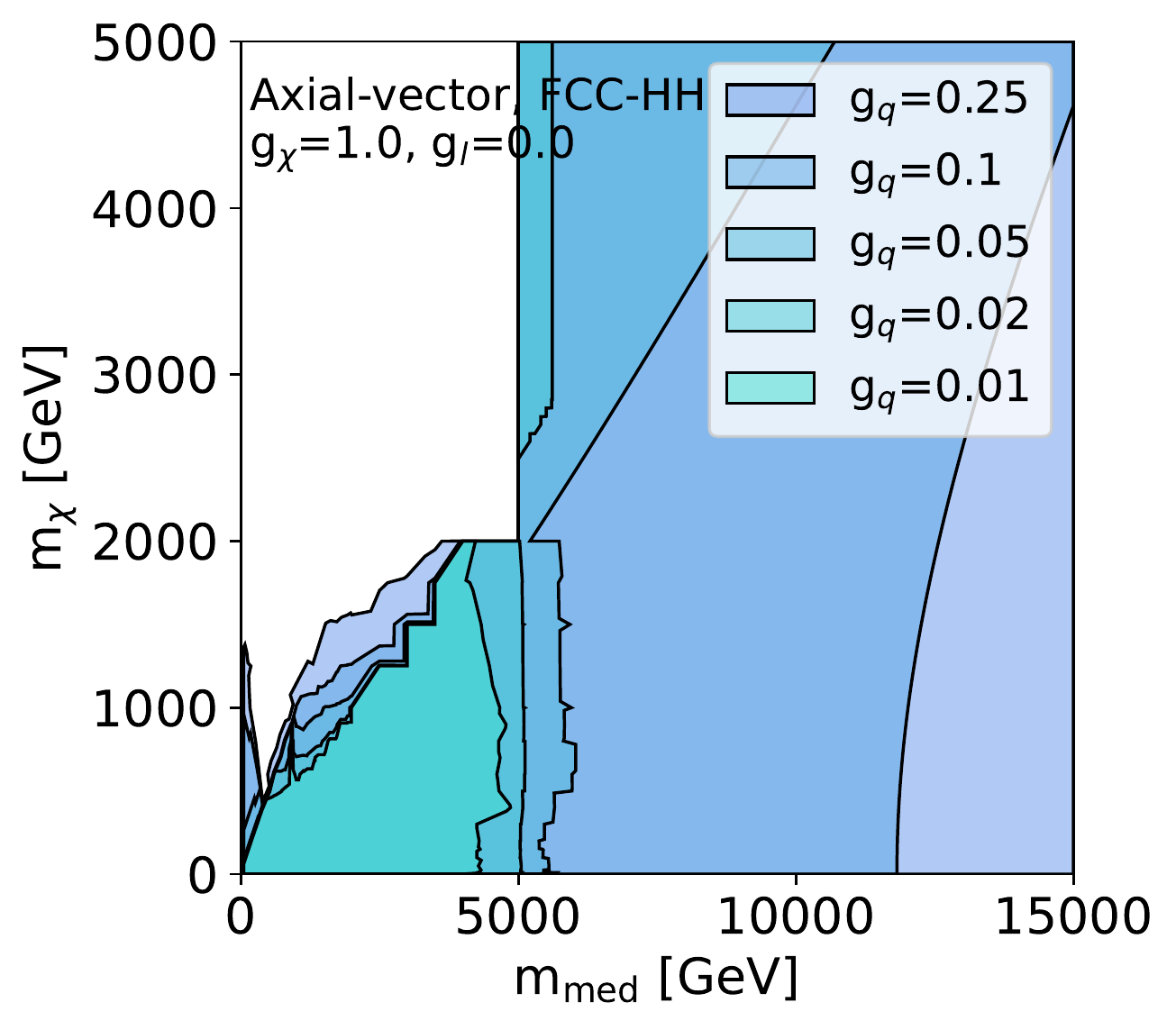}
         \caption{Varying $g_q$ with fixed $g_l=0.0$ and $g_\chi=1.0$}
         \label{subfig:axial-fcc-gqvariations1}
     \end{subfigure}
     \hfill
    \begin{subfigure}[b]{0.49\textwidth}
         \centering
         \includegraphics[width=\textwidth]{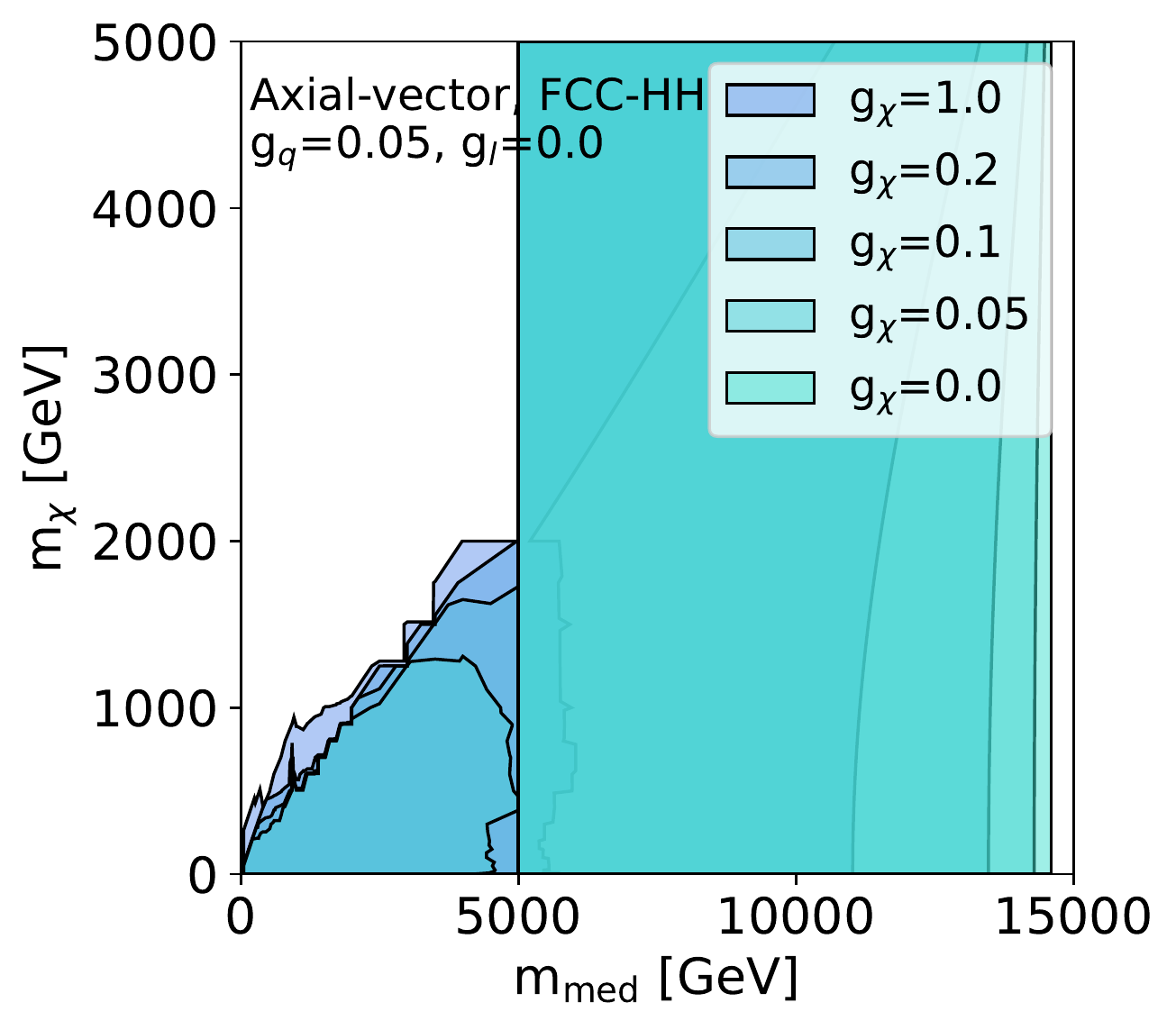}
         \caption{Varying $g_\chi$ with fixed $g_q=0.05$ and $g_l=0.0$}
         \label{subfig:axial-fcc-gqvariations3}
     \end{subfigure}
    \caption{FCC-hh projected exclusions, combined across analyses, for varying coupling values and an axial-vector mediator.}
    \label{fig:fcc-hh-massmass-combined-ax}
\end{figure}

\begin{figure}
    \centering
     \begin{subfigure}[b]{0.49\textwidth}
         \includegraphics[width=\textwidth]{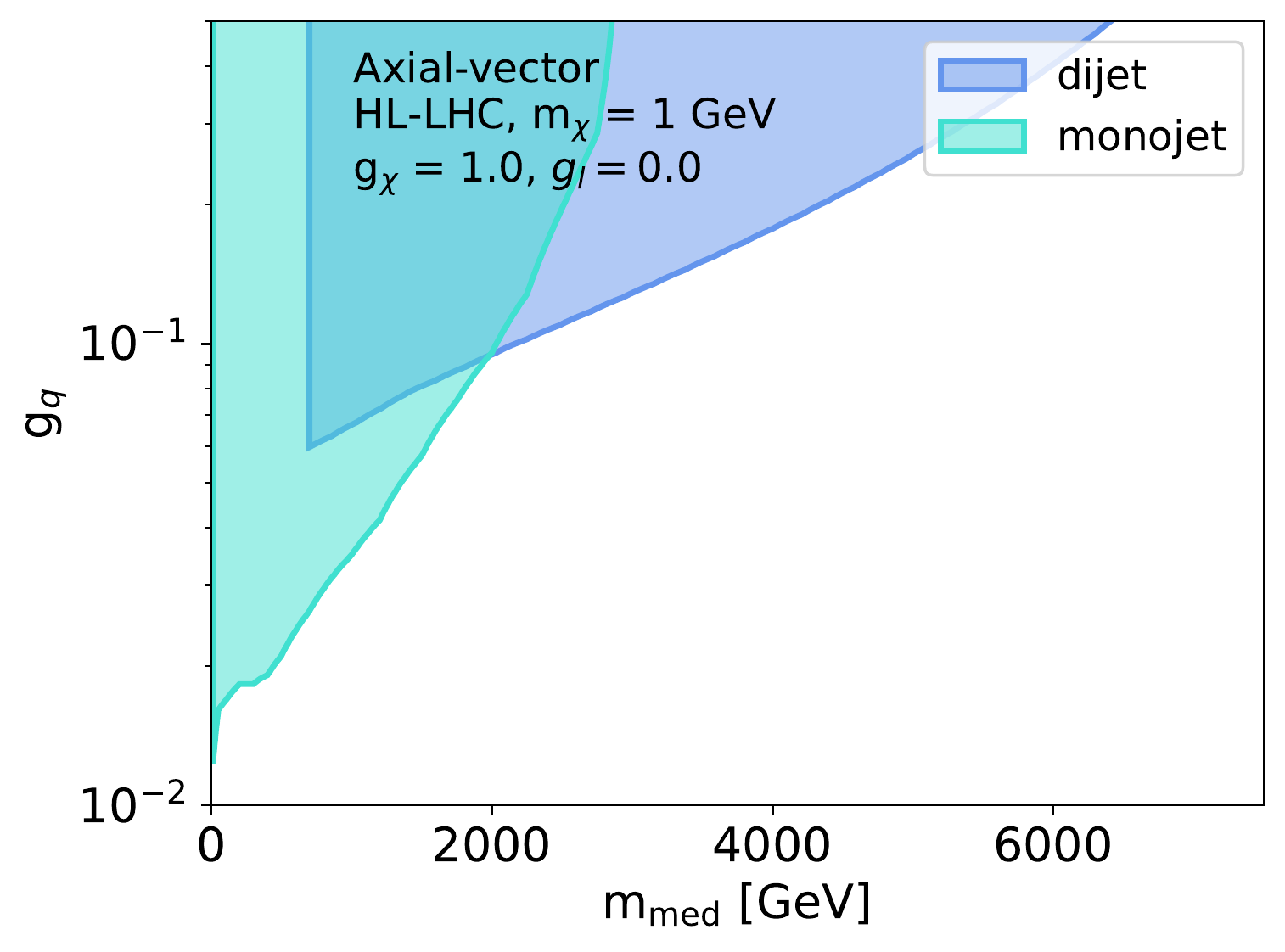}
         \caption{}
         \label{subfig:gqscan-dmLight-ax}
     \end{subfigure}
     \begin{subfigure}[b]{0.49\textwidth}
         \includegraphics[width=\textwidth]{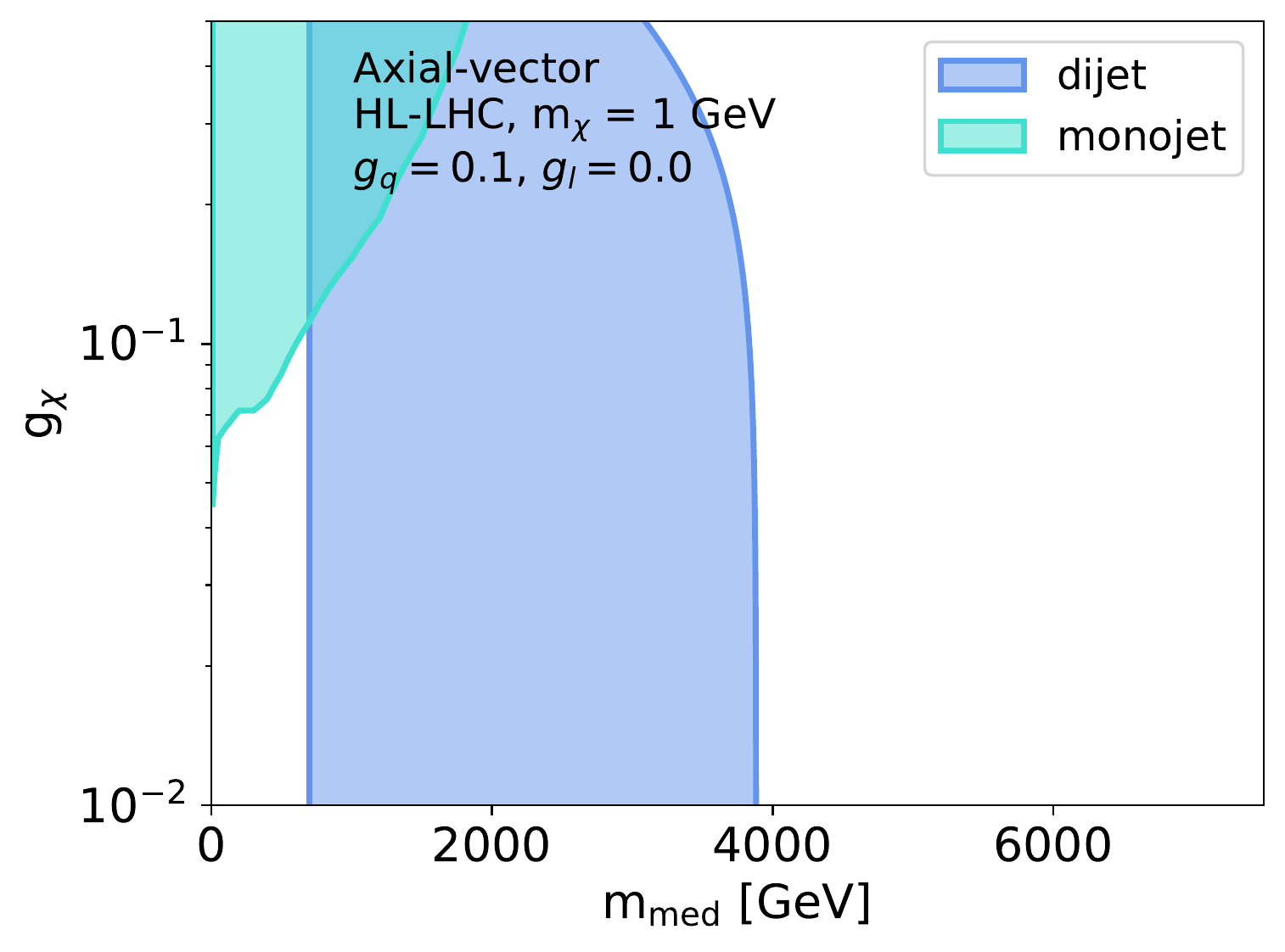}
         \caption{}
         \label{subfig:gchiscan-dmLight-ax}
     \end{subfigure}     
     
    \begin{subfigure}[b]{0.49\textwidth}
         \includegraphics[width=\textwidth]{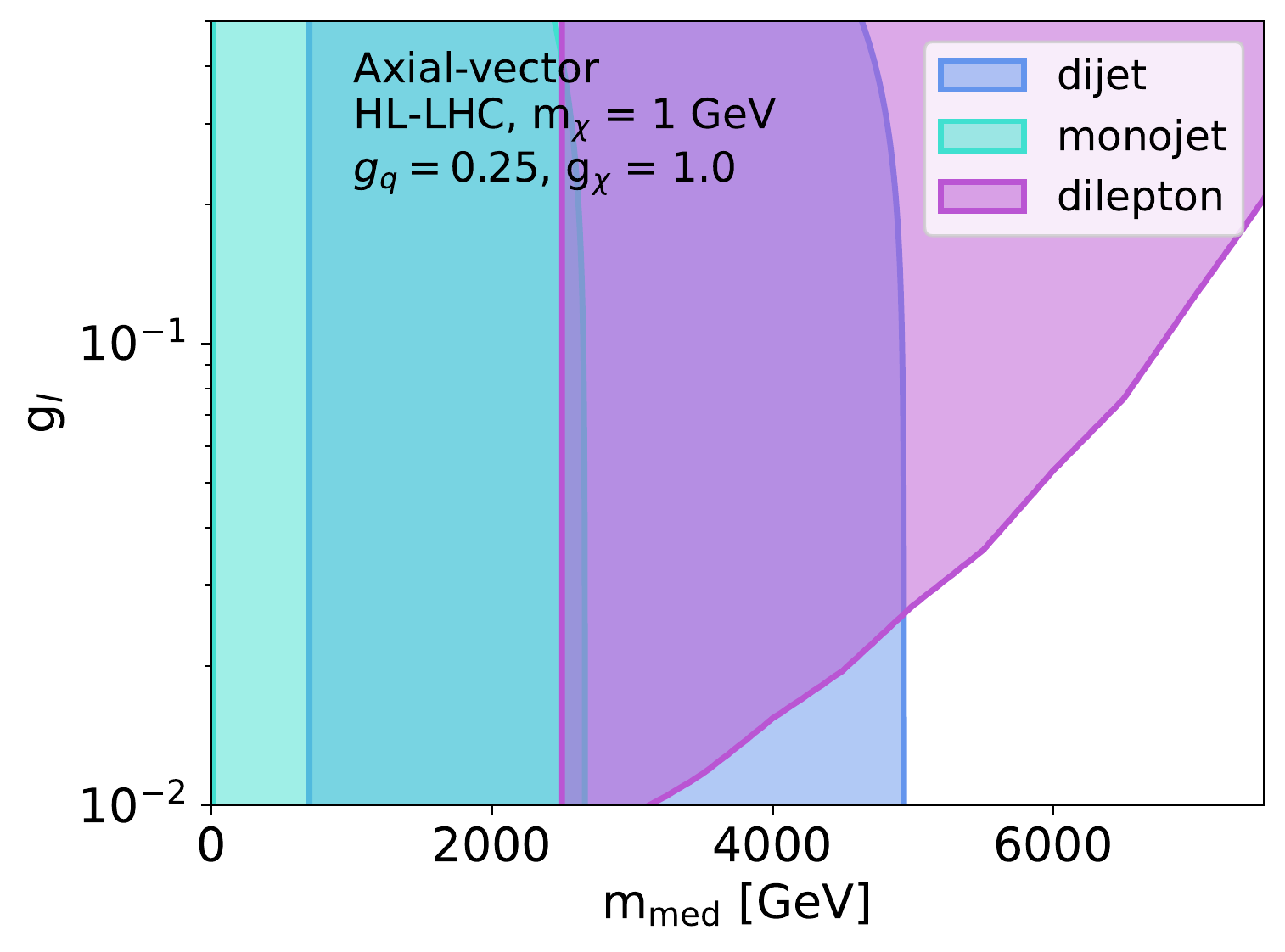}
         \caption{}
         \label{subfig:glscan-dmLight-ax}
     \end{subfigure}    
         \caption{Projected exclusion limits on the couplings $g_q$ (\subref{subfig:gqscan-dmLight-ax}), $g_\chi$ (\subref{subfig:gchiscan-dmLight-ax}), and $g_l$ (\subref{subfig:glscan-dmLight-ax}) for an axial-vector mediator at the HL-LHC. The result is shown as a function of the mediator mass $m_{med}$; the mass of the DM candidate is fixed to 1 GeV in all cases. The coupling on the $y$ axis is varied while the other two couplings are fixed: in~(\subref{subfig:gqscan-dmLight-ax}), $g_\chi$=1.0 and $g_l$=0.0; in~(\subref{subfig:gchiscan-dmLight-ax}), $g_q$=0.1 and $g_l$=0.0; and in~\subref{subfig:glscan-dmLight-ax}, $g_q$=0.25 and $g_\chi$=1.0.}
         \label{fig:couplinglimits-hl-lhc-allanalyses-ax}
\end{figure}

\begin{figure}
    \centering
     \begin{subfigure}[b]{0.49\textwidth}
         \includegraphics[width=\textwidth]{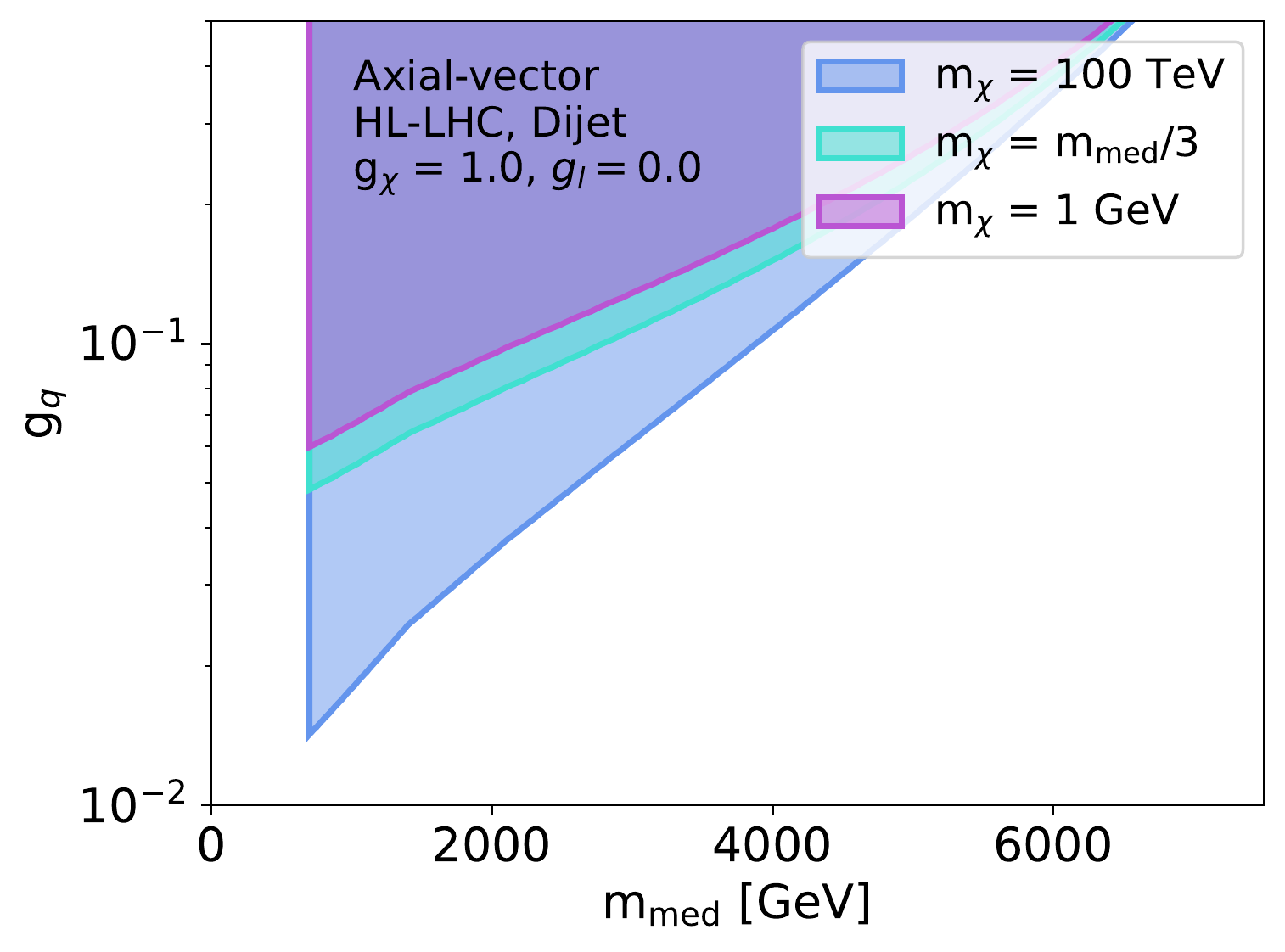}
         \caption{}
         \label{subfig:gqscan-comparedm-ax}
     \end{subfigure}
     \begin{subfigure}[b]{0.49\textwidth}
         \includegraphics[width=\textwidth]{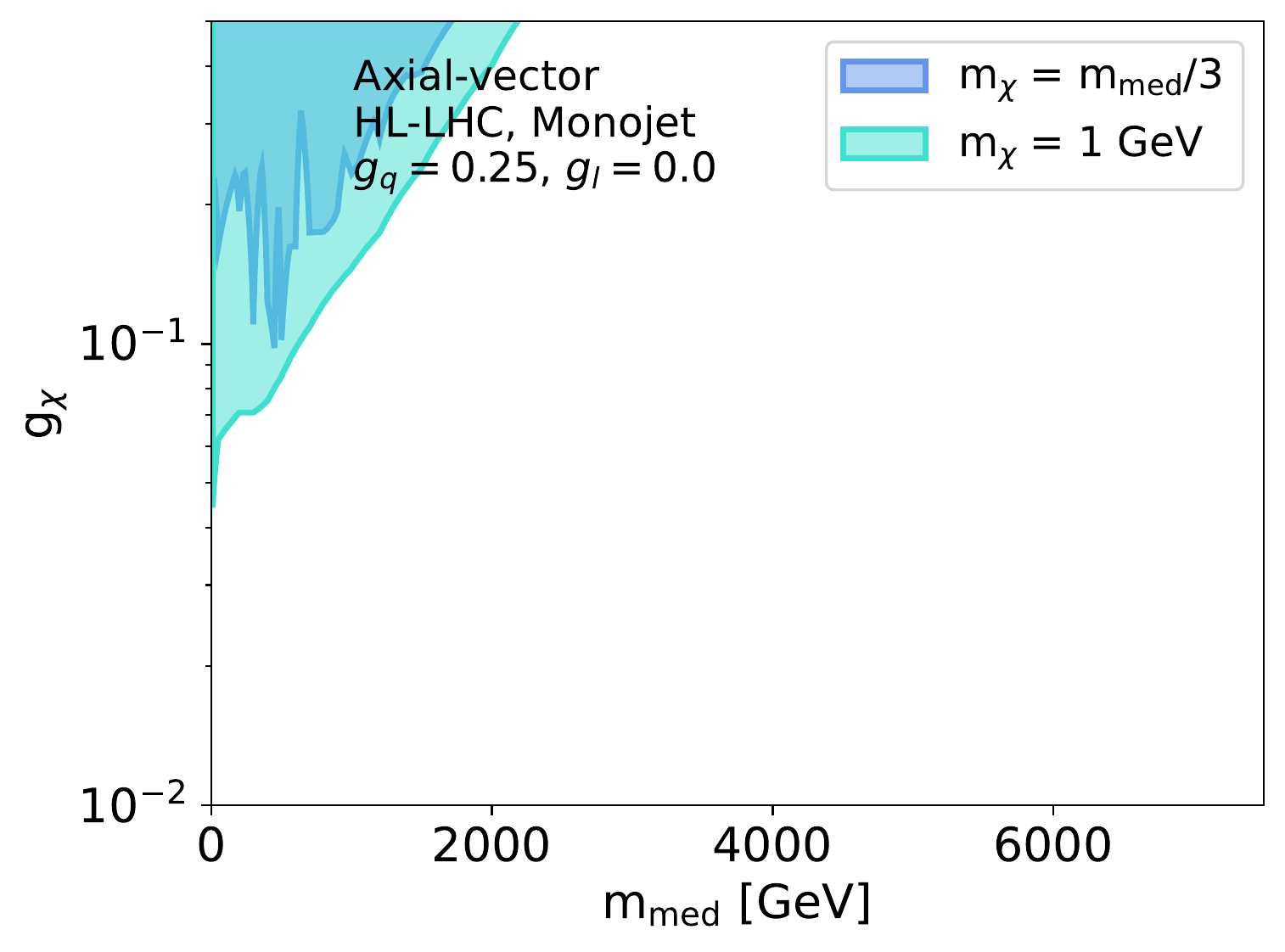}
         \caption{}
         \label{subfig:gchiscan-comparedm-ax}
     \end{subfigure}     
         \caption{An illustration of the impact of the choice of DM particle mass on the coupling exclusion limits. Limits set by the dijet analysis on $g_q$ with $g_\chi=1.0$, $g_l=0.0$ are shown in (\subref{subfig:gqscan-comparedm-ax}). Limits set by the monojet analysis on $g_\chi$ with $g_q=0.25$, $g_l=0.0$ are shown in (\subref{subfig:gchiscan-comparedm-ax}). All limits are displayed axial-vector mediators and HL-LHC projected exclusions. The difference is small between a dark matter mass of 1~GeV and of $1/3$ of the mediator mass. Decoupling dark matter entirely has a larger effect on the strength of the dijet limit, while eliminating any constraints from monojet by definition.}
         \label{fig:couplinglimits-hl-lhc-dmmass-ax}
\end{figure}

\begin{figure}[htb!]
    \centering
     \begin{subfigure}[b]{0.49\textwidth}
         \includegraphics[width=\textwidth]{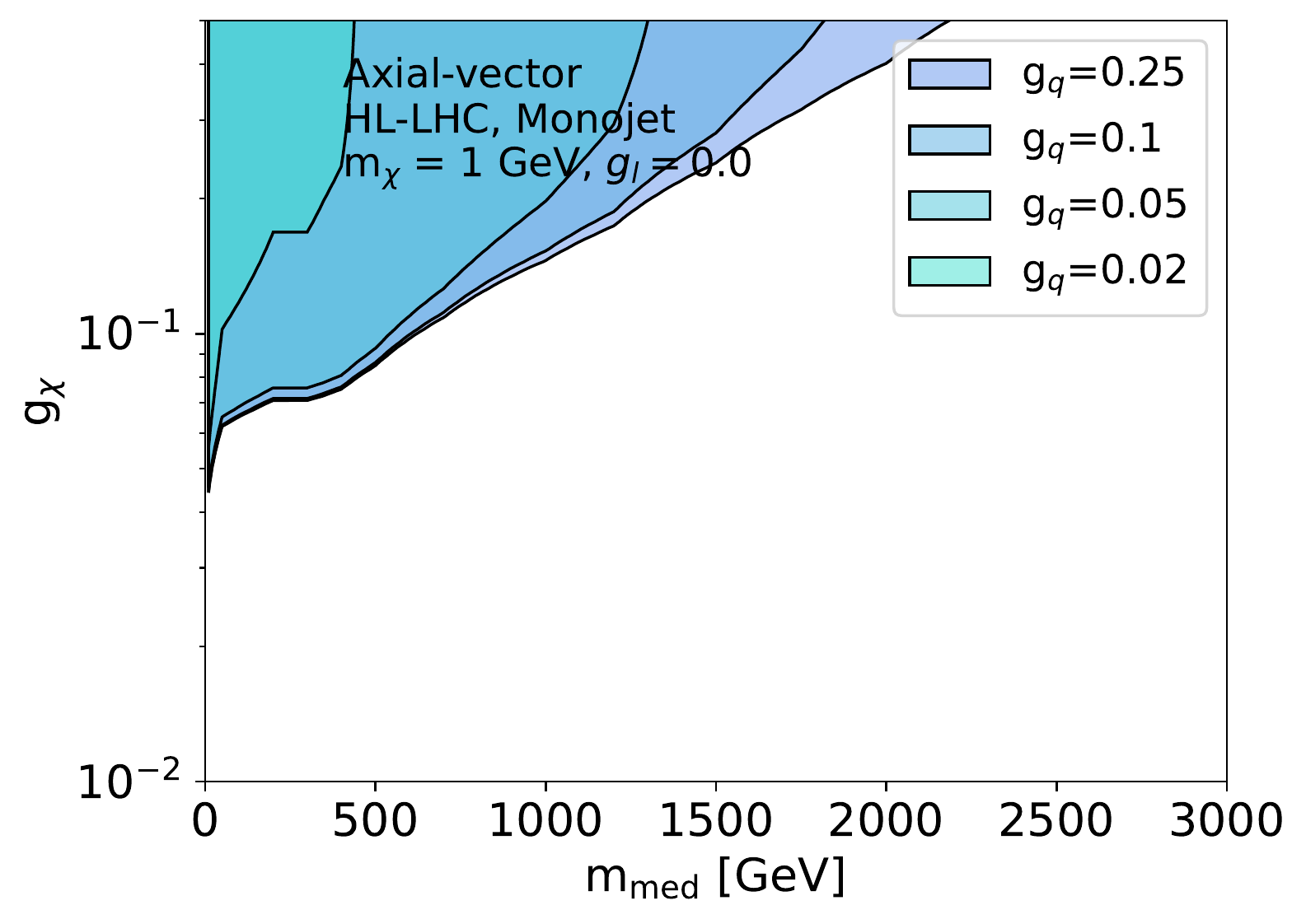}
         \caption{Monojet limits on $g_\chi$}
         \label{subfig:gchiscan-fixedmdm-monojet-ax}
     \end{subfigure}
     \begin{subfigure}[b]{0.49\textwidth}
         \includegraphics[width=\textwidth]{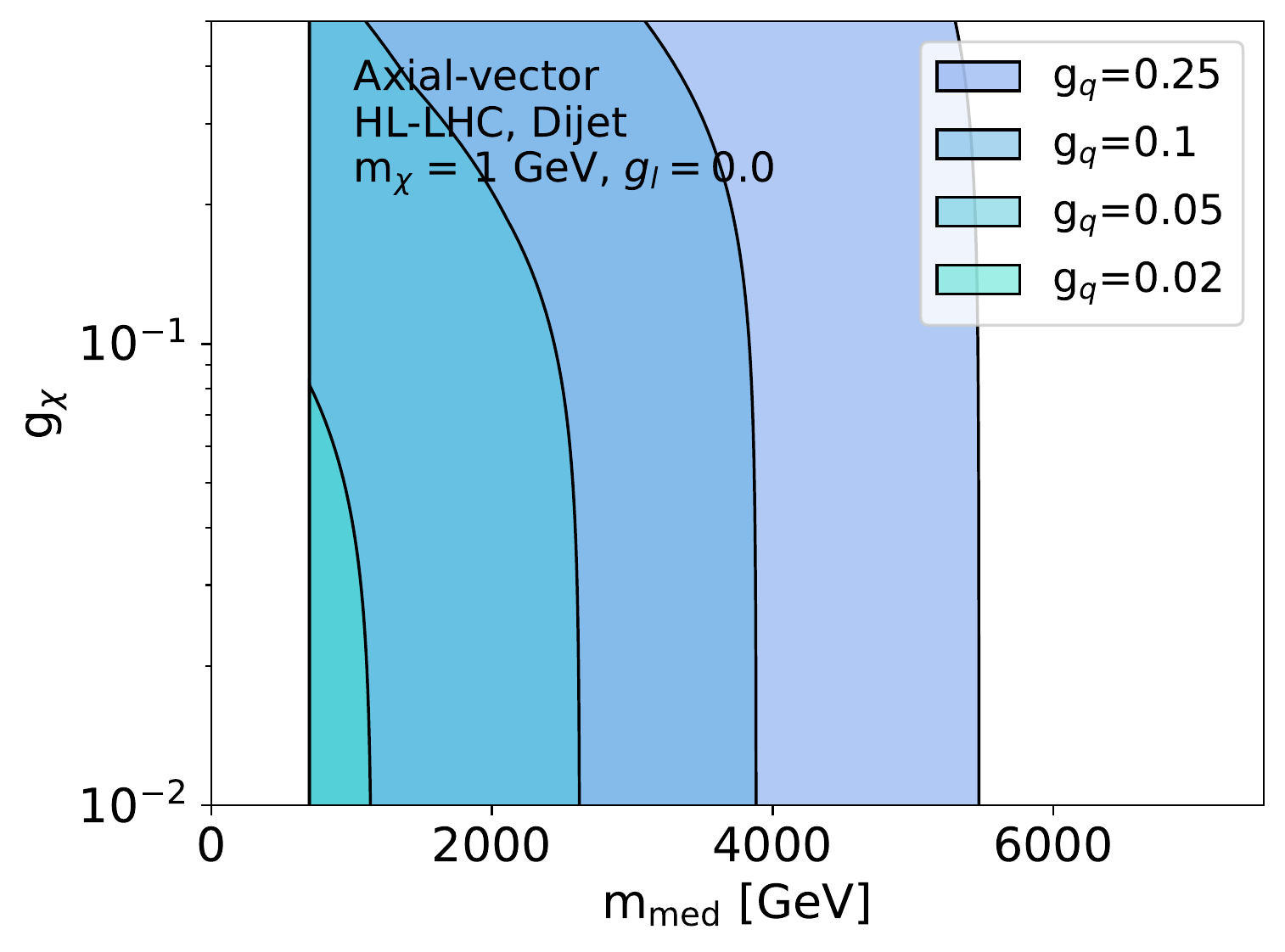}
         \caption{Dijet limits on $g_\chi$}
         \label{subfig:gchiscan-fixedmdm-dijet-ax}
     \end{subfigure}     
     
     \begin{subfigure}[b]{0.49\textwidth}
         \includegraphics[width=\textwidth]{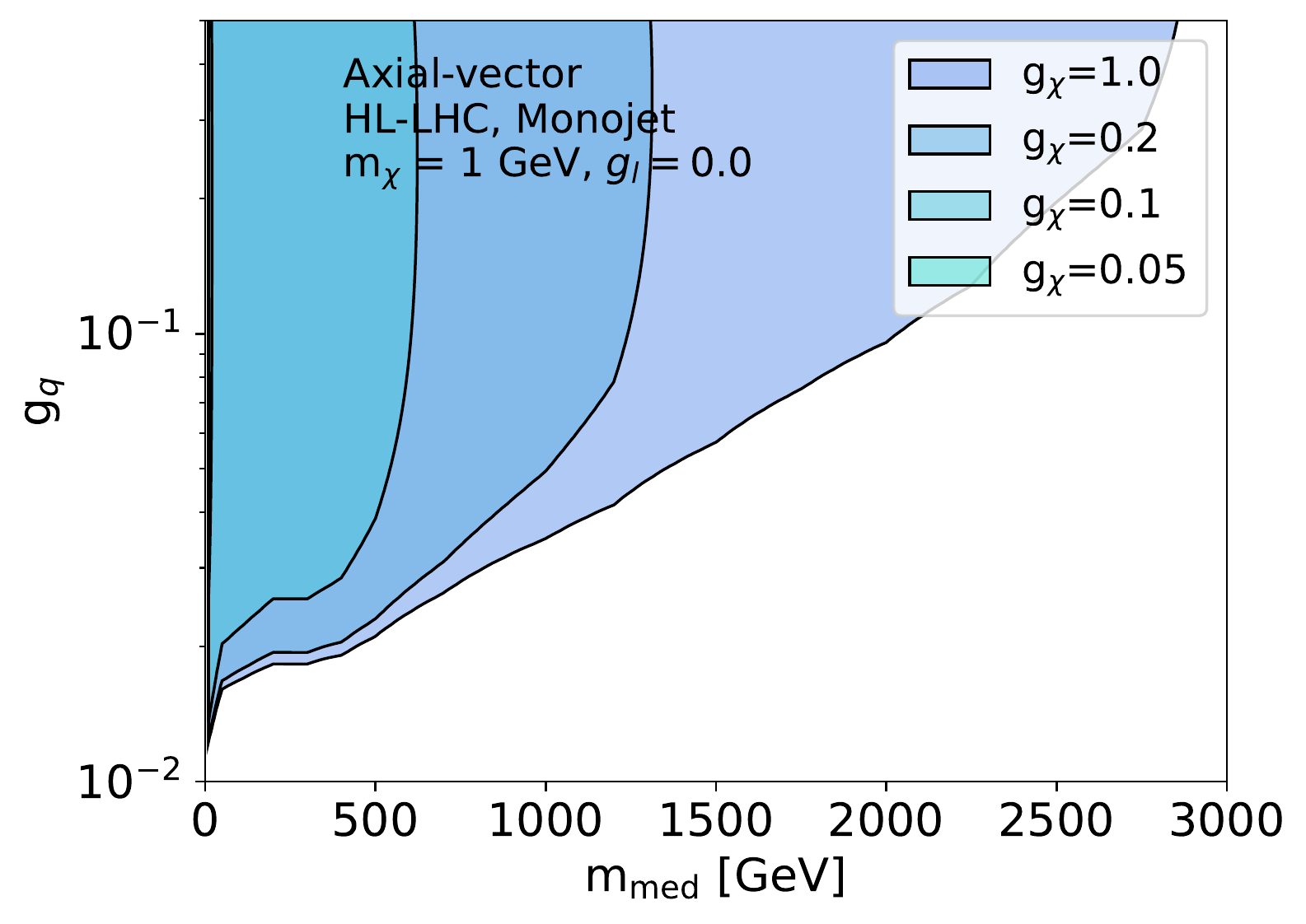}
         \caption{Monojet limits on $g_q$}
         \label{subfig:gqscan-fixedmdm-monojet-ax}
     \end{subfigure}
     \begin{subfigure}[b]{0.49\textwidth}
         \includegraphics[width=\textwidth]{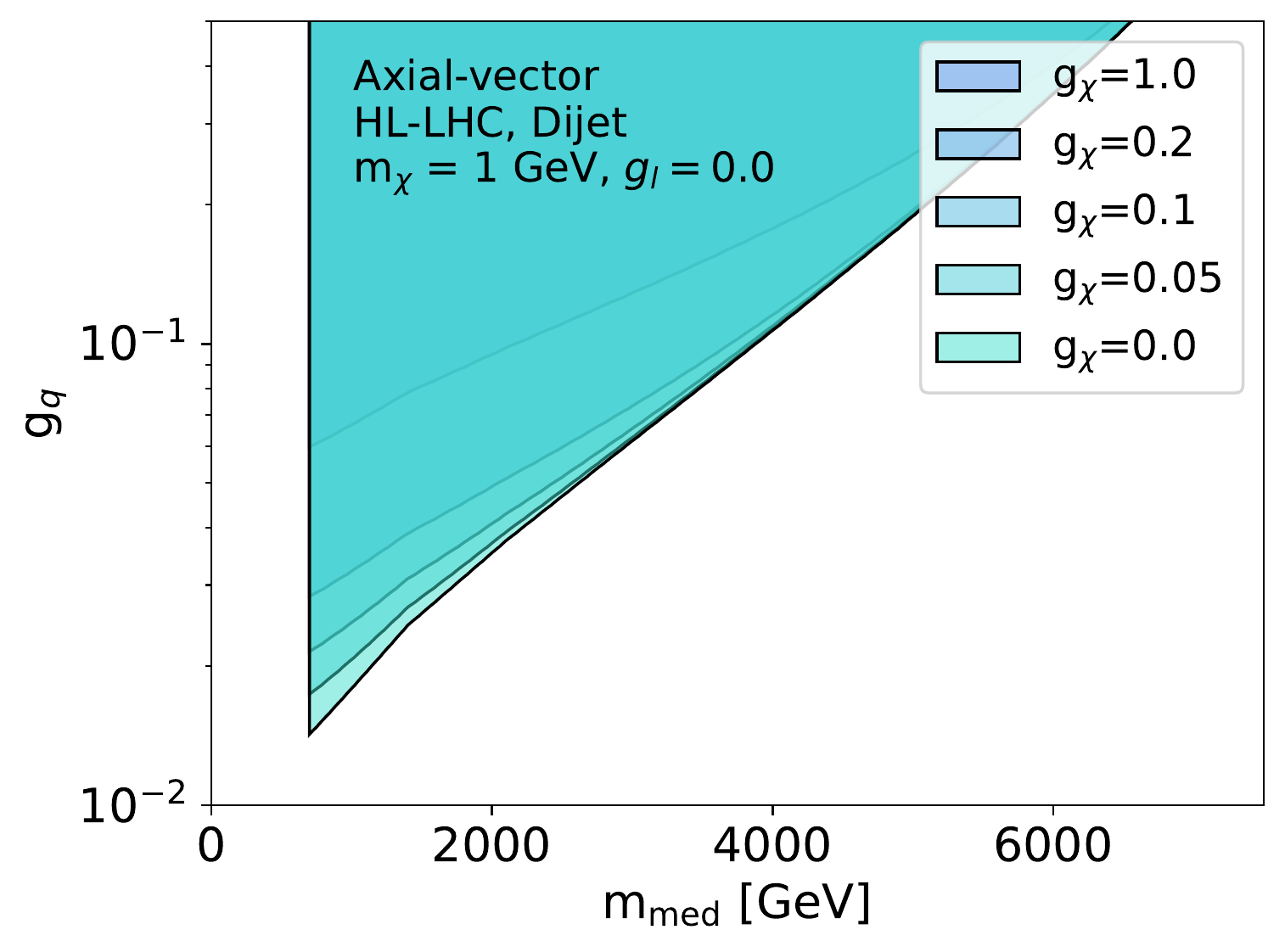}
         \caption{Dijet limits on $g_q$}
         \label{subfig:gqscan-fixedmdm-dijet-ax}
     \end{subfigure}     
         \caption{Limits on $g_\chi$ (\subref{subfig:gchiscan-fixedmdm-monojet-ax},~\subref{subfig:gchiscan-fixedmdm-dijet-ax}) and $g_q$ (\subref{subfig:gqscan-fixedmdm-monojet-ax},~\subref{subfig:gqscan-fixedmdm-dijet-ax}) set by the HL-LHC monojet and dijet analyses as a function of axial-vector mediator mass. The mediator has no coupling to leptons. The mass of the dark matter particle is fixed to $m_\chi=1$~GeV.}
         \label{fig:couplinglimits-hl-lhc-fixedmdm-couplingscan-ax}
\end{figure}

\begin{figure}[htb!]
    \centering
     \begin{subfigure}[b]{0.49\textwidth}
         \includegraphics[width=\textwidth]{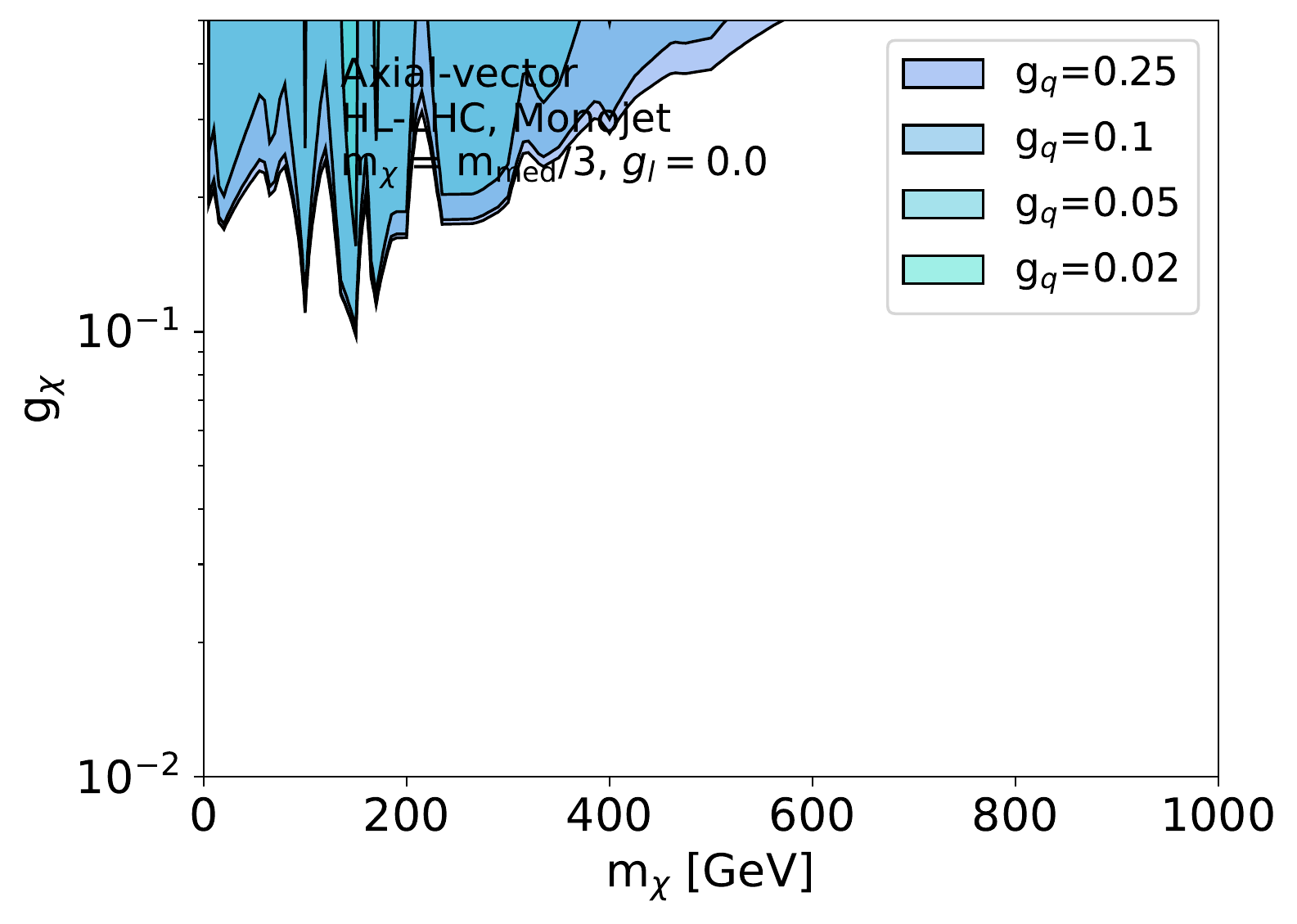}
         \caption{}
         \label{subfig:gchiscan-fixedmmed-monojet-ax}
     \end{subfigure}
     \begin{subfigure}[b]{0.49\textwidth}
         \includegraphics[width=\textwidth]{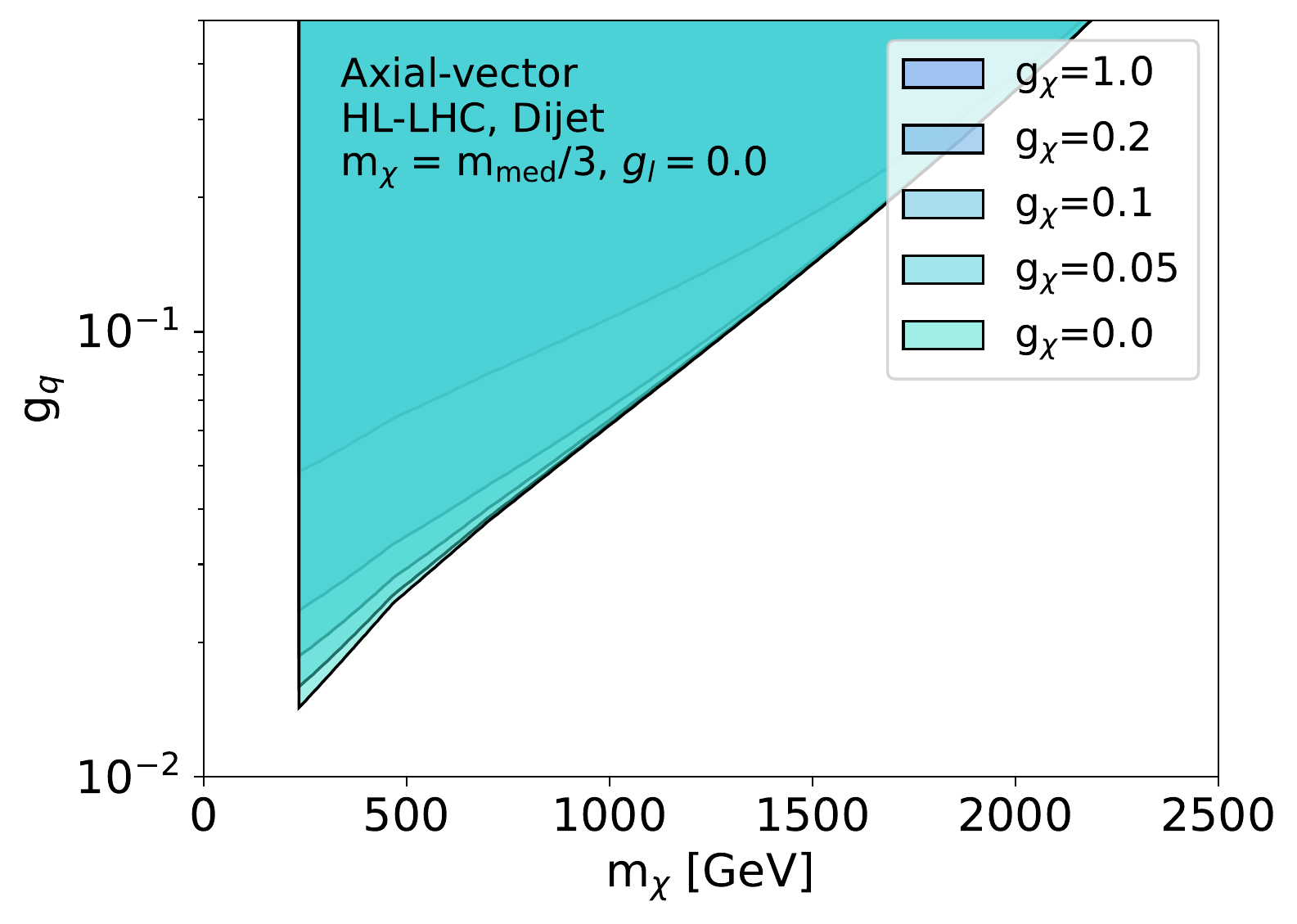}
         \caption{}
         \label{subfig:gqscan-fixedmmed-dijet-ax}
     \end{subfigure}     
         \caption{Coupling limits as a function of $m_\chi$ with axial-vector mediator mass fixed to $m_\mathrm{med} = 3 m_\chi$. Monojet constraints on $g_\chi$ with varying $g_q$ values are shown in (\subref{subfig:gchiscan-fixedmmed-monojet-ax}) while dijet constraints on $g_q$ with varying $g_\chi$ values are shown in (\subref{subfig:gqscan-fixedmmed-dijet-ax}).}
         \label{fig:couplinglimits-hl-lhc-fixedmmed-ax}
\end{figure}

\begin{figure}[htb!]
    \centering
     \begin{subfigure}[b]{0.49\textwidth}
         \includegraphics[width=\textwidth]{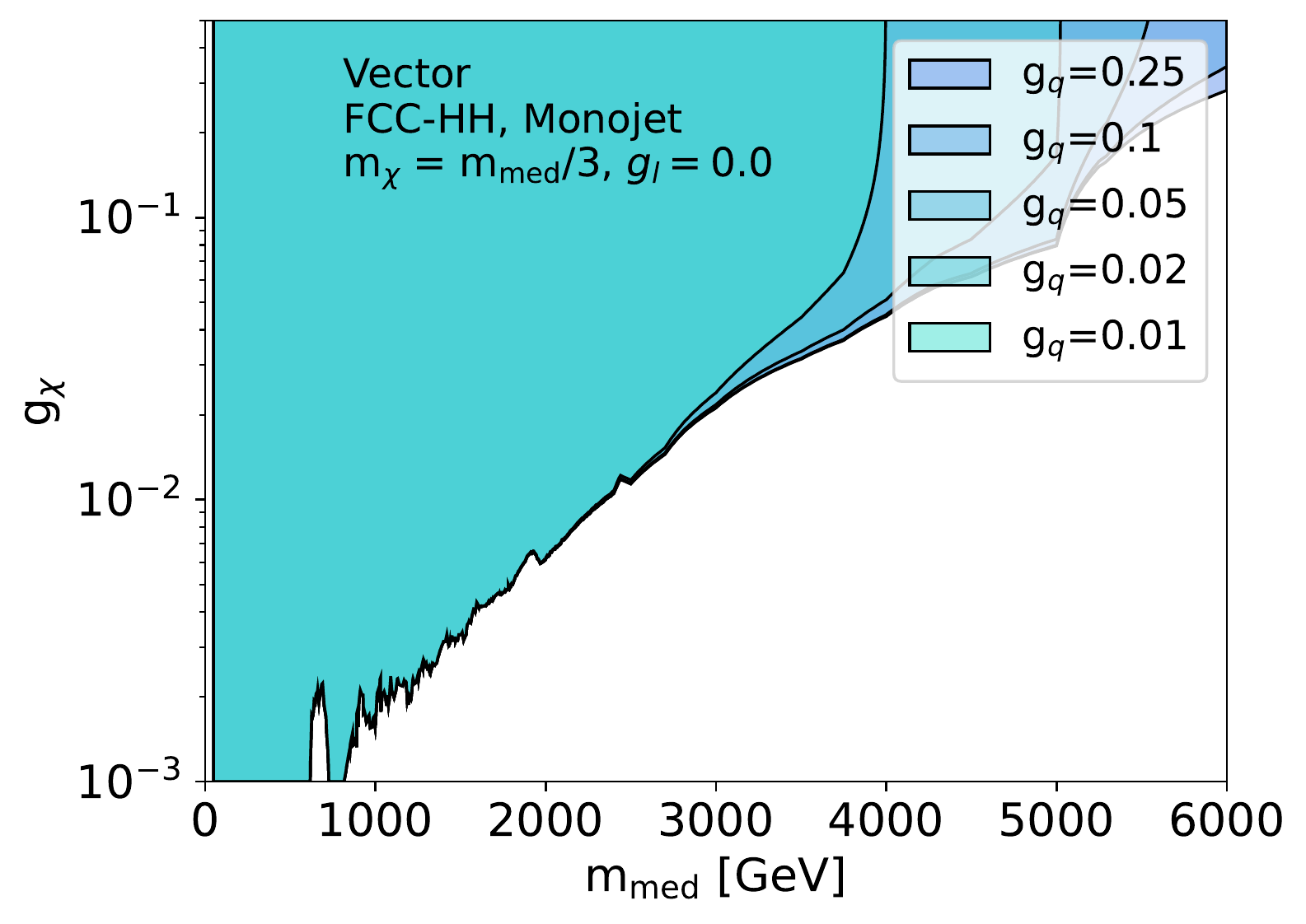}
         \caption{Monojet limits on $g_\chi$}
         \label{subfig:gchiscan-fixedmdm-monojet-fcc-dplike}
     \end{subfigure}
     \begin{subfigure}[b]{0.49\textwidth}
         \includegraphics[width=\textwidth]{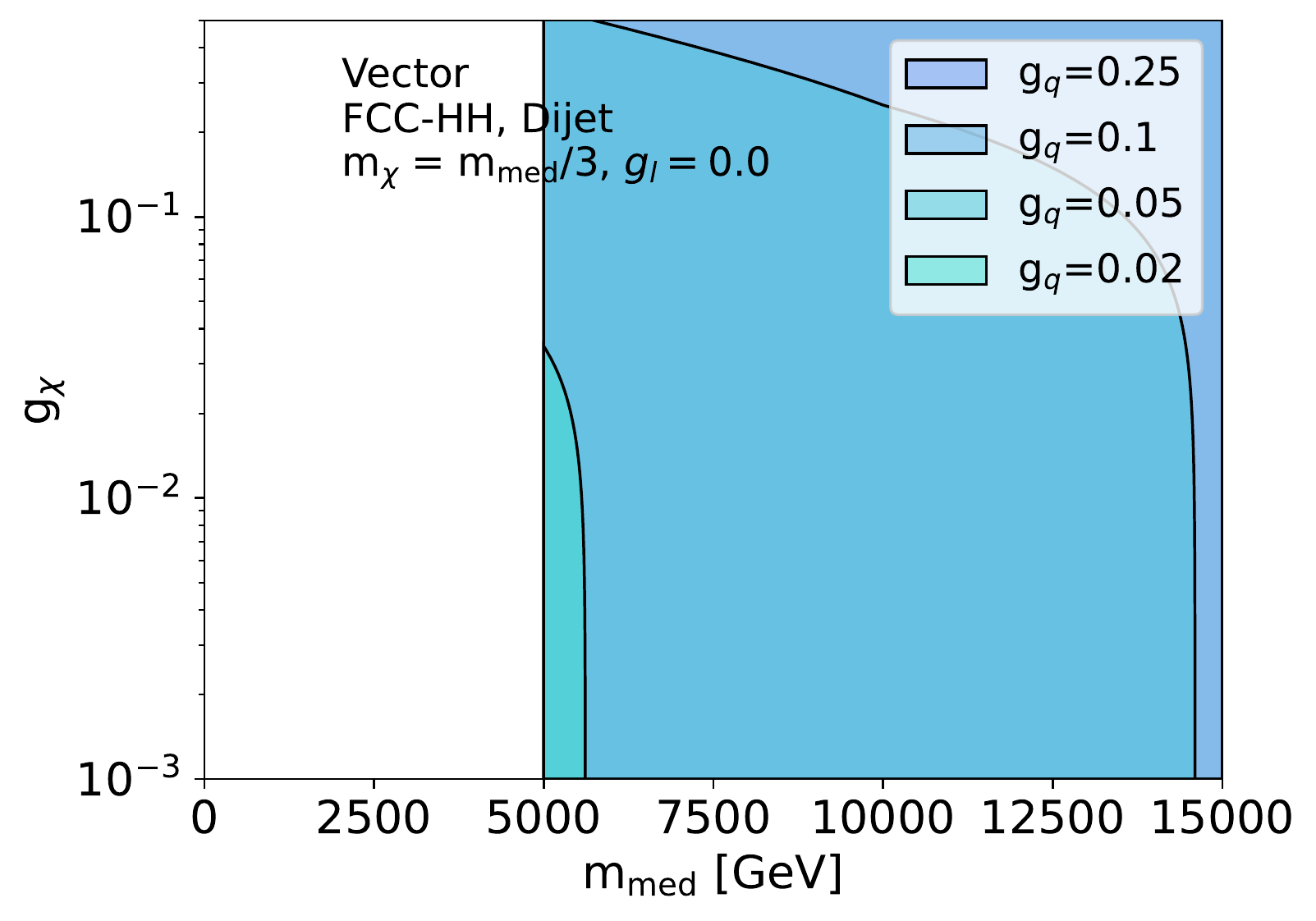}
         \caption{Dijet limits on $g_\chi$}
         \label{subfig:gchiscan-fixedmdm-dijet-fcc-dplike}
     \end{subfigure}     
     
     \begin{subfigure}[b]{0.49\textwidth}
         \includegraphics[width=\textwidth]{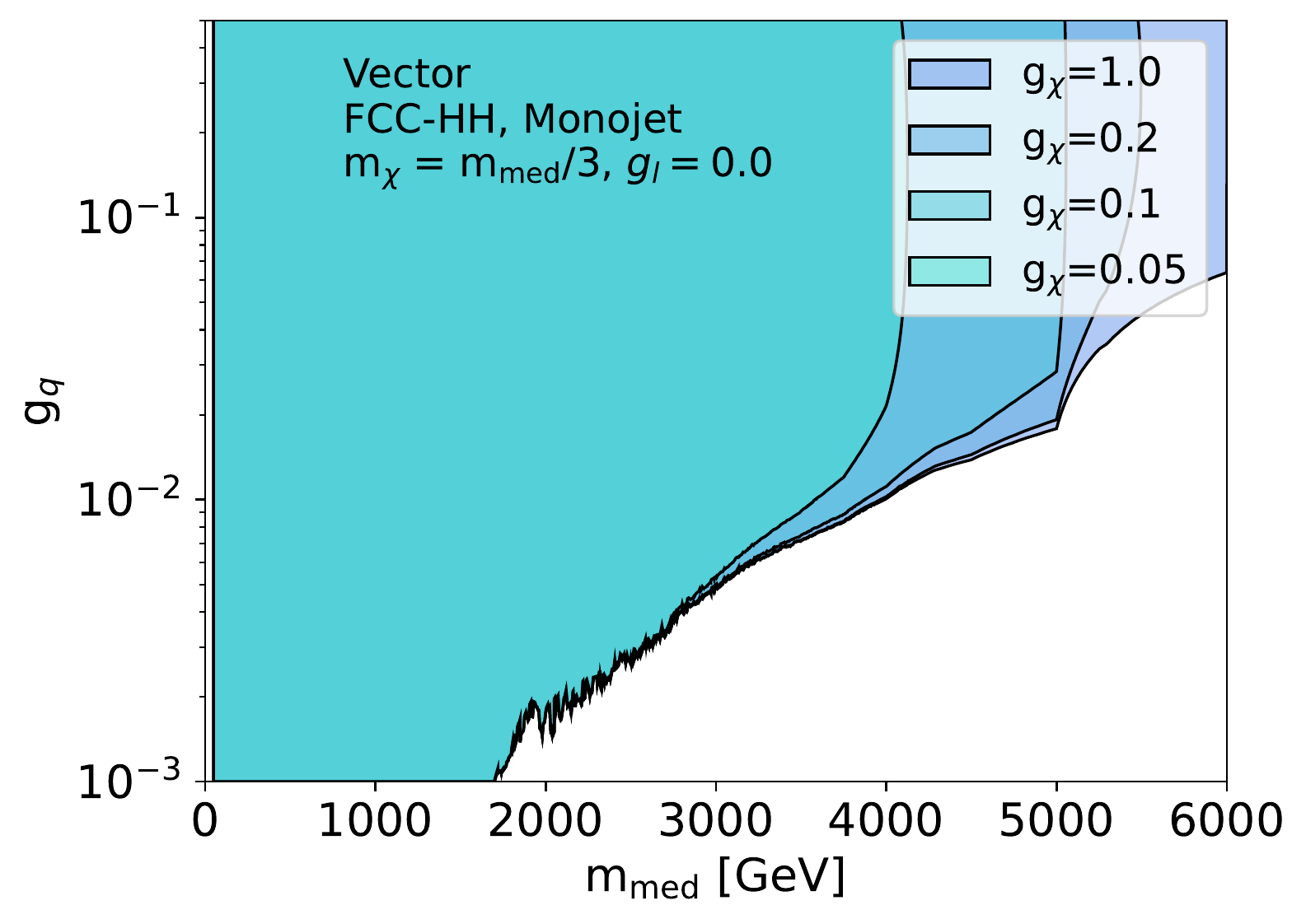}
         \caption{Monojet limits on $g_q$}
         \label{subfig:gqscan-fixedmdm-monojet-fcc-dplike}
     \end{subfigure}
     \begin{subfigure}[b]{0.49\textwidth}
         \includegraphics[width=\textwidth]{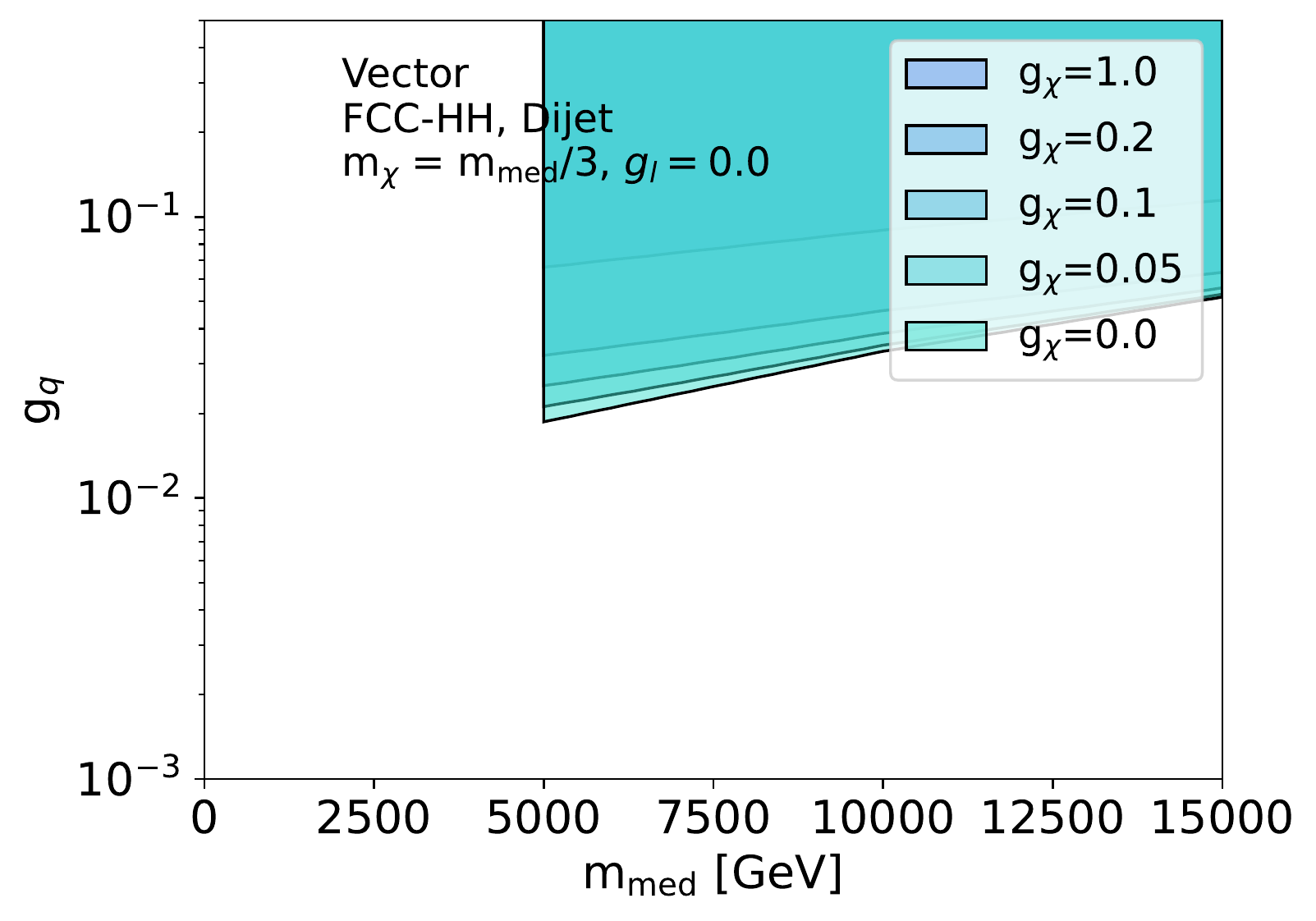}
         \caption{Dijet limits on $g_q$}
         \label{subfig:gqscan-fixedmdm-dijet-fcc-dplike}
     \end{subfigure}       
         \caption{Coupling limits as a function of mediator mass for a vector mediator at FCC-hh. Dark matter mass is fixed to $m_\chi = m_\mathrm{med}/3$.}
         \label{fig:couplinglimits-fcc-hh-various-dplike}
\end{figure}



\begin{figure}[htp]
     \centering
     \begin{subfigure}[b]{0.8\textwidth}
         \centering
         \includegraphics[width=\textwidth]{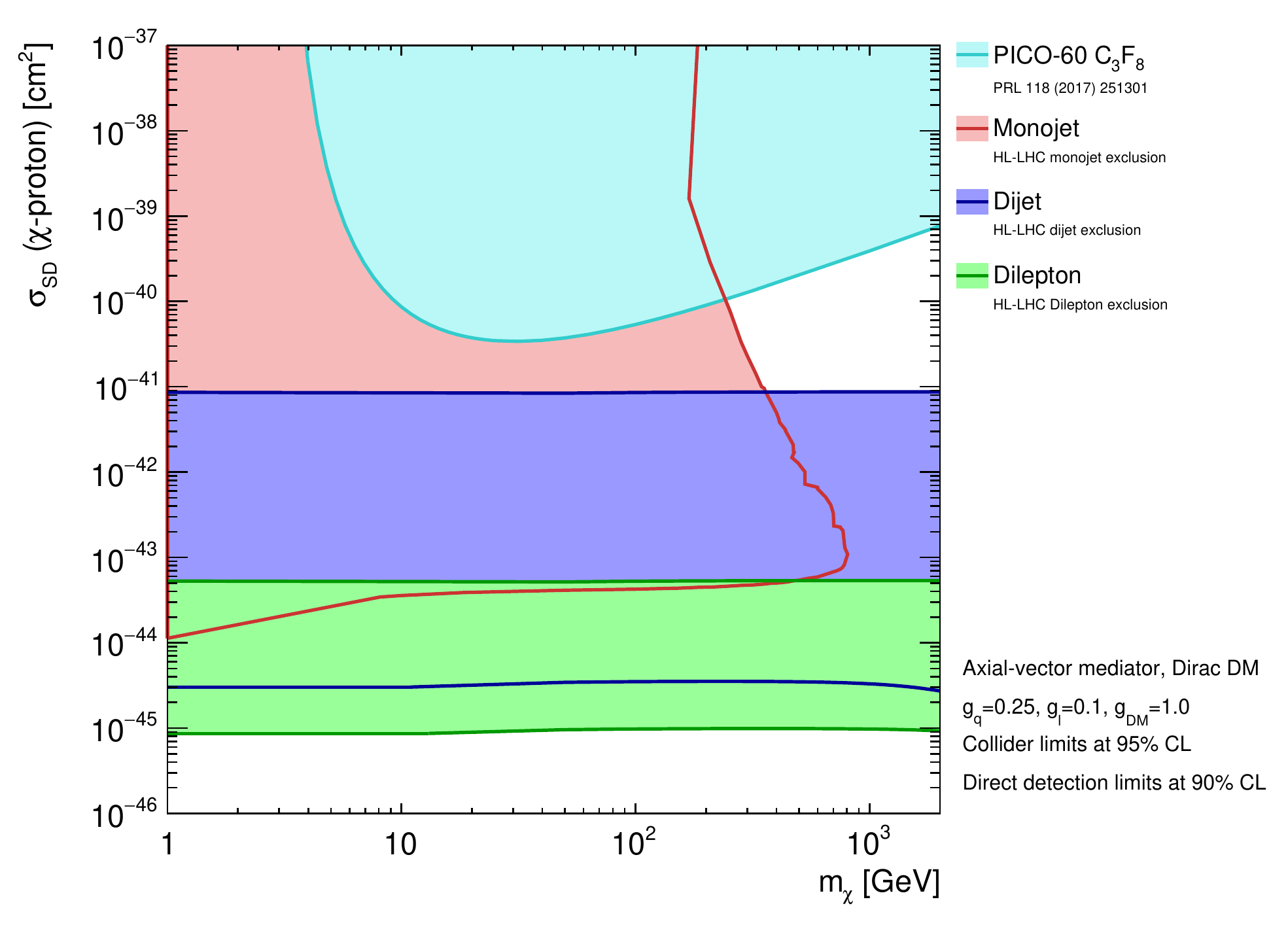}
         \caption{$g_q=0.25$, $g_{\chi}=1.0$, $g_l=0.1$}
         \label{subfig:dd-hl-lhc-axial-1}
     \end{subfigure}

     \begin{subfigure}[b]{0.8\textwidth}
         \centering
         \includegraphics[width=\textwidth]{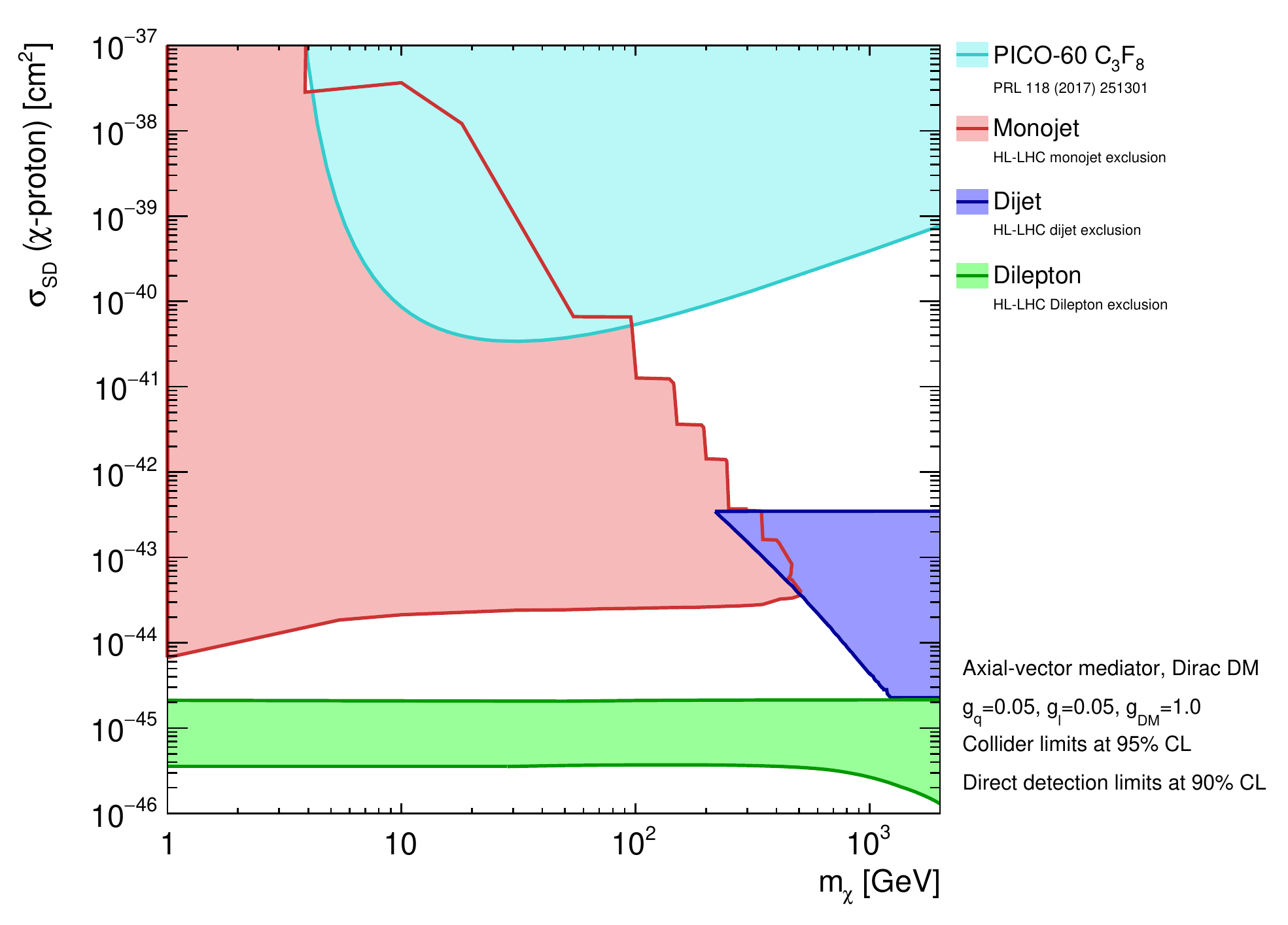}
         \caption{$g_q=0.05$, $g_{\chi}=1.0$, $g_l=0.05$}
         \label{subfig:dd-hl-lhc-axial-2}
     \end{subfigure}
        \caption{Comparison of projected exclusions from HL-LHC with constraints from current direct-detection experiments on the spin-dependent DM–proton scattering cross section in the context of the axial-vector simplified model.}
        \label{fig:hl-lhc-dd-separate-sd}     
\end{figure}

\begin{figure}[htp]
     \centering
     \begin{subfigure}[b]{0.8\textwidth}
         \centering
         \includegraphics[width=\textwidth]{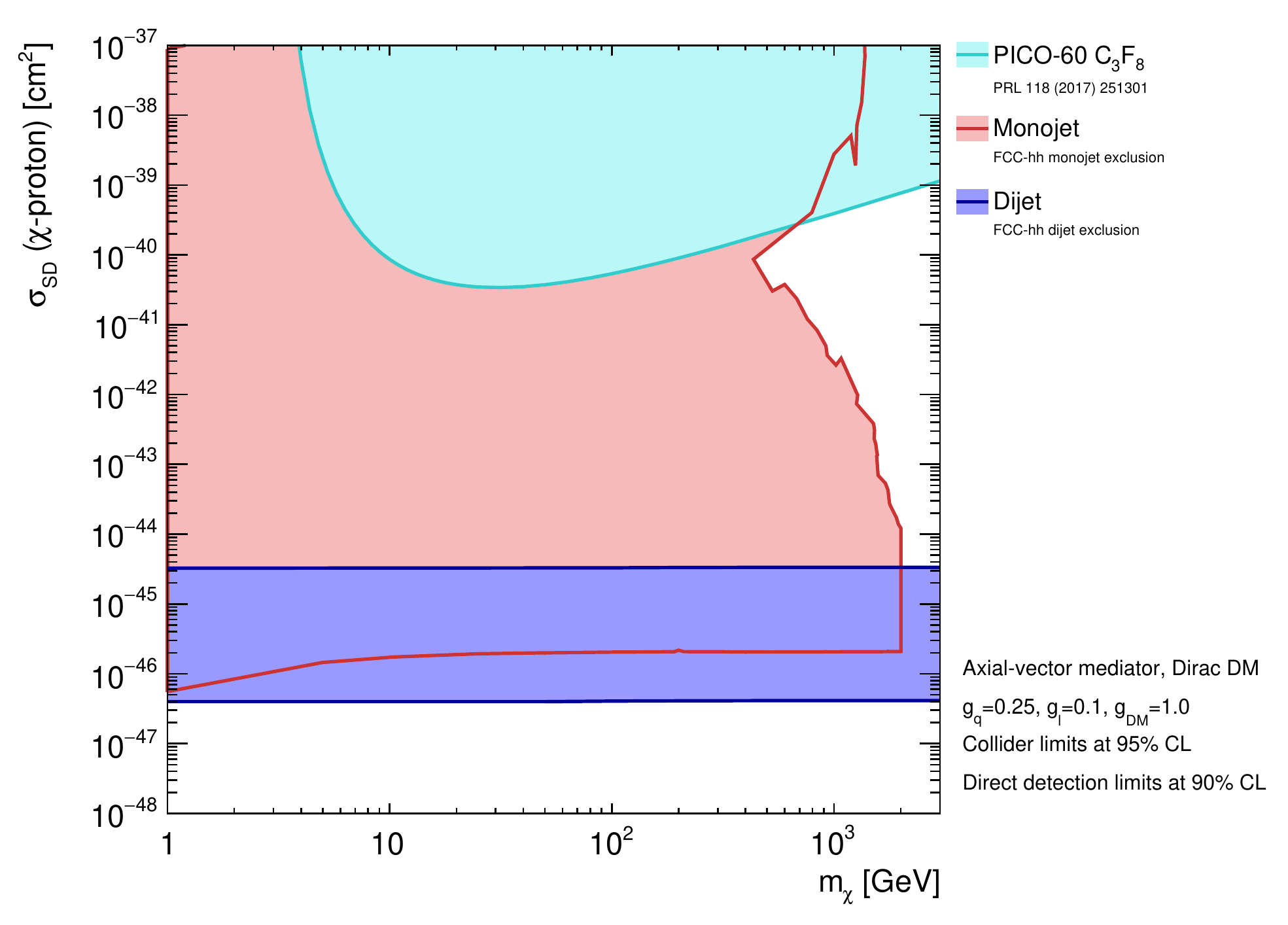}
         \caption{$g_q=0.25$, $g_{\chi}=1.0$, $g_l=0.1$}
         \label{subfig:dd-fcc-hh-axial-1}
     \end{subfigure}

     \begin{subfigure}[b]{0.8\textwidth}
         \centering
         \includegraphics[width=\textwidth]{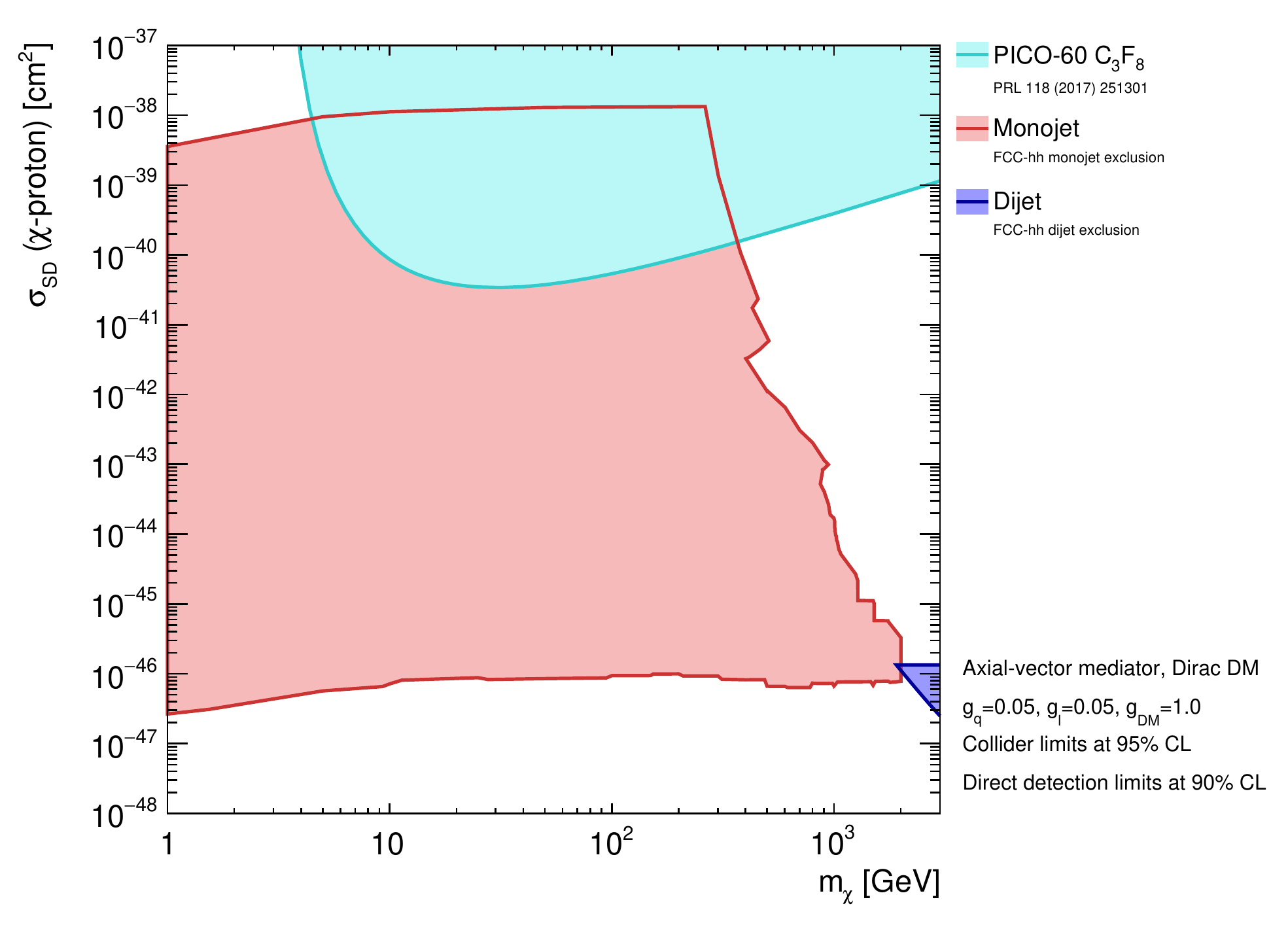}
         \caption{$g_q=0.05$, $g_{\chi}=1.0$, $g_l=0.05$}
         \label{subfig:dd-fcc-hh-axial-2}
     \end{subfigure}
        \caption{Comparison of projected exclusions from FCC-hh with constraints from current direct-detection experiments on the spin-dependent DM–proton scattering cross section in the context of the axial-vector simplified model.}
        \label{fig:fcc-hh-dd-separate-sd}     
\end{figure}

\begin{figure}[htp]
     \centering
     \begin{subfigure}[b]{0.49\textwidth}
         \centering
         \includegraphics[width=\textwidth]{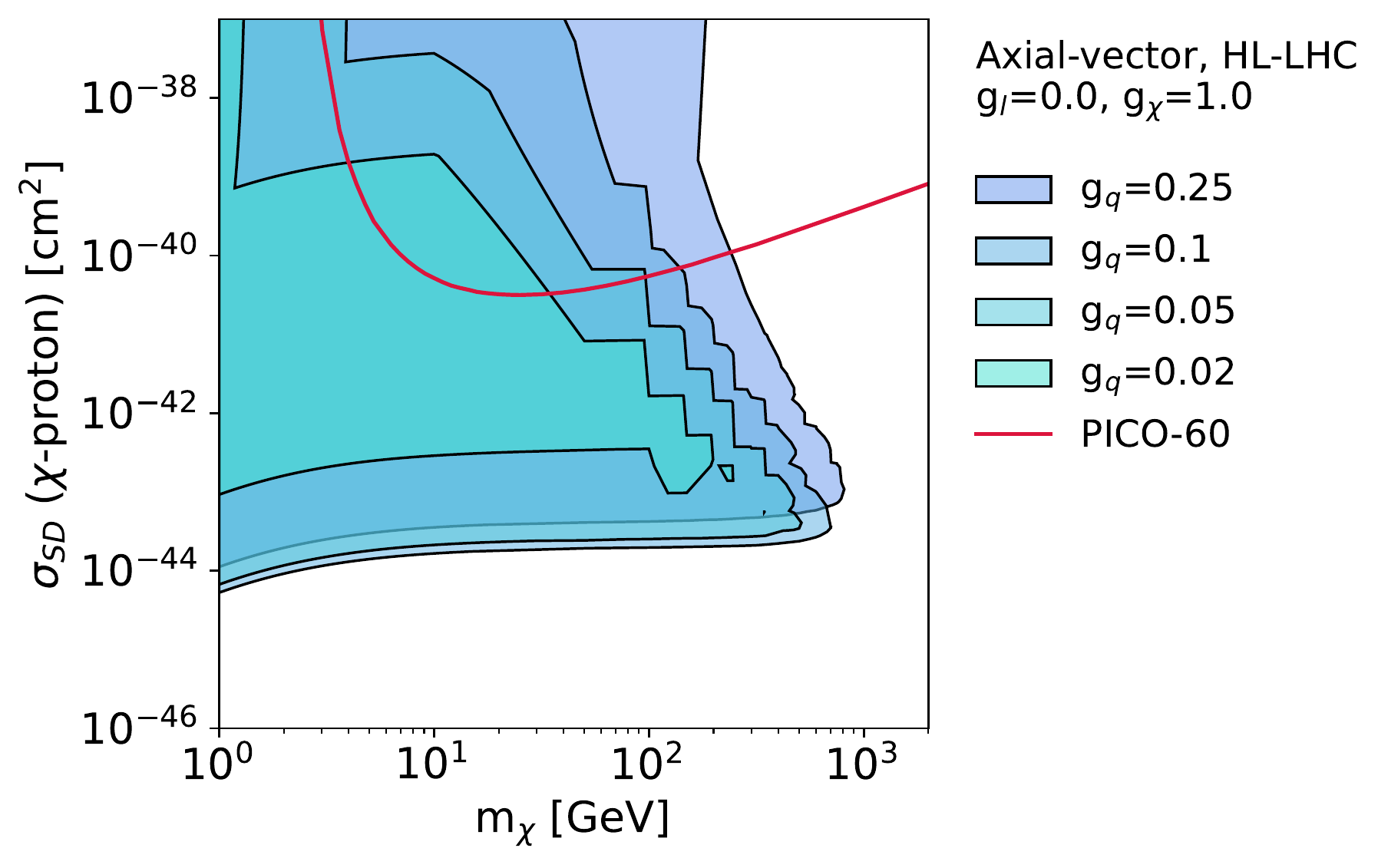}
         \caption{Monojet analysis}
         \label{subfig:couplingscan-dd-monojet-axp}
     \end{subfigure}
     \hfill
     \begin{subfigure}[b]{0.49\textwidth}
         \centering
         \includegraphics[width=\textwidth]{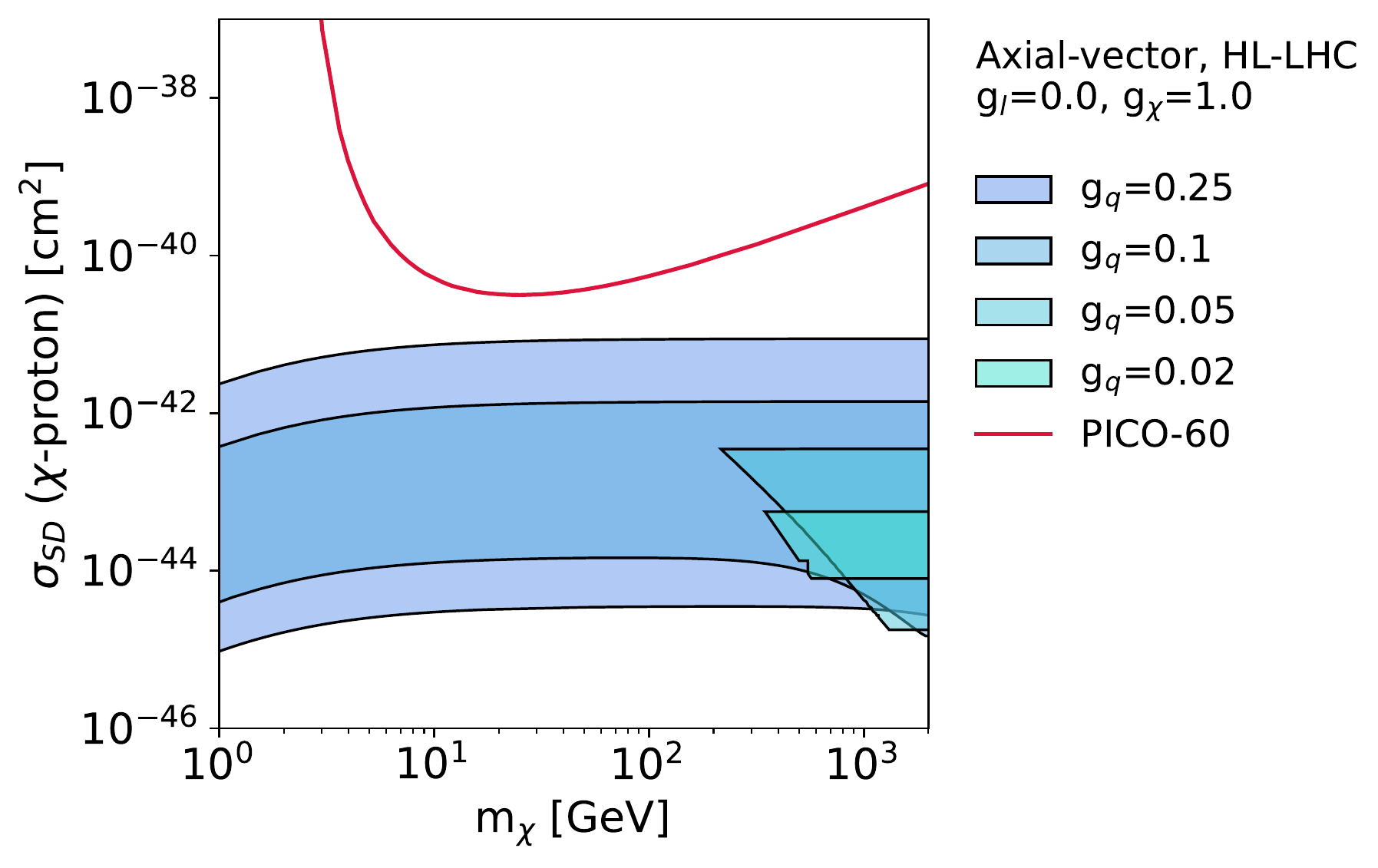}
         \caption{Dijet analysis}
         \label{subfig:couplingscan-dd-dijet-axp}
     \end{subfigure}
    \caption{Effects on the HL-LHC exclusion limits in $\sigma_{\mathrm{SD}}$ for the monojet (\subref{subfig:couplingscan-dd-monojet-axp}) and dijet (\subref{subfig:couplingscan-dd-dijet-axp}) signatures  when varying the $g_q$ coupling. The dark matter coupling is held fixed to $g_{\mathrm{DM}}=1$; there is no coupling to leptons. Results are shown for an axial-vector mediator. Limits on the DM-proton interaction cross section from existing direct-detection experiments are shown for context.}
    \label{fig:couplingscan-dd-axp}
\end{figure}

\begin{figure}[htp]
     \centering
     \begin{subfigure}[b]{0.49\textwidth}
         \centering
         \includegraphics[width=\textwidth]{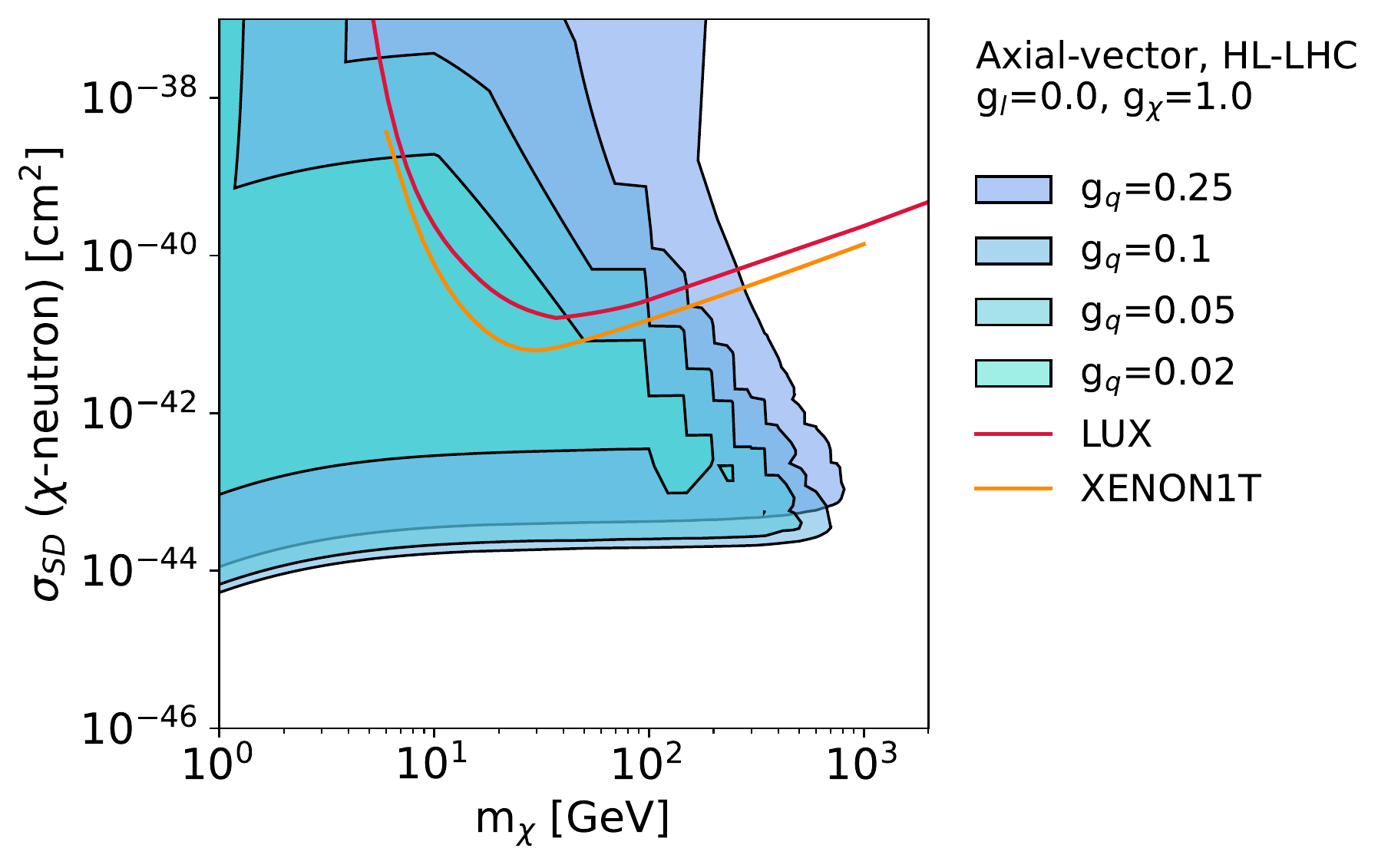}
         \caption{Monojet analysis}
         \label{subfig:couplingscan-dd-monojet-axn}
     \end{subfigure}
     \hfill
     \begin{subfigure}[b]{0.49\textwidth}
         \centering
         \includegraphics[width=\textwidth]{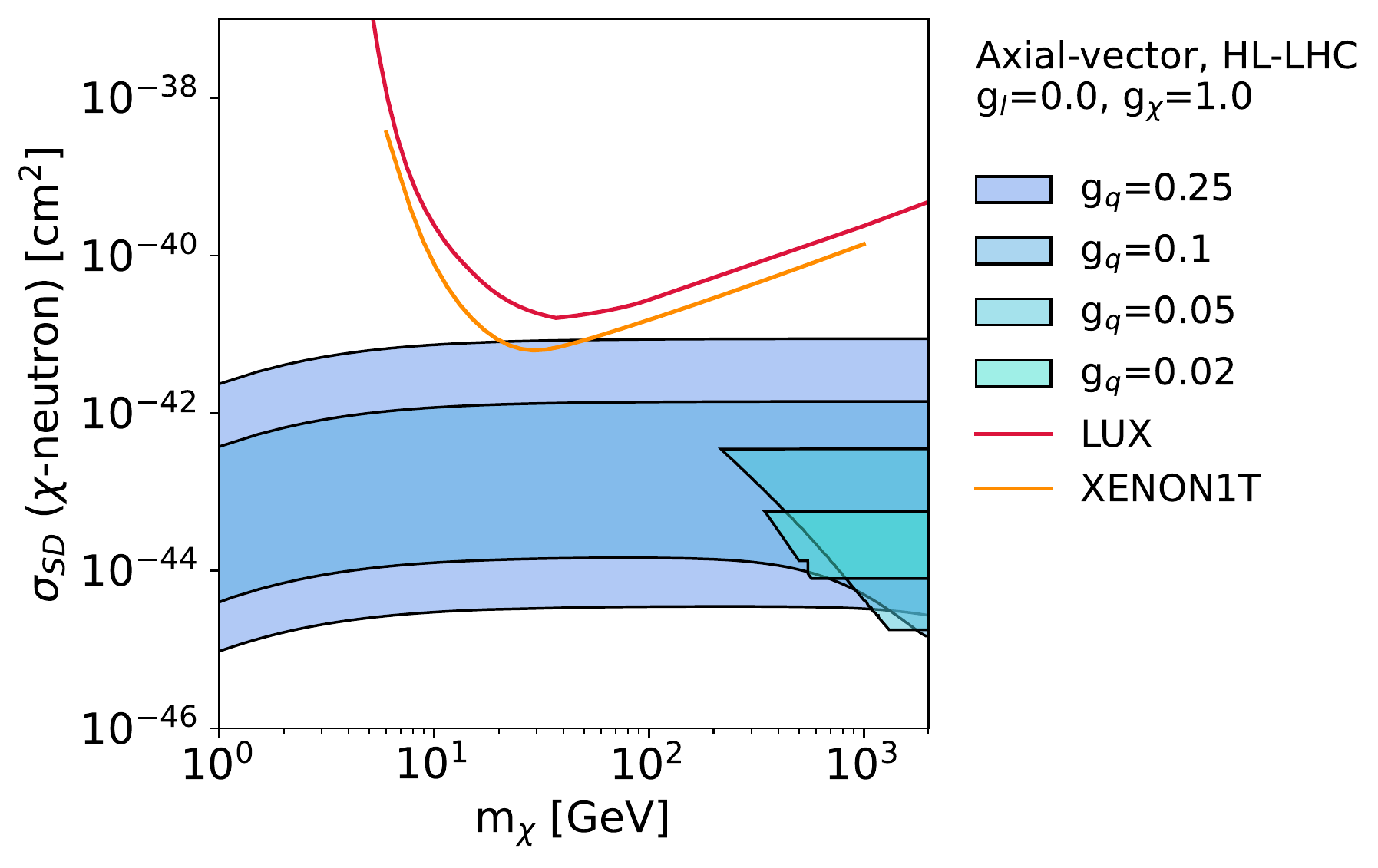}
         \caption{Dijet analysis}
         \label{subfig:couplingscan-dd-dijet-axn}
     \end{subfigure}
    \caption{Effects on the HL-LHC exclusion limits in $\sigma_{\mathrm{SD}}$ for the monojet (\subref{subfig:couplingscan-dd-monojet-axn}) and dijet (\subref{subfig:couplingscan-dd-dijet-axn}) signatures  when varying the $g_q$ coupling. The dark matter coupling is held fixed to $g_{\mathrm{DM}}=1$; there is no coupling to leptons. Results are shown for an axial-vector mediator. Limits on the DM-neutron interaction cross section from existing direct-detection experiments are shown for context.}
    \label{fig:couplingscan-dd-axn}
\end{figure}

\begin{figure}[htp]
     \centering
     \begin{subfigure}[b]{0.49\textwidth}
         \centering
         \includegraphics[width=\textwidth]{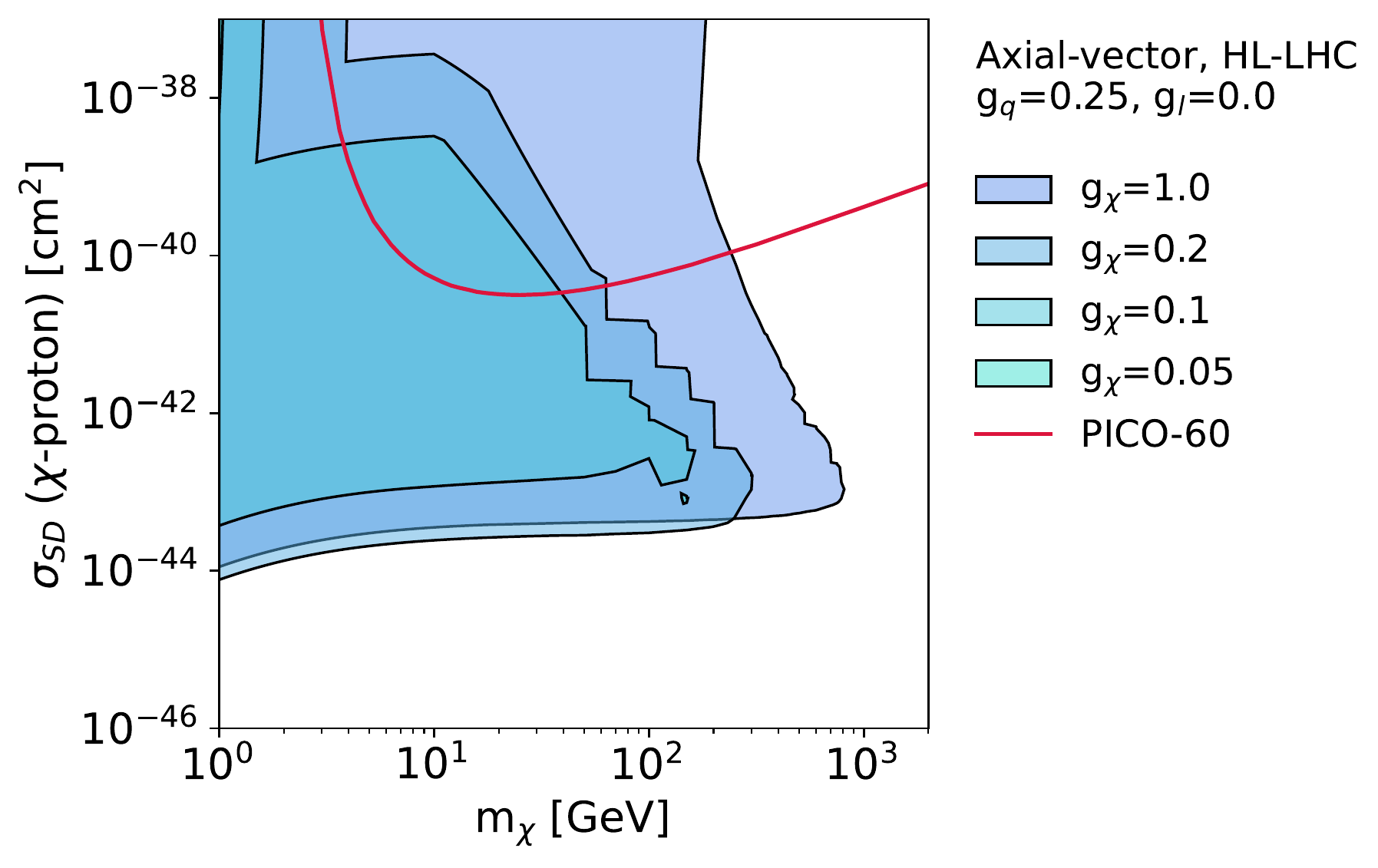}
         \caption{Monojet analysis}
         \label{subfig:couplingscan-dd-gdm-monojet-axp}
     \end{subfigure}
     \hfill
     \begin{subfigure}[b]{0.49\textwidth}
         \centering
         \includegraphics[width=\textwidth]{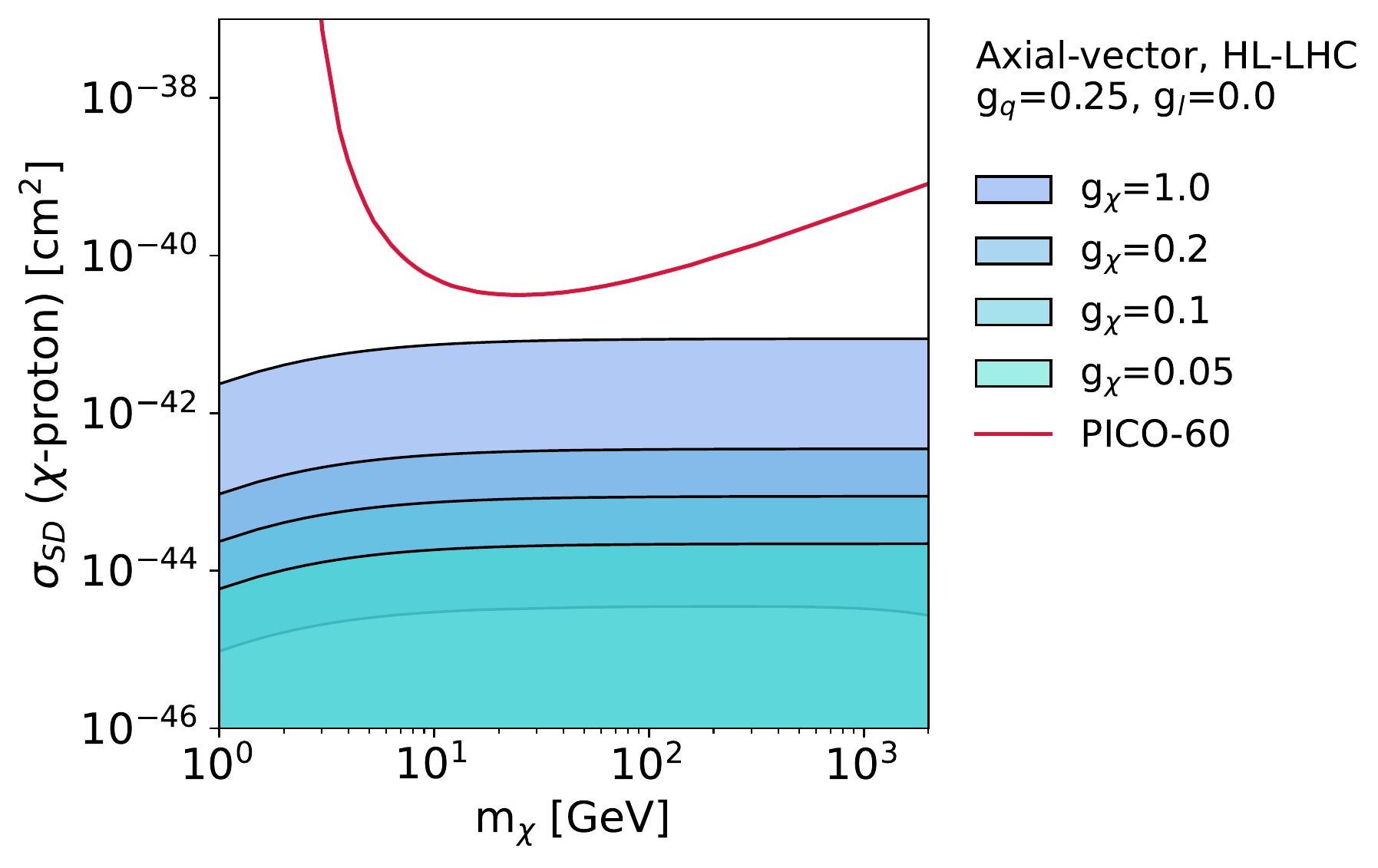}
         \caption{Dijet analysis}
         \label{subfig:couplingscan-dd-gdm-dijet-axp}
     \end{subfigure}
    \caption{Effects on the HL-LHC exclusion limits in $\sigma_{\mathrm{SD}}$ for the monojet (\subref{subfig:couplingscan-dd-gdm-monojet-axp}) and dijet (\subref{subfig:couplingscan-dd-gdm-dijet-axp}) signatures  when varying the $g_\chi$ coupling. The coupling to quarks is held fixed to $g_q=0.25$; there is no coupling to leptons. Results are shown for an axial-vector mediator. Limits on the DM-proton interaction cross section from existing direct-detection experiments are shown for context.}
    \label{fig:couplingscan-dd-gdm-axp}
\end{figure}

\begin{figure}[htp]
     \centering
     \begin{subfigure}[b]{0.49\textwidth}
         \centering
         \includegraphics[width=\textwidth]{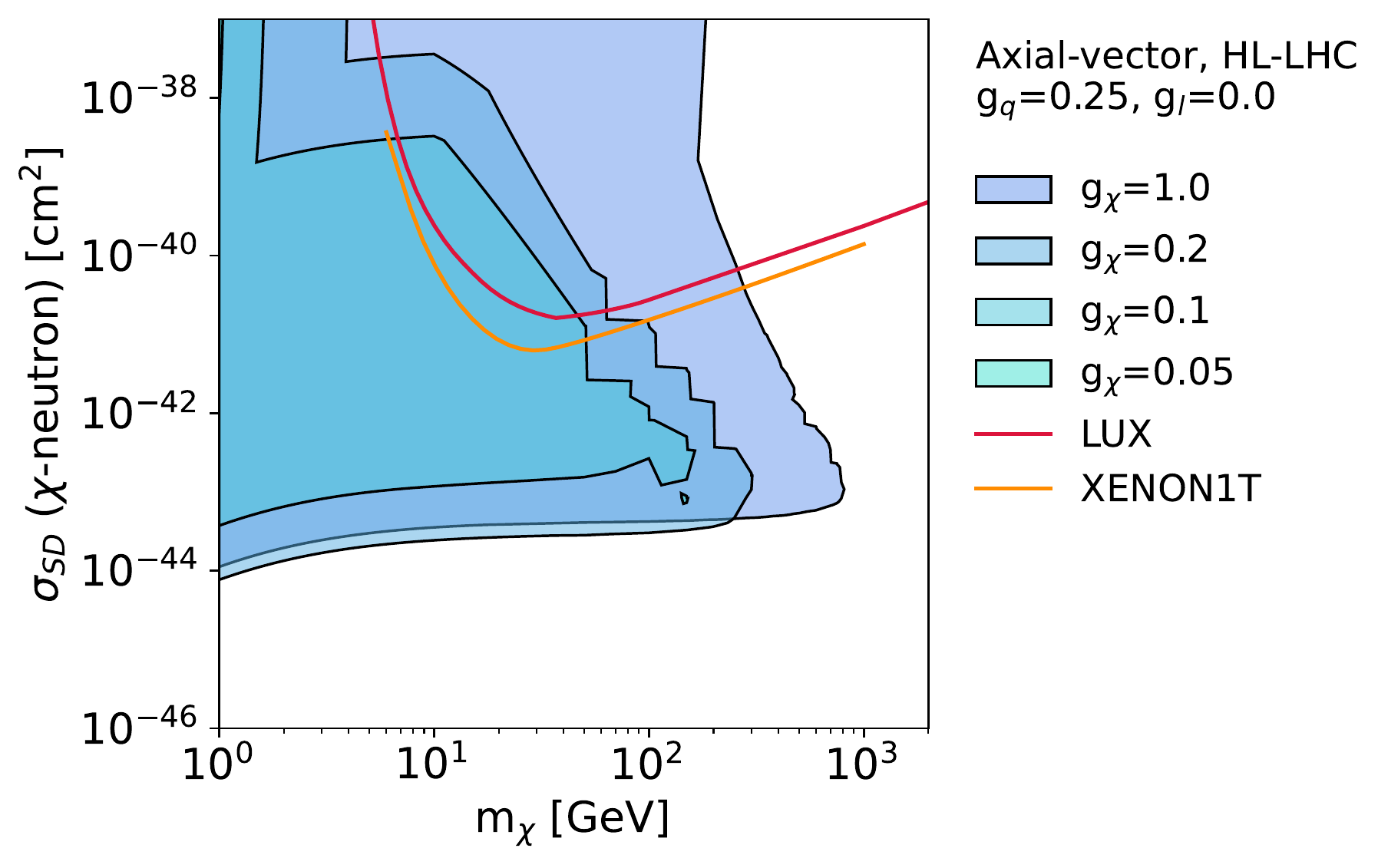}
         \caption{Monojet analysis}
         \label{subfig:couplingscan-dd-gdm-monojet-axn}
     \end{subfigure}
     \hfill
     \begin{subfigure}[b]{0.49\textwidth}
         \centering
         \includegraphics[width=\textwidth]{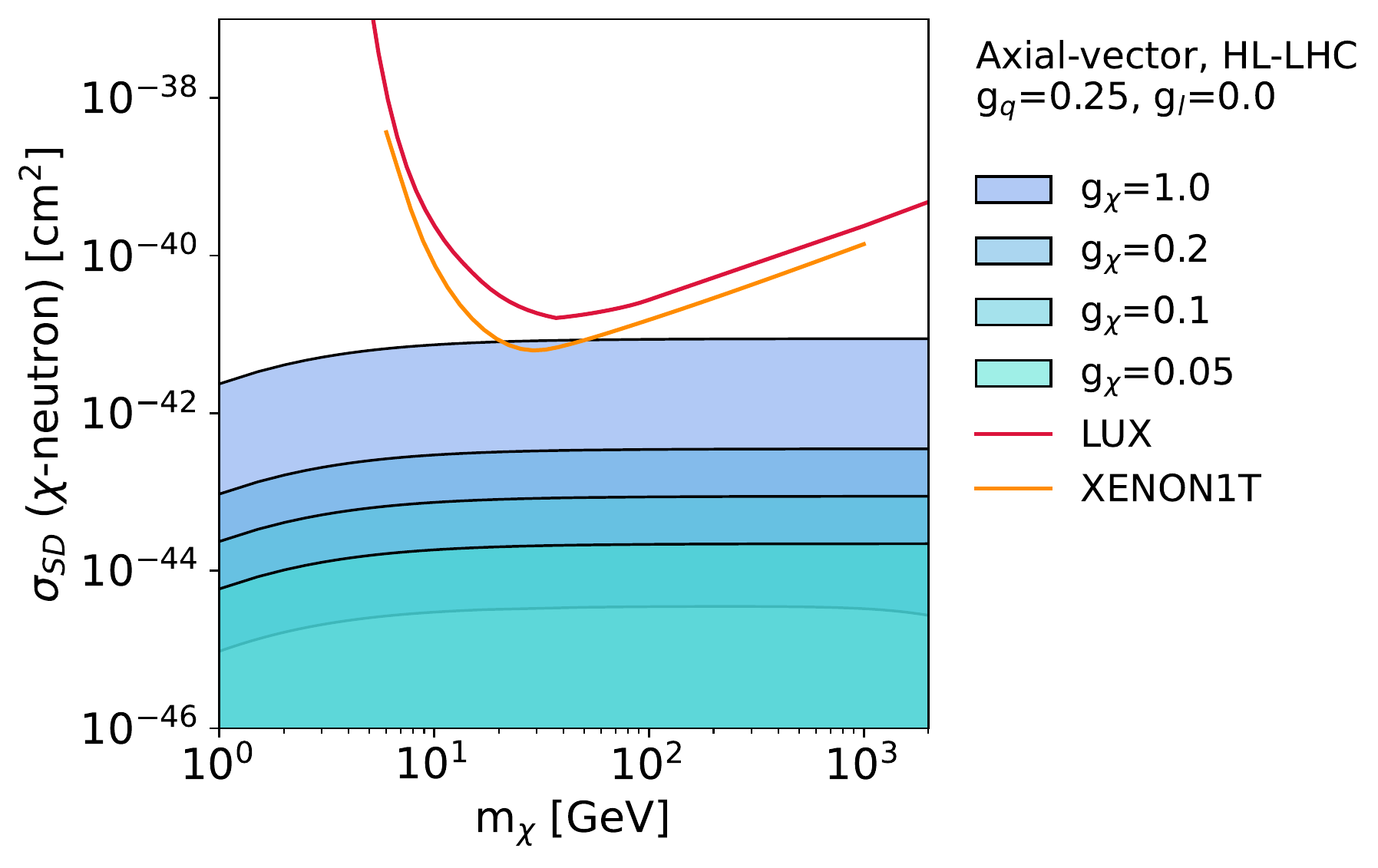}
         \caption{Dijet analysis}
         \label{subfig:couplingscan-dd-gdm-dijet-axn}
     \end{subfigure}
    \caption{Effects on the HL-LHC exclusion limits in $\sigma_{\mathrm{SD}}$ for the monojet (\subref{subfig:couplingscan-dd-gdm-monojet-axn}) and dijet (\subref{subfig:couplingscan-dd-gdm-dijet-axn}) signatures  when varying the $g_\chi$ coupling. The coupling to quarks is held fixed to $g_q=0.25$; there is no coupling to leptons. Results are shown for an axial-vector mediator. Limits on the DM-neutron interaction cross section from existing direct-detection experiments are shown for context.}
    \label{fig:couplingscan-dd-gdm-axn}
\end{figure}

\begin{figure}[htp]
     \centering
     \begin{subfigure}[b]{0.49\textwidth}
         \centering
         \includegraphics[width=\textwidth]{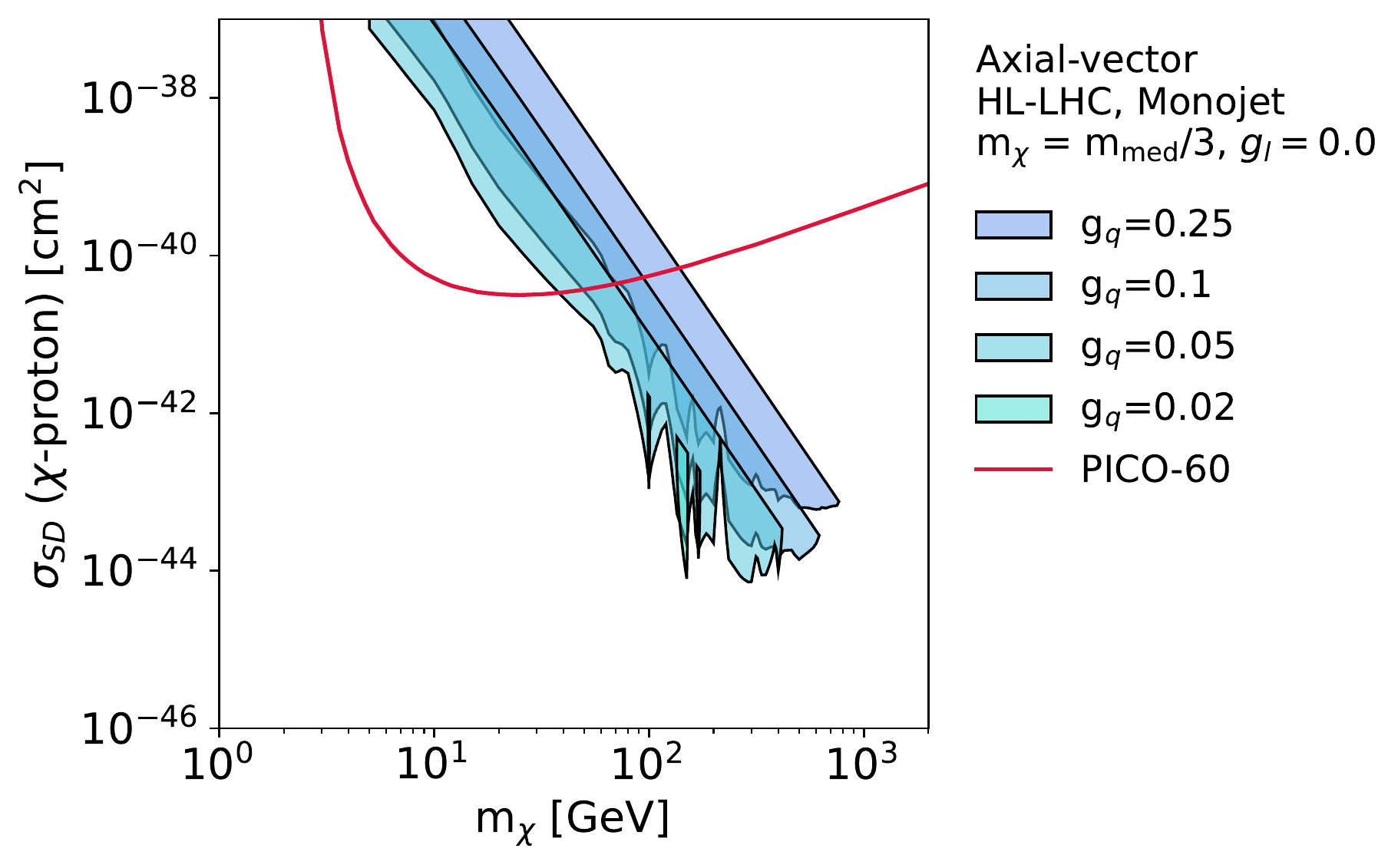}
         \caption{Monojet analysis}
         \label{subfig:couplingscan-dd-gq-monojet-fixedmmed-axp}
     \end{subfigure}
     \hfill
     \begin{subfigure}[b]{0.49\textwidth}
         \centering
         \includegraphics[width=\textwidth]{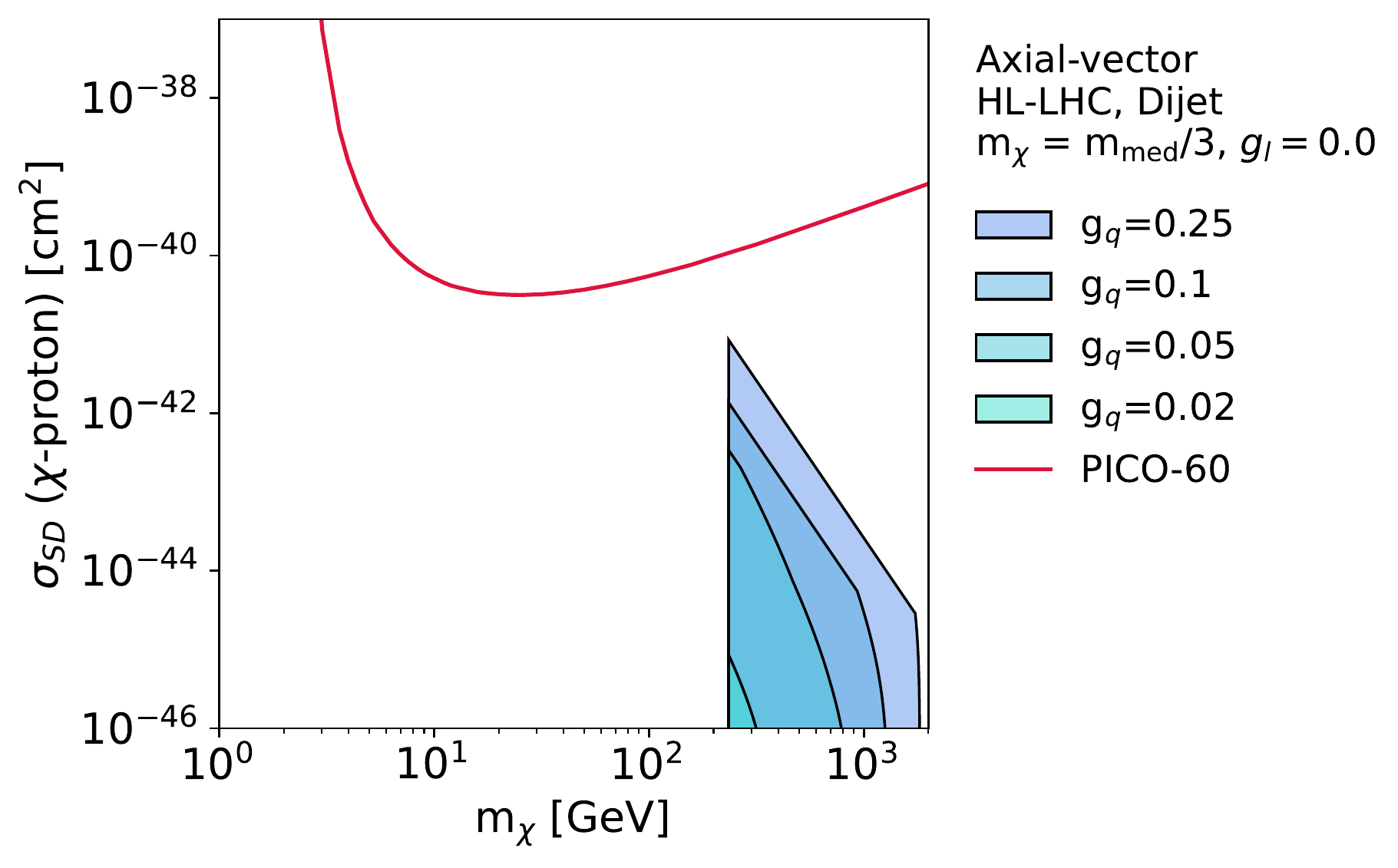}
         \caption{Dijet analysis}
         \label{subfig:couplingscan-dd-gq-dijet-fixedmmed-axp}
     \end{subfigure}
    \caption{Effects on the HL-LHC exclusion limits in $\sigma_{\mathrm{SD}}$ for the monojet (\subref{subfig:couplingscan-dd-gdm-monojet-fixedmmed-axp}) and dijet (\subref{subfig:couplingscan-dd-gdm-dijet-fixedmmed-axp}) signatures when varying the $g_q$ coupling. The mass of the mediator is fixed to $m_\mathrm{med} = 3 m_\chi$; there is no coupling to leptons. Limits on the DM-proton interaction cross section from existing direct-detection experiments are shown for context.}
    \label{fig:couplingscan-dd-gq-fixedmmed-axp}
\end{figure}

\begin{figure}[htp]
     \centering
     \begin{subfigure}[b]{0.49\textwidth}
         \centering
         \includegraphics[width=\textwidth]{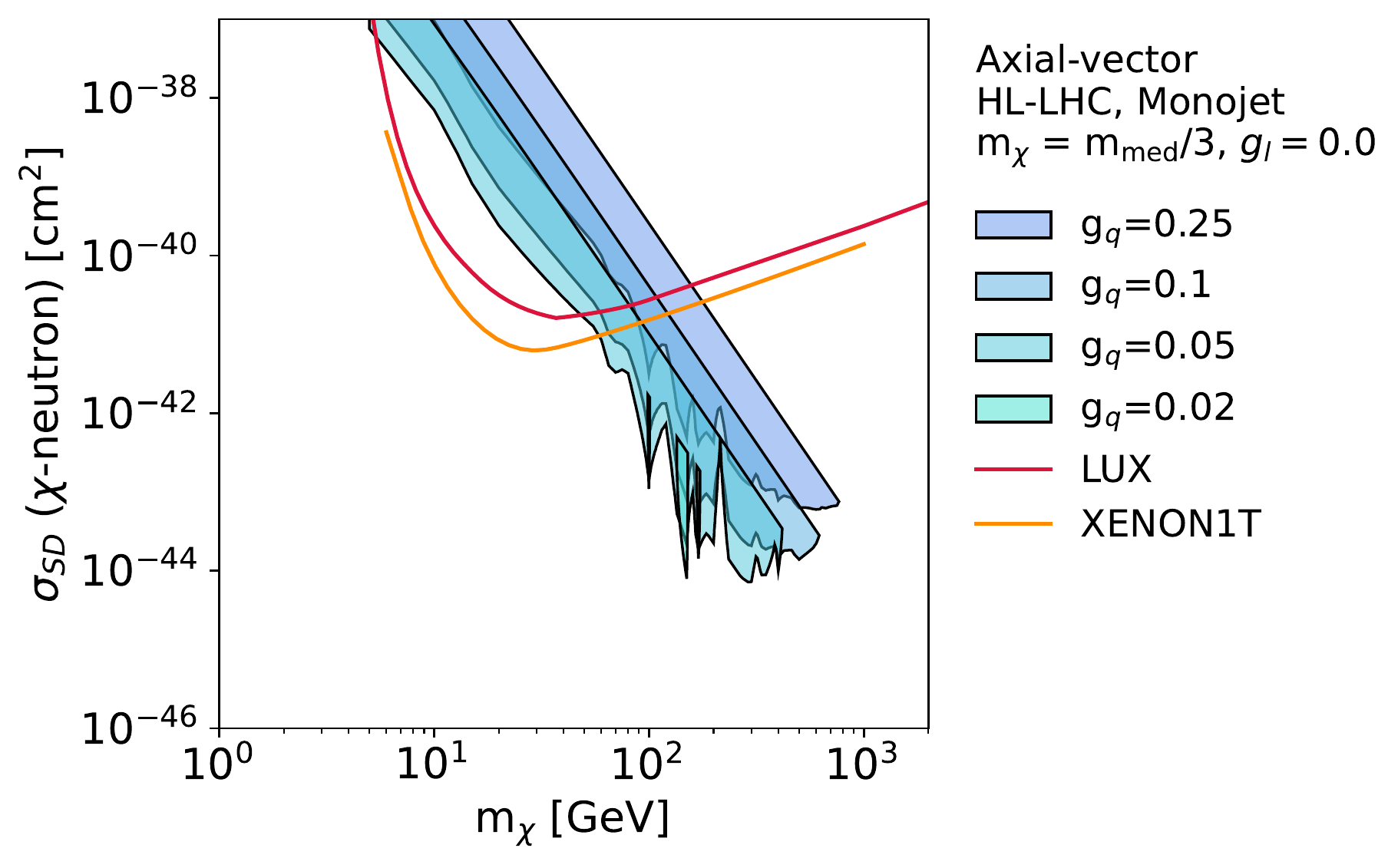}
         \caption{Monojet analysis}
         \label{subfig:couplingscan-dd-gq-monojet-fixedmmed-axn}
     \end{subfigure}
     \hfill
     \begin{subfigure}[b]{0.49\textwidth}
         \centering
         \includegraphics[width=\textwidth]{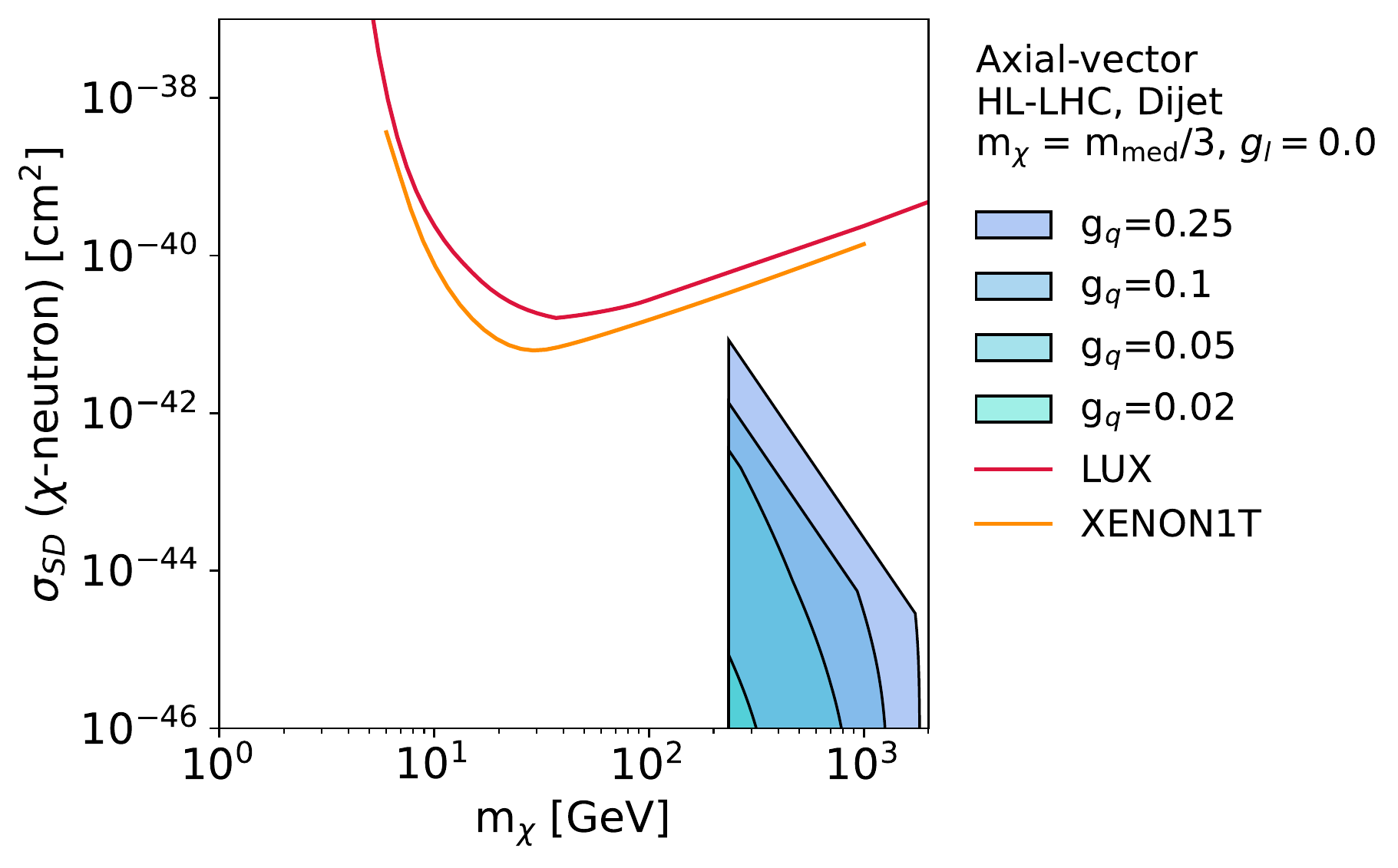}
         \caption{Dijet analysis}
         \label{subfig:couplingscan-dd-gq-dijet-fixedmmed-axn}
     \end{subfigure}
    \caption{Effects on the HL-LHC exclusion limits in $\sigma_{\mathrm{SD}}$ for the monojet (\subref{subfig:couplingscan-dd-gdm-monojet-fixedmmed-axn}) and dijet (\subref{subfig:couplingscan-dd-gdm-dijet-fixedmmed-axn}) signatures when varying the $g_q$ coupling. The mass of the mediator is fixed to $m_\mathrm{med} = 3 m_\chi$; there is no coupling to leptons. Limits on the DM-neutron interaction cross section from existing direct-detection experiments are shown for context.}
    \label{fig:couplingscan-dd-gq-fixedmmed-axn}
\end{figure}

\begin{figure}[htp]
     \centering
     \begin{subfigure}[b]{0.49\textwidth}
         \centering
         \includegraphics[width=\textwidth]{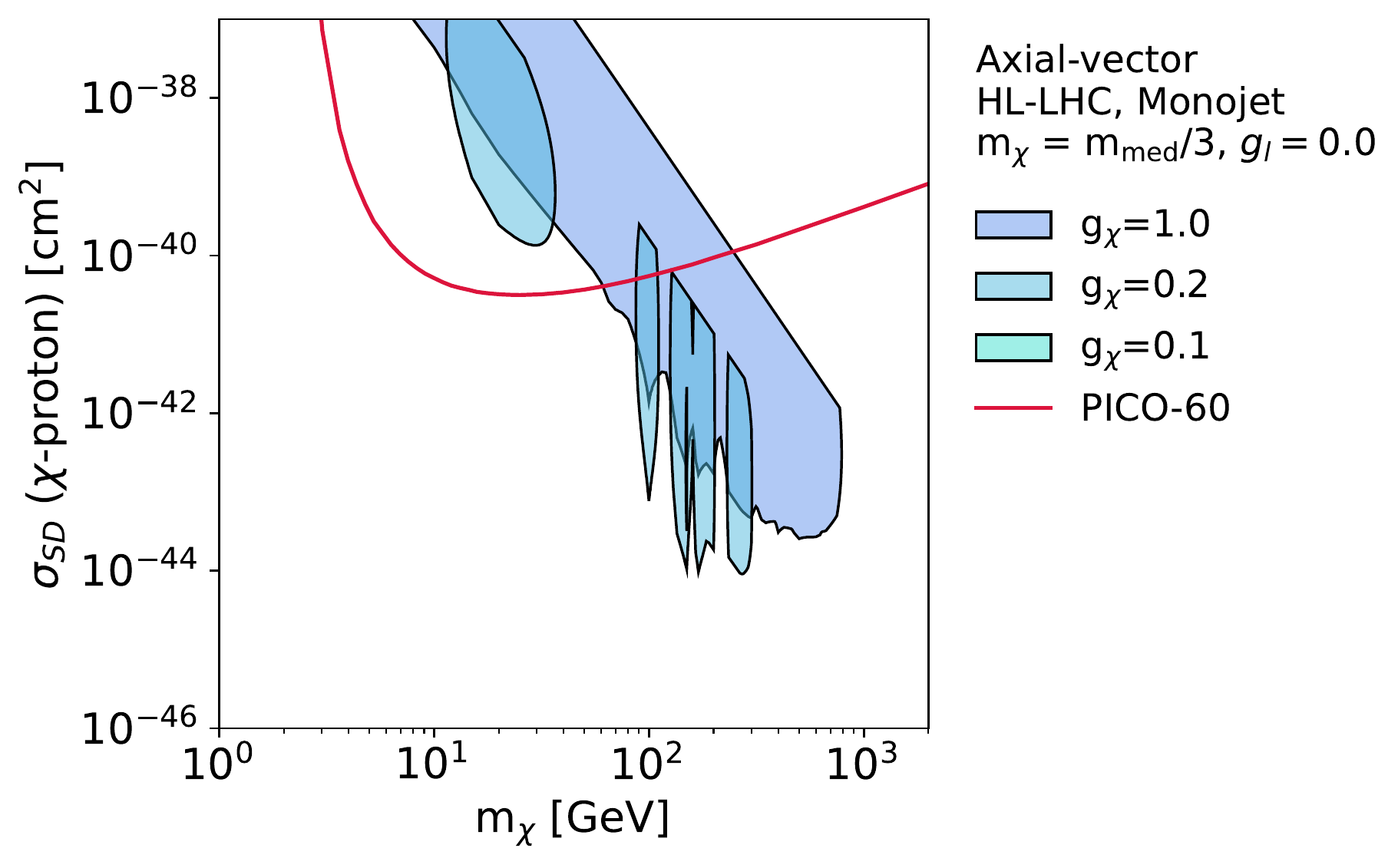}
         \caption{Monojet analysis, $g_\chi$ explicit}
         \label{subfig:couplingscan-dd-gdm-monojet-fixedmmed-axp}
     \end{subfigure}
     \hfill
     \begin{subfigure}[b]{0.49\textwidth}
         \centering
         \includegraphics[width=\textwidth]{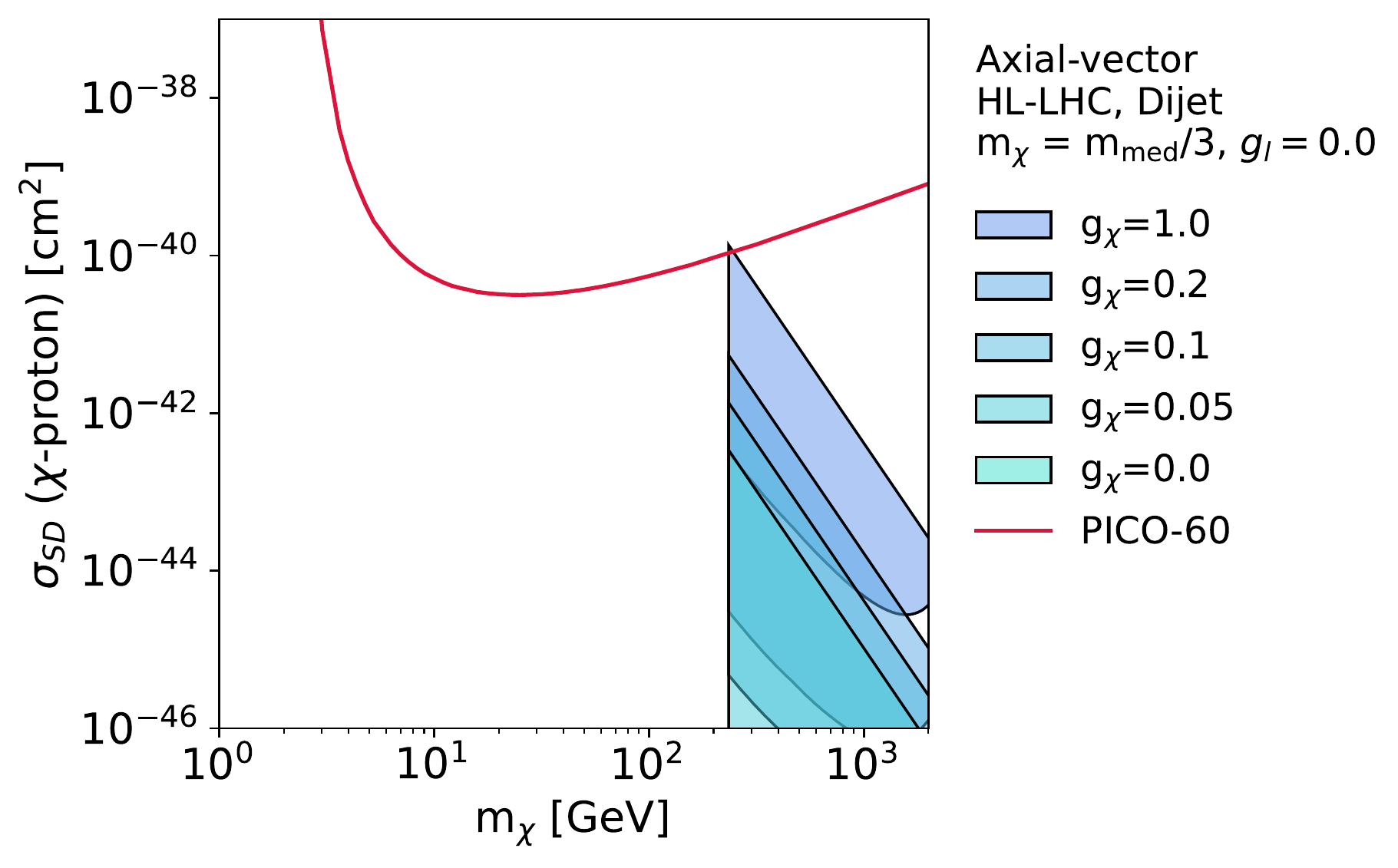}
         \caption{Dijet analysis, $g_\chi$ explicit}
         \label{subfig:couplingscan-dd-gdm-dijet-fixedmmed-axp}
     \end{subfigure}
    \caption{Effects on the HL-LHC exclusion limits in $\sigma_{\mathrm{SD}}$ for the monojet (\subref{subfig:couplingscan-dd-gdm-monojet-fixedmmed-axp}) and dijet (\subref{subfig:couplingscan-dd-gdm-dijet-fixedmmed-axp}) signatures when varying the $g_\chi$ coupling. An axial vector mediator with mass fixed to $m_\mathrm{med} = 3 m_\chi$ is shown; there is no coupling to leptons. Limits on the DM-proton interaction cross section from existing direct-detection experiments are shown for context.}
    \label{fig:couplingscan-dd-gdm-fixedmmed-axp}
\end{figure}

\begin{figure}[htp]
     \centering
     \begin{subfigure}[b]{0.49\textwidth}
         \centering
         \includegraphics[width=\textwidth]{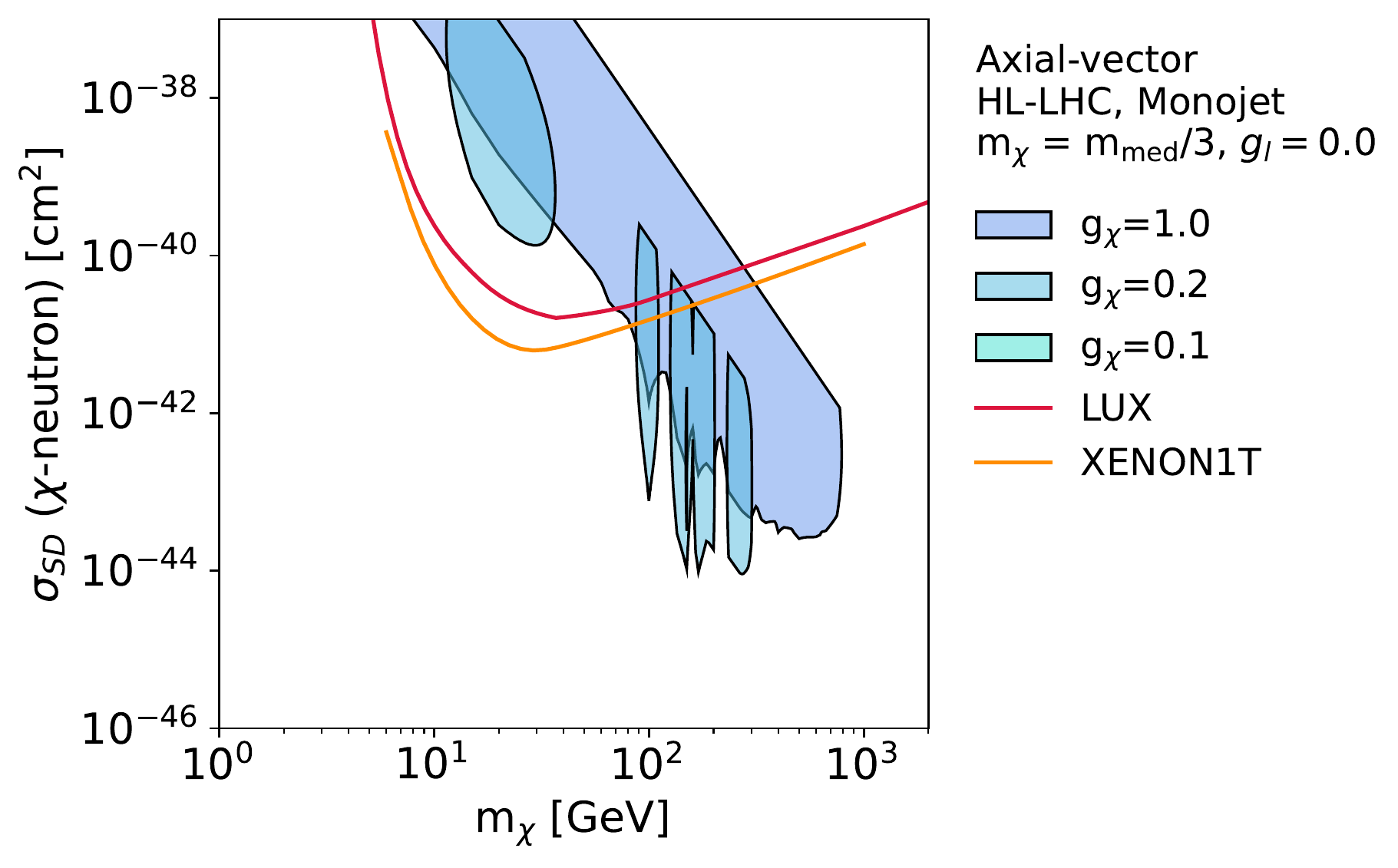}
         \caption{Monojet analysis, $g_\chi$ explicit}
         \label{subfig:couplingscan-dd-gdm-monojet-fixedmmed-axn}
     \end{subfigure}
     \hfill
     \begin{subfigure}[b]{0.49\textwidth}
         \centering
         \includegraphics[width=\textwidth]{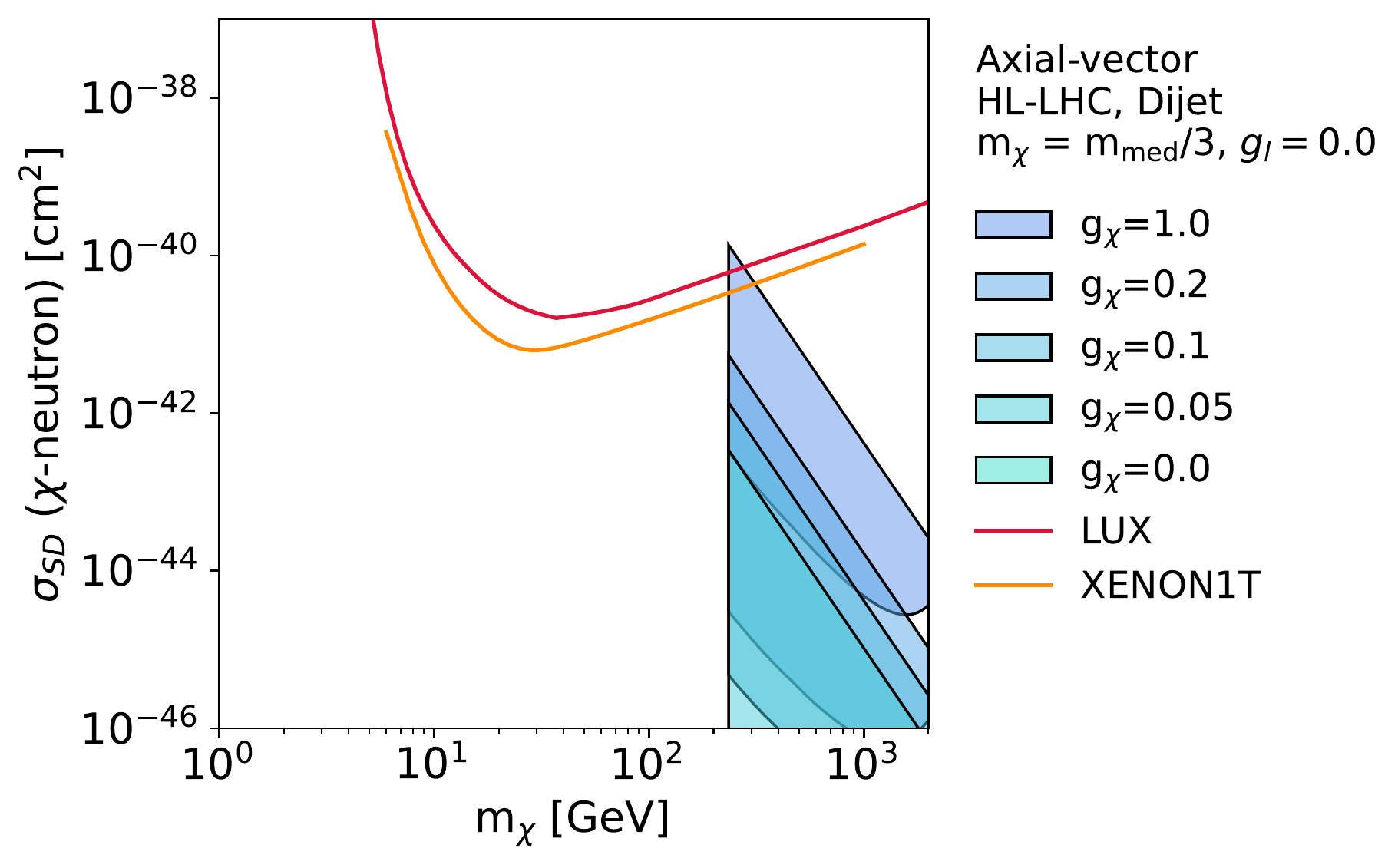}
         \caption{Dijet analysis, $g_\chi$ explicit}
         \label{subfig:couplingscan-dd-gdm-dijet-fixedmmed-axn}
     \end{subfigure}
    \caption{Effects on the HL-LHC exclusion limits in $\sigma_{\mathrm{SD}}$ for the monojet (\subref{subfig:couplingscan-dd-gdm-monojet-fixedmmed-axn}) and dijet (\subref{subfig:couplingscan-dd-gdm-dijet-fixedmmed-axn}) signatures when varying the $g_\chi$ coupling. An axial vector mediator with mass fixed to $m_\mathrm{med} = 3 m_\chi$ is shown; there is no coupling to leptons. Limits on the DM-neutron interaction cross section from existing direct-detection experiments are shown for context.}
    \label{fig:couplingscan-dd-gdm-fixedmmed-axn}
\end{figure}

\end{document}